%% file: main.tex
\documentclass[12pt,a4paper,fleqn]{mybook}

\usepackage{dev}
\usepackage[tbtags]{amsmath}
\usepackage{amssymb}
\usepackage{mathrsfs}

\input{myheadings}

\newcommand{\ui}{\,\mathrm{i}\;\!}
\newcommand{\ud}{\mathrm{d}}
\newcommand{\ue}{\mathrm{e}}
\newcommand{\uD}{\mathrm{D}}
\newcommand{\uU}{\mathrm{U}}
\newcommand{\uO}{\mathrm{O}}
\newcommand{\op}[1]{\mathsf{#1}}
\newcommand{\opgr}[1]{{#1}}

\newcommand{\qop}[1]{\widehat{#1}}
\newcommand{\qopq}{\qop{q}}
\newcommand{\qopp}{\qop{p}}
\newcommand{\qopx}{\qop{x}}
\newcommand{\qopX}{\qop{X}}
\newcommand{\qopA}{\qop{A}}
\newcommand{\qoppi}{\qop{\pi}}
\newcommand{\qopR}{\qop{R}}

\newcommand{\xnc}[1]{\widehat{#1}}
\newcommand{\ncx}{\xnc{x}}

\newcommand{\ncp}{\xnc{p}}
\newcommand{\ncpartial}{\xnc{\partial}}
\newcommand{\nceth}{\xnc{\eth}}
\newcommand{\ncnabla}{\xnc{\nabla}}
\newcommand{\opid}{\mathbf{1}}
\newcommand{\opP}{\mathcal{P}}
\newcommand{\opH}{\mathcal{H}}
\newcommand{\opJ}{\mathcal{J}}
\newcommand{\opG}{\mathcal{G}}
\newcommand{\opD}{\mathcal{D}}
\newcommand{\opK}{\mathcal{K}}
\newcommand{\opF}{\mathcal{F}}
\newcommand{\opA}{\mathcal{A}}
\newcommand{\opB}{\mathcal{B}}
\newcommand{\opQ}{\mathcal{Q}}
\newcommand{\ncopP}{\xnc{\mathcal{P}}}
\newcommand{\ncopH}{\xnc{\mathcal{H}}}
\newcommand{\ncopJ}{\xnc{\mathcal{J}}}
\newcommand{\ncopG}{\xnc{\mathcal{G}}}
\newcommand{\ncopD}{\xnc{\mathcal{D}}}
\newcommand{\ncopK}{\xnc{\mathcal{K}}}
\newcommand{\ncopS}{\xnc{\mathcal{S}}}
\newcommand{\scopP}{\bar{\mathcal{P}}}
\newcommand{\scopH}{\bar{\mathcal{H}}}
\newcommand{\scopJ}{\bar{\mathcal{J}}}
\newcommand{\scopG}{\bar{\mathcal{G}}}
\newcommand{\scopD}{\bar{\mathcal{D}}}
\newcommand{\scopK}{\bar{\mathcal{K}}}

\newcommand{\lagdens}{\mathscr{L}}
\newcommand{\hamdens}{\mathscr{H}}
\newcommand{\weakequals}{\approx}
\newcommand{\real}{\mathbb{R}}
\newcommand{\complex}{\mathbb{C}}
\newcommand{\integer}{\mathbb{Z}}
\newcommand{\weyl}{\xnc{\mathcal{W}}}

\newcommand{\refcite}[1]{\cite{#1}}
\newcommand{\refscite}[1]{\cite{#1}}

\renewcommand{\vec}[1]{\mathbf{#1}}
\newcommand{\fillblank}{\textsf}

\begin{document}


\setlength{\parskip}{1.0ex plus 0.5ex minus 0.5ex}
\frontmatter
\include{titlepage}

\include{cert_from_supervisor}

\include{to_my_parents}

\include{acknowledgements}

\include{list_of_publications}

\include{part_main}

\include{toc}


\mainmatter
\include{chap_intro}

\include{chap_memb}

\include{chap_current}

\include{chap_anomalous}

\include{chap_lorentz}

\include{chap_deform}

\include{chap_conclu}

\backmatter
\include{thebibliography}



\end{document}

%% file: myheadings.tex
\usepackage{fancyhdr}
\renewcommand{\chaptermark}[1]{         
  \markboth{\chaptername\ \thechapter.\ \ #1}{}} %
\renewcommand{\headrulewidth}{0.3pt}    

\makeatletter
\def\cleardoublepage{\clearpage\if@twoside \ifodd\c@page\else%
    \hbox{}%
    \thispagestyle{empty}
    \newpage%
    \if@twocolumn\hbox{}\newpage\fi\fi\fi}
\makeatother


%% file: titlepage.tex

\begin{titlepage}
\begin{center}

\hrulefill\\
\texttt{Thesis defended on July 29, 2008}\\[-0.2cm]
\hrulefill

\Huge
\textbf{Aspects of noncommutativity in field theory, strings and membranes}

\vfill

\normalsize
{\Large Thesis submitted for the degree of}\\[2.2ex]
\textbf{\Large Doctor of Philosophy (Science)}\\[2ex]
{\Large of}\\[2ex]
\textbf{\Large Jadavpur University, Kolkata}

\vfill

{\Large June 2007}

\vfill

\textbf{{\Large Kuldeep Kumar}}\\[2ex]
{\large \mbox{Satyendra Nath Bose National Centre for Basic Sciences}}\\
{\large JD Block, Sector 3, Salt Lake, Kolkata 700098, India}

\end{center}
\end{titlepage}


%% file: cert_from_supervisor.tex

\chapter*{Certificate from the supervisor}
\thispagestyle{empty}


This is to certify that the thesis entitled \fillblank{`Aspects of noncommutativity in field theory, strings and membranes'} submitted by \fillblank{Kuldeep Kumar}, who got his name registered on \fillblank{June 29, 2004} for the award of \fillblank{Ph.D.~(Science)} degree of \fillblank{Jadavpur University}, is absolutely based upon his own work under the supervision of \fillblank{Professor Rabin Banerjee} at \fillblank{S.N.~Bose National Centre for Basic Sciences, Kolkata, India}, and that neither this thesis nor any part of it has been submitted for any degree/diploma or any other academic award anywhere before.

\vspace{3.0cm}

\hfill \begin{tabular}{@{}l@{}}
\fillblank{Rabin Banerjee}\\
Professor\\
S.N.~Bose National Centre for Basic Sciences\\
JD Block, Sector 3, Salt Lake\\
Kolkata 700098, India
\end{tabular}


%% file: to_my_parents.tex

\chapter*{}
\thispagestyle{empty}


{\Large To my beloved parents}


\vfill

\begin{flushright}
{\dn
{\large mAtA kn\4{\qva} Ept\? jo}\\
EjnA{\qva} Em{\qva}jo i\-(\7{T} Etkr \7{p}jAyA
}
\end{flushright}

\vspace{2cm}


%% file: acknowledgements.tex

\chapter*{Acknowledgements}
\thispagestyle{empty}


As a sense of fulfilment at the completion of this phase of academic endeavour, I wish to express my gratitude to all those who made this thesis possible.

It has been my privilege to work under the able guidance of my revered thesis advisor, Prof.~Rabin Banerjee. His insights into various problems and insistence on clarity have been most useful and inspiring. I express my deep sense of gratitude to him for his patience, persistence and his prompt and sincere help whenever I needed it most.

I extend my sincere thanks to Dr.~Biswajit Chakraborty for many fruitful, enthusiastic and illuminating discussions, academic or otherwise. He has always motivated me to do my best. I am grateful to him for always being there to help me in all matters. Thanks are also due to Prof.~Abhijit Mookerjee for his help and care on many occasions during my stay here.

It is my pleasure to thank many friends of the Centre for their cooperation and help. I have had a very nice time with Tomy, Aftab, Ankush, Prasad, Soumen, Sunandan, Anirban, Arindam and Saurav over all these years. They made my stay here an experience I cherish much.

I gratefully acknowledge financial support from the Council of Scientific and Industrial Research, Government of India, during the period of this work.

I owe thanks to my many teachers. Shri Prem Das, Shri Bali Ram, Shri Kartar Chand, Shri Ratan Lal, Shri Somraj Shastri, Smt.~Lalita, Shri R.K.~Chopra, Smt.~Krishna Kanwar, Shri Ram S.~Anand, Shri Harnam S.~Choudhary, Shri K.S.~Pathania, Shri S.C.~Dutta, Shri L.R.~Sharma, Shri K.L.~Verma and Dr.~I.S.~Minhas are still fresh in my memory.

I am fortunte to have many friends and well-wishers. Rajinder, Anoop, Som Dutt, Pradeep and Anup deserve special mention for their helpful attitude and emotional support. I cannot forget Jeewan for his time-to-time help in every way.

My sincerest thanks go to my parents. Although my father himself is no more educated than high school and my mother is an illiterate, they motivated me to pursue higher studies, in spite of many unfavourable circumstances. They have been unshakeable pillars of support, helping me fight and win every battle. Their faith in my abilities has been my strength---my real emotions are much beyond the pages of this thesis. I am also indebted to my sisters and brothers-in-law,\ Trishla--Purshotam, Kanta--Baldev, Shobha--Sharwan, Sushma--Parshottam and Sunita--Ram\! Singh,\ for their unflinching support. I send my love and gratitude to my nephews and nieces,\ Ravinder, Arnu, Monu, Ruma, Neenu, Varinder, Reenu, Pammu, Sannu, Raju, Mittu, Nishu and Sannu. I record my gratitude to my wife, Rachna, for her love and cooperation.

Finally, and most importantly, I thank Him for all His blessings and kindness.


%% file: list_of_publications.tex

\chapter*{List of publications}
\thispagestyle{empty}


\begin{minipage}{13cm}
\begin{enumerate} \raggedright
\item \textbf{Membrane and noncommutativity}\\
  Rabin Banerjee, Biswajit Chakraborty and Kuldeep Kumar\\
  {\em Nucl.\ Phys.} B 668 (2003) 179
\item \textbf{Noncommutative gauge theories and Lorentz symmetry}\\
  Rabin Banerjee, Biswajit Chakraborty and Kuldeep Kumar\\
  {\em Phys.\ Rev.} D 70 (2004) 125004
\item \textbf{Maps for currents and anomalies in noncommutative gauge theories}\\
  Rabin Banerjee and Kuldeep Kumar\\
  {\em Phys.\ Rev.} D 71 (2005) 045013
\item \textbf{Seiberg--Witten maps and commutator anomalies in noncommutative electrodynamics}\\
  Rabin Banerjee and Kuldeep Kumar\\
  {\em Phys.\ Rev.} D 72 (2005) 085012
\item \textbf{Deformed relativistic and nonrelativistic symmetries on canonical noncommutative spaces}\\
  Rabin Banerjee and Kuldeep Kumar\\
  {\em Phys.\ Rev.} D {75} (2007) {045008}
\end{enumerate}
\end{minipage}


%% file: part_main.tex

\chapter*{}
\pagenumbering{roman}
\thispagestyle{empty}

\begin{center}
\uppercase{Aspects of noncommutativity in\\ field theory, strings and membranes}
\end{center}


%% file: toc.tex

\tableofcontents


%% file: chap_intro.tex

\chapter{\label{chap:intro}Introduction}


\section{Noncommutative spaces}

It was noticed a long time ago that various properties of sets of points can be restated in terms of properties of certain commutative rings of functions over those sets. In particular, this observation proved to be extremely fruitful in algebraic geometry and has led to tremendous progress in this subject over the past few decades. In these developments the concept of a point in a space is secondary and overshadowed by the algebraic properties of the (sheaves of) rings of functions on those spaces. This idea also underlies noncommutative geometry, a new direction in mathematics initiated by the French mathematician Alain Connes \cite{Connes:1994}. The central thesis is that the usual notion of a `space'---a set with some extra structure---is inadequate in many interesting cases and that coordinates may profitably be replaced by a noncommutative algebra.

One important source of inspiration for noncommutative geometry is quantum physics. It has been known since the heroic days of quantum mechanics that ordinary concepts of classical mechanics and symplectic geometry do not apply to the subatomic world. In order to understand the physical phenomena taking place at the atomic scale, one needs to replace the concepts of classical geometry by other, noncommutative structures. The classical observables---continuous functions on phase space---are replaced by algebras of operators, in general unbounded, on the Hilbert space of states or quantum observables. In Dirac's parlance, c-numbers get replaced by q-numbers. This procedure is called quantisation.

The simplest example is that of a flat space $\real^2$ which is the phase space of a particle moving in one dimension. After quantisation, the coordinates $q$ and $p$ of a point in $\real^2$ are replaced by operators $\qopq$ and $\qopp$ which obey the Heisenberg--Born--Jordan commutation relation
\begin{equation}
[\qopq, \qopp] = \ui \hbar
\end{equation}
where $\hbar$ is a fundamental constant of nature, Planck's constant. Explicitly, one takes $\qopq \psi(x) = x \psi(x)$, $\qopp \psi(x) = -\ui \hbar (\ud / \ud x) \psi(x)$. This quantisation procedure results in a structure which can be thought of as a noncommutative deformation of a classical phase space. Heisenberg's uncertainty principle implies that there is no natural concept of a point on this quantum deformed phase space: all we have is a nonabelian algebra of `functions on the noncommutative plane'. One can also quantise classical phase spaces with more complicated geometry.

The idea of extension of noncommutativity to the coordinates was first suggested by Heisenberg as a possible solution for removing the infinite quantities of field theories before the renormalisation procedure was developed and had gained acceptance. The first paper on the subject was published in 1947 by Hartland Snyder \cite{Snyder:1946qz}. The success of renormalisation theory however drained interest from the subject for some time. In 1980s noncommutative geometry was studied and developed by mathematicians, most notably Alain Connes \cite{Connes:1994}. The notion of differential structure was generalised to a noncommutative setting. This led to an operator-algebraic description of noncommutative spacetimes and a Yang--Mills theory on a noncommutative torus was developed.

The recent interest by the particle physics community was driven by a paper by Nathan Seiberg and Edward Witten \cite{Seiberg:1999vs}. They argued in the context of string theory that the coordinate functions of the endpoints of open strings constrained to a D-brane in the presence of a constant Neveu--Schwartz $B$-field---equivalent to a constant magnetic field on the brane---would satisfy the noncommutative algebra. The implication is that a quantum field theory on noncommutative spacetime can be interpreted as a low-energy limit of the theory of open strings.

Another possible motivation for the noncommutativity of spacetime was presented by Sergio Doplicher, Klaus Fredenhagen and John Roberts \cite{Doplicher:1994tu}. According to general relativity, when the energy density grows sufficiently large, a black hole is formed. On the other hand according to the Heisenberg's uncertainty principle, a measurement of a spacetime separation causes an uncertainty in momentum inversely proportional to the separation. Thus energy of scale corresponding to the uncertainty in momentum is localised in the system within a region corresponding to the uncertainty in position. When the separation is small enough, the Schwarzschild radius of the system is reached and a black hole is formed, preventing any information to escape the region. Thus a lower limit is introduced for the measurement of length. A sufficient condition for preventing the gravitational collapse can be expressed as a form of uncertainty relation for the coordinates. This relation in turn can be derived from a nontrivial commutation relation for the coordinates.

Voiculescu's free probability theory \cite{Voiculescu:1996} is another example of a noncommutative structure motivated by physics applications. Here the concept of probability space is replaced by a noncommutative structure leading to noncommuting random variables. One of the main results of this theory, Voiculescu's central limit theorem, yields the Wigner semicircle law, which arises in the theory of random matrices. Related fields of quantum ergodic theory and quantum information theory have recently been the focus of a great deal of attention. They play a pivotal role in the emerging field of quantum computation.

\enlargethispage{-2ex} Just as in the quantisation of a classical phase space, a noncommutative space is defined by replacing the local coordinates $x^i$ of $\real^D$ by Hermitian operators $\qopx^i$ obeying the commutation relations
\begin{equation}\label{noncommalg}
\left[ \qopx^i,\qopx^j \right] = \ui \theta^{ij}.
\end{equation}
The $\qopx^i$ then generate a noncommutative algebra of operators. Within the framework of canonical quantisation, Weyl introduced an elegant prescription for associating a quantum operator to a classical function of the phase-space variables \cite{Weyl:1931}. This technique provides a systematic way to describe noncommutative spaces in general and to study field theories defined thereon. Weyl quantisation provides a one-to-one correspondence between the algebra of fields on $\real^D$ and this ring of operators, and it may be thought of as an analogue of the operator-state correspondence of local quantum field theory. Although we shall deal with the commutators \eqref{noncommalg} with constant $\theta^{ij}$, Weyl quantisation also works for more general commutation relations.\footnote{The most common explicit realisations of the noncommutative nature of spacetime coordinates are: a canonical structure
\[
[ \qopx^i, \qopx^j ] = \ui \theta^{ij}, \qquad \theta^{ij} \in \complex,
\]
a Lie-algebra structure
\[
[ \qopx^i, \qopx^j ] = \ui C^{ij}{}_k \qopx^k, \qquad C^{ij}{}_k \in \complex,
\]
and a quantum-space structure \cite{Wess-1999, Cerchiai:1998eg, Wess:1990vh, Schmidke:1991mr}
\[
\qopx^i \qopx^j = q^{-1} \qopR^{ij}{}_{kl} \qopx^k \qopx^l, \qquad \qopR^{ij}{}_{kl} \in \complex.
\]
We shall restrict to the (commonly studied) canonical structure in this thesis.} Given a function $f(x)$, we may interpret it as the coordinate-space representation of the \emph{Weyl operator} $\weyl[f]$. The Weyl operator $\weyl[f]$ is Hermitian if $f(x)$ is real-valued. If we now consider the product of two Weyl operators $\weyl[f]$ and $\weyl[g]$ corresponding to functions $f(x)$ and $g(x)$ then $\weyl[f]\,\weyl[g] = \weyl[f\star g]$ with the star product defined as\footnote{A somewhat detailed discussion about star product and Weyl operator can be found in \refcite{Szabo:2001kg}.}
\begin{equation}\label{star.dcurr}
(f\star g)(x) = \left. \exp\left(\frac{\mathrm{i}}{2}\theta^{\alpha\beta}\partial_{\alpha}\partial'_{\beta} \right)f(x)g(x')\right|_{x'=x},
\end{equation}
where $\partial'_{\beta} \equiv {\partial}/{\partial x'^{\beta}}$. This star-product is associative but noncommutative, and is defined for constant $\theta$. For $\theta=0$ it reduces to the ordinary product of functions. It is a particular example of a star product which is normally defined in deformation quantisation \cite{Bayen:1977ha}.

Therefore, the spacetime noncommutativity may be encoded through ordinary products in the noncommutative $C^*$-algebra of Weyl operators, or equivalently through the deformation of the product of the commutative $C^*$-algebra of functions on spacetime to the noncommutative star product.


\section{Emergence of noncommutativity}

Although it seems that noncommutative geometry is quite a pure mathematical subject, noncommutativity does emerge in some definite limits of string theory. The string-theory origin of noncommutativity is very similar to the coordinate noncommutativity in the lowest Landau level---both rely on the presence of a strong background field. Let us first describe the Landau problem  \cite{Landau:1977} briefly\footnote{See \refscite{Dunne:1989hv, Dunne:1992ew} for a modern account.}, as it is an important physically realised example of noncommuting coordinates. Consider a charged particle of mass $m$ moving in the plane $\vec x=(x^1,x^2)$ and in the presence of a constant, perpendicular magnetic field of magnitude $B$. The classical Lagrangian of the system is
\begin{equation}\label{2}
L = \frac{m}{2} \dot{\vec x}^{\,2} - e \dot{\vec x}\cdot\vec A
\end{equation}
where $e$ is the particle charge and  $\vec{A}$ is the corresponding vector potential.

The quantum Hamiltonian is
\begin{equation}\label{2.5}
\widehat{H} = \frac{1}{2m} \qoppi^i \qoppi^i,
\end{equation}
where $\qoppi^{i} = m  \dot{\qopx}^i = \qopp^i - e \qopA^i$ are the physical momenta and $\qopp^{i}$ are\ the canonical momenta. We notice that the canonical momenta commute, while the physical momenta satisfy the commutation relation
\begin{equation}\label{2.7}
\left[ \qoppi^i, \qoppi^j \right] = \ui \hbar e B \varepsilon^{ij}.
\end{equation}
It is useful to define, in analogy with the classical case, the center-of-orbit operator, whose components are given by
\begin{equation}\label{2.8}
\qopX^i = \qopx^{i} - \frac{\ui}{eB} \qoppi^i.
\end{equation}
These components can be shown to satisfy the commutation relation
\begin{equation}\label{2.9}
\left[ \qopX^{i},\qopX^{j} \right] = - \ui \frac{\hbar\epsilon^{ij}} {eB} = \ui \theta^{ij},
\end{equation}
where $\theta^{ij}=\left(  -\hbar/eB\right)  \varepsilon^{ij}$. While $[\qopx^{i}, \qopx^{j}]=0$, the $\qopX^{i}$ are not allowed to commute due to the presence of the term containing the magnetic field.

Now we consider the strong-magnetic-field limit. In this case, the system is projected onto the lowest Landau level. A rigorous prescription of how to work in this limit, which is achieved by solving the constraint $\qoppi^i \weakequals 0$\ (using a projection technique), may be found in \refcite{Dunne:1989hv}. On heuristic grounds, one can understand the projection onto the lowest Landau level as a process where the particles have their kinetic degrees of freedom frozen and are confined into their respective orbit centers \cite{Banerjee:2001zi}. The particle-coordinate observables in this limit clearly satisfy \eqref{2.9} as a consequence of the coincidence between $\qopX^{i}$ and $\qopx^{i}$.

Usually the relation \eqref{2.9} is achieved in the literature by dropping the kinetic term directly from the Lagrangian \eqref{2}. If we write the vector potential as $\vec{A} = (0,Bx,0),$ and consider the $B \rightarrow \infty$ or $m \rightarrow 0$ limits, we can discard the kinetic term and write the Lagrangian as
\begin{equation}\label{3}
L = e B x^1 \dot{x}^2.
\end{equation}
In this Lagrangian, the $x^1$ and $x^2$ variables are canonically conjugate, and their respective quantum operators satisfy a commutation relation identical to \eqref{2.9}:
\begin{equation}\label{4}
\left[ \qopx^{i}, \qopx^{j} \right] = -\frac{i\hbar}{eB} \varepsilon^{ij}.
\end{equation}
The limit $m \rightarrow 0$ with fixed $B$ is actually the projection of the quantum-mechanical spectrum of this system onto the lowest Landau level. (The mass gap between Landau levels is $B/m$.) The same projection can be done in the limit $B \rightarrow \infty$ of strong magnetic field with fixed mass $m$.

The canonical noncommutativity originating from string theory in \refcite{Seiberg:1999vs} is based on an approximation which is similar to the one of the lowest Landau level just described. Consider open bosonic strings moving in a flat Euclidean space with metric $G_{\mu\nu}$ in the presence of a constant Neveu--Schwarz $B$-field and with D$p$-branes. The $B$-field probed by the open strings is equivalent to a constant magnetic field on the branes, and it can be gauged away in the directions transverse to the D$p$-brane world-volume. The world-sheet action is
\begin{equation}\label{5}
\begin{split}
S &= \frac{T}{2} \int_{\Sigma} \left( G_{\mu\nu} \partial_{\eta}X^{\mu} \partial^{\eta}X^{\nu}
- \frac{\ui}{T} B_{\mu\nu} \varepsilon^{\eta\lambda} \partial_{\eta}X^{\mu} \partial_{\lambda}X^{\nu} \right)\\
&= \frac{T}{2} \int_{\Sigma} G_{\mu\nu} \partial_{\eta}X^{\mu} \partial^{\eta} X^{\nu} - \frac{\ui}{2} \int_{\partial\Sigma} B_{\mu\nu} X^{\mu} \partial_{t}X^{\nu} ,
\end{split}
\end{equation}
where $T$ is the string tension, $\Sigma$ is the string world-sheet, $\partial_{t}$ is a tangential derivative along the world-sheet boundary $\partial\Sigma$ and the $X^{\mu}$ is the embedding function of the strings into flat spacetime. If we consider the limit $G_{\mu\nu} \sim (1/4 \pi^2 T^2) \rightarrow 0,$ keeping $B_{\mu\nu}$ fixed \cite{Seiberg:1999vs}, the bulk kinetic terms of \eqref{5} vanish. The world-sheet theory in this limit is topological. All that remains are the boundary degrees of freedom, which are governed by the action
\begin{equation}\label{6}
S = -\frac{\ui}{2} \int_{\partial\Sigma} B_{\mu\nu} X^{\mu} \partial_{t}X^{\nu}.
\end{equation}
If one regards \eqref{6} as a one-dimensional action, and ignores the fact that the $X^{\mu}\left(t\right)$ are the endpoints of a string, then it can be considered as analogous to the action corresponding to the Lagrangian of the Landau problem \eqref{3}. Under the approximation being considered, the $X^{\mu}\left(t\right)$ may be regarded as operators satisfying the canonical commutation relation
\begin{equation}\label{7}
\left[ \widehat{X}^{\mu}, \widehat{X}^{\nu} \right] = \left( \frac{\ui}{B} \right)^{\mu\nu}.
\end{equation}
The emergence of noncommutativity in the context of string theory will be further illustrated in Chapter \ref{chap:membrane}.


\section{Structure of the thesis}

This thesis, based on the work reported in \refscite{Banerjee:2003tk, Banerjee:2004ev, Banerjee:2005yy, Banerjee:2005zq, Banerjee:2006db}, is devoted to the study of certain aspects of noncommutativity in field theory, strings and membranes.

We start with a review, based on \refcite{Banerjee:2002ky}, of occurrence of noncommutativity in the context of an open string. Then we analyse the dynamics of an open membrane, both for the free case and when it is coupled to a background three-form, whose boundary is attached to $p$-branes. The role of boundary conditions and constraints in the Nambu--Goto and Polyakov formulations is studied. The low-energy approximation that effectively reduces the membrane to an open string is examined. Noncommutative features of the boundary string coordinates, where the cylindrical membrane is attached to the D$p$-branes, are revealed by algebraic consistency arguments and not by treating boundary conditions as primary constraints, as is usually done. The exact form of the noncommutative algebra is obtained in the low-energy limit. This is the subject matter of Chapter \ref{chap:membrane}.

In Chapter \ref{chap:current} we already take a noncommutative spacetime and proceed to see its implications. The Seiberg--Witten map, which provides an alternative method of studying noncommutative gauge theories by recasting these in terms of their commutative equivalents, is discussed. Here we derive maps relating currents and their divergences in nonabelian $\mathrm{U}(N)$ noncommutative gauge theory with the corresponding expressions in the ordinary (commutative) description. For the $\mathrm{U}(1)$ theory in the slowly-varying-field approximation, these maps are also seen to connect the star-gauge-covariant anomaly in the noncommutative theory with the standard Adler--Bell--Jackiw anomaly in the commutative version. For arbitrary fields, derivative corrections to the maps are explicitly computed up to $\uO(\theta^{2})$.

The aim of Chapter \ref{chap:anomalous} is to exploit the Seiberg--Witten maps for fields and currents in a $\uU(1)$ gauge theory relating the noncommutative and commutative (usual) descriptions to obtain the $\uO(\theta)$ structure of the commutator anomalies in noncommutative electrodynamics. These commutators involve the (covariant) current--current algebra and the (covariant) current--field algebra. We also establish the compatibility of the anomalous commutators with the noncommutative covariant anomaly through the use of certain consistency conditions derived here.

One feature of noncommutative field theories is the violation of Lorentz invariance. This issue is discussed in Chapter \ref{chap:lorentz}. Here we explicitly derive, following a Noether-like approach, the criteria for preserving Poincar\'e invariance in noncommutative gauge theories. Using these criteria we discuss the various spacetime symmetries in such theories. The analysis is performed in both the commutative as well as noncommutative descriptions and a compatibility between the two is also established.

Although the noncommutativity of spacetime coordinates is taken as the signature for violation of Lorentz invariance, it has been shown that the relativistic invariance can be retained in the sense of twisted Poincar\'e invariance of the theory \cite{Wess:2003da, Chaichian:2004za, Chaichian:2004yh}. Chapter \ref{chap:deform} is devoted to the study of general \emph{deformed} conformal-Poincar\'e (Galilean) symmetries consistent with relativistic (nonrelativistic) canonical noncommutative spaces. In either case we obtain deformed generators containing arbitrary free parameters, which close to yield new algebraic structures. A particular choice of these parameters reproduces the undeformed algebra. The structures of the deformed generators in both the coordinate and momentum representations are derived. Notably, the deformations in the momentum representation drop out for the specific choice of parameters leading to the undeformed algebra. The modified coproduct rules and the associated Hopf algebra are also obtained. We also show that for the choice of parameters leading to the undeformed algebra, the deformations are represented by twist functions.

Finally, in Chapter \ref{chap:discussions} we summarise the important results.


%% file: chap_memb.tex

\chapter{\label{chap:membrane}Strings, membranes and noncommutativity}


An intriguing connection between string theory, noncommutative geometry and noncommutative (as well as ordinary) Yang--Mills theory was revealed in \refcite{Seiberg:1999vs}. The study of open string, in the presence of a background Neveu--Schwarz two-form field $B_{\mu\nu}$, leads to a noncommutative  structure which manifests in the noncommutativity at the endpoints of the string which are attached to D-branes. Different approaches have been adopted to obtain this result.

Over the last decade string theory has been gradually replaced by M-theory as the most natural candidate for a fundamental description of nature. While a complete definition of M-theory is yet to be given, it is believed that the five perturbatively consistent string theories are different phases of this theory. With the replacement of string theory by M-theory, the string itself has lost its position as the main candidate for the fundamental degree of freedom. Instead, higher-dimensional extended objects like membranes are being considered. (For a review of the theory of membranes, see \refscite{Hoppe:2002km, Taylor:2001vb}.) Indeed it is known that membrane and five-brane occur naturally in eleven-dimensional supergravity, which is argued to be the low-energy limit of M-theory. Also, string theory is effectively described by the low-energy dynamics of a system of branes. For instance, the membrane of M-theory may be `wrapped' around the compact direction of radius $R$ to become the fundamental string of type-IIA string theory, in the limit of vanishing radius.

With the shift in focus from string theory to M-theory, there has been a flurry of activity in analysing noncommutativity in membranes, specifically when an open membrane that couples to a three-form, ends on a D-brane \cite{Bergshoeff:2000jn, Ardalan:1998ce, Ardalan:1999av, Kawamoto:2000zt, Das:2001mg, Tezuka:2002wn}. In this chapter we further this investigation, but with a new perspective and methodology, as explained below.

The study of noncommutative properties in membranes is more involved than the analogous study in the string case since the equations to be solved are nonlinear. Naturally, in contrast to the string situation, the results could be obtained only under some approximations. It is useful to recapitulate how noncommutativity is derived in either the string coupled to the two-form or the membrane coupled to the three-form. There are nontrivial boundary conditions which are incompatible with the basic Poisson brackets of the theory. These boundary conditions are considered as primary constraints in the algorithm of Dirac's constrained Hamiltonian dynamics \cite{Kawamoto:2000zt, Das:2001mg, Tezuka:2002wn, Chu:1998qz, Romero:2002vg}. The primary constraints lead to secondary constraints. Noncommutativity is manifested through the occurrence of nontrivial Dirac brackets. The brackets are found to be gauge dependent, but there is no gauge where it can be made to vanish.

An alternative approach to deal with noncommutativity in strings was advocated in \refcite{Banerjee:2002ky} where, contrary to other approaches, the boundary conditions are not interpreted as primary constraints. The noncommutative algebra emerges from a set of consistency requirements. It is rather similar in spirit to the original analysis of \refcite{HansonReggeTeitelboim:1976} where a modified algebra, involving the periodic delta function instead of the usual one, was found for the coordinates and their conjugate momenta, in the example of the free Nambu--Goto string.

In this chapter we adopt the same strategy to the membrane model. We discuss both the Nambu--Goto and Polyakov forms of action, although noncommutativity is explicitly considered only in the latter formulation. The similarities or otherwise in the analysis of the two actions are illuminated. Analogous to the set of orthonormal gauge-fixing conditions given for the free Nambu--Goto string \cite{Banerjee:2002ky, HansonReggeTeitelboim:1976}, we derive a set of quasi-orthonormal gauge conditions for the free Nambu--Goto membrane. Just as the orthonormal gauge in the Nambu--Goto string corresponds to the conformal gauge in the Polyakov string, we find out the analogue of the quasi-orthonormal gauge in the Polyakov membrane. It corresponds to a choice of the metric that leads to equations of motion that can be explicitly solved in the light-front coordinates \cite{Taylor:2001vb}. The structure and implications of the boundary conditions in the two formulations have been elaborated. In the Nambu--Goto case, the conditions involve the velocities that cannot be inverted so that a phase-space formulation is problematic. Only by fixing a gauge is it possible to get hold of a phase-space description. In the Polyakov type, on the other hand, the boundary condition is expressible in phase-space variables without the need of any gauge choice. This is because the metric itself is regarded as an independent field. In this sense, therefore, there is no qualitative difference between string and membrane boundary conditions, since even in the Nambu--Goto string, a gauge fixing is required for writing the boundary conditions in terms of phase-space variables. We thus differ from \refcite{Das:2001mg} where it is claimed that it is imperative in the membrane case, as opposed to the string case, to gauge-fix, in order to express the boundary conditions in phase-space coordinates, as a first step in the Hamiltonian formalism.

The mandatory gauge fixing in the Nambu--Goto membrane, as  we shall show, converts the reparametrisation-invariant (first-class) system into a second-class one, necessitating the use of Dirac brackets. This involves the inversion of highly nonlinear expressions, so that approximations become essential to make any progress. Hence we avoid this formulation in favour of the Polyakov version, where gauge fixing is not mandatory.

A detailed constrained Hamiltonian analysis of the free bosonic Polyakov membrane naturally leads to three restrictions on the world-volume metric. These are found to be identical to those obtained by counting the independent degrees of freedom. Unlike the case of the classical string where there are three components of the metric and three continuous symmetries (two diffeomorphism symmetries and one scale symmetry), leading to a complete specification of the metric by gauge fixing, for the membrane there are six independent metric components and only three diffeomorphism symmetries. Thus only three restrictions on the metric can be imposed. Interestingly, the restrictions usually put in by hand \cite{Taylor:2001vb} to perform calculations in the light-front coordinates are obtained directly in our Hamiltonian formalism. This gauge fixing is only partial in the sense that the nontrivial gauge generating first-class constraints remain unaffected. Effectively, therefore, it is a gauge-independent Hamiltonian formalism. We show that the boundary string coordinates corresponding to the membrane--D$p$-brane system (i.e., when the boundary of the open membrane is attached to $p$-branes) satisfy the usual Poisson algebra without any noncommutativity. By imposing further gauge conditions, it is possible to simulate a situation where the cylindrical membrane is wrapped around a circle of vanishing radius so that the open membrane passes over to an open string. The boundary conditions of the membrane reduce to the well-known Neumann boundary conditions of the string in the conformal gauge, just as the membrane metric reduces to the conformal metric of the Polyakov string.

Next, the interacting membrane in the presence of a constant three-form tensor potential is discussed. Proceeding in a gauge-independent manner, it is shown that, contrary to the free theory, the boundary string coordinates must be noncommutative. This is shown from certain algebraic conditions. However, in contrast to the string case where it was possible to solve  these equations \cite{Banerjee:2002ky}, here an explicit solution is prevented from the nonlinear structure. Nevertheless, by passing to the low-energy limit (wrapping the membrane on a circle of vanishingly small radius), the explicit form of the noncommutativity in an open string, whose endpoints are attached to a D-brane, are reproduced.

Section \ref{sec:str}, which deals with a brief discussion of noncommutativity in an open string, is a summary of the essential results of \refcite{Banerjee:2002ky}. In section \ref{sec:memb-ng} the free Nambu--Goto membrane is discussed and the form of the quasi-orthonormal gauge conditions, which act as the analogue of the orthonormal gauge conditions in the Nambu--Goto string \cite{HansonReggeTeitelboim:1976}, is derived. The role of the boundary conditions in maintaining stability of the membrane is discussed. The free Polyakov membrane is considered in section \ref{sec:memb-poly}, where its detailed constrained Hamiltonian account is given. The complete form of the energy--momentum tensor is derived. All components of this tensor are written as a linear combination of the constraints. This is a generalisation of the string case since even though Weyl symmetry is absent in the membrane, the energy--momentum tensor has a (weakly) vanishing trace; namely, it vanishes only on the constraint shell. The brackets for the free theory with a cylindrical topology for the membrane 
yield the expected Poisson algebra without any noncommutativity. The low-energy limit where the membrane is approximated by the string, is discussed in 
in the last part of this section.
Section \ref{sec:memb-interact} gives an analysis of the interacting theory. General algebraic requirements enforce a noncommutativity of the boundary coordinates of the membrane, which are attached to the $p$-branes. No gauge fixing or approximation is needed to reveal this noncommutativity. The explicit structure of the algebra is once again computed in the low-energy approximation, when the result agrees with the conformal-gauge expression for the noncommutativity among the coordinates of the endpoints of the string attached to D-branes.


\section{\label{sec:str}Noncommutativity in open string}

We begin by summarising the essential results of \refcite{Banerjee:2002ky} that will be used for an easy comparison of our results of open membrane with those of open string.


\subsection{Free Polyakov string}

The free Polyakov string action is
\begin{equation} \label{901}
S_\mathrm{P} = -\frac{T_\mathrm{s}}{2} \int \ud \tau \ud\sigma \sqrt{-g} g^{ij} \partial_i X^{\mu} \partial_j X_{\mu}
\equiv \int_{\Sigma} \ud^2 \sigma \lagdens, \qquad i,j = 0,1,
\end{equation}
where $T_\mathrm{s}$ stands for string tension, $\tau$ and $\sigma$ are the usual world-sheet parameters and $g_{ij}$, up to a Weyl factor, is the induced metric $h_{ij} = \partial_i X^{\mu} \partial_j X_{\mu}$ on the world-sheet. $X^{\mu}$ are the string coordinates in the $D$-dimensional Minkowskian target space with metric $\text{diag}(-1,1,1,\cdots,1)$. This action has the usual Poincar\'e, Weyl and diffeomorphism invariances. Both $X^{\mu}$ and $g_{ij}$ are regarded as independent dynamical variables \cite{VanHolten:2001nj}. The canonical momenta are
\begin{equation} \label{903}
\Pi_{\mu} = \frac{\partial \lagdens}{\partial (\partial_0 X^{\mu})} = - T_\mathrm{s} \sqrt{-g} \partial^0 X_{\mu}, \qquad 
\pi_{ij} = \frac{\partial \lagdens}{\partial (\partial_0 g^{ij})} = 0.
\end{equation}
It is clear that while $\Pi_{\mu}$ are genuine momenta, $\pi_{ij} \weakequals 0$ are the primary constraints of the theory. To determine the secondary constraints one can either follow the traditional Dirac's Hamiltonian approach or just read them off from the equation obtained by varying $g_{ij}$, since this is basically a Lagrange multiplier. This imposes the vanishing of the symmetric energy--momentum tensor:
\begin{equation} \label{904}
T_{ij} = \frac{2}{\sqrt{-g}} \frac{\delta S_\mathrm{P}}{\delta g^{ij}}
= - T_\mathrm{s} \partial_i X^{\mu} \partial_j X_{\mu} + \frac{T_\mathrm{s}}{2} g_{ij} g^{kl} \partial_k X^{\mu} \partial_l X_{\mu}
\weakequals 0.
\end{equation}
Because of the Weyl invariance, the energy--momentum tensor is traceless:
\begin{equation} \label{905}
{T^i}_i = g^{ij} T_{ij} = 0
\end{equation}
so that only two components of $T_{ij}$ are independent. These components, which are the constraints of the theory, are given by
\begin{gather}
\label{906}
\chi_1 \equiv 2T_\mathrm{s} g T^{00} = - 2T_\mathrm{s} T_{11} = \Pi^2 + T_\mathrm{s}^2 h_{11} \weakequals 0, \\
\label{907}
\chi_2 \equiv \sqrt{-g} T^0_1 = \Pi_{\mu} \partial_1 X^{\mu} \weakequals 0.
\end{gather}
The canonical Hamiltonian density obtained from Eq.~\eqref{901} by a Legendre transformation is given by
\begin{equation} \label{908}
\hamdens_C = \sqrt{-g} {T^0}_0
= \frac{\sqrt{-g}}{2 T_\mathrm{s} g_{11}} \chi_1 + \frac{g_{01}}{g_{11}} \chi_2,
\end{equation}
which, as expected, turns out to be a linear combination of the constraints. The boundary condition written in terms of phase-space variables is given by
\begin{equation} \label{911}
\left[ T_\mathrm{s} \partial_1 X^{\mu} + \sqrt{-g} g^{01} \Pi^{\mu} \right]_{\sigma=0,\pi} = 0,
\end{equation}
where the string parameters are in the region $-\infty \leq \tau \leq +\infty$, $0 \leq \sigma \leq \pi$. This boundary condition is incompatible with the first of the basic Poisson brackets:
\begin{gather}
\label{912}
\left\{ X^{\mu}(\tau,\sigma), \Pi_{\nu}(\tau,\sigma') \right\}
= \delta^{\mu}_{\nu} \delta(\sigma-\sigma'), \\
\label{913}
\left\{ g_{ij}(\tau,\sigma), \pi^{kl}(\tau,\sigma') \right\}
= \tfrac{1}{2} (\delta^k_i \delta^l_j + \delta^l_i \delta^k_j) \delta(\sigma -\sigma').
\end{gather}
From the basic Poisson brackets it is easy to generate a first-class (involutive) algebra:
\begin{equation} \label{914}
\begin{aligned}
&\left\{ \chi_1(\sigma), \chi_1(\sigma') \right\}
= 4T_\mathrm{s}^2 \left[ \chi_2(\sigma ) + \chi_2(\sigma ') \right] \partial_1 \delta(\sigma-\sigma'), \\
&\left\{ \chi_2(\sigma), \chi_1(\sigma') \right\}
= \left[ \chi_1(\sigma) + \chi_1(\sigma') \right] \partial_1 \delta(\sigma-\sigma'), \\
&\left\{ \chi_2(\sigma), \chi_2(\sigma') \right\}
= \left[ \chi_2(\sigma) + \chi_2(\sigma') \right] \partial_1 \delta(\sigma-\sigma').
\end{aligned}
\end{equation}
The constraints $\chi_1$ and $\chi_2$ generate the diffeomorphism transformations.

The boundary condition \eqref{911} is not a constraint in the Dirac sense, since it is applicable only at the boundary. Thus, there has to be an appropriate modification in the Poisson brackets to incorporate this condition. This is not unexpected and occurs, for instance, in the example of a free scalar field $\phi(x)$ in $1+1$ dimensions, subjected to periodic boundary condition of period, say, $2\pi$. There the Poisson bracket between the field $\phi(t,x)$ and its conjugate momentum $\pi(t,x)$ is given by
\begin{equation} \label{915}
\left\{ \phi(t,x), \pi(t,y) \right\} = \delta_\mathrm{p}(x-y),
\end{equation}
which is obtained automatically if one starts with the canonical harmonic-oscillator algebra for each mode in the Fourier space. Here
\begin{equation}\label{st403}
\delta_{\mathrm{p}}(x-x') = \frac{1}{2\pi} \sum_{n\in {\integer}} e^{\ui n (x-x')}
\end{equation}
is the periodic delta function of period $2\pi$

Before discussing the mixed condition \eqref{911}, let us consider the simpler Neumann-type condition. Since the string coordinates $X^{\mu}(\tau,\sigma)$ transform as a world-sheet scalar under its reparametrisation, it is more convenient to get
back to scalar field $\phi(t,x)$ defined on $(1+1)$-dimensional spacetime, but with the periodic boundary condition replaced by Neumann boundary condition,
\begin{equation} \label{916}
\left. \partial_x \phi \right|_{\sigma=0,\pi} = 0,
\end{equation}
at the endpoints of a 1-dimensional box of compact size, i.e., of length $\pi$.
Correspondingly, the $\delta_\mathrm{p}$ appearing there in the Poisson bracket \eqref{915}---consistent with periodic boundary condition---has to be replaced now with a suitable `delta function' incorporating Neumann boundary condition, rather than periodic boundary condition.

The following usual property of delta function is also satisfied by $\delta_\mathrm{p}(x-x')$:
\begin{equation} \label{917}
\int_{-\pi}^{+\pi} \ud x' \delta_\mathrm{p}(x-x') f(x') = f(x)
\end{equation}
for any periodic function $f(x) = f(x+2\pi)$ defined in the interval $[-\pi, +\pi]$. Restricting to the case of even and odd functions, $f_\pm(-x) = \pm f_\pm(x)$, the above integral reduces to
\begin{equation} \label{918}
\int_0^{\pi} \ud x' \Delta_{\pm}(x,x') f_\pm(x') = f_\pm(x),
\end{equation}
where
\[
\Delta_{\pm}(x',x)= \delta_\mathrm{p}(x'-x)\pm \delta_\mathrm{p}(x'+x)
\]
Since any function $\phi(x)$ defined in the interval $[0,\pi ]$ can be regarded as a part of an even or odd function $f_\pm(x)$ defined in the interval $[-\pi,\pi]$, both $\Delta_{\pm}(x,x')$ act as delta functions defined in half of the interval at the right, i.e., $[0,\pi]$ as follows from Eq.~\eqref{918}. It is still not clear which of these $\Delta_{\pm}(x,x')$ functions should replace $\delta_\mathrm{p}(x-x')$ in the Poisson-bracket relation. We now consider the Fourier decomposition of an arbitrary function $f(x)$ satisfying periodic boundary condition $f(x) = f(x+2\pi)$:
\begin{equation} \label{921}
f(x) = \sum_{n \in \integer} f_n e^{\ui nx}.
\end{equation}
Clearly,
\[
f'(0) = \ui \sum_{n>0} n(f_n-f_{-n}), \qquad
f'(\pi) = \ui \sum_{n>0} (-1)^n n(f_n-f_{-n}).
\]
Now for even and odd functions, the Fourier coefficients are related as $f_{-n} = \pm f_n$ so that Neumann boundary condition $f'(0) = f'(\pi) = 0$ is satisfied if and only if $f(x)$ is even. Therefore, one has to regard the scalar field $\phi(x)$ defined in the interval $[0,\pi ]$ and subjected to  Neumann boundary condition \eqref{916} as a part of an even periodic function $f_+(x)$ defined in the extended interval $[-\pi, +\pi]$. It thus follows that the appropriate Poisson bracket for the scalar theory is given by $\{ \phi(t,x), \pi(t,x') \} = \Delta_+(x,x')$. It is straightforward to generalise it to the string case as
\begin{equation} \label{922}
\left\{ X^{\mu}(\tau,\sigma), \Pi_{\nu}(\tau,\sigma') \right\}
= \delta^{\mu}_{\nu} \Delta_+(\sigma,\sigma'),
\end{equation}
the Lorentz indices playing the role of `isospin' indices, as viewed from the world-sheet. We observe also that the other brackets
\begin{gather} \label{923}
\left\{ X^{\mu}(\tau,\sigma), X^{\nu}(\tau,\sigma') \right\} = 0, \\
\label{924}
\left\{ \Pi^{\mu}(\tau,\sigma), \Pi^{\nu}(\tau,\sigma') \right\} = 0
\end{gather}
are consistent with the boundary conditions and hence remain unchanged.

The mixed condition \eqref{911} is compatible with the modified brackets \eqref{922} and \eqref{924}, but not with \eqref{923}. Therefore, let us make an ansatz,
\begin{equation} \label{925}
\left\{ X^{\mu}(\tau,\sigma), X^{\nu}(\tau,\sigma') \right\}
= C^{\mu\nu}(\sigma,\sigma'),
\end{equation}
where
\begin{equation} \label{926}
C^{\mu\nu}(\sigma,\sigma') = -C^{\nu\mu}(\sigma',\sigma).
\end{equation}
Imposing the boundary condition \eqref{911} on this algebra, one gets
\begin{equation} \label{927}
\left. \partial'_1 C^{\mu\nu}(\sigma,\sigma') \right|_{\sigma'=0,\pi}
= \left. \partial_1 C^{\mu\nu}(\sigma,\sigma') \right|_{\sigma=0,\pi}
= \sqrt{-g} g^{01} \eta^{\mu \nu} \Delta_+(\sigma,\sigma').
\end{equation}
For a restricted class of metrics that satisfy $\partial_1 g_{ij} = 0$ it is possible to give a quick solution of this equation as
\begin{equation} \label{928}
C^{\mu\nu}(\sigma,\sigma')
= \sqrt{-g} g^{01} \eta^{\mu\nu} \left[ \Theta(\sigma,\sigma') - \Theta(\sigma',\sigma) \right].
\end{equation}
This noncommutativity can be made to vanish in gauges like conformal gauge, where $g^{01}=0$, thereby restoring the usual commutative structure. The essential structure of the involutive algebra \eqref{914} is still preserved, but with $\delta(\sigma-\sigma')$ replaced by $\Delta_+(\sigma,\sigma')$.


\subsection{Interacting Polyakov string}

The Polyakov action for a bosonic string moving in the presence of a constant background Neveu--Schwarz two-form field $B_{\mu \nu}$ is given by
\begin{equation} \label{441}
S_\mathrm{P} = -\frac{T_\mathrm{s}}{2} \int \ud \tau \ud \sigma \left( \sqrt{-g} g^{ij} \partial_i X^{\mu} \partial_j X_{\mu} + e \varepsilon^{ij} B_{\mu \nu} \partial_i X^{\mu} \partial_j X^{\nu} \right),
\end{equation}
where a `coupling constant' $e$ has been introduced. A usual canonical analysis leads to the following set of primary first-class constraints:
\begin{gather}
\label{442}
g T^{00} = \tfrac{1}{2} \left[ (\Pi_{\mu} + e B_{\mu\nu} \partial_1 X^{\nu}) (\Pi^{\mu} + e B^{\mu \nu} \partial_1 X_{\nu}) + T_\mathrm{s}^2 h_{11} \right]
 \weakequals 0, \\
\label{443}
\sqrt{-g} {T^0}_1 = \Pi_{\mu} \partial_1 X^{\mu} \weakequals 0, 
\end{gather}
where
\begin{equation} \label{444}
\Pi_{\mu} = -T_\mathrm{s} \left[ \sqrt{-g} \partial^0 X_{\mu} + e B_{\mu \nu} \partial_1 X^{\nu} \right]
\end{equation}
is the momentum conjugate to $X^{\mu}$. The boundary condition written in terms of phase-space variables is
\begin{equation} \label{448}
\left[ \partial_1 X_{\mu} + \Pi^{\rho} (NM^{-1})_{\rho\mu} \right]_{\sigma=0,\pi} = 0,
\end{equation}
where
\begin{equation} \label{447}
{M^{\rho}}_{\mu} = T_\mathrm{s} \left[ {\delta^{\rho}}_{\mu} - e^2 B^{\rho\nu} B_{\nu\mu} \right], \qquad
N_{\nu\mu} = \frac{g_{01}}{\sqrt{-g}} \eta_{\nu\mu} + e B_{\nu\mu}.
\end{equation}
The $\{X^\mu,\Pi_\nu\}$ Poissson bracket is the same as that of the free string whereas considering the general structure \eqref{925} and exploiting the above boundary condition, one obtains
\begin{equation} \label{449}
\left. \partial_1 C_{\mu\nu}(\sigma,\sigma') \right|_{\sigma=0,\pi}
= \left. (NM^{-1})_{\nu\mu} \Delta_+(\sigma,\sigma') \right|_{\sigma=0,\pi}.
\end{equation}
As in the free case, restricting to the class of metrics satisfying $\partial_1 g_{ij} = 0$, the above equation has a solution
\begin{equation} \label{450}
\begin{split}
C_{\mu\nu}(\sigma,\sigma')
&= \tfrac{1}{2} (NM^{-1})_{(\nu\mu )} \left[ \Theta(\sigma,\sigma') - \Theta(\sigma',\sigma) \right] \\
&\quad {} + \tfrac{1}{2} (NM^{-1})_{[\nu\mu]} \left[ \Theta(\sigma,\sigma') + \Theta(\sigma',\sigma) - 1 \right],
\end{split}
\end{equation}
where $(NM^{-1})_{(\nu \mu)}$ the symmetric and $(NM^{-1})_{[\nu \mu]}$ the antisymmetric part of $(NM^{-1})_{\nu \mu}$.
The modified algebra is gauge dependent; it depends on the choice of the metric. However, there is no choice for which the noncommutativity vanishes. To show this, note that the origin of the noncommutativity is the presence of non-vanishing $N_{\nu\mu }$ in the boundary condition \eqref{448}. Vanishing $N_{\nu\mu }$ would make $B_{\mu\nu}$ and $\eta_{\mu\nu}$ proportional which obviously cannot happen, as the former is an antisymmetric and the latter is a symmetric tensor. Hence, noncommutativity will persist for any choice of world-sheet metric $g_{ij}$. Specially interesting are the expressions for noncommutativity at the boundaries:
\begin{equation} \label{451}
\begin{aligned}
&C_{\mu\nu}(0,0) = -C_{\mu\nu}(\pi,\pi) = -\tfrac{1}{2} (NM^{-1})_{[\nu\mu ]}, \\
&C_{\mu\nu}(0,\pi) = -C_{\mu\nu}(\pi,0) = -\tfrac{1}{2} (NM^{-1})_{(\nu\mu)}.
\end{aligned}
\end{equation}


\section{\label{sec:memb-ng}Free Nambu--Goto membrane}

A dynamical membrane moving in $D-1$ spatial dimensions sweeps out a three-dimensional world-volume in $D$-dimensional spacetime. We use a metric with signature $(-,+,+,\cdots,+)$ in the target space whose indices are $\mu,\nu = 0,1,2,\ldots,D-1$. We can locally choose a set of three coordinates $\sigma^i$, $i = 0,1,2$, on the world-volume to parameterise it. We shall sometime use the notation $\tau = \sigma^0$ and the indices $a,b,\ldots$ to describe `spatial' coordinates $\sigma^a$, $a = 1,2$, on the membrane world-volume. In such a coordinate system, the motion of the membrane through spacetime is described by a set of $D$ functions $X^{\mu}(\sigma^0,\sigma^1,\sigma^2)$ which are the membrane coordinates in the target space.

Although we are going to study the noncommutativity through the Polyakov action, we find it convenient to briefly discuss the Nambu--Goto action also. The Nambu--Goto analysis will be just an extension of the string case considered in \refcite{HansonReggeTeitelboim:1976}. The Nambu--Goto action for a membrane moving in flat spacetime is given by the integrated proper volume swept out by the membrane:
\begin{equation} \label{memb101}
S_{\mathrm{NG}} = -T \int_{\Sigma} \ud^3 \sigma \sqrt{-h}
\equiv \int_{\Sigma} \ud^3 \sigma \lagdens_{\mathrm{NG}} \left( X^{\mu},\partial_i X^{\mu} \right),
\end{equation}
where $T$ is a constant which can be interpreted as the membrane tension and $h = \det h_{ij}$ with
\begin{equation} \label{memb102}
h_{ij} = \partial_i X^{\mu} \partial_j X_{\mu}
\end{equation}
being the induced metric on $(2+1)$-dimensional world-volume, which is nothing but the pullback of the flat spacetime metric on this three-dimensional sub-manifold. This induced metric, however, does not have the status of an independent field in the world-volume; it is rather determined through the embedding fields $X^{\mu}$. The Lagrangian density is $\lagdens_{\mathrm{NG}} = -T \sqrt{-h}$. The Euler--Lagrange equation is given by
\begin{equation} \label{103}
\partial_i \left( \sqrt{-h} h^{ij} \partial_j X^{\mu} \right) = 0 \, ,
\end{equation}
while the boundary conditions are given by
\begin{equation} \label{memb104}
\left. \mathcal{P}^a_{\mu} \right|_{\partial\Sigma}
= \left. -T \sqrt{-h} \partial^a X_{\mu} \right|_{\partial\Sigma} = 0 \, ,
\end{equation}
where
\begin{equation} \label{105}
\mathcal{P}^i_{\mu} = \frac{\partial \lagdens_{\mathrm{NG}}}{\partial (\partial_i X^{\mu})}
= -T \sqrt{-h} \partial^i X_{\mu}
\end{equation}
and $\partial \Sigma$ represents the boundary. The components $\mathcal{P}^0_{\mu} \equiv \Pi_{\mu}$ are the canonical momenta conjugate to $X^{\mu}$. Using this, the Euler--Lagrange equation~\eqref{103} can be rewritten as
\begin{equation} \label{106}
\partial_0 \Pi^{\mu} + \partial_a \mathcal{P}^{a\mu} = 0 \, .
\end{equation}
It can be seen easily that the theory admits the primary constranits
\begin{gather}
\label{memb107}
\psi \equiv \Pi^{2} + T^{2} \bar{h} \weakequals 0 \, , \\
\label{108}
\phi_{a} \equiv \Pi_{\mu}\partial_{a}X^{\mu} \weakequals 0 \, , \qquad a = 1,2 \, ,
\end{gather}
where  $\Pi^{2} \equiv \Pi^{\mu}\Pi_{\mu}$ and $ \bar{h}=\det h_{ab}
= h_{11}h_{22}-(h_{12})^{2}$. These constraints are first-class since the
brackets between them vanish weakly:
\begin{equation} \label{1109}
\begin{aligned}
&\begin{aligned}
\left\{\psi\left(\tau,\vec{\sigma}\right),\psi\left(\tau,\vec{\sigma}'\right)\right\}
&= 4T^{2}\left[ \left\{h_{22}\left(\tau,\vec{\sigma}\right)\partial_{1}\delta\left(\vec{\sigma}\!-\!\vec{\sigma}'\right)-h_{12}\left(\tau,\vec{\sigma}\right)\partial_{2}\delta\left(\vec{\sigma}\!-\!\vec{\sigma}'\right)\right\}\phi_{1}\left(\tau,\vec{\sigma}\right)\right. \\
&\quad {} +\left\{h_{11}\left(\tau,\vec{\sigma}\right)\partial_{2}\delta\left(\vec{\sigma}\!-\!\vec{\sigma}'\right)-h_{12}\left(\tau,\vec{\sigma}\right)\partial_{1}\delta\left(\vec{\sigma}\!-\!\vec{\sigma}'\right)\right\}\phi_{2}\left(\tau,\vec{\sigma}\right) \\
&\quad {} -\left\{h_{22}\left(\tau,\vec{\sigma}'\right)\partial'_{1}\delta\left(\vec{\sigma}\!-\!\vec{\sigma}'\right)-h_{12}\left(\tau,\vec{\sigma}'\right)\partial'_{2}\delta\left(\vec{\sigma}\!-\!\vec{\sigma}'\right)\right\}\phi_{1}\left(\tau,\vec{\sigma}'\right) \\
&\quad \left. {} -\left\{h_{11}\left(\tau,\vec{\sigma}'\right)\partial'_{2}\delta\left(\vec{\sigma}\!-\!\vec{\sigma}'\right)-h_{12}\left(\tau,\vec{\sigma}'\right)\partial'_{1}\delta\left(\vec{\sigma}\!-\!\vec{\sigma}'\right)\right\}\phi_{2}\left(\tau,\vec{\sigma}'\right)\right] \\
&\weakequals 0,
\end{aligned} \\
&\left\{\phi_{a}\left(\tau,\vec{\sigma}\right),\phi_{b}\left(\tau,\vec{\sigma}'\right)\right\}
= \phi_{b}\left(\tau,\vec{\sigma}\right)\partial_{a}\delta\left(\vec{\sigma}\!-\!\vec{\sigma}'\right)-\phi_{a}\left(\tau,\vec{\sigma}'\right)\partial'_{b}\delta\left(\vec{\sigma}\!-\!\vec{\sigma}'\right)\weakequals0, \\
&\left\{\psi\left(\tau,\vec{\sigma}\right),\phi_{a}\left(\tau,\vec{\sigma}'\right)\right\}
= 2\psi\left(\tau,\vec{\sigma}\right)\partial_{a}\delta\left(\vec{\sigma}\!-\!\vec{\sigma}'\right)+\partial_{a}\psi\,\delta\left(\vec{\sigma}\!-\!\vec{\sigma}'\right)\weakequals 0,
\end{aligned}
\end{equation}
where $\partial'_{a} \equiv \frac{\partial}{\partial \sigma'^{a}}$.

The canonical world-volume energy--momentum tensor density{\footnote{$\lagdens_{\mathrm{NG}}$ transforms as a scalar \emph{density} under diffeomorphism.}} can be obtained through Noether theorem:
\begin{equation}
{[\theta_{C}]^{i}}_{j} = \frac{\partial\lagdens_{\mathrm{NG}}}{\partial(\partial_{i} X^{\mu})}\partial_{j}X^{\mu}-{\delta^{i}}_{j}\lagdens_{\mathrm{NG}}.
\label{1110}
\end{equation}
In particular, ${[\theta_{C}]^{0}}_{0}=0$, ${[\theta_{C}]^{0}}_{a} = \phi_{a}\weakequals0$, ${[\theta_{C}]^{a}}_{0}=0$ and ${[\theta_{C}]^{a}}_{b}=0$. We notice that the canonical Hamiltonian density, ${\mathcal H}_{C}={[\theta_{C}]^{0}}_{0}$, obtained by Legendre transformation, vanishes strongly. Since the canonical energy--momentum tensor density is first-class, we may add to it a linear combination of first-class constraints with tensor-valued coefficients to write down the total energy--momentum tensor density as
\begin{equation} \label{1111}
{\theta^{i}}_{j} = {U^{i}}_{j}\psi+{V^{ai}}_{j}\phi_{a} \weakequals 0.
\end{equation}
The generators of $\tau$- and $\sigma^{a}$-translations are
\begin{equation}
H_T = \int \ud^{2}\sigma{\theta^{0}}_{0}, \qquad 
H_a = \int \ud^{2}\sigma{\theta^{0}}_{a}.
\label{1112}
\end{equation}
As one can easily see, there are no secondary constraints. The Hamilton's equation $\dot{X}^{\mu} = \{X^{\mu}, H_{T}\}$ gives $\partial_{0}X^{\mu} = 2{U^{0}}_{0}\Pi^{\mu}+{V^{a0}}_{0}\partial_{a}X^{\mu}$, which reproduces the definition of momenta $\Pi^{\mu}$ for the following choice of ${U^{0}}_{0}$ and ${V^{a0}}_{0}$:
\begin{equation}
{U^{0}}_{0} = \frac{\sqrt{-h}}{2T\bar{h}}, \qquad
{V^{a0}}_{0} = -\frac{hh^{0a}}{\bar{h}}= \bar{h}^{ab}h_{0b},
\label{1113}
\end{equation}
where $\bar{h}^{ab}\, (\neq h^{ab}$, which is obtained by chopping off first row and first column from $h^{ij}$, the inverse of $h_{ij})$ is the inverse of $h_{ab}$ in the two-dimensional subspace. The other equation, $\dot{\Pi}^{\mu} = \{\Pi^{\mu}, H_{T}\}$, reproduces the Euler--Lagrange equation \eqref{103} whereas $\partial_{a}X^{\mu} = \{X^{\mu}, H_{a}\}$ gives $\partial_{a}X^{\mu} = 2{U^{0}}_{a}\Pi^{\mu}+{V^{b0}}_{a}\partial_{b}X^{\mu}$, which is satisfied for
\begin{equation}
{U^{0}}_{a}=0, \qquad
{V^{b0}}_{a}= {\delta^{b}}_{a}.
\label{1114}
\end{equation}

Coming to the conserved Poincar\'e generators in the target space, the translational generator is given by
\[
P_{\mu} = \int \ud^{2}\sigma \Pi_{\mu}, 
\]
and the angular-momentum generator by
\[
M^{\mu\nu} = \int \ud^{2}\sigma\left(X^{\mu}\Pi^{\nu}-X^{\nu}\Pi^{\mu}\right).
\]
As can be easily checked, these generators generate appropriate Poincar\'e transformations. The above analysis can be generalised in a straightforward manner to an arbitrary $p$-brane.

There is an interesting implication of the boundary conditions \eqref{memb104}. For a cylindrical membrane with $\sigma^{1} \in [0, \pi]$, $\sigma^{2} \in [0, 2\pi)$, $\sigma^{2}$ representing the compact direction, the boundary condition is written as
\[
\left.\mathcal{P}^{1}_{\mu}\right|_{\sigma^{1}=0,\pi}=\left.-T\sqrt{-h}
\partial^{1}X_{\mu}\right|_{\sigma^{1}=0,\pi} = 0.
\]
Squaring the above equation, we get
\begin{equation}
\left.hh^{11}\right|_{\sigma^{1}=0,\pi}
= \left[h_{00}h_{22}-(h_{02})^{2}\right]_{\sigma^{1}=0,\pi} = 0,
\label{1123}
\end{equation}
which implies
\begin{equation}
h_{00}|_{\sigma^{1}=0,\pi} = \left.\frac{(h_{02})^{2}}{h_{22}}\right|_{\sigma^{1}=0,\pi}.
\label{112376}
\end{equation}
However, $h_{22}$ is strictly positive and cannot vanish at the boundary in order to prevent it from collapsing to a point as the length of the boundary is given by $\int^{2\pi}_{0} \sqrt{h_{22}} \ud \sigma^2$. This indicates that
\[
\left. \dot{X}^{2} \right|_{\sigma^{1}=0,\pi}=h_{00}|_{\sigma^{1}=0,\pi} \geq 0
\]
so that the points on the boundary move along either a space-like or light-like trajectory. If we now demand that the speed of these boundary points should not exceed the speed of light then we must have $h_{02}|_{\sigma^{1}=0,\pi}=0$ in Eq.~\eqref{112376} so that
\[
\left. \dot{X}^{2} \right|_{\sigma^{1}=0,\pi}=h_{00}|_{\sigma^{1}=0,\pi} = 0.
\]
Therefore the boundary points move with the speed of light which is a direct generalisation of the string case where a similar result holds. For a square membrane with $\sigma^{1}, \sigma^{2} \in [0, \pi]$, the boundary conditions \eqref{memb104} are written as
\begin{eqnarray*}
\left.\mathcal{P}^{1}_{\mu}\right|_{\sigma^{1}=0,\pi}\!\!&=&\!\!\left.-T\sqrt{-h}
\partial^{1}X_{\mu}\right|_{\sigma^{1}=0,\pi} = 0, \\
\left.\mathcal{P}^{2}_{\mu}\right|_{\sigma^{2}=0,\pi}\!\!&=&\!\!\left.-T\sqrt{-h}
\partial^{2}X_{\mu}\right|_{\sigma^{2}=0,\pi} = 0.
\end{eqnarray*}
Therefore, in addition to Eq.~\eqref{1123}, we also have
\[
\left.hh^{22}\right|_{\sigma^{2}=0,\pi}
= \left[h_{00}h_{11}-(h_{01})^{2}\right]_{\sigma^{2}=0,\pi} = 0.
\]
Proceeding just as in the case of cylindrical membrane, we find that we must have $h_{02}|_{\sigma^{1}=0,\pi}=0$ and $h_{01}|_{\sigma^{2}=0,\pi}=0$ so that
\[
\dot{X}^{2}|_{\sigma^{1}=0,\pi} = 0 = \dot{X}^{2}|_{\sigma^{2}=0,\pi},
\]
which shows that the boundary points move with the speed of light. Also, since $h_{0a}\weakequals 0$ at the boundary, for both the cylindrical or square topology, it implies that the vector $\partial_0 X^\mu$ is not only null, but also orthogonal to all directions tangent to the membrane world-volume. Hence the boundary points move with the speed of light, perpendicularly to the membrane. This peculiar motion is exactly reminiscent of the string case. The tension in the free membrane would cause it to collapse. This is prevented by the angular momentum generated by the boundary motion, just as the collapse of the free string is thwarted by a similar motion of the string endpoints \cite{Mandelstam:1974br}.


\paragraph{Quasi-orthonormal gauge fixing conditions.} As we shall see now, the membrane case, or any $p$-brane with $p > 1$ for that matter, involves some subtle issues. The first step is to provide  a set of complete gauge fixing conditions.  Taking a cue from the previous analysis  we would like to generalise the condition $h_{0a}\weakequals 0$, so that it holds everywhere, instead of just at the boundary. This is also quite similar in spirit to what is done for implementing the orthonormal gauge in the string case. Indeed, following the string analysis of  \cite{HansonReggeTeitelboim:1976}, we first impose the following gauge fixing conditions:
\begin{gather}
\label{memb109}
\lambda_{\mu} \left( X^{\mu} \left(\tau, \vec{\sigma}\right) - \frac{P^{\mu}\tau}{TA} \right) \weakequals 0,
\\
\label{memb110}
\lambda_{\mu} \left( \Pi^{\mu} \left(\tau, \vec{\sigma}\right) - \frac{P^{\mu}}{A} \right) \weakequals 0,
\end{gather}
where $\vec{\sigma} = (\sigma^{1}, \sigma^{2})$ and $\lambda_{\mu}$ is an arbitrary constant vector and $A$ is taken to be the `parametric area' of the membrane. For example, if the membrane is of square topology with $\sigma^{1}, \sigma^{2} \in [0, \pi]$, it will be $\pi^{2}$ and for cylindrical topology with $\sigma^{1} \in [0, \pi]$, $\sigma^{2} \in [0, 2\pi)$ (membrane periodic along $\sigma^{2}$-direction), it will be $2\pi^{2}$. Clearly, this `parametric area' is not an invariant quantity under two-dimensional diffeomorphism. One can think of the square or cylindrical membrane to be flat at one instant to admit a Cartesian-like coordinate system on the membrane surface which will provide a coordinate chart for it during its future time evolution.

Differentiating Eq.~\eqref{memb109} with respect to $\tau$ and using Eq.~\eqref{memb110}, we get
\begin{equation} \label{memb111}
\lambda \cdot \dot{X} \weakequals \frac{\lambda \cdot P}{TA} 
\weakequals \frac{\lambda \cdot \Pi}{T}.
\end{equation}
Differentiating Eq.~\eqref{memb109} with respect to $\sigma^{a}$, and Eq.~\eqref{memb110} with respect to $\tau$ we get
\begin{gather}
\label{112}
\lambda \cdot \partial_{a} X \weakequals 0,
\\
\label{memb113}
\partial_{0} \left( \lambda \cdot \Pi \right) \weakequals 0.
\end{gather}
Using Eq.~\eqref{memb113}, it follows from the form \eqref{106} of Euler--Lagrange equation that
\begin{equation} \label{memb114}
\partial_{a}\left(\lambda \cdot \mathcal{P}^{a}\right) \weakequals 0.
\end{equation}
Upon contraction with $\lambda^{\mu}$, the boundary conditions \eqref{memb104} give
\begin{equation} \label{memb115}
\left. \lambda \cdot \mathcal{P}^a \right|_{\partial\Sigma} = 0.
\end{equation}
Now we impose an additional gauge fixing condition\footnote{One can generalise this gauge fixing condition~\eqref{memb116} for higher-dimensional hyper-membranes. Any $n$-dimensional divergenceless vector field $A^{a}$, subjected to the boundary condition $A^{a}|_{\partial\Sigma}~=~0$ (just like $\lambda \cdot \mathcal{P}^{a}$ in \eqref{memb114} and \eqref{memb115}) can be expressed as $A^{a}=\varepsilon^{abc_{1}\ldots c_{n-2}}\partial_{b}B_{c_{1}\ldots c_{n-2}}$, where $B_{c_{1}\ldots c_{n-2}}$ are the components of an $(n-2)$-form. Like the Kalb-Ramond gauge fields, these $B$'s have a hierarchy of `gauge symmetries' given by $B~\rightarrow~B'~=~B~+~\ud B_{(n-3)}$, $B_{(n-3)}~\rightarrow~B'_{(n-3)}~=~B_{(n-3)}~+~\ud B_{(n-4)}, \ldots$, so on and so forth, where $B_{(p)}$ is a $p$-form. One can therefore easily see that the demand $A^{a}~=~0$ entails $(n-1)$ additional constraints as there are $(n-1)$ independent components of $B_{(n-2)}$. With two gauge fixing conditions of type \eqref{memb109} and \eqref{memb110}, this gives rise to $(n+1)$ number of independent constraints, which exactly matches with the number of first-class constraints of the type \eqref{memb107} and \eqref{108} of the theory. For the special case of $n~=~2$, $A^{a}~=~\varepsilon^{ab}\partial_{b}B$, where $B$ is now a pseudo-scalar. Clearly the demand $A^{a}~=~0$ is equivalent to the gauge fixing condition \eqref{memb116}. For the case $n~=~3$, $A^{a}~=~\varepsilon^{abc}\partial_{b}B_{c}$ so that 3-vector is expressed as a curl of another 3-vector, in a standard manner, having only two transverse degrees of freedom; the longitudinal one having been eliminated through the above mentioned gauge transformation.}
\begin{equation}
\varepsilon^{ab} \partial_{a}\left(\lambda \cdot \mathcal{P}_{b}\right) 
\weakequals 0.
\label{memb116}
\end{equation}
Thus, we have
from Eqs.~\eqref{memb114} and \eqref{memb116} both the divergence and curl vanishing for
the vector field $(\lambda \cdot \mathcal{P}^{a})$ in the 2-dimensional
membrane, which is also subjected to the boundary conditions \eqref{memb115}.
We thus have
\begin{equation}
\lambda \cdot \mathcal{P}^{a} = 0 \qquad \forall \sigma^a.
\label{memb117}
\end{equation}
In view of Eq.~\eqref{112}, we have $\lambda \cdot \Pi \weakequals -T\sqrt{-h}h^{00}(\lambda \cdot \partial_{0}X)$, which, using Eq.~\eqref{memb111} gives
\begin{equation}
h^{00}\sqrt{-h} \weakequals -1.
\label{memb118}
\end{equation}
Using Eqs.~\eqref{memb111} and \eqref{112}, Eq.~\eqref{memb117} gives $h^{0a} \weakequals 0$
which implies
\begin{equation}
h_{0a} \weakequals 0, \qquad
h^{00} \weakequals \frac{1}{h_{00}}.
\label{memb119}
\end{equation}
From Eqs.~\eqref{memb118} and \eqref{memb119} it follows that
\begin{equation}
h_{00} + \bar{h} \weakequals 0.
\label{memb120}
\end{equation}
The term quasi-orthonormality in this case means that the time-like vector $\partial_{0}$ is orthogonal to the space-like vectors $\partial_{a}$, which follows from Eq.~\eqref{memb119}. However, the two space-like directions $\partial_{1}$ and $\partial_{2}$ need not be orthogonal to each other. Also note that by replacing $\tau \rightarrow \alpha \tau$, $\alpha$ a constant number, in Eq.~\eqref{memb109}, the normalisation condition \eqref{memb120} will change to $h_{00}+\alpha^{2}\bar{h} \weakequals 0$.

Using the quasi-orthonormal conditions \eqref{memb119} and \eqref{memb120}, the
Lagrangian density becomes $\lagdens_\mathrm{NG} \weakequals -T\bar{h} \weakequals Th_{00} \weakequals \frac{T}{2}\left(h_{00}-\bar{h}\right)$. The effective action thus becomes
\begin{equation}
S^{\mathrm{eff}} = \frac{T}{2} \int_{\Sigma} \ud^3 \sigma \left[h_{00}-h_{11}h_{22}+(h_{12})^{2}\right],
\label{12006081}
\end{equation}
which gives the equation of motion:
\begin{equation}
\partial_{0}\partial_{0}X_{\mu} + \partial_{1}\left(h_{12}\partial_{2}X_{\mu}-h_{22}\partial_{1}X_{\mu}\right) + \partial_{2}\left(h_{12}\partial_{1}X_{\mu}-h_{11}\partial_{2}X_{\mu}\right) = 0.
\label{12006082}
\end{equation}

Note that the quasi-orthonormal conditions \eqref{memb119} and \eqref{memb120} do not correspond to any gauge conditions themselves as they contain time derivatives. Actually they follow as a consequence of the conditions \eqref{memb109}, \eqref{memb110} and \eqref{memb116} which are to be regarded as gauge fixing conditions. These gauge conditions, when imposed, render the first-class constraints \eqref{memb107} and \eqref{108} of the theory into second-class as can be seen from their non-vanishing Poisson-bracket structure. Therefore, Nambu--Goto formalism requires the evaluation of Dirac brackets where these constraints are implemented strongly. As we shall see subsequently, in the Polyakov formulation the constraints \eqref{memb107} and \eqref{108} are not rendered into second-class and we can avoid the detailed calculation of Dirac brackets.

It is possible to draw a parallel between the quasi-orthonormal gauge discussed here and the usual orthonormal gauge in Nambu--Goto string, which is the analogue of the conformal gauge in the Polyakov string. In the latter case the equations of motion linearise reducing to the D'Alembert equations. This is possible because the gauge choice induces a net of coordinates that form a locally orthonormal system \cite{Rebbi:1974tc}. For the membrane, the invariances are insufficient to make such a choice and the best that we could do was to provide a quasi-orthonormal system. It is however amusing to note that if we forced an orthonormal choice, so that $h_{0a}\weakequals 0$ is supplemented with $h_{12}\weakequals 0$ and $h_{11}=h_{22}\weakequals 1$, then the equation of motion \eqref{12006082} indeed simplifies to the D'Alembert equation. This provides an alternative way of looking at the quasi-orthonormality.

If we do not impose quasi-orthonormality, it is highly nontrivial, if not totally impossible, to express  the boundary conditions~(\ref{memb104}) in terms of phase-space variables because the canonical momentum $\Pi_{\mu}~=~\mathcal{P}^{0}_{\mu}$
\eqref{105}, which can be re-expressed as
\[
\Pi_{\mu} = \frac{T\bar{h}}{\sqrt{-h}}\left(\eta_{\mu\nu}-\partial_{a}
X_{\mu}\bar{h}^{ab}\partial_{b}X_{\nu}\right)\partial_{0}X^{\nu}
\]
involves a projection operator given by the expression within the parentheses in the above equation. The velocity terms appear both in the right of the projection operator and in $\sqrt{-h}$ appearing in the denominator. This makes the inversion of the above equation to write the velocities in terms of momenta highly nontrivial. Nevertheless, all this simplifies drastically in the quasi-orthonormal gauge to enable us to simplify the above expression to
\begin{equation}
\Pi_{\mu} = T\partial_{0}X_{\mu}
\label{12006084}
\end{equation}
so that the boundary condition \eqref{memb104} is now expressible in terms of phase-space variables as
\[
\left.\left(h_{22}\partial_{1}X_{\mu}-h_{12}\partial_{2}X_{\mu}\right)\Pi^{2}\right|_{\sigma^{1}=0,\pi}=0.
\]

Finally we notice that the parameters ${U^{0}}_{0}$ and ${V^{a0}}_{0}$ given by Eq.~\eqref{1113} simplify in this gauge to
\begin{equation}
{U^{0}}_{0} = \frac{1}{2T}, \qquad {V^{a0}}_{0} = 0
\label{12006083}
\end{equation}
while ${U^{0}}_{a}$ and ${V^{b0}}_{a}$ given by Eq.~\eqref{1114} remain unchanged. Now the generators of $\tau$ and $\sigma^{a}$ translations \eqref{1112} become
\begin{equation}
H_{T} = \frac{1}{2T} \int \ud^{2} \sigma \psi, \qquad 
H_{a} = \int \ud^{2} \sigma \phi_{a}.
\label{121}
\end{equation}
It is straightforward to reproduce the action \eqref{12006081} by performing an inverse Legendre transformation. Computing the Poisson bracket of $X_{\mu}(\tau, \vec{\sigma})$ with the above $H_{T}$, the Hamilton's equation $\partial_{0}X_{\mu} = \{X_{\mu}, H_{T}\}$ gives Eq.~\eqref{12006084}, the definition of momenta in this gauge. Then,
\[
S^{\mathrm{eff}} = \int_\Sigma \ud^3 \sigma \,\Pi_\mu\partial_0 X^\mu - \int \ud \tau\, H_T
\]
just yields \eqref{12006081}. The other equation, $\partial_{0}\Pi_{\mu} = \{\Pi_{\mu}, H_{T}\}$, reproduces Eq.~\eqref{12006082}, which is the Euler--Lagrange equation following from the effective action \eqref{12006081}. 

Notice that the values of ${U^{0}}_{0}$ and ${V^{a0}}_{0}$ are gauge dependent. The particular values given by Eq.~\eqref{12006083} correspond to our quasi-orthonormal gauge. Had we chosen a different gauge, we would have obtained different values for these parameters. On the contrary, the parameters ${U^{0}}_{a}$ and ${V^{b0}}_{a}$ are gauge independent. This is consistent with the symmetries of the problem. There are three reparametrisation invariances, so that three parameters among these $U$'s and $V$'s must be gauge dependent, manifesting these symmetries. Since the reparametrization invariances govern the time evolution of the system,  the gauge dependent parameters are given by ${U^{0}}_{0}$ and ${V^{a0}}_{0}$, while the others are gauge independent.


\section{\label{sec:memb-poly}Free Polyakov membrane}

The Polyakov action for the bosonic membrane is \cite{Taylor:2001vb}
\begin{equation} \label{301}
S_\mathrm{P} = -\frac{T}{2} \int_{\Sigma} \ud^3 \sigma \sqrt{-g} \left( g^{ij} \partial_i X^{\mu} \partial_j X_{\mu} - 1 \right),
\end{equation}
where an auxiliary metric $g_{ij}$ on the membrane world-volume has been introduced and will be given the status of an independent field variable in the enlarged configuration space. The final term $(-1)$ inside the parentheses does not appear in the analogous string theory action. A consistent set of equations can be obtained only by taking the `cosmological' constant to be $-1$. Indeed, the equations of motion following from  the action \eqref{301} but with arbitrary cosmological constant $\Lambda$ are
\begin{gather}
\label{302}
\partial_i \left( \sqrt{-g} g^{ij} \partial_j X^{\mu} \right) = 0, \\
\label{303}
h_{ij} = \frac{1}{2} g_{ij} \left( g^{kl} h_{kl} + \Lambda \right)
\end{gather}
while the boundary conditions are 
\begin{equation} \label{304}
\left. \partial^a X^{\mu} \right|_{\partial \Sigma} = 0.
\end{equation}
Equation \eqref{303} can now be satisfied if and only if we identify $g_{ij}$ with $h_{ij}$:
\begin{equation} \label{305}
g_{ij} = h_{ij} \equiv \partial_i X^{\mu} \partial_j X_{\mu},
\end{equation}
for the case $\Lambda = -1$ so that the action \eqref{301} reduces to the Nambu--Goto action \eqref{memb101}. The canonical momenta corresponding to the fields $X^{\mu}$ and $g_{ij}$ are
\begin{gather}
\label{306}
\Pi_{\mu} = \frac{\partial \lagdens} {\partial \dot{X}^{\mu}} = -T \sqrt{-g} \partial^0 X_{\mu}, \\
\label{307}
\pi^{ij} = \frac{\partial \lagdens}{\partial \dot{g}_{ij}} = 0.
\end{gather}
Clearly, $\pi^{ij} \weakequals 0$ represent primary constraints of the theory. The canonical Hamiltonian density is 
\begin{equation} \label{308}
\begin{split}
\hamdens_C &= \Pi_{\mu} \partial_0 X^{\mu} - \lagdens \\
&= \frac{\sqrt{-g}}{2T\bar{g}} \Pi^2 - \frac{gg^{0a}}{\bar{g}} \Pi_{\mu} \partial_a X^{\mu} + \frac{T\sqrt{-g}}{2\bar{g}} \left(g_{22} h_{11} + g_{11} h_{22} - 2 g_{12} h_{12} - \bar{g} \right).
\end{split}
\end{equation}
Therefore, the total Hamiltonian is written as
\begin{equation} \label{309}
H_T = \int \ud^2 \sigma \left( \hamdens_C + \lambda_{ij} \pi^{ij} \right),
\end{equation}
where $\lambda_{ij}$ are arbitrary Lagrange multipliers. Conserving the constraint $\pi^{00} \weakequals 0$ with time, $\dot{\pi}^{00} = \left\{ \pi^{00}, H_T \right\} \weakequals 0$, we get
\begin{equation} \label{3010}
\Omega_1 \equiv \Pi^2 + T^2 \left( g_{22} h_{11} + g_{11} h_{22} - 2 g_{12} h_{12} - \bar{g} \right) \weakequals 0.
\end{equation}
Similarly, conserving other primary constraints with time, we get
\begin{gather}
\label{3011}
\begin{split}
\Omega_2 &\equiv \frac{\sqrt{-g}}{4T{\bar{g}}^2} \left( 2g_{22} - \bar{g} g^{11} \right) \left\{ \Pi^2 + T^2 \left( g_{22} h_{11} + g_{11} h_{22} - 2g_{12} h_{12} \right) \right\} \\
&\quad {} - \frac{gg_{22}}{\bar{g}^{2}} g^{0a} \Pi_{\mu} \partial_a X^{\mu} - \frac{g_{02}}{\bar{g}} \Pi_{\mu} \partial_2 X^{\mu} - \frac{T\sqrt{-g}}{4\bar{g}} \left( 2h_{22} - \bar{g} g^{11} \right) \\
&\weakequals 0,
\end{split}
\\
\label{3012}
\begin{split}
\Omega_3 &\equiv \frac{\sqrt{-g}}{4T{\bar{g}}^2} \left( 2g_{11} - \bar{g} g^{22} \right) \left\{ \Pi^2 + T^2 \left( g_{22} h_{11} + g_{11} h_{22} - 2g_{12} h_{12} \right) \right\} \\
&\quad {} - \frac{gg_{11}}{\bar{g}^{2}} g^{0a} \Pi_{\mu} \partial_a X^{\mu} - \frac{g_{01}}{\bar{g}} \Pi_{\mu} \partial_1 X^{\mu} - \frac{T\sqrt{-g}}{4\bar{g}} \left( 2h_{11} - \bar{g} g^{22} \right) \\
&\weakequals 0,
\end{split}
\\
\label{3013}
\begin{split}
\Omega_4 &\equiv -\frac{\sqrt{-g}}{2T{\bar{g}}^2} \left( 2g_{12} + \bar{g} g^{12} \right) \left\{ \Pi^2 + T^2 \left( g_{22} h_{11} + g_{11} h_{22} - 2g_{12} h_{12} \right) \right\} \\
&\quad {} + \frac{2g g_{12}}{\bar{g}^2} g^{0a} \Pi_{\mu} \partial_a X^{\mu} + \frac{g_{02}}{\bar{g}} \Pi_{\mu} \partial_1 X^{\mu} + \frac{g_{01}}{\bar{g}} \Pi_{\mu} \partial_2 X^{\mu} + \frac{T\sqrt{-g}}{2\bar{g}} \left( 2h_{12} + \bar{g} g^{12} \right) \\
&\weakequals 0,
\end{split}
\\
\label{3014}
\begin{split}
\Omega_5 &\equiv -\frac{\sqrt{-g}g^{01}}{2T} \left\{ \Pi^2+ T^2 \left( g_{22} h_{11} + g_{11} h_{22} - 2g_{12} h_{12} - \bar{g} \right)\right\} \\
&\quad {} - g_{22} \Pi_{\mu} \partial_1 X^{\mu} + g_{12} \Pi_{\mu} \partial_2 X^{\mu} \\
&\weakequals 0,
\end{split}
\\
\label{3015}
\begin{split}
\Omega_6 &\equiv -\frac{\sqrt{-g} g^{02}}{2T} \left\{ \Pi^2+ T^2 \left( g_{22} h_{11} + g_{11} h_{22} - 2g_{12} h_{12} - \bar{g} \right) \right\} \\
&\quad {} - g_{11} \Pi_{\mu} \partial_2 X^{\mu} + g_{12} \Pi_{\mu} \partial_1 X^{\mu} \\
& \weakequals 0.
\end{split}
\end{gather}
The above constraints appear to have a complicated form. Also, their connection with the constraints obtained in the Nambu--Goto formalism, is not particularly transparent. To bring the constraints into a more tractable form and to illuminate this connection, it is desirable to express them by the following combinations:
\begin{gather}
\label{30101}
\Omega_1 = \psi - T^2 \bar{\chi} \weakequals 0,
\\
\label{30112}
\Omega_2 = \frac{\sqrt{-g}}{4T{\bar{g}}^2} \left( 2g_{22} - \bar{g} g^{11} \right) \Omega_1 - \frac{gg_{22}}{\bar{g}^2} g^{0a} \phi_{a} - \frac{g_{02}}{\bar{g}} \phi_2 + \frac{T\sqrt{-g}}{2\bar{g}} \chi_{22} \weakequals 0,
\\
\label{30123}
\Omega_3 = \frac{\sqrt{-g}}{4T{\bar{g}}^2} \left( 2g_{11} - \bar{g} g^{22} \right) \Omega_1 - \frac{gg_{11}}{\bar{g}^{2}} g^{0a} \phi_a - \frac{g_{01}}{\bar{g}} \phi_1 + \frac{T\sqrt{-g}}{2\bar{g}} \chi_{11} \weakequals 0,
\\
\label{30134}
\Omega_4 = -\frac{\sqrt{-g}}{2T{\bar{g}}^2} \left( 2g_{12} + \bar{g} g^{12} \right) \Omega_1 + \frac{2gg_{12}}{\bar{g}^{2}} g^{0a} \phi_a + \frac{g_{02}}{\bar{g}} \phi_1 + \frac{g_{01}}{\bar{g}} \phi_2 - \frac{T\sqrt{-g}}{\bar{g}} \chi_{12} \weakequals 0,
\\
\label{30145}
\Omega_5 = -\frac{\sqrt{-g}g^{01}}{2T}\Omega_1 - g_{22} \phi_1 + g_{12} \phi_2 \weakequals 0,
\\
\label{30156}
\Omega_6 = -\frac{\sqrt{-g}g^{02}}{2T}\Omega_1 - g_{11} \phi_2 + g_{12} \phi_1 \weakequals 0,
\end{gather}
where
\begin{gather}
\label{310}
\psi \equiv \Pi^2 + T^2 \bar{h} \weakequals 0, \\
\label{311}
\phi_a \equiv \Pi_{\mu} \partial_a X^{\mu} \weakequals 0, \\
\label{312} 
\chi_{ab} \equiv g_{ab} - h_{ab} \weakequals 0
\end{gather}
and $\bar{\chi} = \chi_{11} \chi_{22} - (\chi_{12})^2$. As all the constraints $\Omega$'s appearing in Eqs.~\eqref{30101}--\eqref{30156} are combinations of $\psi$, $\phi_a$ and $\chi_{ab}$ in Eqs.~\eqref{310}--\eqref{312}, we can treat these $\psi$, $\phi_a$ and $\chi_{ab}$ as an alternative set of secondary constraints. These constraints along with the primary constraints \eqref{307}, $\pi^{ij} \weakequals 0$, constitute the complete set of constraints of the theory. This is because the canonical Hamiltonian density \eqref{308} can be expressed as a combination of constraints in the following manner:
\begin{equation}
\label{313}
\hamdens_C = \frac{\sqrt{-g}}{2T\bar{g}} \psi - \frac{gg^{0a}}{\bar{g}} \phi_a - \frac{T\sqrt{-g}}{2\bar{g}} \bar{\chi} \weakequals 0
\end{equation}
and the non-vanishing Poisson brackets between the constraints of the theory are
\begin{equation} \label{314}
\begin{aligned}
&\left\{\psi(\tau, \vec{\sigma}), \chi_{ab}(\tau, \vec{\sigma}')\right\}
\weakequals 2\left(\partial_{a}\Pi_{\mu}\partial_{b}X^{\mu}+\partial_{b} \Pi_{\mu}\partial_{a}X^{\mu}\right) \delta \left( \vec{\sigma}- \vec{\sigma}' \right), \\
&\begin{aligned}
\left\{\phi_{a}(\tau, \vec{\sigma}), \chi_{bc}(\tau, \vec{\sigma}')\right\}
&= h_{ab}(\tau, \vec{\sigma}')\partial'_{c}\delta\left(\vec{\sigma} -\vec{\sigma}'\right) + h_{ac}(\tau, \vec{\sigma}')\partial'_{b}\delta \left(\vec{\sigma} - \vec{\sigma}'\right) \\
&\quad {} + \left(\partial_{b}X^{\mu}\partial_{c}\partial_{a}X_{\mu} + \partial_{c}X^{\mu}\partial_{b}\partial_{a}X_{\mu}\right)\delta \left(\vec{\sigma}-\vec{\sigma}'\right),
\end{aligned} \\
&\left\{\pi^{ab}(\tau, \vec{\sigma}), \chi_{cd}(\tau, \vec{\sigma}')\right\}
= -\frac{1}{2} \left(\delta^{a}_{c}\delta^{b}_{d}+\delta^{a}_{d}\delta^{b}_{c}\right) \delta \left(\vec{\sigma} - \vec{\sigma}'\right), 
\end{aligned}
\end{equation}
while the weakly vanishing brackets are the same as given by \eqref{1109}. As far as the rest of the brackets are concerned, it is trivial to see that they vanish strongly. Thus, as it appears, none of the constraints except $\pi^{0i}$ in the set is first-class. But we have not yet extracted the maximal number of first-class constraints from the set \eqref{307}, \eqref{310}--\eqref{312} by constructing appropriate linear combinations of the constraints. However, it is highly nontrivial to find such a linear combination in the present case as one can see from the complicated structure of the Poisson brackets given above in \eqref{314}. Nevertheless, one can bypass such an elaborate procedure to extract the first-class constraints from the given set by noting that the complete set of constraints can be split into two sectors. In one sector we retain $\psi, \phi_a $ and $\pi^{0i}$, which are first-class among themselves, while the other sector contains the canonically conjugate pairs $\chi_{ab}$ and $\pi^{ab}$.  This allows an iterative computation of the Dirac brackets \cite{GitmanTyutin:1990}; namely, it is possible to eliminate this set completely by calculating the Dirac brackets within this sector. The brackets of the other constraints are now computed with respect to these Dirac brackets. Obviously $\psi$, $\phi_a, $ will have vanishing brackets with $\pi^{ab}$, $\chi_{cd}$. Moreover, the original first-class algebra among $\psi$ and $\phi_a$ will be retained. This follows from the fact that  the Dirac constraint matrix involving $\pi^{ab}$ and $\chi_{cd}$ has entries only in the off-diagonal pieces, while $\psi$ and $\phi_a$ have non-vanishing contributions coming just from the bracket with one of them; i.e., $\chi_{cd}$ (see \eqref{314}). The Dirac brackets of $\psi$ and $\phi_{a}$ are thus identical to their Poisson brackets, satisfying the same algebra as in the Nambu--Goto case.

We are therefore left with the first-class constraints $\psi \weakequals 0$, $\phi_{a} \weakequals 0$ and $\pi^{0i} \weakequals 0$. At this stage, we note that the constraints $\pi^{0i} \weakequals 0$ are analogous to $\pi^{0} \weakequals 0$ in free Maxwell theory, where $\pi^{0}$ is canonical conjugate to $A_{0}$. Consequently, the time evolution of $g_{0i}$ is arbitrary as follows from the Hamiltonian \eqref{309}. Therefore, we can set
\begin{equation}
g_{0a} = 0, \qquad g_{00} = -\bar{h},
\label{316}
\end{equation}
as new gauge fixing conditions.\footnote{We cannot set $g_{00} = 0$ as it will make the metric singular. We therefore set $g_{00} = -\bar{h}$ to make it match with the corresponding condition \eqref{memb120} in Nambu--Goto case.} With that ($g_{0a},\pi^{0a}$) and ($g_{00},\pi^{00}$) are discarded from the phase-space. This is again analogous to the arbitrary time evolution of $A_{0}$ in Maxwell theory, where we can set $A_{0}=0$ as a gauge fixing condition and discard the pair ($A_{0},\pi^{0}$) from the phase-space altogether.

These gauge fixing conditions \eqref{316} are the counterpart of the quasi-orthonormal conditions \eqref{memb119} and \eqref{memb120} in the Nambu--Goto case. However, unlike the Nambu--Goto case, these second-class constraints \eqref{316} do not render the residual first-class constraints of the theory, viz. $\psi \weakequals 0$ and $\phi_{a} \weakequals 0$ into second-class constraints. Therefore, they represent partial gauge fixing conditions. This stems from the fact that $g_{0i}$ were still regarded as independent field variables in the configuration space whereas $g_{ab}$ have already been strongly identified with $h_{ab}$ \eqref{312}. We therefore note that the calculation of the Dirac brackets is not necessary in Polyakov formulation. This motivates us to study the noncommutativity vis-\`a-vis the modified brackets $\{X^{\mu},X^{\nu}\}$ in the simpler Polyakov version. For that we shall first consider the free theory in the next section.

Let us now make some pertinent observations about the structure of the symmetric form of energy--momentum tensor, which is obtained by functionally differentiating the action with respect to the metric. The various components of this tensor are given by
\begin{gather}
\label{317}
\begin{split}
T_{00} &= \frac{g_{00}}{2T\bar{g}} \psi + \frac{2g\sqrt{-g}}{{\bar{g}^{2}}} g^{0a} \phi_{a} + \left( \frac{1}{g^{00}} - g_{00} \right) \frac{\Pi^{2}}{T\bar{g}} \\
&\quad {} -\frac{Tg^{2}}{\bar{g}^{2}} \left[ (g^{01})^{2} h_{11} + (g^{02})^{2} h_{22} + 2g^{01} g^{02} h_{12} \right] - \frac{Tg_{00}}{2\bar{g}} \bar{\chi},
\end{split} \\
\label{318}
\begin{split}
T_{01} &= -\frac{g_{01}}{2T\bar{g}} \psi - \frac{\sqrt{-g}}{\bar{g}} \phi_{1} \\
&\quad {} + \frac{T}{\bar{g}} \left[ \left(g_{02}h_{11}+g_{01}h_{12}\right) \chi_{12} - g_{02} h_{12} \chi_{11} - g_{01} h_{11} \chi_{22} -\frac{g_{01}}{2} \bar{\chi} \right],
\end{split} \\
\label{319}
\begin{split}
T_{02} &= -\frac{g_{02}}{2T\bar{g}} \psi - \frac{\sqrt{-g}}{\bar{g}} \phi_{2} \\
&\quad {} + \frac{T}{\bar{g}} \left[ \left(g_{01}h_{22}+g_{02}h_{12}\right) \chi_{12} - g_{01} h_{12} \chi_{22} - g_{02} h_{22} \chi_{11} - \frac{g_{02}}{2} \bar{\chi} \right],
\end{split} \\
\label{320}
T_{ab} = -\frac{g_{ab}}{2T\bar{g}} \psi + T \chi_{ab} + \frac{Tg_{ab}}{2\bar{g}} \bar{\chi} - \frac{Tg_{ab}}{\bar{g}} \left( g_{22} \chi_{11} + g_{11} \chi_{22} - 2g_{12} \chi_{12} \right).
\end{gather}
Unlike the case of string \cite{Banerjee:2002ky}, the component $T_{00}$ cannot be written in terms of constraints of the theory. However, the other components can be expressed in terms of these constraints, of which $\chi_{ab}$ are second-class and have already been put strongly to  zero by using Dirac brackets, so that the form of $T_{0a}$ and $T_{ab}$ simplifies to
\begin{eqnarray*}
T_{0a}\!\!&=&\!\!-\frac{g_{0a}}{2T\bar{g}}\psi-\frac{\sqrt{-g}}{\bar{g}}\phi_{a}, \\
T_{ab}\!\!&=&\!\!-\frac{g_{ab}}{2T\bar{g}}\psi.
\end{eqnarray*}
However, for $T_{00}$ we have to make use of the gauge conditions \eqref{316}, which hold strongly as was discussed earlier, to enable us to write $T_{00}=-\frac{1}{2T}\psi$. Let us now compare it with Nambu--Goto case. First we notice that ${\theta^{i}}_{j}$ appearing in Eq.~\eqref{1111} is not a tensor itself but it is a tensor density. The corresponding tensor is $\frac{1}{\sqrt{-g}}{\theta^{i}}_{j}$. In quasi-orthonormal gauge we have $\sqrt{-g}{T^{0}}_{0}={\theta^{0}}_{0}=\frac{1}{2T}\psi$, which reproduces the canonical Hamiltonian density \eqref{313} in this gauge. Also, in this gauge we have $\sqrt{-g}{T^{0}}_{a}=\phi_{a}$, which matches with ${\theta^{0}}_{a}$ in quasi-orthonormal gauge. This also provides a direct generalisation of the string case \cite{Banerjee:2002ky}. Although, unlike the string case, the Weyl symmetry is absent in the membrane case, we still have a vanishing trace, albeit weakly, of the energy--momentum tensor:
\[
{T^{i}}_{i} = - \frac{1}{2T\bar{h}}\psi \weakequals 0.
\]


\paragraph{Brackets for a free theory.}
Here we consider a cylindrical topology for the membrane which is taken to be periodic along $\sigma^{2}$-direction, i.e., $\sigma^{2}\in [0, 2\pi)$ and $\sigma^{1} \in [0, \pi]$. Following the example of string case \cite{Banerjee:2002ky}, we write down the first version of the brackets as:
\begin{equation}
\{X^{\mu}(\tau ,\vec{\sigma}),\Pi_{\nu}(\tau ,\vec{\sigma}')\}
= \delta^{\mu}_{\nu}\Delta_{+}(\sigma^{1}, \sigma'^{1})\delta_{\mathrm{p}}(\sigma^{2}\!-\!\sigma'^{2}),
\label{401} 
\end{equation}
and the other brackets vanishing.\footnote{The $\{X^{\mu},\Pi_{\nu}\}$ brackets are not affected as we implemented the second-class constraints and the gauge fixing conditions strongly in the preceding section. They are the only surviving phase-space variables as $g_{ij}$ have lost their independent status.} Here
\begin{equation}
\delta_{\mathrm{p}}(\sigma-\sigma') = \frac{1}{2\pi} \sum_{n\in {\integer}} e^{\ui n (\sigma-\sigma')}
\label{403}
\end{equation}
is the periodic delta function of period $2\pi$ which satisfies
\begin{equation}
\int_{-\pi}^{+\pi} \ud \sigma'\delta_{\mathrm{p}}(\sigma-\sigma')f(\sigma') = f(\sigma)
\label{404}
\end{equation}
for any periodic function $f(\sigma)=f(\sigma+2\pi )$ defined in the interval $[-\pi, +\pi ]$; and if, in addition, $f(\sigma)$ is taken to be an even function in the interval $[-\pi, +\pi ]$, then the above integral \eqref{404} reduces to
\begin{equation}
\int_0^{\pi} \ud \sigma'\Delta_{+}(\sigma,\sigma')f(\sigma') = f(\sigma),
\label{405}
\end{equation}
where
\begin{eqnarray}
\Delta_{+}(\sigma,\sigma') \!\!&=&\!\! \delta_{\mathrm{p}}(\sigma-\sigma')+\delta_{\mathrm{p}}(\sigma+\sigma')
\nonumber \\
&=&\!\! \frac{1}{\pi}+\frac{1}{\pi}\sum_{n \neq 0}\cos(n\sigma)\cos(n\sigma').
\label{406}
\end{eqnarray}
This structure of the brackets is, however, consistent only with Neumann boundary conditions along $\sigma^{1}$-direction. On the other hand, we have a mixed boundary condition \eqref{304} which can be expressed in terms of phase-space variables as
\begin{equation}
\left[g_{22}T\partial_{1}X^{\mu} + \sqrt{-g}g^{01}\Pi^{\mu}
- g_{12}T\partial_{2}X^{\mu}\right]_{\sigma^{1}=0,\pi} = 0.
\label{407}
\end{equation}
We notice that in Nambu--Goto formulation it was necessary to fix gauge in order to express the boundary condition in terms of phase-space variables. However, this is not the case with Polyakov formulation since $g_{ij}$ are taken to be independent fields. Using the strongly valid equations \eqref{312} and the gauge fixing conditions \eqref{316}, this simplifies further to
\begin{equation}
\left[\partial_{2}X^{\nu}\partial_{2}X_{\nu}\partial_{1}X^{\mu}
- \partial_{1}X^{\nu}\partial_{2}X_{\nu}\partial_{2}X^{\mu}
\right]_{\sigma^{1}=0,\pi} = 0.
\label{408}
\end{equation}
Although we are using the gauge \eqref{316}, the nontrivial gauge generating first-class constraints \eqref{310} and \eqref{311} will be retained in the gauge-independent analysis both here and in the interacting case. As we see, the above boundary condition is nontrivial in nature and involves both the $\partial_{1}$ and $\partial_{2}$ derivatives. But, since the coordinates and momenta are not related at the boundary, we do not require to postulate a non-vanishing $\{X^{\mu}, X^{\nu}\}$ bracket as in the case of free string in conformal gauge \cite{Banerjee:2002ky}. Therefore, the free membrane theory, like its string counterpart, does not exhibit noncommutativity in the boundary coordinates.


\paragraph{Low-energy limit.}
Finally, we would like to see how the results in the free membrane theory go over to those of free string theory in the limit of small radius for the cylindrical membrane.

The cylindrical membrane is usually taken to propagate in an 11-dimensional compactified target space ${\real}^{9-p} \times M^{p} \times S^{1} \times I$, where $M^{p}$ is a $p$-dimensional flat Minkowski spacetime and $I$ is an interval with finite length. There exist at the boundaries of $I$ two $p$-branes on which an open membrane can end. And the topology of the $p$-branes is given by $M^{p} \times S^{1}$. Also, the cylindrical membrane is assumed to wrap around this $S^{1}$. The radius of this circle is supposed to be very small so that in the low-energy limit the target space effectively goes over to 10-dimensional  ${\real}^{9-p} \times M^{p} \times I$ and the cylindrical membrane goes over to the open string.

At this stage, we choose further gauge fixing conditions:
\begin{equation}
X^{0} = \tau, \qquad X^{2} = \sigma^{2}R,
\label{410}
\end{equation}
where we have introduced $R$ to indicate the radius of the cylindrical membrane and $X^{2}$ represents the compact dimension $S^{1}$.\footnote{In \cite{Kawamoto:2000zt}, another gauge fixing condition $X^{1}=\sigma^{1}$, ($\pi$ being the length of the cylindrical membrane) has been used. But we notice that imposition of this gauge fixing condition would be inconsistent with the boundary condition \eqref{408} since, for $\mu = 1$, it yields a topology changing condition (cylinder $\rightarrow$ sphere), $R^{2}|_{\sigma^{1}=0,\pi}=0$, which is clearly unacceptable. Therefore, the choice \eqref{410} does not allow us to choose $X^{1}=\sigma^{1}$ as well, which is not needed either for our purpose.} Before choosing the gauge conditions \eqref{410}, the $\tau$ and $\sigma^{a}$ translations were generated by the constraints $\frac{1}{2T}\psi$ and $\phi_{a}$ respectively, just as in the Nambu--Goto case \eqref{121}. Now we have
\begin{gather*}
\left\{ \psi \left(\tau, \vec{\sigma}\right), X^0 \left(\tau, \vec{\sigma}'\right) - \tau \right\}
= {} - 2 \Pi^0 \left(\tau, \vec{\sigma}\right) \Delta_{+}(\sigma^{1},\sigma'{^1}) \delta_{\mathrm{p}}(\sigma^{2},\sigma'{^2}),
\\
\begin{split}
\left\{ \phi_{2} \left(\tau, \vec{\sigma}\right), X^{2} \left(\tau, \vec{\sigma}'\right) - \sigma'^{2} R \right\}
&= {} - \partial_{2} X^{2} \left(\tau, \vec{\sigma}\right) \Delta_{+}(\sigma^{1},\sigma'{^1}) \delta_{\mathrm{p}}(\sigma^{2},\sigma'{^2}) \\
&\weakequals {} - R \Delta_{+}(\sigma^{1},\sigma'{^1}) \delta_{\mathrm{p}}(\sigma^{2},\sigma'{^2}),
\end{split}
\end{gather*}
whereas
\begin{gather*}
\left\{ \phi_{1} \left(\tau, \vec{\sigma}\right), X^{0} \left(\tau, \vec{\sigma}'\right) - \tau \right\}
= {} - \partial_{1} X^{0} \left(\tau, \vec{\sigma}\right) \Delta_{+}(\sigma^{1},\sigma'{^1}) \delta_{\mathrm{p}}(\sigma^{2},\sigma'{^2}) \weakequals 0, \\
\left\{ \phi_{1} \left(\tau, \vec{\sigma}\right), X^{2} \left(\tau, \vec{\sigma}'\right) - \sigma'^{2} R \right\}
= {} - \partial_{1} X^{2} \left(\tau, \vec{\sigma}\right) \Delta_{+}(\sigma^{1},\sigma'{^1}) \delta_{\mathrm{p}}(\sigma^{2},\sigma'{^2}) \weakequals 0.
\end{gather*}
Thus, the (partial) gauge fixing conditions \eqref{410} take care of the world-volume diffeomorphism generated by $\psi$ and $\phi_{2}$ in the sense that these constraints are rendered into second-class while the diffeomorphism generated by $\phi_{1}$ is still there.

Coming back to the low-energy limit, we would like to show that the $\sigma^{2}$ dependence of all the fields except $X^{2}$ itself drops out effectively in the gauge \eqref{410}. To motivate it, let us consider the case of a free massless scalar field defined on a space with one compact dimension of ignorable size. Let the space be $M^{p} \times S^{1}$, where $M^{p}$ is a $p$-dimensional Minkowski spacetime taken to be flat for simplicity and $S^{1}$ is a circle of radius $R$ which is very small. We take $\theta \in [0,2\pi)$ to be the angle coordinate coorresponding to this circle so that the metric is given by $\ud s^{2} = \eta_{\mu\nu} \ud x^{\mu} \ud x^{\nu} = \eta_{\mu'\nu'} \ud x^{\mu'} \ud x^{\nu'} + R^{2} \ud \theta^{2}$ with $\mu, \nu$ ranging from 0 to $p$ and $\mu', \nu'$ from 0 to $(p-1)$. The action is
\[
S = -\frac{1}{2} \int \ud^p x \ud \theta \partial_{\mu} \phi \partial^{\mu} \phi.
\]
Separating the index corresponding to the compact dimension, we rewrite it as
\[
S = -\frac{1}{2} \int \ud^p x \ud \theta \left( \partial_{\mu'} \phi \partial^{\mu'} \phi + \frac{1}{R^{2}} \partial_{\theta} \phi \partial_{\theta} \phi \right).
\]
Substituting the Fourier expansion
\[
\phi(x,\theta)
= \frac{1}{\sqrt{2 \pi}} \sum_{n \in {\integer}} \phi_{(n)}(x) e^{\ui n \theta}, \qquad
\phi_{(-n)} = \phi_{(n)}^{*}
\]
in the action and integrating out the compact dimension, we get
\[
S\rightarrow S'= -\frac{1}{2} \int \ud^p x\sum_{n\in {\integer}}\left(\partial_{\mu'}\phi_{(n)}\partial^{\mu'}\phi^{*}_{(n)}+\frac{n^{2}}{R^{2}}\phi_{(n)}\phi^{*}_{(n)}\right).
\]
Thus the Fourier coefficients represent a whole tower of effective massive complex scalar fields of mass $\sim n/R$ in a lower-dimensional non-compact spacetime. These masses are usually of the Planck order if $R$ is of the order of Planck length and are therefore ignored in the low-energy regime. Equivalently, one ignores the $\theta$ dependece of the field $\phi$. This can also be understood from physical considerations. In the low-energy limit, the associated wavelengths are very large as compared to $R$ so that variation of the field along the circle is ignorable.

Now the membrane goes over to string in the low-energy regime when the circle $S^{1}$ effectively disappears in the limit $R \rightarrow 0$. So the field theory living in the membrane world-volume is expected to correspond to the field theory living on string world-sheet. To verify this, let us substitute the Fourier expansion of the world-volume fields $X^{\mu}(\tau, \sigma^{1}, \sigma^{2})$ around $\sigma^{2}$:
\begin{equation} \label{414}
X^{\mu}(\tau, \sigma^{1}, \sigma^{2}) = \frac{1}{\sqrt{2 \pi}} \sum_{n \in {\integer}} X^{\mu}_{(n)}(\tau, \sigma^{1}) e^{\ui n \sigma^{2}}, \qquad
X^{\mu}_{(-n)} = {X^{\mu}_{(n)}}^{*}
\end{equation}
in the Poisson bracket \eqref{401} to find that the Fourier coefficients $X^{\mu}_{(n)}(\tau, \sigma^{1})$ satisfy
\begin{equation}
\{X^{\mu}_{(n)}(\tau, \sigma^{1}),\Pi_{\nu}^{(m)}(\tau, \sigma'^{1})\}
= \delta^{\mu}_{\nu}\delta^{-m}_{n}\Delta_{+}(\sigma^{1},\sigma'^{1}).
\label{413}
\end{equation}
As in the case of free scalar field discussed above, the Fourier coefficients $X^{\mu}_{(0)}(\tau, \sigma^{1})$ will represent the effective (real) fields in the string world-sheet satisfying
\begin{equation} \label{415}
\{ X^{\mu}_{(0)}(\tau, \sigma^{1}), \Pi_{\nu}^{(0)}(\tau, \sigma'^{1}) \}
= \delta^{\mu}_{\nu} \Delta_{+}(\sigma^{1}, \sigma'^{1}),
\end{equation}
which reproduces the Poisson bracket for string. The sub/superscript (0) will be dropped now onwards for convenience. Using $\partial_{2} X^{\mu} = R \delta^{\mu}_2$, the boundary condition \eqref{408} gives
\begin{equation} \label{411}
\left. \partial_{1} X^{\mu} \right|_{\sigma^{1}=0,\pi} = 0
\end{equation}
so that we recover the boundary condition for free string in conformal gauge.\footnote{Actually, we do not get \eqref{411} directly, rather it is accompanied by a pre-factor $R^{2}$. However, this equation is not satisfied trivially if $R\rightarrow 0$, as this limit should not be taken literally in a mathematical sense. This just means that $R$ should be taken to have a very small nonzero value and presumably should be of the order of Planck length, as we have mentioned earlier.}

Now we would like to show how the gauge fixed world-volume membrane metric \eqref{316} reduces to the world-sheet string metric in conformal gauge. For that we first note that the components of the metric tensor in a matrix form can be written as
\[
\{g_{ij}\}
=\begin{pmatrix}
g_{00} & g_{01} & g_{02} \\
g_{01} & g_{11} & g_{12} \\
g_{02} & g_{12} & g_{22} \end{pmatrix}
=\begin{pmatrix}
g_{00} & 0 & 0 \\
0 & h_{11} & 0 \\
0 & 0 & R^2 \end{pmatrix},
\]
where we have made use of the first gauge fixing condition in \eqref{316} and by now the strongly valid equations \eqref{312}. Clearly, this matrix becomes singular in the limit $R \rightarrow 0$ taken in a proper mathematical sense. It must therefore correspond to a two-dimensional surface embedded in three-dimensional world-volume. The metric corresponding to it can be easily obtained by chopping off the last row and last column in the above three-dimensional metric to get $\left(\begin{smallmatrix}g_{00} & 0 \\ 0 & h_{11}\end{smallmatrix}\right)$. Now, we make use of the second gauge fixing condition in \eqref{316} to replace $g_{00}$ by $(-\bar{h})$. However, this $\bar{h}$ can be simplified further using the gauge \eqref{410} to get $R^{2}h_{11}$ so that the above $2 \times 2$ matrix becomes $h_{11}\left(\begin{smallmatrix}-R^2 & 0 \\ 0 & 1\end{smallmatrix}\right)$ and the diagonal elements get identified up to a scale factor. It can now be put in the standard form, ${\rm diag}\,(-1,1)$, up to an overall Weyl factor, by replacing the second condition in \eqref{316} by $g_{00}=-\alpha^{2} \bar{h}$ and choosing $\alpha$ suitably. We also notice that using $h_{0a}=0$, the Nambu--Goto action for the membrane becomes
\[
S_\mathrm{NG} = -T \int \ud^3 \sigma \sqrt{-h_{00}\bar{h}},
\]
which using the gauge conditions \eqref{410} and integrating out $\sigma^{2}$, reduces to the Nambu--Goto action for string in orthonormal gauge:
\[
S_\mathrm{NG} \rightarrow S'_\mathrm{NG} = -2\pi RT \int \ud^2 \sigma \sqrt{-h_{00}h_{11}}.
\]
This also shows that the string tension is $\sim TR$ if the original membrane tension is given by $T$.  Actually one takes the limit $R\rightarrow 0$ together with the membrane tension $T\rightarrow \infty $ in such a way that their product $(TR)$ is finite. Such a limit was earlier discussed from other considerations in \cite{Lindstrom:1989xa}.


\section{\label{sec:memb-interact}Interacting Polyakov membrane}

The Polyakov action for a  membrane moving in the presence of a constant antisymmetric background field $A_{\mu\nu\rho}$ is
\begin{equation} \label{501}
S_\mathrm{P} = -\frac{T}{2} \int_{\Sigma} \ud^3 \sigma \left[ \sqrt{-g} \left( g^{ij} \partial_i X^{\mu} \partial_j X_{\mu} -1 \right) + \frac{e}{3} \varepsilon^{ijk} \partial_i X^{\mu} \partial_j X^{\nu} \partial_k X^{\rho} A_{\mu\nu\rho} \right],
\end{equation}
where we have introduced a coupling constant $e$.\footnote{As it stands, the interaction term involving the three-form field $A_{\mu\nu\rho}$ in \eqref{501} is not gauge invariant under the transformation $A \rightarrow A + \ud \Lambda$, where $\Lambda$ is a two-form field. One can, however, make it gauge invariant by adding a surface term $2e\int_{\partial\Sigma}B$, where $B$ is a two-form undergoing the compensating gauge transformation $B \rightarrow B - \Lambda$. But, using Stoke's theorem, this gets combined to a single integral over the world-volume as $\int_{\Sigma}(A+\ud B)$ so that $(A+\ud B)$ is gauge invariant as a whole. In the action \eqref{501}, $A$ is taken to correspond to this gauge invariant quantity by absorbing $\ud B$ in $A$.} The equations of motion are
\begin{gather}
\label{502}
\partial_i \left( \sqrt{-g} g^{ij} \partial_j X_{\mu} + \frac{e}{2} \varepsilon^{ijk} \partial_j X^{\nu} \partial_k X^{\rho} A_{\mu\nu\rho} \right) = 0,
\\
\label{503}
g_{ij} = h_{ij} \equiv \partial_i X^{\mu} \partial_j X_{\mu}. 
\end{gather}
We note that the second equation does not change from the free case $(e=0)$ despite the presence of interaction term as this term is topological in nature and does not involve the metric $g_{ij}$. The canonical momenta are
\begin{gather}
\label{504}
\Pi_{\mu} = \frac{\partial \lagdens}{\partial \dot{X}^{\mu}}
= -T \left( \sqrt{-g} \partial^0 X_{\mu} + \frac{e}{2} \varepsilon^{ab} \partial_a X^{\nu} \partial_b X^{\rho} A_{\mu\nu\rho} \right),
\\
\label{505}
\pi^{ij} = \frac{\partial \lagdens}{\partial \dot{g}_{ij}} = 0.
\end{gather}
For convenience, we define
\begin{equation} \label{506}
\widetilde{\Pi}_{\mu} \equiv \Pi_{\mu} + \frac{eT}{2} \varepsilon^{ab} \partial_a X^{\nu} \partial_b X^{\rho} A_{\mu\nu\rho}
= -T \sqrt{-g} \partial^0 X_{\mu}.
\end{equation}
Proceeding just as in the free case, the structure of the Hamiltonian density $\hamdens_C$ and the set of constraints is obtained just by replacing ${\Pi}_{\mu} \rightarrow \widetilde{\Pi}_{\mu}$, so that we are finally left with the following first-class constraints:
\begin{gather}
\label{507}
\psi \equiv \widetilde{\Pi}^2 + T^2 \bar{h} \weakequals 0,
\\
\label{508}
\phi_a \equiv \widetilde{\Pi}_{\mu} \partial_a X^{\mu} \weakequals 0
\end{gather}
and, as argued in the free case, we adopt the same gauge fixing conditions \eqref{316}.

For a cylindrical membrane periodic along $\sigma^{2}$-direction with $\sigma^{1} \in [0, \pi]$, $\sigma^{2} \in [0, 2\pi)$, the boundary condition is given by
\begin{equation} \label{509}
\left[ \sqrt{-g} \partial^1 X_{\mu} + e \partial_2 X^{\nu} \partial_0 X^{\rho} A_{\mu\nu\rho} \right]_{\sigma^{1}=0,\pi} = 0,
\end{equation}
which when expressed in terms of phase-space variables looks as
\begin{equation} \label{510}
\begin{split}
&\big[ g_{22} T \partial_1 X_{\mu} - g_{12} T \partial_2 X_{\mu} + \sqrt{-g} g^{01} \Pi_{\mu} \\
&\qquad {} + e \left( \Pi^{\rho} + e T \partial_1 X^{\lambda} \partial_2 X^{\kappa} {A^{\rho}}_{\lambda\kappa} \right) \partial_2 X^{\nu} A_{\mu\nu\rho} \big]_{\sigma^{1}=0,\pi} = 0.
\end{split}
\end{equation}
As in the free case, here also we use the strongly valid equations \eqref{312} and the gauge fixing conditions \eqref{316} so that the above boundary condition simplifies to
\begin{equation} \label{511}
\begin{split}
&\big[ T \partial_{2} X^{\nu} \partial_{2} X_{\nu} \partial_{1} X_{\mu} - T \partial_{1} X^{\nu} \partial_{2} X_{\nu} \partial_{2} X_{\mu} \\
&\qquad {} + e \left( \Pi^{\rho} + e T \partial_{1} X^{\lambda} \partial_{2} X^{\kappa} {A^{\rho}}_{\lambda\kappa} \right)
\partial_{2} X^{\nu} A_{\mu\nu\rho} \big]_{\sigma^{1}=0,\pi} = 0.
\end{split}
\end{equation}
Here we notice that the above boundary condition involves both phase-space coordinates $X^{\mu}$ and $\Pi_{\nu}$. Using the brackets of the free theory to compute the Poisson bracket of the left-hand side of above equation with $X_{\sigma}(\tau, \vec{\sigma}')$, we find that it does not vanish. The boundary condition is therefore not compatible with the brackets of the free theory. Thus, we have to postulate a non-vanishing $\{X^{\mu}, X^{\nu}\}$ bracket.{\footnote {In the case of free Polyakov string also, the incompatibility of the boundary condition with the basic Poisson brackets forces us to postulate  a non-vanishing $\{X^{\mu},X^{\nu}\}$. However, in contrast to the interacting string, this bracket vanishes in a particular gauge---the conformal gauge.}} For that we make an ansatz:
\begin{equation} \label{512}
\left\{ X_{\mu}(\tau,\vec{\sigma}), X_{\nu}(\tau,\vec{\sigma}') \right\}
= \mathcal{C}_{\mu \nu}(\vec{\sigma},\vec{\sigma}') = C_{\mu \nu}(\sigma^{1},\sigma'^{1}) \delta_{P}(\sigma^{2}-\sigma'^{2})
\end{equation}
with
\begin{equation} \label{513}
C_{\mu \nu}(\sigma^{1},\sigma'^{1}) = -C_{\nu \mu}(\sigma'^{1},\sigma^{1}).
\end{equation}
and the $\{X^{\mu},\Pi_{\nu}\}$ bracket is taken to be the same as in the free case---Eq.~\eqref{401}. At this stage, we note that the boundary condition \eqref{511}, if bracketted with $X_{\sigma}(\tau, \vec{\sigma}')$, yields at the boundary
\begin{equation} \label{514}
\begin{split}
&\left[ T \partial_{2} X^{\nu} \partial_{2} X_{\nu} \delta_{\mu}^{\lambda} - T \partial_{2} X^{\lambda} \partial_{2} X_{\mu} + e^{2} T \partial_{2} X^{\kappa} \partial_{2} X^{\nu} A_{\mu\nu\rho} {A^{\rho\lambda}}_{\kappa} \right] \partial_{1} {\mathcal{C}}_{\lambda\sigma} (\vec{\sigma},\vec{\sigma}') \\
&\quad {} + \Big[ 2T \partial_{1} X_{\mu} \partial_{2} X^{\kappa} - T \partial_{1} X^{\nu} \partial_{2} X_{\nu} \delta_{\mu}^{\kappa} - T \partial_{2} X_{\mu} \partial_{1} X^{\kappa} + e \Pi_{\rho} {A_{\mu}}^{\kappa\rho} \\
&\quad {} + e^{2} T \partial_{1} X^{\lambda} \partial_{2} X^{\nu} \left( A_{\mu\nu\rho} {{A^{\rho}}_{\lambda}}^{\kappa} + {{A_{\mu}}^{\kappa}}_{\rho} {A^{\rho}}_{\lambda\nu} \right) \Big] \partial_{2} {\mathcal{C}}_{\kappa\sigma} (\vec{\sigma},\vec{\sigma}') \\
&= e \partial_{2} X^{\nu} A_{\mu\nu\sigma} \Delta_{+}(\sigma^{1}, \sigma'^{1}) \delta_{P}(\sigma^{2}\!-\!\sigma'^{2}),
\end{split}
\end{equation}
which involves both $\partial_{1}{\mathcal{C}}$ and $\partial_{2}{\mathcal{C}}$ and leads to a contradiction if we put ${\mathcal{C}}_{\mu\nu}(\vec{\sigma},\vec{\sigma}')=0$. This is another way of seeing that there must be a noncommutativity in the membrane coordinates. However, there is no contradiction with ${\mathcal{C}}_{\mu\nu}(\vec{\sigma},\vec{\sigma}')~=~0$ provided $A_{\mu\nu\rho}~=~0$, thereby implying that there is no noncommutativity in the free theory.

Because of the nonlinearity in the above equation, it is problematic to find an exact solution. It should however be stressed that the above relation has been derived in a general (gauge-independent) manner. At this point there does not seem to be any compelling reason to choose a particular gauge to simplify this equation further to enable an exact solution. Nonlinearity would, in all probability, prevent this.  This is in contrast to the string case where the analysis naturally leads to a class of light-cone gauges where the corresponding equation was solvable \cite{Banerjee:2002ky}.
However, by taking recourse to the low-energy approximation, we show that the results for the string case are recovered. To this end, we substitute the expansion \eqref{414} in \eqref{512} to get
\begin{equation} \label{5131}
\{ X^{\mu}_{(n)}(\tau, \sigma^{1}), X^{\nu}_{(m)}(\tau, \sigma'^{1}) \}
= \delta_{n,-m} C^{\mu\nu}(\sigma^{1},\sigma'^{1}).
\end{equation}
But again, as in the free case, we retain only the real fields $X^{\mu}_{(0)}(\tau, \sigma^{1})\equiv X^{\mu}(\tau, \sigma^{1})$ when we consider the low-energy regime. Using the gauge fixing conditions \eqref{410}, the boundary condition \eqref{511} reduces to
\begin{equation} \label{515}
\left[ (TR) \partial_{1} X_{\mu} - e \Pi^{\rho} A_{\mu\rho2} - e^{2} (TR) \partial_{1} X_{\lambda} {A^{\rho\lambda}}_{2} A_{\mu\rho2} \right]_{\sigma^{1}=0,\pi}=0.
\end{equation}
Here, $X^{\mu}$ and $\Pi_{\nu}$ can be taken to correspond to $X^{\mu}_{(0)}(\tau, \sigma^{1})$ and $\Pi_{\nu}^{(0)}(\tau, \sigma^{1})$ respectively. Thus we recover the boundary condition of the string theory in conformal gauge with the correspondence $TR\leftrightarrow T_{\mathrm{s}}$ and $A_{\mu\nu2}\leftrightarrow B_{\mu\nu}$, where $T_{\mathrm{s}}$ is the effective (string) tension and $B_{\mu\nu}$ is the 2-form background field appearing in the string theory \cite{Banerjee:2002ky}. Now taking the Poisson bracket of the boundary condition \eqref{515} with $X_{\sigma}(\tau, \sigma^{1})$, the low-energy effective real fields, one gets for $\mu \neq 2$ the following differential condition satisfied by $C_{\mu\sigma}$ at the boundary
\begin{equation} \label{516}
\left. T_{\mathrm{s}} \left( \delta_{\mu}^{\lambda} - e^{2} A_{\mu\rho2} {A^{\rho\lambda}}_{2} \right) \partial_{1} C_{\lambda\sigma}(\sigma^{1},\sigma'^{1}) \right|_{\sigma^{1}=0,\pi}
= \left. e A_{\sigma\mu2} \Delta_{+}(\sigma^{1},\sigma'^{1}) \right|_{\sigma^{1}=0,\pi},
\end{equation}
which just reproduces the corresponding equation in string theory---see Eq.~\eqref{449}. We therefore obtain the noncommutativity:
\begin{equation} \label{517}
\begin{split}
C_{\mu\nu}(\sigma^{1},\sigma'^{1})
&= \tfrac{1}{2} (NM^{-1})_{(\nu \mu )} [ \Theta(\sigma^{1},\sigma'^{1}) - \Theta(\sigma'^{1},\sigma^{1}) ] \\
&\quad\, {}+ \tfrac{1}{2} (NM^{-1})_{[\nu \mu]} [ \Theta(\sigma^{1},\sigma'^{1}) + \Theta(\sigma'^{1},\sigma^{1})-1 ],
\end{split}
\end{equation}
where $N_{\nu\sigma} = e A_{\nu\sigma2}$ and ${M^{\lambda}}_{\mu} = T_\mathrm{s} ( \delta^{\lambda}_{\mu} - e^2 A_{\mu\rho2} {A^{\rho\lambda}}_2 )$, while 
\begin{equation} \label{518}
\Theta(\sigma^{1},\sigma'^{1}) = \frac{\sigma^1}{\pi} + \frac{1}{\pi} \sum_{n\neq 0} \frac{1}{n} \sin(n\sigma^{1}) \cos(n\sigma'^{1})
\end{equation}
being the generalised step function which satisfies
\begin{equation} \label{519}
\partial_1 \Theta(\sigma^{1},\sigma'^{1}) = \Delta_{+}(\sigma^{1},\sigma'^{1}).
\end{equation}
It has the property
\begin{gather*}
\Theta(\sigma^{1},\sigma'^{1}) = 1 \qquad \text{for } \sigma^{1} > \sigma'^{1}, \\
\Theta(\sigma^{1},\sigma'^{1}) = 0 \qquad \text{for } \sigma^{1} < \sigma'^{1}.
\end{gather*}


%% file: chap_current.tex

\chapter{\label{chap:current}Maps for currents and anomalies in noncommutative gauge theories}


The occurrence of noncommutativity was discussed in the previous chapter. The study of an open string in the presence of a background two-form field led to a noncommutative structure which manifests in the noncommutativity at the endpoints of the string which are attached to D-branes. In the same way, for membrane interacting with a three-form potential a nontrivial algebraic relation revealed the occurrence of noncommutativity independent of any gauge or any approximation. Now we already take such a noncommutative structure and proceed to see its implications.

There are two approaches to noncommutative field theory.\footnote{A commuting (ordinary) field theory is a field theory defined on ordinary commuting space and a noncommutative field theory is a field theory in which the coordinates do not commute.} One is in terms of the star-products which we discussed in the beginning. However, it is difficult to have local observables in this formulation. Local quantities in noncommutative field theory are gauge variant and no gauge invariant meaning can be assigned to their profiles. Nonlocal, integrated, expressions can be gauge invariant (in the noncommutative electrodynamics, for example, the action is gauge invariant) but in ordinary theory we deal with local quantities and we would like to compare these local quantites to corresponding quantities in the noncommutative theory.

A way out of this difficulty is provided by Seiberg and Witten's observation that the noncommuting gauge theory may be equivalently described by a commuting (usual) gauge theory that is formulated in terms of ordinary (not star) products of commuting variables, together with an explicit dependence on $\theta^{\alpha\beta}$, which acts as a constant `background'.

The Seiberg--Witten map \cite{Seiberg:1999vs} replaces the noncommuting vector potential by a function of a commuting potential and of $\theta$;  i.e., the former is viewed as a function of the latter. The relationship between the two follows from the requirement of stability against gauge transformations: a noncommuting gauge transformation of the noncommuting gauge potential should be equivalent to a commuting gauge transformation on the commuting vector potential on which the noncommuting potential depends. We shall discuss this map in section \ref{sec:current-SW}. Maps for the matter sector \cite{Bichl:2001gu, Rivelles:2002ez, Yang:2004vd, Banerjee:2004rs} as well as for currents and energy--momentum tensors \cite{Banerjee:2003vc} also exist in the literature.

An intriguing issue is the validity of such classical maps at the quantum level. Studies in this direction \cite{Jurco:2000fs, Kaminsky:2003qq, Kaminsky:2003mn} have principally focussed on extending the purported classical equivalence of Chern--Simons theories (in $2+1$ dimensions) in different descriptions \cite{Grandi:2000av, Banerjee:2003ce} to the quantum formulation.

In this chapter, we provide an alternative approach to study these quantum aspects by relating the current-divergence anomalies in the noncommutative and commutative pictures through a Seiberg--Witten-type map. Taking a cue from \refcite{Banerjee:2003vc}, we first derive a map connecting the star-gauge-covariant current in the noncommutative gauge theory with the gauge-invariant current in the $\theta$-expanded gauge theory. From this relation, a mapping between the (star-) covariant divergence of the covariant current and the ordinary divergence of the invariant current in the two descriptions, respectively, is deduced. We find that ordinary current-conservation in the $\theta$-expanded theory implies covariant conservation in the original noncommutative theory, and vice versa. The result is true irrespective of the choice of the current to be vector or axial vector. This is also to be expected on classical considerations.

The issue is quite nontrivial for a quantum treatment due to the occurrence of current-divergence anomalies for axial (chiral) currents. Since the star-gauge-covariant anomaly is known \cite{Ardalan:2000cy, Gracia-Bondia:2000pz} and the gauge-invariant anomaly in the $\theta$-expanded theory, which is in fact identical to the ordinary Adler--Bell--Jackiw anomaly (ABJ anomaly) \cite{Brandt:2003fx}, is also known, it is possible to test the map by inserting these expressions. We find that the classical map does not hold in general. However, if we confine to a slowly-varying-field approximation\footnote{This approximation is also used in \refcite{Seiberg:1999vs} to show the equivalence of Dirac--Born--Infeld actions (DBI actions) in the two descriptions.}, then there is a remarkable set of simplifications and the classical map holds. We also give a modified map, that includes the derivative corrections, which is valid for arbitrary field configurations.

After briefly summarising the standard Seiberg--Witten map in section \ref{sec:current-SW}, the map for currents and their divergences is derived in section \ref{sec:current-map}. Here the treatment is for the nonabelian gauge group $\mathrm{U}(N)$. In section \ref{sec:current-abelian}, we discuss the map for anomalous currents and their divergences. The abelian $\mathrm{U}(1)$ theory is considered and results are given up to $\uO(\theta^{2})$. As already mentioned, the map for the axial anomalies (in two and four dimensions) holds in the slowly-varying-field limit. A possible scheme is discussed whereby further higher-order results are confirmed. Especially, $\uO(\theta^{3})$ computations are done in some detail. In section \ref{sec:current-conclu} we briefly discuss the implications of this analysis on the definition of effective actions.


\section{\label{sec:current-SW}The Seiberg--Witten map}

We shall now briefly review the salient features of the Seiberg--Witten map. The ordinary Yang--Mills action is given by
\begin{equation}\label{S-YM}
S_{\mathrm{YM}} = -\frac{1}{4}\int\!\mathrm{d}^{4}x\,\mathrm{Tr}\left(F_{\mu\nu}F^{\mu\nu}\right),
\end{equation}
where the nonabelian field strength is defined as
\begin{equation}\label{F.dcurr}
F_{\mu\nu}=\partial_{\mu}A_{\nu}-\partial_{\nu}A_{\mu}-\mathrm{i}[A_{\mu},A_{\nu}]
\end{equation}
in terms of the Hermitian U($N$) gauge fields $A_{\mu}(x)$. The noncommutativity of spacetime is characterised by the algebra
\begin{equation}\label{theta.d}
\left[x^{\alpha},x^{\beta}\right]_{\star} \equiv x^{\alpha}\star x^{\beta}-x^{\beta}\star x^{\alpha} = \mathrm{i}\theta^{\alpha\beta},
\end{equation}
with $\theta^{\alpha\beta}$ real, constant and antisymmetric, and the star product as defined in Eq.~\eqref{star.dcurr}.\footnote{It is perhaps worthwhile to mention here that the star product also appears in other instances, for example, in the context of charged fluids in an intense magnetic field \cite{Jackiw:2001dj}.} In noncommutative spacetime, the usual multiplication of functions is replaced by the star product. The Yang--Mills theory is generalised to
\begin{equation}\label{S-hat-YM}
\xnc{S}_{\mathrm{YM}} = -\frac{1}{4}\int\!\mathrm{d}^{4}x\,\mathrm{Tr}\left(\xnc{F}_{\mu\nu}\star\xnc{F}^{\mu\nu}\right)
\end{equation}
with the noncommutative field strength
\begin{equation}\label{F-hat.dcurr}
\xnc{F}_{\mu\nu}=\partial_{\mu}\xnc{A}_{\nu}-\partial_{\nu}\xnc{A}_{\mu}-\mathrm{i}\left[\xnc{A}_{\mu},\xnc{A}_{\nu}\right]_{\star}.
\end{equation}
This theory reduces to the conventional U($N$) Yang--Mills theory for $\theta \rightarrow 0$.

To first order in $\theta$, it is possible to relate the variables in the noncommutative spacetime with those in the usual one by the classical maps \cite{Seiberg:1999vs}
\begin{gather}
\label{A-hat.m}
\xnc{A}_{\mu} = A_{\mu}-\frac{1}{4}\theta^{\alpha\beta}\{A_{\alpha},\partial_{\beta}A_{\mu}+F_{\beta\mu}\}+\uO(\theta^{2}),\\
\label{F-hat.m}
\xnc{F}_{\mu\nu} = F_{\mu\nu}+\frac{1}{4}\theta^{\alpha\beta}\left(2\{F_{\mu\alpha},F_{\nu\beta}\}-\{A_{\alpha},\mathrm{D}_{\beta}F_{\mu\nu}+\partial_{\beta}F_{\mu\nu}\}\right)+\uO(\theta^{2}),
\end{gather}
where the bracketed expressions denote the anticommutator and $\mathrm{D}_{\beta}$ denotes the covariant derivative as defined below in Eq.~\eqref{A-hat.gtcurr}. A further map among gauge parameters,
\begin{equation}\label{lmda-hat.m}
\xnc{\lambda}=\lambda+\frac{1}{4}\theta^{\alpha\beta}\left\{\partial_{\alpha}\lambda,A_{\beta}\right\}+\uO(\theta^{2}),
\end{equation}
ensures the stability of gauge transformations
\begin{gather}
\label{A-hat.gtcurr}
\delta_{\lambda}A_{\mu} = \partial_{\mu}\lambda+\mathrm{i}\left[\lambda,A_{\mu}\right] \equiv \mathrm{D}_{\mu}\lambda,\\
\label{A.gtcurr}
\xnc{\delta}_{\xnc{\lambda}}\xnc{A}_{\mu} = \partial_{\mu}\xnc{\lambda}+\mathrm{i}\left[\xnc{\lambda},\xnc{A}_{\mu}\right]_{\star} \equiv \xnc{\mathrm{D}}_{\mu}\star\xnc{\lambda}.
\end{gather}
That is, if two ordinary gauge fields $A_{\mu}$ and $A'_{\mu}$ are equivalent by an ordinary gauge transformation, then the corresponding noncommutative gauge fields, $\xnc{A}_{\mu}$ and $\xnc{A'}_{\mu}$, will also be gauge-equivalent by a noncommutative gauge transformation. It may be noted that the map \eqref{F-hat.m} is a consequence of the map \eqref{A-hat.m} and the definition \eqref{F-hat.dcurr} of the noncommutative field strength. The field strengths $F_{\mu\nu}$ and $\xnc{F}_{\mu\nu}$ transform covariantly under the usual and the star-gauge transformations, respectively:
\begin{equation}\label{F&F-hat.gt}
\delta_{\lambda}F_{\mu\nu} = \mathrm{i}\left[\lambda,F_{\mu\nu}\right],\qquad
\xnc{\delta}_{\xnc{\lambda}}\xnc{F}_{\mu\nu} = \mathrm{i}\left[\xnc{\lambda},\xnc{F}_{\mu\nu}\right]_{\star}.
\end{equation}

The gauge fields $A_{\mu}(x)$ may be expanded in terms of the Lie-algebra generators $T^{a}$ of U($N$) as $A_{\mu}(x)=A_{\mu}^{a}(x)T^{a}$. These generators satisfy
\begin{equation}\label{T.p}
\left[T^{a},T^{b}\right] = \mathrm{i}f^{abc}T^{c}, \qquad
\left\{T^{a},T^{b}\right\}=d^{abc}T^{c}, \qquad
\mathrm{Tr}\left(T^{a}T^{b}\right)=\delta^{ab}.
\end{equation}
We shall take the structure functions $f^{abc}$ and $d^{abc}$ to be, respectively, totally antisymmetric and totally symmetric. The Yang--Mills action \eqref{S-YM} can now be rewritten as\footnote{A lower gauge index is equivalent to a raised one---whether a gauge index appears as a superscript or as a subscript is a matter of notational convenience.}
\begin{equation}\label{S-YM2}
S_{\mathrm{YM}} = -\frac{1}{4}\int\!\mathrm{d}^{4}x\,F_{\mu\nu}^{a}F^{\mu\nu}_{a},
\end{equation}
where
\begin{equation}\label{F.d2}
F_{\mu\nu}^{a}=\partial_{\mu}A_{\nu}^{a}-\partial_{\nu}A_{\mu}^{a}+f^{abc}A_{\mu}^{b}A_{\nu}^{c}.
\end{equation}
In view of relations \eqref{T.p}, the maps \eqref{A-hat.m}--\eqref{lmda-hat.m} can also be written as
\begin{gather}
\label{A-hat.m2}
\xnc{A}_{\mu}^{c} = A_{\mu}^{c}-\frac{1}{4}\theta^{\alpha\beta}d^{abc}A_{\alpha}^{a}\left(\partial_{\beta}A_{\mu}^{b}+F_{\beta\mu}^{b}\right) + \uO(\theta^{2}),\\
\label{F-hat.m2}
\xnc{F}_{\mu\nu}^{c} = F_{\mu\nu}^{c}+\frac{1}{2}\theta^{\alpha\beta}d^{abc}\left(F_{\mu\alpha}^{a}F_{\nu\beta}^{b}-A_{\alpha}^{a}\partial_{\beta}F_{\mu\nu}^{b}+\frac{1}{2}f^{bde}A_{\alpha}^{a}A_{\beta}^{e}F_{\mu\nu}^{d}\right) + \uO(\theta^{2}),\\
\label{lmda-hat.m2}
\xnc{\lambda}^{c} = \lambda^{c}+\frac{1}{4}\theta^{\alpha\beta}d^{abc}\partial_{\alpha}\lambda^{a}A_{\beta}^{b} + \uO(\theta^{2}),
\end{gather}
and the gauge transformations \eqref{A.gtcurr}--\eqref{F&F-hat.gt} as
\begin{gather}
\label{A.gt2}
\delta_{\lambda}A_{\mu}^{a} = \partial_{\mu}\lambda^{a}+f^{abc}A_{\mu}^{b}\lambda^{c},\\
\label{F.gt2}
\delta_{\lambda}F_{\mu\nu}^{a} = f^{abc}F_{\mu\nu}^{b}\lambda^{c},\\
\label{Ahat.gt2}\begin{split}
\xnc{\delta}_{\xnc{\lambda}}\xnc{A}_{\mu}^{a} &= \partial_{\mu}\xnc{\lambda}^{a}+\frac{\mathrm{i}}{2}d^{abc}\left[\xnc{\lambda}^{b},\xnc{A}_{\mu}^{c}\right]_{\star}-\frac{1}{2}f^{abc}\left\{\xnc{\lambda}^{b},\xnc{A}_{\mu}^{c}\right\}_{\star}\\
&= \partial_{\mu}\xnc{\lambda}^{a}+f^{abc}\xnc{A}_{\mu}^{b}\xnc{\lambda}^{c}+\frac{1}{2}\theta^{\alpha\beta}d^{abc}\partial_{\alpha}\xnc{A}_{\mu}^{b}\partial_{\beta}\xnc{\lambda}^{c} + \uO(\theta^{2}),\end{split}\\
\label{Fhat.gt2}\begin{split}
\xnc{\delta}_{\xnc{\lambda}}\xnc{F}_{\mu\nu}^{a} &= \frac{\mathrm{i}}{2}d^{abc}\left[\xnc{\lambda}^{b},\xnc{F}_{\mu\nu}^{c}\right]_{\star}-\frac{1}{2}f^{abc}\left\{\xnc{\lambda}^{b},\xnc{F}_{\mu\nu}^{c}\right\}_{\star}\\
&= f^{abc}\xnc{F}_{\mu\nu}^{b}\xnc{\lambda}^{c}+\frac{1}{2}\theta^{\alpha\beta}d^{abc}\partial_{\alpha}\xnc{F}_{\mu\nu}^{b}\partial_{\beta}\xnc{\lambda}^{c} + \uO(\theta^{2}).
\end{split}
\end{gather}


\section{\label{sec:current-map}Map for nonabelian currents: classical aspects}

In order to discuss noncommutative gauge theories with sources, it is essential to have a map for the sources also, so that a complete transition between noncommutative gauge theories and the usual ones is possible. Such a map was first briefly discussed in \refcite{Banerjee:2003vc} for the abelian case. We consider the nonabelian case in this section.

Let the noncommutative action be defined as
\begin{equation}\label{S-hatcurr}
\xnc{S}(\xnc{A},\xnc{\psi})=\xnc{S}_{\mathrm{YM}}(\xnc{A})+\xnc{S}_{\mathrm{M}}(\xnc{\psi},\xnc{A}),
\end{equation}
where $\xnc{\psi}_{\alpha}$ are the charged matter fields. The equation of motion for $\xnc{A}_{\mu}^{a}$ is\footnote{We mention that the noncommutative gauge field $\xnc{A}_{\mu}$ is in general an element of the enveloping algebra of the gauge group. Only for specific cases, as for instance the considered case of $\mathrm{U}(N)$ gauge symmetry, it is Lie-algebra valued.}
\begin{equation}\label{eom-ncurr}
\frac{\delta\xnc{S}_{\mathrm{YM}}}{\delta\xnc{A}_{\mu}^{a}}=\xnc{\mathrm{D}}_{\nu}\star\xnc{F}^{\nu\mu}_{a}=-\xnc{J}^{\mu}_{a},
\end{equation}
where
\begin{equation}\label{J-hat.dcurr}
\xnc{J}^{\mu}_{a}=\left.\frac{\delta\xnc{S}_{\mathrm{M}}}{\delta\xnc{A}_{\mu}^{a}}\right|_{\xnc{\psi}}.
\end{equation}
Equation \eqref{eom-ncurr} shows that $\xnc{J}^{\mu}_{a}$ transforms covariantly under the star-gauge transformation:
\begin{equation}\label{sgt2curr}
\xnc{\delta}_{\xnc{\lambda}}\xnc{J}^{\mu} = -\mathrm{i}\left[\xnc{J}^{\mu},\xnc{\lambda}\right]_{\star}, \qquad
\xnc{\delta}_{\xnc{\lambda}}\xnc{J}^{\mu}_{a} = f^{abc}\xnc{J}^{\mu}_{b}\xnc{\lambda}^{c}+\frac{1}{2}\theta^{\alpha\beta}d^{abc}\partial_{\alpha}\xnc{J}^{\mu}_{b}\partial_{\beta}\xnc{\lambda}^{c}+ \uO(\theta^{2}).
\end{equation}
Also, it satisfies the noncommutative covariant conservation law
\begin{equation}\label{DhJh3}
\xnc{\mathrm{D}}_{\mu}\star\xnc{J}^{\mu}_{a}=0,
\end{equation}
which may be seen from Eq.~\eqref{eom-ncurr} by taking the noncommutative covariant divergence.

The use of Seiberg--Witten map in the action \eqref{S-hatcurr} gives its $\theta$-expanded version in commutative space:
\begin{equation}\label{S-thetacurr}
\xnc{S}(\xnc{A},\xnc{\psi})\rightarrow S^{\theta}(A,\psi)=S^{\theta}_{\mathrm{YM}}(A)+S^{\theta}_{\mathrm{M}}(\psi,A),
\end{equation}
where $S^{\theta}_{\mathrm{YM}}(A)$ contains all terms involving $A_{\mu}^{a}$ only, and is given by
\begin{equation}\label{S-th-YM}
S^{\theta}_{\mathrm{YM}} = -\frac{1}{4}\int\!\mathrm{d}^{4}x\left[F_{\mu\nu}^{a}F^{\mu\nu}_{a}+\theta^{\alpha\beta}d^{abc}F^{\mu\nu}_{a}\left(F_{\mu\alpha}^{b}F_{\nu\beta}^{c}+\frac{1}{4}F_{\beta\alpha}^{b}F_{\mu\nu}^{c}\right) + \uO(\theta^{2})\right],
\end{equation}
also, we have dropped a boundary term in order to express it solely in terms of the field strength. The equation of motion following from the action \eqref{S-thetacurr} is
\begin{equation}\label{eom-thcurr}
\frac{\delta S^{\theta}_{\mathrm{YM}}}{\delta A_{\mu}^{a}}=-J^{\mu}_{a},
\end{equation}
where
\begin{equation}\label{J.dcurr}
J^{\mu}_{a}=\left.\frac{\delta S^{\theta}_{\mathrm{M}}}{\delta A_{\mu}^{a}}\right|_{\psi}.
\end{equation}
Expectedly, from these relations, it follows that $J^{\mu}_{a}$ transforms covariantly,
\begin{equation}\label{gt6}
\delta_{\lambda}J^{\mu} = -\mathrm{i}\left[J^{\mu}, \lambda\right], \qquad
\delta_{\lambda}J^{\mu}_{a} = f^{abc}J^{\mu}_{b}\lambda^{c},
\end{equation}
and satisfies the covariant conservation law
\begin{equation}\label{DJ3}
\mathrm{D}_{\mu}J^{\mu}_{a} = 0.
\end{equation}

Now the application of Seiberg--Witten map on the right-hand side of Eq.~\eqref{J-hat.dcurr} yields the relation between $\xnc{J}^{\mu}_{a}$ and $J^{\mu}_{a}$:
\begin{equation}\label{J-hat.m}
\xnc{J}^{\mu}_{a}(x) = \int\!\mathrm{d}^{4}y\left[\left.\frac{\delta S^{\theta}_{\mathrm{M}}}{\delta A_{\nu}^{c}(y)}\right|_{\psi}\frac{\delta A_{\nu}^{c}(y)}{\delta \xnc{A}_{\mu}^{a}(x)}+\left.\frac{\delta S^{\theta}_{\mathrm{M}}}{\delta \psi_{\alpha}^{c}(y)}\right|_{A}\frac{\delta \psi_{\alpha}^{c}(y)}{\delta \xnc{A}_{\mu}^{a}(x)}\right] = \int\!\mathrm{d}^{4}y\,J^{\nu}_{c}(y)\frac{\delta A_{\nu}^{c}(y)}{\delta \xnc{A}_{\mu}^{a}(x)},
\end{equation}
where the second term obtained in the first step has been dropped on using the equation of motion for $\psi_{\alpha}^{a}$.

We consider Eq.~\eqref{J-hat.m} as a closed form for the map among the sources. To get its explicit structure, the map \eqref{A-hat.m2} among the gauge potentials is necessary. Since the map \eqref{A-hat.m2} is a classical result, the map for the sources obtained in this way is also classical.

Let us next obtain the explicit form of this map up to first order in $\theta$. Using the map (\ref{A-hat.m2}) and its inverse,
\begin{equation}\label{A-in.m2}
A_{\mu}^{c}=\xnc{A}_{\mu}^{c}+\frac{1}{4}\theta^{\alpha\beta}d^{abc}\xnc{A}_{\alpha}^{a}\left(\partial_{\beta}\xnc{A}_{\mu}^{b}+\xnc{F}_{\beta\mu}^{b}\right) + \uO(\theta^{2}),
\end{equation}
we can compute the functional derivative
\begin{equation}\label{fund}\begin{split}
\frac{\delta A_{\nu}^{c}(y)}{\delta \xnc{A}_{\mu}^{a}(x)} &= \delta^{\mu}_{\nu}\delta^{ac}\delta(x-y)\\
&\quad {}+\frac{1}{4}\theta^{\alpha\beta}\delta^{\mu}_{\nu}\left[2d^{abc}A_{\alpha}^{b}(y)\partial_{\beta}^{y}\delta(x-y)+d^{edc}f^{bad}A_{\alpha}^{e}(y)A_{\beta}^{b}(y)\delta(x-y)\right]\\
&\quad {}-\frac{1}{4}\theta^{\alpha\mu}\left[d^{abc}A_{\alpha}^{b}(y)\partial_{\nu}^{y}\delta(x-y)\right.\\
&\qquad\qquad\quad \left.{}+\left(d^{abc}\partial_{\alpha}^{y}A_{\nu}^{b}(y)+d^{abc}F_{\alpha\nu}^{b}(y)-d^{edc}f^{dab}A_{\alpha}^{e}(y)A_{\nu}^{b}(y)\right)\delta(x-y)\right]\\
&\quad {} + \uO(\theta^{2}),\end{split}
\end{equation}
where $\partial_{\beta}^{y}$ stands for ${\partial}/{\partial y^{\beta}}$. Putting this in Eq.~\eqref{J-hat.m}, we get
\begin{equation}\label{J-hat.m2'}\begin{split}
\xnc{J}^{\mu}_{a} &= J^{\mu}_{a}-\frac{1}{2}\theta^{\alpha\beta}\left[d^{abc}\partial_{\beta}\left(A_{\alpha}^{b}J^{\mu}_{c}\right)-\frac{1}{2}d^{edc}f^{bad}A_{\alpha}^{e}A_{\beta}^{b}J^{\mu}_{c}\right]\\
&\quad {}-\frac{1}{2}\theta^{\alpha\mu}\left[d^{abc}F_{\alpha\nu}^{b}J^{\nu}_{c}-\frac{1}{2}\left(d^{cad}f^{dbe}+d^{bcd}f^{dae}\right)A_{\alpha}^{b}A_{\nu}^{e}J^{\nu}_{c}-\frac{1}{2}d^{abd}A_{\alpha}^{b}\partial_{\nu}J^{\nu}_{d}\right] + \uO(\theta^{2}).\end{split}
\end{equation}
Since $\mathrm{D}_{\nu}J^{\nu}_{a} \equiv \partial_{\nu}J^{\nu}_{a}-f^{abc}J^{\nu}_{b}A_{\nu}^{c}$, we can use Eq.~\eqref{DJ3} to substitute
\begin{equation}\label{xxx}
\partial_{\nu}J^{\nu}_{d}=f^{dce}J^{\nu}_{c}A_{\nu}^{e}
\end{equation}
in the last term on the right-hand side of Eq.~\eqref{J-hat.m2'} to obtain
\begin{equation}\label{J-hat.m2}
\xnc{J}^{\mu}_{a}=J^{\mu}_{a}-\frac{1}{2}\theta^{\alpha\beta}\left[d^{abc}\partial_{\beta}\left(A_{\alpha}^{b}J^{\mu}_{c}\right)-\frac{1}{2}d^{edc}f^{bad}A_{\alpha}^{e}A_{\beta}^{b}J^{\mu}_{c}\right]
-\frac{1}{2}\theta^{\alpha\mu}d^{abc}F_{\alpha\nu}^{b}J^{\nu}_{c} + \uO(\theta^{2}),
\end{equation}
where we have used the identity
\begin{equation}\label{df.i}
d^{abd}f^{dce}+d^{bcd}f^{dae}+d^{cad}f^{dbe} = 0.
\end{equation}
As a simple yet nontrivial consistency check, we show the stability of the map under gauge transformations. Under the ordinary gauge transformations given by Eqs.~\eqref{A.gt2} and \eqref{F.gt2}, and using the covariant transformation law \eqref{gt6} for $J^{\mu}_{a}$, the right-hand side of Eq.~\eqref{J-hat.m2} transforms as
\begin{equation}\label{J-hat.gt1}\begin{split}
\delta_{\lambda}\xnc{J}^{\mu}_{a} &= f^{abc}J^{\mu}_{b}\lambda^{c}
-\frac{1}{2}\theta^{\alpha\beta}\bigg[d^{abc}\partial_{\beta}J^{\mu}_{c}\partial_{\alpha}\lambda^{b}+d^{cdb}f^{bea}\partial_{\beta}\left(A_{\alpha}^{d}J^{\mu}_{c}\lambda^{e}\right)\\
&\qquad\qquad\qquad\qquad\quad {}+\frac{1}{2}\left(d^{ecb}f^{bda}-d^{cdb}f^{bea}\right)A_{\alpha}^{d}J^{\mu}_{c}\partial_{\beta}\lambda^{e}\\
&\qquad\qquad\qquad\qquad\quad {}+\frac{1}{2}d^{gcd}\left(f^{abe}f^{edh}+f^{dae}f^{ebh}\right)A_{\alpha}^{g}A_{\beta}^{b}J^{\mu}_{c}\lambda^{h}\bigg]\\
&\quad {}+\frac{1}{2}\theta^{\alpha\mu}d^{cdb}f^{bae}F_{\alpha\nu}^{d}J^{\nu}_{c}\lambda^{e} + \uO(\theta^{2}),\end{split}
\end{equation}
where we have used the relation \eqref{df.i}. On the other hand, using the maps \eqref{lmda-hat.m2} and \eqref{J-hat.m2}, and the identity
\begin{equation}\label{ff.i}
f^{abe}f^{edh}+f^{bde}f^{eah}+f^{dae}f^{ebh}=0,
\end{equation}
the right-hand side of the second relation in Eq.~\eqref{sgt2curr} reproduces the right-hand side of Eq.~\eqref{J-hat.gt1}. Hence,
\begin{equation}\label{J-hat.gt2}
\xnc{\delta}_{\xnc{\lambda}}\xnc{J}^{\mu}_{a} = \delta_{\lambda}\xnc{J}^{\mu}_{a},
\end{equation}
thereby proving the stability of the map \eqref{J-hat.m2} under the gauge transformations. This statement is equivalent to the usual notion of stability which ensures that the star-gauge-transformed noncommutative current is mapped to the usual-gauge-transformed ordinary current, as may be verified by performing a Taylor expansion of the right-hand side of $\xnc{J}^{\mu}_{a}(J, A)+\xnc{\delta}_{\xnc{\lambda}}\xnc{J}^{\mu}_{a}(J, A) = \xnc{J}^{\mu}_{a}(J+\delta_{\lambda}J, A+\delta_{\lambda}A)$ and comparing both sides.\footnote{Exactly the same thing happens when discussing the stability of the map \eqref{A-hat.m} for the potentials.} 

It is worthwhile to mention that the use  of Eq.~\eqref{xxx} in obtaining the map \eqref{J-hat.m2} is crucial to get the correct transformation property of $\xnc{J}^{\mu}_{a}$. This is because issues of gauge covariance and covariant conservation are not independent. In an ordinary abelian gauge theory, for example, current conservation and gauge invariance are related. Likewise, in the nonabelian case, covariant conservation and gauge covariance are related. This intertwining property is a peculiarity of the mapping among the sources and is not to be found in the mapping among the potentials or the field strengths. 

From these results, it is possible to give a map for the covariant derivatives of the currents. We recall that
\begin{equation}\label{D-hJ-h}\begin{split}
\xnc{\mathrm{D}}_{\mu}\star\xnc{J}^{\mu}_{a} &= \partial_{\mu}\xnc{J}^{\mu}_{a}+\frac{\mathrm{i}}{2}d^{abc}\left[\xnc{J}^{\mu}_{b},\xnc{A}_{\mu}^{c}\right]_{\star}-\frac{1}{2}f^{abc}\left\{\xnc{J}^{\mu}_{b},\xnc{A}_{\mu}^{c}\right\}_{\star}\\
&= \partial_{\mu}\xnc{J}^{\mu}_{a}+f^{abc}\xnc{A}_{\mu}^{b}\xnc{J}^{\mu}_{c}+\frac{1}{2}\theta^{\alpha\beta}d^{abc}\partial_{\alpha}\xnc{A}_{\mu}^{b}\partial_{\beta}\xnc{J}^{\mu}_{c} + \uO(\theta^{2}),\end{split}
\end{equation}
which, using the maps \eqref{A-hat.m2} and \eqref{J-hat.m2}, gives
\begin{equation}\label{D-hJ-h2}
\xnc{\mathrm{D}}_{\mu}\star\xnc{J}^{\mu}_{a} = \mathrm{D}_{\mu}J^{\mu}_{a}-\frac{1}{2}\theta^{\alpha\beta}\left[d^{abc}\partial_{\beta}\left(A_{\alpha}^{b}\mathrm{D}_{\mu}J^{\mu}_{c}\right)-\frac{1}{2}d^{edc}f^{bad}A_{\alpha}^{e}A_{\beta}^{b}\mathrm{D}_{\mu}J^{\mu}_{c}\right] + \uO(\theta^{2}),
\end{equation}
where we have used the Jacobi identities \eqref{df.i} and \eqref{ff.i}, and the relation \eqref{xxx}. Thus we see that covariant conservation of the ordinary current, $\mathrm{D}_{\mu}J^{\mu}_{a}=0$, implies that $\xnc{J}^{\mu}_{a}$ given by Eq.~\eqref{J-hat.m2} indeed satisfies the noncommutative covariant conservation law, $\xnc{\mathrm{D}}_{\mu}\star\xnc{J}^{\mu}_{a}=0$. This is also to be expected from classical notions.

At this point, an intriguing issue arises. Is it possible to use Eq.~\eqref{D-hJ-h2} to relate the anomalies in the different descriptions? Indeed the analysis presented for the vector current can be readily taken over for the chiral current. Classically everything would be fine since the relevant currents are both conserved. At the quantum level, however, the chiral currents are not conserved. We would like to ascertain whether the relation \eqref{D-hJ-h2} is still valid by substituting the relevant chiral anomalies in place of $\xnc{\mathrm{D}}_{\mu}\star\xnc{J}^{\mu}_{a}$ and $\mathrm{D}_{\mu}J^{\mu}_{a}$. Since the main aspects get highlighted for the abelian theory itself, we confine to this case, and present a detailed analysis in the remainder of this chapter.


\section{\label{sec:current-abelian}Map for abelian currents: classical and quantum aspects}

Some discussion on the use of the map \eqref{D-hJ-h2}, in the abelian case, for relating anomalies up to $\uO(\theta)$ was earlier given in \refcite{Banerjee:2003vc}. In order to gain a deeper understanding, it is essential to consider higher orders in $\theta$. Keeping this in mind, we present a calculation up to $\uO(\theta^{2})$ for two- and four-dimensional theories.

The maps to the second order in $\theta$ in the abelian case are given by \cite{Fidanza:2001qm}
\begin{gather}
\label{A.m}\begin{split}
\xnc{A}_{\mu} &= A_{\mu}-\frac{1}{2}\theta^{\alpha\beta}A_{\alpha}\left(\partial_{\beta}A_{\mu}+F_{\beta\mu}\right)\\
&\quad {}+\frac{1}{6}\theta^{\alpha\beta}\theta^{\kappa\sigma}A_{\alpha}\left[\partial_{\beta}\left(A_{\kappa}\partial_{\sigma}A_{\mu}+2A_{\kappa}F_{\sigma\mu}\right)+F_{\beta\kappa}\left(\partial_{\sigma}A_{\mu}+2F_{\sigma\mu}\right)\right] + \uO(\theta^{3}),\end{split}\\
\label{107}\begin{split}
\xnc{F}_{\mu\nu} &= F_{\mu\nu}-\theta^{\alpha\beta}\left(A_{\alpha}\partial_{\beta}F_{\mu\nu}+F_{\mu\alpha}F_{\beta\nu}\right)\\
&\quad {}+\frac{1}{2}\theta^{\alpha\beta}\theta^{\kappa\sigma}\left[A_{\alpha}\partial_{\beta}\left(A_{\kappa}\partial_{\sigma}F_{\mu\nu}+2F_{\mu\kappa}F_{\sigma\nu}\right)+F_{\beta\kappa}\left(A_{\alpha}\partial_{\sigma}F_{\mu\nu}+2F_{\mu\alpha}F_{\sigma\nu}\right)\right]\\
&\quad {} + \uO(\theta^{3}),\end{split}\\
\label{102}
\xnc{\lambda} = \lambda-\frac{1}{2}\theta^{\alpha\beta}A_{\alpha}\partial_{\beta}\lambda+\frac{1}{6}\theta^{\alpha\beta}\theta^{\kappa\sigma}A_{\alpha}\left[\partial_{\beta}\left(A_{\kappa}\partial_{\sigma}\lambda\right)+F_{\beta\kappa}\partial_{\sigma}\lambda\right] + \uO(\theta^{3}),
\end{gather}
which ensure the stability of gauge transformations
\begin{gather}
\label{1040}
\xnc{\delta}_{\xnc{\lambda}}\xnc{A}_{\mu} = \xnc{\mathrm{D}}_{\mu}\star\xnc{\lambda}
\equiv\partial_{\mu}\xnc{\lambda}+\mathrm{i}\left[\xnc{\lambda},\xnc{A}_{\mu}\right]_{\star}
= \partial_{\mu}\xnc{\lambda}+\theta^{\alpha\beta}\partial_{\alpha}\xnc{A}_{\mu}\partial_{\beta}\xnc{\lambda} + \uO(\theta^{3}),\\
\label{104}
\delta_{\lambda}A_{\mu} = \partial_{\mu}\lambda.
\end{gather}
Analogous to the nonabelian theory, the map for currents is consistent with the requirements that while the current $J^{\mu}$ is gauge-invariant and satisfies the ordinary conservation law, $\partial_{\mu}J^{\mu}=0$, the current $\xnc{J}^{\mu}$ is star-gauge-covariant and satisfies the noncommutative covariant conservation law, $\xnc{\mathrm{D}}_{\mu}\star\xnc{J}^{\mu}=0$. Now the currents $J^{\mu}$ and $\xnc{J}^{\mu}$ are related by the abelian version of Eq.~\eqref{J-hat.m} \cite{Banerjee:2003vc},
\begin{equation}\label{109}
\xnc{J}^{\mu}(x) = \int\!\mathrm{d}^{4}yJ^{\nu}(y)\frac{\delta A_{\nu}(y)}{\delta\xnc{A}_{\mu}(x)},
\end{equation}
which, using the map \eqref{A.m} and its inverse,
\begin{equation}\label{110}\begin{split}
A_{\mu} &= \xnc{A}_{\mu}+\frac{1}{2}\theta^{\alpha\beta}\xnc{A}_{\alpha}\left(\partial_{\beta}\xnc{A}_{\mu}+\xnc{F}_{\beta\mu}\right)\\
&\quad {}+\frac{1}{6}\theta^{\alpha\beta}\theta^{\kappa\sigma}\xnc{A}_{\alpha}\left[\frac{1}{2}\partial_{\beta}\left(\xnc{A}_{\kappa}\partial_{\sigma}\xnc{A}_{\mu}-\xnc{A}_{\kappa}\xnc{F}_{\sigma\mu}\right)+\frac{1}{2}\xnc{F}_{\beta\kappa}\left(\partial_{\sigma}\xnc{A}_{\mu}+5\xnc{F}_{\sigma\mu}\right)\right.\\
&\qquad\qquad\qquad\qquad {}+\left.\frac{3}{2}\left(2\xnc{A}_{\kappa}\partial_{\beta}\xnc{F}_{\sigma\mu}+\partial_{\beta}\xnc{A}_{\kappa}\partial_{\sigma}\xnc{A}_{\mu}+\partial_{\beta}\xnc{A}_{\kappa}\xnc{F}_{\sigma\mu}\right)\right] + \uO(\theta^{3}),\end{split}
\end{equation}
yields the explicit $\uO(\theta^{2})$ form of the source map:
\begin{equation}\label{J-hat2.m}\begin{split}
\xnc{J}^{\mu} &= J^{\mu}-\theta^{\alpha\beta}\left(A_{\alpha}\partial_{\beta}J^{\mu}-\frac{1}{2}F_{\alpha\beta}J^{\mu}\right)+\theta^{\mu\alpha}F_{\alpha\beta}J^{\beta}\\
&\quad {}+\frac{1}{2}\theta^{\alpha\beta}\theta^{\kappa\sigma}\partial_{\alpha}\left(A_{\kappa}F_{\beta\sigma}J^{\mu}-A_{\beta}A_{\kappa}\partial_{\sigma}J^{\mu}+\frac{1}{2}A_{\beta}F_{\kappa\sigma}J^{\mu}\right)-\theta^{\alpha\beta}\theta^{\kappa\mu}\partial_{\alpha}\left(A_{\beta}F_{\kappa\nu}J^{\nu}\right)\\
&\quad {} + \uO(\theta^{3}),\end{split}
\end{equation}
where we have used $\partial_{\mu}J^{\mu}=0$ to simplify the integrand.\footnote{This is essential to ensure the stability of map \eqref{J-hat2.m} under appropriate gauge transformations. A similar manipulation was needed for getting the nonabelian expression \eqref{J-hat.m2}.} The above map, up to $\uO(\theta)$, was earlier given in \refcite{Banerjee:2003vc}. Now let us check explicitly the stability under the gauge transformations. Under the ordinary gauge transformation, $\delta_{\lambda}A_{\mu}=\partial_{\mu}\lambda$, $\delta_{\lambda}F_{\mu\nu}=0$, and $\delta_{\lambda}J^{\mu}=0$. Hence the right-hand side of Eq.~\eqref{J-hat2.m} transforms as
\begin{equation}\label{113}\begin{split}
\delta_{\lambda}\xnc{J}^{\mu} &= \theta^{\alpha\beta}\partial_{\alpha}J^{\mu}\partial_{\beta}\lambda+\theta^{\alpha\beta}\theta^{\mu\kappa}\partial_{\alpha}\left(F_{\kappa\nu}J^{\nu}\right)\partial_{\beta}\lambda\\
&\quad {}+\frac{1}{2}\theta^{\alpha\beta}\theta^{\kappa\sigma}\left[2\partial_{\beta}\partial_{\sigma}\left(A_{\kappa}J^{\mu}\right)\partial_{\alpha}\lambda-\partial_{\beta}\left(A_{\kappa}\partial_{\sigma}\lambda\right)\partial_{\alpha}J^{\mu}\right] + \uO(\theta^{3}).\end{split}
\end{equation}
On the other hand,
\begin{equation}\label{1131}
\xnc{\delta}_{\xnc{\lambda}}\xnc{J}^{\mu} = \mathrm{i}\left[\xnc{\lambda},\xnc{J}^{\mu}\right]_{\star} = \theta^{\alpha\beta}\partial_{\alpha}\xnc{J}^{\mu}\partial_{\beta}\xnc{\lambda} + \uO(\theta^{3}).
\end{equation}
Next, using the maps \eqref{102} and \eqref{J-hat2.m} in the above equation, one finds that the right-hand side of Eq.~\eqref{113} is reproduced. Hence,
\begin{equation}\label{11319}
\xnc{\delta}_{\xnc{\lambda}}\xnc{J}^{\mu} = \delta_{\lambda}\xnc{J}^{\mu},
\end{equation}
thereby proving the gauge-equivalence, as observed earlier. Furthermore, using the maps \eqref{A.m} and \eqref{J-hat2.m}, the covariant divergence of $\xnc{J}^{\mu}$,
\begin{equation}\label{DJ}
\xnc{\mathrm{D}}_{\mu}\star\xnc{J}^{\mu} = \partial_{\mu}\xnc{J}^{\mu}+\mathrm{i}\left[\xnc{J}^{\mu},\xnc{A}_{\mu}\right]_{\star} = \partial_{\mu}\xnc{J}^{\mu}-\theta^{\alpha\beta}\partial_{\alpha}\xnc{J}^{\mu}\partial_{\beta}\xnc{A}_{\mu} + \uO(\theta^{3}),
\end{equation}
can be expressed as
\begin{equation}\label{115}\begin{split}
\xnc{\mathrm{D}}_{\mu}\star\xnc{J}^{\mu} &= \partial_{\mu}J^{\mu}+\theta^{\alpha\beta}\partial_{\alpha}\left(A_{\beta}\partial_{\mu}J^{\mu}\right)
+\frac{1}{2}\theta^{\alpha\beta}\theta^{\kappa\sigma}\partial_{\alpha}\left[A_{\kappa}F_{\beta\sigma}\partial_{\mu}J^{\mu}-A_{\beta}\partial_{\sigma}\left(A_{\kappa}\partial_{\mu}J^{\mu}\right)\right]\\
&\quad {} + \uO(\theta^{3}),\end{split}
\end{equation}
where each term on the right-hand side involves $\partial_{\mu}J^{\mu}$, so that the covariant conservation of $\xnc{J}^{\mu}$ follows from the ordinary conservation of $J^{\mu}$. This is the abelian analogue of Eq.~\eqref{D-hJ-h2}, but valid up to $\uO(\theta^{2})$.

We are now in a position to discuss the mapping of anomalies. Since the maps have been obtained for the gauge currents, the anomalies refer to chiral anomalies found in chiral gauge theories. Moreover, we implicitly assume a regularisation which preserves vector-current conservation so that the chiral anomaly $\partial_\mu[\bar\psi\gamma^\mu\{(1+\gamma_5)/{2}\}\psi]$ is proportional to the usual ABJ anomaly $\partial_{\mu} J^{\mu}_{5}$ \cite{Banerjee:1985ti}. The first step is to realise that the standard ABJ anomaly \cite{Adler:1969gk, Bell:1969ts} is not modified in $\theta$-expanded gauge theory \cite{Brandt:2003fx}. In other words,
\begin{equation}\label{anom.d}
\mathscr{A} = \partial_{\mu}J^{\mu}_{5} = \frac{1}{16\pi^{2}}\varepsilon_{\mu\nu\lambda\rho}F^{\mu\nu}F^{\lambda\rho}
\end{equation}
still holds. The star-gauge-covariant anomaly is just given by a standard deformation of the above result \cite{Ardalan:2000cy, Gracia-Bondia:2000pz}:
\begin{equation}\label{anom-cov.d}
\xnc{\mathscr{A}} = \xnc{\mathrm{D}}_{\mu}\star\xnc{J}^{\mu}_{5} = \frac{1}{16\pi^{2}}\varepsilon_{\mu\nu\lambda\rho}\xnc{F}^{\mu\nu}\star\xnc{F}^{\lambda\rho}.
\end{equation}
The expected map for anomalies, obtained by a lift from the classical result \eqref{115}, follows as
\begin{equation}\label{anom.m}
\xnc{\mathscr{A}} = \mathscr{A}+\theta^{\alpha\beta}\partial_{\alpha}\left(A_{\beta}\mathscr{A}\right)
+\frac{1}{2}\theta^{\alpha\beta}\theta^{\kappa\sigma}\partial_{\alpha}\left[A_{\kappa}F_{\beta\sigma}\mathscr{A}-A_{\beta}\partial_{\sigma}\left(A_{\kappa}\mathscr{A}\right)\right] + \uO(\theta^{3}).
\end{equation}

Let us digress a bit on this map. The starting point is the classical map \eqref{J-hat2.m} with the vector current replaced by the axial one. Although current conservation is used to derive the map \eqref{J-hat2.m}, the analysis still remains valid since the axial current is also classically conserved. Also, as discussed earlier, the retention of the term proportional to the divergence of the current would spoil the stability of the gauge transformations, which must hold irrespective of whether the current is vector or axial. From the map \eqref{J-hat2.m} one is led to the relation \eqref{115}. Now we would like to see whether this classical map persists even at the quantum level, written in the form \eqref{anom.m}. As far as gauge-transformation properties are concerned, it is obviously compatible since the anomalies in the different descriptions transform exactly as the corresponding currents. Corrections, if any, would thus entail only gauge-invariant terms, involving the field tensor $F_{\mu\nu}$. We now prove that the relation \eqref{anom.m} is indeed valid for the slowly-varying-field approximation, which was also essential for demonstrating the equivalence of DBI actions \cite{Seiberg:1999vs}. Later on we shall compute the corrections that appear for arbitrary field configurations. In the slowly-varying-field approximation, since derivatives on $\xnc{F}^{\mu\nu}$ can be ignored, the star product in Eq.~\eqref{anom-cov.d} is dropped. Using the map \eqref{107}, we write this expression as
\begin{equation}\label{120-1}\begin{split}
\xnc{\mathscr{A}} &= \frac{1}{16\pi^{2}}\varepsilon_{\mu\nu\lambda\rho}\xnc{F}^{\mu\nu}\xnc{F}^{\lambda\rho}\\
&= \frac{1}{16\pi^{2}}\varepsilon_{\mu\nu\lambda\rho}\bigg[F^{\mu\nu}F^{\lambda\rho} + \theta^{\alpha\beta}\left\{A_{\beta}\partial_{\alpha}\left(F^{\mu\nu}F^{\lambda\rho}\right)-2F^{\mu\nu}{F^{\lambda}}_{\alpha}{F_{\beta}}^{\rho}\right\}\\
&\qquad\qquad\qquad {}+ \theta^{\alpha\beta}\theta^{\kappa\sigma}\bigg\{\frac{1}{2}A_{\alpha}\partial_{\beta}\left[A_{\kappa}\partial_{\sigma}\left(F^{\mu\nu}F^{\lambda\rho}\right)\right]+\frac{1}{2}A_{\alpha}F_{\beta\kappa}\partial_{\sigma}\left(F^{\mu\nu}F^{\lambda\rho}\right)\\
&\qquad\qquad\qquad\qquad\qquad\quad {}+2A_{\alpha}\partial_{\beta}\left(F^{\mu\nu}{F^{\lambda}}_{\kappa}{F_{\sigma}}^{\rho}\right)+2F^{\mu\nu}{F^{\lambda}}_{\alpha}F_{\beta\kappa}{F_{\sigma}}^{\rho}\\
&\qquad\qquad\qquad\qquad\qquad\quad {}+{F^{\mu}}_{\alpha}{F_{\beta}}^{\nu}{F^{\lambda}}_{\kappa}{F_{\sigma}}^{\rho}\bigg\} + \uO(\theta^{3})\bigg].\end{split}
\end{equation}
Next, using the identities \cite{Banerjee:2003ce}
\begin{gather}
\label{118}
\varepsilon_{\mu\nu\lambda\rho}\theta^{\alpha\beta}\left[F^{\mu\nu}F^{\lambda\rho}F_{\alpha\beta}+4F^{\mu\nu}{F^{\lambda}}_{\alpha}{F_{\beta}}^{\rho}\right] = 0,\\
\label{119}\begin{split}
\varepsilon_{\mu\nu\lambda\rho}\theta^{\alpha\beta}\theta^{\kappa\sigma}&\bigg[{F^{\mu}}_{\alpha}{F_{\beta}}^{\nu}{F^{\lambda}}_{\kappa}{F_{\sigma}}^{\rho}+2F^{\mu\nu}{F^{\lambda}}_{\alpha}F_{\beta\kappa}{F_{\sigma}}^{\rho}\\
&\quad {}+\frac{1}{2}F^{\mu\nu}{F^{\lambda}}_{\kappa}{F_{\sigma}}^{\rho}F_{\alpha\beta}+\frac{1}{4}F^{\mu\nu}F^{\lambda\rho}F_{\alpha\kappa}F_{\sigma\beta}\bigg] = 0,\end{split}
\end{gather}
and the usual Bianchi identity, we can write down
\begin{equation}\label{120}\begin{split}
\xnc{\mathscr{A}} &= \frac{1}{16\pi^{2}}\varepsilon_{\mu\nu\lambda\rho}\bigg[F^{\mu\nu}F^{\lambda\rho}+\theta^{\alpha\beta}\partial_{\alpha}\left(A_{\beta}F^{\mu\nu}F^{\lambda\rho}\right)\\
&\qquad\qquad\qquad {}+\frac{1}{2}\theta^{\alpha\beta}\theta^{\kappa\sigma}\partial_{\alpha}\Big\{A_{\kappa}F_{\beta\sigma}F^{\mu\nu}F^{\lambda\rho}-A_{\beta}\partial_{\sigma}\left(A_{\kappa}F^{\mu\nu}F^{\lambda\rho}\right)\Big\} + \uO(\theta^{3})\bigg].\end{split}
\end{equation}
The identities \eqref{118} and \eqref{119} are valid in four dimensions and, in fact, hold not only for just $F^{\mu\nu}$ but for any antisymmetric tensor, in particular, for $\xnc{F}^{\mu\nu}$ also. This gives a definite way for obtaining the identity \eqref{119} starting form \eqref{118}. The identity \eqref{119} may be obtained from the identity \eqref{118} by doing the replacement $F^{\mu\nu}\rightarrow\xnc{F}^{\mu\nu}$ followed by using the map \eqref{107} and retaining $\uO(\theta^{2})$ terms. Alternatively, one can check it by explicitly carrying out all the summations. Now substituting for the anomaly \eqref{anom.d} on the right-hand side of Eq.~\eqref{120}, we indeed get back our expected anomaly map \eqref{anom.m}.

It is easy to show that the map \eqref{anom.m} is equally valid in two dimensions\footnote{Contrary to the four-dimensional example, the map holds for arbitrary fields. This is because the anomaly does not involve any (star) product of fields and hence the slowly-varying-field approximation becomes redundant.}, in which case,
\begin{equation}\label{an-nc-2d}
\mathscr{A}_{2\mathrm{d}} = \partial_{\mu}J^{\mu}_{5} = \frac{1}{2\pi}\varepsilon_{\mu\nu}F^{\mu\nu},\qquad
\xnc{\mathscr{A}}_{2\mathrm{d}} = \xnc{\mathrm{D}}_{\mu}\star\xnc{J}^{\mu}_{5} = \frac{1}{2\pi}\varepsilon_{\mu\nu}\xnc{F}^{\mu\nu}.
\end{equation}
It follows from the map \eqref{107} for the field strength that
\begin{equation}\label{an2-2d}\begin{split}
\xnc{\mathscr{A}}_{2\mathrm{d}} &= \frac{1}{2\pi}\varepsilon_{\mu\nu}\xnc{F}^{\mu\nu}\\
&= \frac{1}{2\pi}\varepsilon_{\mu\nu}\bigg[F^{\mu\nu}-\theta^{\alpha\beta}\left(A_{\alpha}\partial_{\beta}F^{\mu\nu}+{F^{\mu}}_{\alpha}{F_{\beta}}^{\nu}\right)\\
&\qquad\qquad {}+\frac{1}{2}\theta^{\alpha\beta}\theta^{\kappa\sigma}\left\{A_{\alpha}\partial_{\beta}\left(A_{\kappa}\partial_{\sigma}F^{\mu\nu}\right)+A_{\alpha}F_{\beta\kappa}\partial_{\sigma}F^{\mu\nu}\right.\\
&\qquad\qquad\qquad\qquad\quad \left.{}+2A_{\alpha}\partial_{\beta}\left({F^{\mu}}_{\kappa}{F_{\sigma}}^{\nu}\right)+2{F^{\mu}}_{\alpha}F_{\beta\kappa}{F_{\sigma}}^{\nu}\right\} + \uO(\theta^{3})\bigg].\end{split}
\end{equation}
In two dimensions, we have the identities
\begin{gather}
\label{id-2d}
\varepsilon_{\mu\nu}\theta^{\alpha\beta}\left(F_{\alpha\beta}F^{\mu\nu}+2{F^{\mu}}_{\alpha}{F_{\beta}}^{\nu}\right) = 0,\\
\label{id2-2d}
\varepsilon_{\mu\nu}\theta^{\alpha\beta}\theta^{\kappa\sigma}\left(F_{\alpha\kappa}F_{\sigma\beta}F^{\mu\nu}+F_{\alpha\beta}{F^{\mu}}_{\kappa}{F_{\sigma}}^{\nu}+4{F^{\mu}}_{\alpha}F_{\beta\kappa}{F_{\sigma}}^{\nu}\right) = 0,
\end{gather}
which are the analogue of the identities \eqref{118} and \eqref{119}. Likewise, these identities hold for any antisymmetric second-rank tensor, and the second identitiy can be obtained from the first by replacing the usual field strength by the noncommutative field strength and then using the Seiberg--Witten map. Using these identities, Eq.~\eqref{an2-2d} can be rewritten as
\begin{equation}\label{an3-2d}\begin{split}
\xnc{\mathscr{A}}_{2\mathrm{d}} = \frac{1}{2\pi}\varepsilon_{\mu\nu}&\bigg[F^{\mu\nu}+\theta^{\alpha\beta}\partial_{\alpha}\left(A_{\beta}F^{\mu\nu}\right)\\
&\quad {}+\frac{1}{2}\theta^{\alpha\beta}\theta^{\kappa\sigma}\partial_{\alpha}\left\{A_{\kappa}F_{\beta\sigma}F^{\mu\nu}-A_{\beta}\partial_{\sigma}\left(A_{\kappa}F^{\mu\nu}\right)\right\} + \uO(\theta^{3})\bigg],\end{split}
\end{equation}
which, substituting for the usual anomaly on the right-hand side, reproduces the map \eqref{anom.m} for the two-dimensional case. 

For arbitrary fields, the derivative corrections to the map in the four-dimensional case are next computed. Now the noncommutative anomaly takes the form
\begin{equation}\label{120-d}\begin{split}
\xnc{\mathscr{A}} &= \frac{1}{16\pi^{2}}\varepsilon_{\mu\nu\lambda\rho}\xnc{F}^{\mu\nu}\star\xnc{F}^{\lambda\rho}\\
&= \frac{1}{16\pi^{2}}\varepsilon_{\mu\nu\lambda\rho}\bigg[F^{\mu\nu}F^{\lambda\rho}+\theta^{\alpha\beta}\partial_{\alpha}\left(A_{\beta}F^{\mu\nu}F^{\lambda\rho}\right)\\
&\qquad\qquad\qquad {}+\frac{1}{2}\theta^{\alpha\beta}\theta^{\kappa\sigma}\partial_{\alpha}\Big\{A_{\kappa}F_{\beta\sigma}F^{\mu\nu}F^{\lambda\rho}-A_{\beta}\partial_{\sigma}\left(A_{\kappa}F^{\mu\nu}F^{\lambda\rho}\right)\Big\}\bigg]\\
&\quad {}-\frac{1}{128\pi^{2}}\varepsilon_{\mu\nu\lambda\rho}\theta^{\alpha\beta}\theta^{\kappa\sigma}\partial_{\alpha}\partial_{\kappa}F^{\mu\nu}\partial_{\beta}\partial_{\sigma}F^{\lambda\rho} + \uO(\theta^{3}).\end{split}
\end{equation}
The last term is the new piece added to Eq.~\eqref{120}. Thus, the map \eqref{anom.m} gets modified as
\begin{equation}\label{anomaly4-m}\begin{split}
\xnc{\mathscr{A}} &= \mathscr{A}+\theta^{\alpha\beta}\partial_{\alpha}\left(A_{\beta}\mathscr{A}\right)
+\frac{1}{2}\theta^{\alpha\beta}\theta^{\kappa\sigma}\partial_{\alpha}\left[A_{\kappa}F_{\beta\sigma}\mathscr{A}-A_{\beta}\partial_{\sigma}\left(A_{\kappa}\mathscr{A}\right)\right]\\
&\quad {}-\frac{1}{128\pi^{2}}\varepsilon_{\mu\nu\lambda\rho}\theta^{\alpha\beta}\theta^{\kappa\sigma}\partial_{\alpha}\left(\partial_{\kappa}F^{\mu\nu}\partial_{\beta}\partial_{\sigma}F^{\lambda\rho}\right) + \uO(\theta^{3}).\end{split}
\end{equation}
This is reproduced by including a derivative correction to the classical map \eqref{J-hat2.m} for currents:
\begin{equation}\label{J-hat5.m}
\begin{split}
\xnc{J}^{\mu}_{5} &= J^{\mu}_{5}-\theta^{\alpha\beta}\left(A_{\alpha}\partial_{\beta}J^{\mu}_{5}-\frac{1}{2}F_{\alpha\beta}J^{\mu}_{5}\right)+\theta^{\mu\alpha}F_{\alpha\beta}J^{\beta}_{5}\\
&\quad {}+\frac{1}{2}\theta^{\alpha\beta}\theta^{\kappa\sigma}\partial_{\alpha}\left(A_{\kappa}F_{\beta\sigma}J^{\mu}_{5}-A_{\beta}A_{\kappa}\partial_{\sigma}J^{\mu}_{5}+\frac{1}{2}A_{\beta}F_{\kappa\sigma}J^{\mu}_{5} \right)-\theta^{\alpha\beta}\theta^{\kappa\mu}\partial_{\alpha}\left(A_{\beta}F_{\kappa\nu}J^{\nu}_{5}\right)\\
&\quad {}+\frac{1}{128\pi^{2}}\varepsilon_{\sigma\nu\lambda\rho}\theta^{\alpha\beta}\theta^{\kappa\mu}\partial_{\alpha}F^{\sigma\nu}\partial_{\kappa}\partial_{\beta}F^{\lambda\rho} + \uO(\theta^{3}),
\end{split}
\end{equation}
with the correction term given at the end. It is straightforward to see the contribution of this derivative term. Since this is an $\uO(\theta^2)$ term and we are restricting ourselves to the second order itself, taking its noncommutative covariant derivative amounts to just taking its ordinary partial derivative. Then taking into account the antisymmetric nature of $\theta^{\kappa\mu}$ it immediately yields the corresponding term in Eq.~\eqref{anomaly4-m}. We therefore interpret this term as a quantum correction for correctly mapping anomalies for arbitrary fields.

It is to be noted that Eq.~\eqref{anomaly4-m} can be put in a form so that the $\theta$-dependent terms are all expressed as a total derivative. This implies
\begin{equation}\label{xx}
\int\!\mathrm{d}^{4}x\,\xnc{\mathrm{D}}_{\mu}\star\xnc{J}^{\mu}_{5} = \int\!\mathrm{d}^{4}x\,\partial_{\mu}J^{\mu}_{5},
\end{equation}
reproducing the familiar equivalence of the integrated anomalies \cite{Banerjee:2003ce, Ardalan:2000cy, Gracia-Bondia:2000pz, Banerjee:2001un}.

We shall now give some useful inverse maps. From maps \eqref{110} and \eqref{J-hat2.m}, the inverse map for the currents follows:
\begin{equation}\label{123}\begin{split}
J^{\mu} &= \xnc{J}^{\mu}+\theta^{\alpha\beta}\left(\xnc{A}_{\alpha}\partial_{\beta}\xnc{J}^{\mu}-\frac{1}{2}\xnc{F}_{\alpha\beta}\xnc{J}^{\mu}\right)-\theta^{\mu\alpha}\xnc{F}_{\alpha\beta}\xnc{J}^{\beta}\\
&\quad {}-\frac{1}{2}\theta^{\alpha\beta}\theta^{\kappa\sigma}\left[\xnc{A}_{\kappa}\partial_{\beta}\xnc{F}_{\sigma\alpha}\xnc{J}^{\mu}-\xnc{A}_{\alpha}\xnc{A}_{\kappa}\partial_{\beta}\partial_{\sigma}\xnc{J}^{\mu}-2\xnc{A}_{\alpha}\partial_{\beta}\xnc{A}_{\kappa}\partial_{\sigma}\xnc{J}^{\mu}-\frac{1}{2}\xnc{A}_{\kappa}\xnc{F}_{\alpha\beta}\partial_{\sigma}\xnc{J}^{\mu}\right.\\
&\qquad\qquad\qquad \left. {}+\frac{3}{2}\xnc{A}_{\alpha}\partial_{\beta}\left(\xnc{F}_{\kappa\sigma}\xnc{J}^{\mu}\right)+\frac{1}{2}\xnc{F}_{\alpha\kappa}\xnc{F}_{\sigma\beta}\xnc{J}^{\mu}-\frac{1}{4}\xnc{F}_{\alpha\beta}\xnc{F}_{\kappa\sigma}\xnc{J}^{\mu}\right]\\
&\quad {}-\theta^{\alpha\beta}\theta^{\kappa\mu}\partial_{\alpha}\left(\xnc{A}_{\beta}\xnc{F}_{\kappa\nu}\xnc{J}^{\nu}\right) + \uO(\theta^{3}).\end{split}
\end{equation}
Taking the ordinary derivative and doing some simplifications yields
\begin{equation}\label{125}
\begin{split}
\partial_{\mu}J^{\mu} &= \xnc{\mathrm{D}}_{\mu}\star\xnc{J}^{\mu}-\theta^{\alpha\beta}\partial_{\alpha}\left[\xnc{A}_{\beta}\left(\xnc{\mathrm{D}}_{\mu}\star\xnc{J}^{\mu}\right)\right] \\
&\quad\, {} +\frac{1}{2}\theta^{\alpha\beta}\theta^{\kappa\sigma}\partial_{\alpha}\partial_{\kappa}\left[\xnc{A}_{\beta}\xnc{A}_{\sigma}\left(\xnc{\mathrm{D}}_{\mu}\star\xnc{J}^{\mu}\right)\right] + \uO(\theta^{3}),
\end{split}
\end{equation}
which may be regarded as the inverse map of \eqref{115}. Indeed, use of this relation reduces the expression on the right-hand side of Eq.~\eqref{115} to that on its left-hand side which shows the consistency of the results. This also proves that the covariant conservation of $\xnc{J}^{\mu}$ implies the ordinary conservation of $J^{\mu}$, as expected.

Likewise, inverting the relation \eqref{107}, we obtain
\begin{equation}\label{127}\begin{split}
F_{\mu\nu} &= \xnc{F}_{\mu\nu}+\theta^{\alpha\beta}\left(\xnc{A}_{\alpha}\partial_{\beta}\xnc{F}_{\mu\nu}+\xnc{F}_{\mu\alpha}\xnc{F}_{\beta\nu}\right)\\
&\quad {}+\theta^{\alpha\beta}\theta^{\kappa\sigma}\left[\xnc{A}_{\alpha}\partial_{\beta}\xnc{A}_{\kappa}\partial_{\sigma}\xnc{F}_{\mu\nu}+\frac{1}{2}\xnc{A}_{\alpha}\xnc{A}_{\kappa}\partial_{\beta}\partial_{\sigma}\xnc{F}_{\mu\nu}+\xnc{A}_{\alpha}\partial_{\beta}\left(\xnc{F}_{\mu\kappa}\xnc{F}_{\sigma\nu}\right)+\xnc{F}_{\mu\alpha}\xnc{F}_{\beta\kappa}\xnc{F}_{\sigma\nu}\right]\\
&\quad {} + \uO(\theta^{3}).\end{split}
\end{equation}
If we now write down the usual anomaly as
\begin{equation}\label{126}
\frac{1}{16\pi^{2}}\varepsilon_{\mu\nu\lambda\rho}F^{\mu\nu}F^{\lambda\rho} = \frac{1}{16\pi^{2}}\varepsilon_{\mu\nu\lambda\rho}\left(F^{\mu\nu}\star F^{\lambda\rho}+\frac{1}{8}\theta^{\alpha\beta}\theta^{\kappa\sigma}\partial_{\alpha}\partial_{\kappa}F^{\mu\nu}\partial_{\beta}\partial_{\sigma}F^{\lambda\rho} + \uO(\theta^{3})\right),
\end{equation}
and use Eq.~\eqref{127} on the right-hand side, we get
\begin{equation}\label{128}\begin{split}
\frac{1}{16\pi^{2}}\varepsilon_{\mu\nu\lambda\rho}F^{\mu\nu}F^{\lambda\rho} &= \frac{1}{16\pi^{2}}\varepsilon_{\mu\nu\lambda\rho}\bigg[\xnc{F}^{\mu\nu}\star\xnc{F}^{\lambda\rho}-\theta^{\alpha\beta}\partial_{\alpha}\left\{\xnc{A}_{\beta}\left(\xnc{F}^{\mu\nu}\star\xnc{F}^{\lambda\rho}\right)\right\}\\
&\qquad\qquad\qquad {}+\frac{1}{2}\theta^{\alpha\beta}\theta^{\kappa\sigma}\partial_{\alpha}\partial_{\kappa}\Big\{\xnc{A}_{\beta}\xnc{A}_{\sigma}\left(\xnc{F}^{\mu\nu}\star\xnc{F}^{\lambda\rho}\right)\Big\}\bigg]\\
&\quad {}+\frac{1}{128\pi^{2}}\varepsilon_{\mu\nu\lambda\rho}\theta^{\alpha\beta}\theta^{\kappa\sigma}\partial_{\alpha}\partial_{\kappa}\xnc{F}^{\mu\nu}\partial_{\beta}\partial_{\sigma}\xnc{F}^{\lambda\rho} + \uO(\theta^{3}),\end{split}
\end{equation}
where we have used the identities \eqref{118} and \eqref{119} with the replacement  $F^{\mu\nu}\rightarrow\xnc{F}^{\mu\nu}$. Thus we have the map for the anomalies:
\begin{equation}\label{129}\begin{split}
\partial_{\mu}J^{\mu}_{5} &= \xnc{\mathrm{D}}_{\mu}\star\xnc{J}^{\mu}_{5}-\theta^{\alpha\beta}\partial_{\alpha}\left[\xnc{A}_{\beta}\left(\xnc{\mathrm{D}}_{\mu}\star\xnc{J}^{\mu}_{5}\right)\right]+\frac{1}{2}\theta^{\alpha\beta}\theta^{\kappa\sigma}\partial_{\alpha}\partial_{\kappa}\left[\xnc{A}_{\beta}\xnc{A}_{\sigma}\left(\xnc{\mathrm{D}}_{\mu}\star\xnc{J}^{\mu}_{5}\right)\right]\\
&\quad {}+\frac{1}{128\pi^{2}}\varepsilon_{\mu\nu\lambda\rho}\theta^{\alpha\beta}\theta^{\kappa\sigma}\partial_{\alpha}\partial_{\kappa}\xnc{F}^{\mu\nu}\partial_{\beta}\partial_{\sigma}\xnc{F}^{\lambda\rho} + \uO(\theta^{3}).\end{split}
\end{equation}
In the slowly-varying-field approximation, the last term drops out. Then it mimics the usual map \eqref{125}. Again, as before, it is possible to find the correction term for arbitrary fields and write down the map for anomalous current as
\begin{equation}\label{130}\begin{split}
J^{\mu}_{5} &= \xnc{J}^{\mu}_{5}+\theta^{\alpha\beta}\left(\xnc{A}_{\alpha}\partial_{\beta}\xnc{J}^{\mu}_{5}-\frac{1}{2}\xnc{F}_{\alpha\beta}\xnc{J}^{\mu}_{5}\right)-\theta^{\mu\alpha}\xnc{F}_{\alpha\beta}\xnc{J}^{\beta}_{5}\\
&\quad {}-\frac{1}{2}\theta^{\alpha\beta}\theta^{\kappa\sigma}\left[\xnc{A}_{\kappa}\partial_{\beta}\xnc{F}_{\sigma\alpha}\xnc{J}^{\mu}_{5}-\xnc{A}_{\alpha}\xnc{A}_{\kappa}\partial_{\beta}\partial_{\sigma}\xnc{J}^{\mu}_{5}-2\xnc{A}_{\alpha}\partial_{\beta}\xnc{A}_{\kappa}\partial_{\sigma}\xnc{J}^{\mu}_{5}-\frac{1}{2}\xnc{A}_{\kappa}\xnc{F}_{\alpha\beta}\partial_{\sigma}\xnc{J}^{\mu}_{5}\right.\\
&\qquad\qquad\qquad \left. {}+\frac{3}{2}\xnc{A}_{\alpha}\partial_{\beta}\left(\xnc{F}_{\kappa\sigma}\xnc{J}^{\mu}_{5}\right)+\frac{1}{2}\xnc{F}_{\alpha\kappa}\xnc{F}_{\sigma\beta}\xnc{J}^{\mu}_{5}-\frac{1}{4}\xnc{F}_{\alpha\beta}\xnc{F}_{\kappa\sigma}\xnc{J}^{\mu}_{5}\right]\\
&\quad {}-\theta^{\alpha\beta}\theta^{\kappa\mu}\partial_{\alpha}\left(\xnc{A}_{\beta}\xnc{F}_{\kappa\nu}\xnc{J}^{\nu}_{5}\right)-\frac{1}{128\pi^{2}}\varepsilon_{\sigma\nu\lambda\rho}\theta^{\alpha\beta}\theta^{\kappa\mu}\partial_{\alpha}\xnc{F}^{\sigma\nu}\partial_{\kappa}\partial_{\beta}\xnc{F}^{\lambda\rho} + \uO(\theta^{3}),\end{split}
\end{equation}
which reproduces Eq.~\eqref{129} correctly. Substituting this map, the expression on the right-hand side of Eq.~\eqref{J-hat5.m} reduces to that on its left-hand side, which shows the consistency of the results.

Now we provide a mapping between modified chiral currents which are anomaly-free but no longer gauge invariant. In the ordinary (commutative) theory, such a modified chiral current may be defined as
\begin{equation}\label{Jcal}
\mathcal{J}^{\mu} = J^{\mu}_{5}-\frac{1}{8\pi^{2}}\varepsilon^{\mu\nu\lambda\rho}A_{\nu}F_{\lambda\rho}.
\end{equation}
By construction, this is anomaly-free ($\partial_{\mu}\mathcal{J}^{\mu}=0$) but no longer gauge-invariant. It is possible to do a similar thing for the noncommutative theory. We rewrite Eq.~\eqref{J-hat5.m} by replacing $J^{\mu}_{5}$ in favour of $\mathcal{J}^{\mu}$. The terms independent of $\mathcal{J}^{\mu}$, including the quantum correction, are then moved to the other side and a new current is defined as
\begin{equation}\label{Jcalhat}
\xnc{\mathcal{J}}^{\mu} = \xnc{J}^{\mu}_{5}+\xnc{X}^{\mu}(\xnc{A}),
\end{equation}
where all $A_{\mu}$-dependent terms lumped in $\xnc{X}^{\mu}$ have been expressed in terms of the noncommutative variables using the Seiberg--Witten map. Thus we have
\begin{equation}\label{111cal}\begin{split}
\xnc{\mathcal{J}}^{\mu} &= \mathcal{J}^{\mu}-\theta^{\alpha\beta}\left(A_{\alpha}\partial_{\beta}\mathcal{J}^{\mu}-\frac{1}{2}F_{\alpha\beta}\mathcal{J}^{\mu}\right)+\theta^{\mu\alpha}F_{\alpha\beta}\mathcal{J}^{\beta}\\
&\quad {}+\frac{1}{2}\theta^{\alpha\beta}\theta^{\kappa\sigma}\partial_{\alpha}\left(A_{\kappa}F_{\beta\sigma}\mathcal{J}^{\mu}-A_{\beta}A_{\kappa}\partial_{\sigma}\mathcal{J}^{\mu}+\frac{1}{2}A_{\beta}F_{\kappa\sigma}\mathcal{J}^{\mu}\right)
-\theta^{\alpha\beta}\theta^{\kappa\mu}\partial_{\alpha}\left(A_{\beta}F_{\kappa\nu}\mathcal{J}^{\nu}\right)\\
&\quad {} + \uO(\theta^{3}).\end{split}
\end{equation}
Since the above equation is structurally identical to Eq.~\eqref{J-hat2.m}, a relation akin to \eqref{115} follows:
\begin{equation}\label{115cal}\begin{split}
\xnc{\mathrm{D}}_{\mu}\star\xnc{\mathcal{J}}^{\mu} &= \partial_{\mu}\mathcal{J}^{\mu}+\theta^{\alpha\beta}\partial_{\alpha}\left(A_{\beta}\partial_{\mu}\mathcal{J}^{\mu}\right)\\
&\quad {}+\frac{1}{2}\theta^{\alpha\beta}\theta^{\kappa\sigma}\partial_{\alpha}\left[A_{\kappa}F_{\beta\sigma}\partial_{\mu}\mathcal{J}^{\mu}-A_{\beta}\partial_{\sigma}\left(A_{\kappa}\partial_{\mu}\mathcal{J}^{\mu}\right)\right] + \uO(\theta^{3}),\end{split}
\end{equation}
which shows that $\partial_{\mu}\mathcal{J}^{\mu}=0$ implies $\xnc{\mathrm{D}}_{\mu}\star\xnc{\mathcal{J}}^{\mu}=0$. We are thus successful in constructing  an anomaly-free current which however does not transform (star-) covariantly. It is the $\xnc{X}^{\mu}$ appearing in Eq.~\eqref{Jcalhat} which spoils the covariance of $\xnc{\mathcal{J}}^{\mu}$.


\paragraph{Higher-order computations.}
Results discussed so far were valid up to $\uO(\theta^{2})$. A natural question that arises is the validity of these results for further higher-order corrections. Here we face a problem. The point is that although the map \eqref{109} for sources is given in a closed form, its explicit structure is dictated by the map involving the potentials. Thus one has to first construct the latter map before proceeding. All these features make higher- (than $\uO(\theta^{2})$) order computations very formidable, if not practically impossible. An alternate approach is suggested, which is explicitly demonstrated by considering $\uO(\theta^{3})$ calculations.

Consider first the two-dimensional example. The star-gauge-covariant anomaly, after an application of the Seiberg--Witten map, is given by
\begin{equation}\label{F-3.m1}
\xnc{\mathscr{A}}_{2\mathrm{d}} = \frac{1}{2\pi}\varepsilon_{\mu\nu}\xnc{F}^{\mu\nu} = {\mathscr{A}}_{2\mathrm{d}}^{(0)}+{\mathscr{A}}_{2\mathrm{d}}^{(1)}+{\mathscr{A}}_{2\mathrm{d}}^{(2)}+{\mathscr{A}}_{2\mathrm{d}}^{(3)} + \uO(\theta^{4}),
\end{equation}
with ${\mathscr{A}}_{2\mathrm{d}}^{(0)}$, ${\mathscr{A}}_{2\mathrm{d}}^{(1)}$ and ${\mathscr{A}}_{2\mathrm{d}}^{(2)}$ respectively being the zeroth-, first- and second-order (in $\theta$) parts already appearing on the right-hand side of Eq.~\eqref{an3-2d}, and
\begin{equation}\label{F-3.m}\begin{split}
{\mathscr{A}}_{2\mathrm{d}}^{(3)} &= - \frac{1}{12\pi}\varepsilon_{\mu\nu}\theta^{\alpha\beta}\theta^{\kappa\sigma}\theta^{\tau\xi}\\
&\qquad \times\left[A_{\alpha}\partial_{\beta}\left\{A_{\kappa}\partial_{\sigma}\left(A_{\tau}\partial_{\xi}F^{\mu\nu}+3{F^{\mu}}_{\tau}{F_{\xi}}^{\nu}\right)+2F_{\sigma\tau}\left(A_{\kappa}\partial_{\xi}F^{\mu\nu}+3{F^{\mu}}_{\kappa}{F_{\xi}}^{\nu}\right)\right\}\right.\\
&\qquad\quad \left.{}+A_{\alpha}F_{\beta\kappa}\partial_{\sigma}\left(A_{\tau}\partial_{\xi}F^{\mu\nu}+3{F^{\mu}}_{\tau}{F_{\xi}}^{\nu}\right)+2F_{\beta\kappa}F_{\sigma\tau}\left(A_{\alpha}\partial_{\xi}F^{\mu\nu}+3{F^{\mu}}_{\alpha}{F_{\xi}}^{\nu}\right)\right],\end{split}
\end{equation}
where the $\uO(\theta^{3})$ contribution to the map \eqref{107} has been taken from \refcite{Fidanza:2001qm}.

Now our objective is to rewrite the $\uO(\theta^{3})$ contribution in a form akin to $\uO(\theta)$ and $\uO(\theta^{2})$ terms; namely, to recast it as something proportional to the commutative anomaly ($\varepsilon_{\mu\nu}F^{\mu\nu}$), and also as a total derivative. Expressing it as a total derivative is necessary to preserve the equality of the integrated anomalies ($\int\!\mathrm{d}^{2}x\,\varepsilon_{\mu\nu}\xnc{F}^{\mu\nu} = \int\!\mathrm{d}^{2}x\,\varepsilon_{\mu\nu}F^{\mu\nu}$) \cite{Banerjee:2004rs, Banerjee:2003vc, Banerjee:2003ce, Ardalan:2000cy, Gracia-Bondia:2000pz}.

The $\uO(\theta^{3})$ contribution may be expressed as
\begin{equation}\label{an-o3.2d}\begin{split}
{\mathscr{A}}_{2\mathrm{d}}^{(3)} &= -\frac{1}{12\pi}\varepsilon_{\mu\nu}\theta^{\alpha\beta}\theta^{\kappa\sigma}\theta^{\tau\xi}\left[A_{\alpha}\partial_{\beta}\left\{A_{\kappa}\partial_{\sigma}\left(A_{\tau}\partial_{\xi}F^{\mu\nu}-\frac{3}{2}F_{\tau\xi}F^{\mu\nu}\right)+2A_{\kappa}F_{\sigma\tau}\partial_{\xi}F^{\mu\nu}\right.\right.\\
&\qquad\qquad\qquad\qquad\qquad\qquad\qquad \left.{}+\frac{3}{4}\left(F_{\kappa\sigma}F_{\tau\xi}-2F_{\kappa\tau}F_{\xi\sigma}\right)F^{\mu\nu}\right\}\\
&\qquad\qquad\qquad\qquad\qquad {}+A_{\alpha}F_{\beta\kappa}\left\{\partial_{\sigma}\left(A_{\tau}\partial_{\xi}F^{\mu\nu}-\frac{3}{2}F_{\tau\xi}F^{\mu\nu}\right)+2F_{\sigma\tau}\partial_{\xi}F^{\mu\nu}\right\}\\
&\qquad\qquad\qquad\qquad\qquad \left.{}-\left(F_{\alpha\tau}F_{\xi\kappa}F_{\sigma\beta}+\frac{1}{8}F_{\alpha\beta}F_{\kappa\sigma}F_{\tau\xi}-\frac{3}{4}F_{\alpha\beta}F_{\kappa\tau}F_{\xi\sigma}\right)F^{\mu\nu}\right],\end{split}
\end{equation}
where, in addition to the identities \eqref{id-2d} and \eqref{id2-2d}, we have also used
\begin{equation}\label{id3-2d}\begin{split}
\varepsilon_{\mu\nu}\theta^{\alpha\beta}\theta^{\kappa\sigma}\theta^{\tau\xi}&\left(F^{\mu\nu}F_{\alpha\tau}F_{\xi\kappa}F_{\sigma\beta}+{F^{\mu}}_{\kappa}{F_{\sigma}}^{\nu}F_{\alpha\tau}F_{\xi\beta}\right.\\
&\left.{}+{F^{\mu}}_{\kappa}F_{\sigma\tau}{F_{\xi}}^{\nu}F_{\alpha\beta}+6{F^{\mu}}_{\alpha}F_{\beta\kappa}F_{\sigma\tau}{F_{\xi}}^{\nu}\right) =0,\end{split}
\end{equation}
which follows from the identity \eqref{id2-2d} by doing the replacement $F^{\mu\nu}\rightarrow\xnc{F}^{\mu\nu}$ followed by exploiting the Seiberg--Witten map and retaining $\uO(\theta^{3})$ terms. We notice that each term on the right-hand side of Eq.~\eqref{an-o3.2d} contains the usual anomaly, as desired. After some algebra, the right-hand side of Eq.~\eqref{an-o3.2d} can be written as a total divergence, which gives us the final improved version of the map \eqref{an3-2d} as
\begin{equation}\label{an5-2d}\begin{split}
\xnc{\mathscr{A}}_{2\mathrm{d}} &= \frac{1}{2\pi}\varepsilon_{\mu\nu}\bigg[F^{\mu\nu}+\theta^{\alpha\beta}\partial_{\alpha}\left(A_{\beta}F^{\mu\nu}\right)+\frac{1}{2}\theta^{\alpha\beta}\theta^{\kappa\sigma}\partial_{\alpha}\left\{A_{\kappa}F_{\beta\sigma}F^{\mu\nu}-A_{\beta}\partial_{\sigma}\left(A_{\kappa}F^{\mu\nu}\right)\right\}\\
&\qquad\qquad\quad {}+\frac{1}{6}\theta^{\alpha\beta}\theta^{\kappa\sigma}\theta^{\tau\xi}\partial_{\alpha}\bigg\{F^{\mu\nu}\bigg(2A_{\tau}F_{\xi\kappa}F_{\sigma\beta}-2A_{\beta}A_{\kappa}\partial_{\sigma}F_{\tau\xi}-\frac{3}{2}A_{\beta}F_{\kappa\tau}F_{\xi\sigma}\\
&\qquad\qquad\qquad\qquad\qquad\qquad\qquad\quad {}+\frac{1}{4}A_{\beta}F_{\kappa\sigma}F_{\tau\xi}-A_{\beta}\partial_{\sigma}\left(A_{\tau}F_{\xi\kappa}\right)-\frac{1}{2}A_{\kappa}F_{\sigma\beta}F_{\tau\xi}\bigg)\\
&\qquad\qquad\qquad\qquad\qquad\qquad\quad {}+\partial_{\xi}F^{\mu\nu}\left[A_{\beta}A_{\kappa}\left(\partial_{\sigma}A_{\tau}+2F_{\sigma\tau}\right)-A_{\tau}\left(A_{\kappa}F_{\beta\sigma}+A_{\beta}F_{\kappa\sigma}\right)\right]\\
&\qquad\qquad\qquad\qquad\qquad\qquad\quad {}+A_{\beta}A_{\kappa}A_{\tau}\partial_{\sigma}\partial_{\xi}F^{\mu\nu}\bigg\} + \uO(\theta^{4})\bigg].\end{split}
\end{equation}
Thus, in two dimensions, the noncommutative anomaly can be written in terms of the usual anomaly at $\uO(\theta^{3})$ also:
\begin{equation}\label{an51-2d}\begin{split}
\xnc{\mathscr{A}}_{2\mathrm{d}} &= \mathscr{A}_{2\mathrm{d}}+\theta^{\alpha\beta}\partial_{\alpha}\left(A_{\beta}\mathscr{A}_{2\mathrm{d}}\right)+\frac{1}{2}\theta^{\alpha\beta}\theta^{\kappa\sigma}\partial_{\alpha}\left\{A_{\kappa}F_{\beta\sigma}\mathscr{A}_{2\mathrm{d}}-A_{\beta}\partial_{\sigma}\left(A_{\kappa}\mathscr{A}_{2\mathrm{d}}\right)\right\}\\
&\quad {}+\frac{1}{6}\theta^{\alpha\beta}\theta^{\kappa\sigma}\theta^{\tau\xi}\partial_{\alpha}\bigg[\mathscr{A}_{2\mathrm{d}}\bigg(2A_{\tau}F_{\xi\kappa}F_{\sigma\beta}-2A_{\beta}A_{\kappa}\partial_{\sigma}F_{\tau\xi}-\frac{3}{2}A_{\beta}F_{\kappa\tau}F_{\xi\sigma}\\
&\qquad\qquad\qquad\qquad\qquad\quad {}+\frac{1}{4}A_{\beta}F_{\kappa\sigma}F_{\tau\xi}-A_{\beta}\partial_{\sigma}\left(A_{\tau}F_{\xi\kappa}\right)-\frac{1}{2}A_{\kappa}F_{\sigma\beta}F_{\tau\xi}\bigg)\\
&\qquad\qquad\qquad\qquad\quad {}+\partial_{\xi}\mathscr{A}_{2\mathrm{d}}\left\{A_{\beta}A_{\kappa}\left(\partial_{\sigma}A_{\tau}+2F_{\sigma\tau}\right)-A_{\tau}\left(A_{\kappa}F_{\beta\sigma}+A_{\beta}F_{\kappa\sigma}\right)\right\}\\
&\qquad\qquad\qquad\qquad\quad {}+A_{\beta}A_{\kappa}A_{\tau}\partial_{\sigma}\partial_{\xi}\mathscr{A}_{2\mathrm{d}}\bigg] + \uO(\theta^{4}).\end{split}
\end{equation}

If the anomalies in four dimensions also satisfy the above map, then clearly we have a general result, valid up to $\uO(\theta^{3})$. Now it will be shown that, in the slowly-varying-field approximation, such a relation indeed holds. We have
\begin{equation}\label{an5-4d}\begin{split}
\lefteqn{\frac{1}{16\pi^{2}}\varepsilon_{\mu\nu\lambda\rho}\xnc{F}^{\mu\nu}\xnc{F}^{\lambda\rho}}\\
&= \frac{1}{16\pi^{2}}\varepsilon_{\mu\nu\lambda\rho}\bigg[F^{\mu\nu}F^{\lambda\rho}+\theta^{\alpha\beta}\partial_{\alpha}\left(A_{\beta}F^{\mu\nu}F^{\lambda\rho}\right)\\
&\qquad\qquad\qquad {}+\frac{1}{2}\theta^{\alpha\beta}\theta^{\kappa\sigma}\partial_{\alpha}\left\{A_{\kappa}F_{\beta\sigma}F^{\mu\nu}F^{\lambda\rho}-A_{\beta}\partial_{\sigma}\left(A_{\kappa}F^{\mu\nu}F^{\lambda\rho}\right)\right\}\\
&\qquad\qquad\qquad {}+\frac{1}{6}\theta^{\alpha\beta}\theta^{\kappa\sigma}\theta^{\tau\xi}\partial_{\alpha}\bigg\{F^{\mu\nu}F^{\lambda\rho}\bigg(2A_{\tau}F_{\xi\kappa}F_{\sigma\beta}-2A_{\beta}A_{\kappa}\partial_{\sigma}F_{\tau\xi}-\frac{3}{2}A_{\beta}F_{\kappa\tau}F_{\xi\sigma}\\
&\qquad\qquad\qquad\qquad\qquad\qquad\qquad\qquad\qquad {}+\frac{1}{4}A_{\beta}F_{\kappa\sigma}F_{\tau\xi}-A_{\beta}\partial_{\sigma}\left(A_{\tau}F_{\xi\kappa}\right)-\frac{1}{2}A_{\kappa}F_{\sigma\beta}F_{\tau\xi}\bigg)\\
&\qquad\qquad\qquad\qquad\qquad\qquad\qquad {}+\partial_{\xi}\left(F^{\mu\nu}F^{\lambda\rho}\right)\left[A_{\beta}A_{\kappa}\left(\partial_{\sigma}A_{\tau}+2F_{\sigma\tau}\right)\right.\\
&\qquad\qquad\qquad\qquad\qquad\qquad\qquad\qquad\qquad\qquad\quad \left.{}
-A_{\tau}\left(A_{\kappa}F_{\beta\sigma}+A_{\beta}F_{\kappa\sigma}\right)\right]\\
&\qquad\qquad\qquad\qquad\qquad\qquad\qquad {}+A_{\beta}A_{\kappa}A_{\tau}\partial_{\sigma}\partial_{\xi}\left(F^{\mu\nu}F^{\lambda\rho}\right)\bigg\} + \uO(\theta^{4})\bigg].\end{split}
\end{equation}\label{id3-4d}
In obtaining this equation, it is necessary to use the identities \eqref{118} and \eqref{119}, and a new one (given below), which follows from the identity \eqref{119} by doing the replacement $F^{\mu\nu}\rightarrow\xnc{F}^{\mu\nu}$ followed by using the Seiberg--Witten map and retaining $\uO(\theta^{3})$ terms:
\begin{equation}\begin{split}
\varepsilon_{\mu\nu\lambda\rho}\theta^{\alpha\beta}\theta^{\kappa\sigma}\theta^{\tau\xi}&\bigg(6{F^{\mu}}_{\alpha}{F_{\beta}}^{\nu}{F^{\lambda}}_{\kappa}F_{\sigma\tau}{F_{\xi}}^{\rho}+6F^{\mu\nu}{F^{\lambda}}_{\alpha}F_{\beta\kappa}F_{\sigma\tau}{F_{\xi}}^{\rho}\\
&\quad {}+F^{\mu\nu}{F^{\lambda}}_{\kappa}{F_{\sigma}}^{\rho}F_{\alpha\tau}F_{\xi\beta}+F^{\mu\nu}{F^{\lambda}}_{\kappa}F_{\sigma\tau}{F_{\xi}}^{\rho}F_{\alpha\beta}\\
&\quad {}+\frac{1}{2}{F^{\mu}}_{\tau}{F_{\xi}}^{\nu}{F^{\lambda}}_{\kappa}{F_{\sigma}}^{\rho}F_{\alpha\beta}+\frac{1}{2}F^{\mu\nu}F^{\lambda\rho}F_{\alpha\tau}F_{\xi\kappa}F_{\sigma\beta}\bigg) = 0.\end{split}
\end{equation}
Obviously, Eq.~\eqref{an5-4d} reproduces the map \eqref{an51-2d}, with $\xnc{\mathscr{A}}_{2\mathrm{d}}$ and $\mathscr{A}_{2\mathrm{d}}$ replaced by the corresponding expressions in four dimensions. This proves our claim.

Starting from the results in two dimensions, it is thus feasible to infer the general structure valid in higher dimensions. This is an outcome of the topological properties of anomalies. Proceeding in this fashion, the map for the anomalies can be extended to higher orders.


\section{\label{sec:current-conclu}Discussion}

To put our results in a proper perspective, let us recall that the Seiberg--Witten maps are classical maps. \emph{A priori}, therefore, it was not clear whether they had any role in the mapping of anomalies which are essentially of quantum origin. The first hint that such a possibility might exist came from Eq.~\eqref{115}, or Eq.~\eqref{anom.m}, where the covariant derivative of the noncommutative covariant current was expressed in terms of the ordinary derivative of the commutative current. Indeed, to put the map in this form was quite nontrivial. Classically, such a map was trivially consistent, since both the covariant divergence in the noncommutative description and the ordinary divergence in the usual (commutative) picture vanish. The remarkable feature, however, was that such a map remained valid even for the quantum case in the slowly-varying-field approximation which was checked explicitly by inserting the familiar anomalies in the different descriptions (the planar anomaly for the noncommutative description and the ABJ anomaly for the commutative case). Incidentally, the slowly-varying-field approximation is quite significant in discussions of the Seiberg--Witten maps. For instance, it was in this approximation that the equivalence of the DBI actions in the noncommutative and the commutative pictures was established \cite{Seiberg:1999vs} through the use of Seiberg--Witten maps.

Our analysis has certain implications for the mapping among the effective actions (for chiral theories) obtained by integrating out the matter degrees of freedom. The point is that the anomalies are the gauge-variations of the effective actions and if the anomalies get mapped then one expects that, modulo local counterterms, the effective actions might get identified, i.e., it suggests that
\begin{equation}\label{zzz}
\xnc{W}\left(\xnc{A}(A)\right) \equiv W(A) + \mbox{local counterterms},
\end{equation}
where $W$ and $\xnc{W}$ denote the effective actions in the commutative and noncommutative formulations, respectively. Taking the gauge-variations (with parameters $\lambda$ and $\xnc{\lambda}$), yields
\begin{equation}\label{xxx2}
\int\!\mathrm{d}^{4}x\,\left(\xnc{\mathrm{D}}_{\mu}\star\xnc{J}^{\mu}_{5}\right)\star\xnc{\lambda} = \int\!\mathrm{d}^{4}x\,\left(\partial_{\mu}J^{\mu}_{5}\right)\lambda+\int\!\mathrm{d}^{4}x\,\left(\partial_{\mu}\Lambda^{\mu}\right)\lambda,
\end{equation}
where
\begin{equation}\label{yyy}
\xnc{J}^{\mu}_{5} = \frac{\delta\xnc{W}}{\delta\xnc{A}_{\mu}}, \quad J^{\mu}_{5} = \frac{\delta W}{\delta A_{\mu}}
\end{equation}
and $\Lambda^{\mu}$ accounts for the ambiguity (local counterterms) in obtaining the effective actions. Now Eq.~\eqref{120-d} expresses the noncommutative anomaly in terms of the commutative variables. Using that result and the Seiberg--Witten map \eqref{102} for the gauge parameter $\xnc{\lambda}$ simplifies the left-hand side of Eq.~\eqref{xxx2}:
\begin{equation}\label{qqq}\begin{split}
\int\!\mathrm{d}^{4}x\,\left(\xnc{\mathrm{D}}_{\mu}\star\xnc{J}^{\mu}_{5}\right)\star\xnc{\lambda} &= \int\!\mathrm{d}^{4}x\,\left(\xnc{\mathrm{D}}_{\mu}\star\xnc{J}^{\mu}_{5}\right)\xnc{\lambda} = \frac{1}{16\pi^{2}}\varepsilon_{\mu\nu\lambda\rho}\int\!\mathrm{d}^{4}x\,\left(\xnc{F}^{\mu\nu}\star\xnc{F}^{\lambda\rho}\right)\xnc{\lambda}\\
&= \frac{1}{16\pi^{2}}\varepsilon_{\mu\nu\lambda\rho}\int\!\mathrm{d}^{4}x\,\left(F^{\mu\nu}F^{\lambda\rho}\right)\lambda+\int\!\mathrm{d}^{4}x\,\left(\partial_{\alpha}\Lambda^{\alpha}\right)\lambda,\end{split}
\end{equation}
where Eq.~\eqref{120-d} and the  map \eqref{102} have been used in the last step, and
\begin{equation}\label{Lambda}
\begin{split}
\Lambda^{\alpha} &= \frac{1}{16\pi^{2}}\varepsilon_{\mu\nu\lambda\rho}\left[\frac{1}{2}\theta^{\alpha\beta}A_{\beta}F^{\mu\nu}F^{\lambda\rho}+\theta^{\alpha\beta}\theta^{\kappa\sigma}\left(\frac{1}{3}A_{\kappa}F_{\beta\sigma}F^{\mu\nu}F^{\lambda\rho}+\frac{1}{6}A_{\beta}\partial_{\kappa}\left(A_{\sigma}F^{\mu\nu}F^{\lambda\rho}\right)\right.\right.\\
&\qquad\qquad\qquad\qquad\qquad\qquad\qquad\qquad\qquad \left.\left.{}-\frac{1}{8}\partial_{\kappa}F^{\mu\nu}\partial_{\beta}\partial_{\sigma}F^{\lambda\rho}\right)\right] + \uO(\theta^{3}),
\end{split}
\end{equation}
thereby proving Eq.~\eqref{xxx2} and establishing the claim \eqref{zzz}.

We further stress, to avoid any confusion, that the relation \eqref{zzz} was not assumed, either explicitly or implicitly, in our calculations.\footnote{Indeed, as already stated, there cannot be any \emph{a priori} basis for such an assumption since the classical Seiberg--Witten map need not be valid for mapping effective actions that take into account loop effects.} Rather, as shown here, our analysis suggested such a relation. Its explicit verification confirms the consistency of our approach. It should be mentioned that the map among anomalies \eqref{anom.m} follows from the map \eqref{J-hat2.m} for currents through a series of algebraic manipulations. This does not depend on the interpretation of the anomaly as gauge-variation of an effective action. If one sticks to this interpretation and furthermore \emph{assumes} the relation \eqref{zzz}, then it might be possible to get a relation, like Eq.~\eqref{xxx2}, involving the integrated version of the products of anomalies and gauge parameters. Our formulation always led to maps involving unintegrated anomalies or currents, which are more fundamental.

We also note that the map \eqref{anom.m} for the unintegrated anomalies, which follows from the basic map \eqref{J-hat2.m} among the currents, was only valid in the slowly-varying-field approximation. The suggested map \eqref{zzz} among the effective actions, on the other hand, led to the map \eqref{xxx2}, involving the integrated anomalies and the gauge parameters, that was valid in general. For the pure integrated anomalies we have the familiar map \eqref{xx} that has been discussed extensively in the literature \cite{Banerjee:2003ce, Ardalan:2000cy, Gracia-Bondia:2000pz, Banerjee:2001un}.


%% file: chap_anomalous.tex

\chapter{\label{chap:anomalous}Commutator anomalies in noncommutative electrodynamics}


The subject of anomalies in gauge theories has been studied extensively in the literature.\footnote{See \cite{Bertlmann:2000, FujikawaSuzuki:2004} for reviews.} Ever since the importance of noncommutative manifolds was realised\footnote{See \cite{Douglas:2001ba, Szabo:2001kg} for recent reviews.}, it has been natural to investigate the structure of anomalies in such a setting. Various results have been reported in this context. In particular, it has been noted \cite{Ardalan:2000qk} that, due to noncommutativity, two different currents can be defined even for a $\uU(1)$ theory. These are the star-gauge-invariant and the star-gauge-covariant currents which are defined according to their distinct gauge-transformation properties. In this chapter we shall be exclusively dealing with the star-gauge-covariant currents. Now the covariant divergence of the star-gauge-covariant axial current reveals an anomaly---this is the star-gauge-covariant anomaly \cite{Ardalan:2000cy, Gracia-Bondia:2000pz, Armoni:2002fh} which is basically the covariant deformation of the usual gauge-invariant ABJ anomaly \cite{Adler:1969gk, Bell:1969ts}.

The next logical step would be to compute the anomalous commutators involving the currents and see their connection with the anomaly, as happens for the commutative description \cite{Adler:1970qb, DeserOthers:1970, Jackiw:1968xt}. The structure of the anomalous commutators in the noncommutative setting, however, is lacking in the literature. This chapter is aimed at investigating this aspect. Here we would like to mention that the computation of noncommutative commutators from loop diagrams following the `Bjorken-limit' approach might not be practically feasible. Even in the ordinary case, the computation of anomalous commutators is much more involved than that of the divergence anomaly.

Based on the various results of the previous chapter, here we provide an approach to obtain the structure of the anomalous commutators in a noncommutative theory. We exploit the maps for fields and currents in a $\uU(1)$ gauge theory in noncommutative and commutative (usual) descriptions~\cite{Seiberg:1999vs, Banerjee:2003vc, Banerjee:2005yy} to express the commutators in the noncommutative theory in favour of their commutative counterparts, where the results are known \cite{Adler:1970qb, DeserOthers:1970}. Using these known results we obtain the explicit structure for the anomalous commutators in the noncommutative theory.

The new results on anomalous commutators in noncommutative electrodynamics are by themselves interesting. Their compatibility with the noncommutative divergence anomalies, exhibited through consistency conditions derived here, further supports our analysis. Moreover the computational method provides a nontrivial application of various Seiberg--Witten maps.

After enumerating the known results for ordinary anomalous commutators in the first part of section \ref{sec:anomalous-commut}, we compute the commutators in the noncommutative theory in the second part. Although we have considered massless quantum electrodynamics (QED) here, the structure of these commutators remains equally valid for the massive case as well. Explicit results are given for the current--current as well as the current--field commutators. The compatibility of our results for these anomalous commutators with the noncommutative covariant anomaly has been established in section \ref{sec:anomalous-consis} through the use of certain consistency conditions. It is known that in the ordinary theory there is a possibility of the presence of additional terms in some of the commutators. Last part of this section deals with the implications of these ambiguities on our scheme.


\section{\label{sec:anomalous-commut}Anomalous commutators}

Our method of computing the commutators is straightforward. The maps connecting the variables in the two descriptions will be used to express the commutators in the noncommutative theory in favour of their commutative counterparts. From a knowledge of the latter the former is easily obtained. We shall restrict to the first order in $\theta$. Let us enumerate the various anomalous commutators in the ordinary theory.


\subsection{Anomalous commutators in the ordinary theory}

We consider massless QED given by the Lagrangian density
\begin{equation}\label{S101}
\mathscr{L} = \ui\bar{\psi}\gamma^{\mu}\partial_{\mu}\psi-\frac{1}{4}F_{\mu\nu}F^{\mu\nu}-\bar{\psi}\gamma^{\mu}\psi A_{\mu},
\end{equation}
where the $(+,-,-,-)$ signature has been used. We shall take $\varepsilon_{0123}=\varepsilon_{123}=1$, $E^{i} = F_{0i}$, $B^{i} = -\varepsilon_{ijk}\partial_{j}A_{k}$ with $i,j,k = 1, 2, 3$. The equations of motion for the fields are
\begin{gather}
\label{Seom1}
\ui\gamma^{\mu}\partial_{\mu}\psi = \gamma^{\mu}\psi A_{\mu},\\
\label{Seom2}
\partial_{\nu}F^{\nu\mu} = J^{\mu},
\end{gather}
where $J^{\mu} = \bar{\psi}\gamma^{\mu}\psi$. The usual current conservation, $\partial_{\mu}J^{\mu} = 0$, follows upon using the equation of motion. The canonical anticommutator relations of the spinor fields are
\begin{equation}\label{Sacr}
\left\{\psi_{\alpha}(\mathbf{x},t), \psi^{\dagger}_{\beta}(\mathbf{y},t)\right\}
= \delta_{\alpha\beta}\delta^{3}(\mathbf{x}-\mathbf{y}),
\end{equation}
with $\alpha, \beta = 1, \ldots, 4$, the labels of the spinor components, and the canonical commutation relations of the photon fields in the Feynman gauge are
\begin{equation}\label{Scr}
\begin{aligned}
&\left[A_{\mu}(\mathbf{x},t), \partial_{0}A_{\nu}(\mathbf{y},t)\right] = -\ui\eta_{\mu\nu}\delta^{3}(\mathbf{x}-\mathbf{y}),\\
&\left[A_{\mu}(\mathbf{x},t), A_{\nu}(\mathbf{y},t)\right] = \left[\partial_{0}A_{\mu}(\mathbf{x},t), \partial_{0}A_{\nu}(\mathbf{y},t)\right] = 0.
\end{aligned}
\end{equation}
It has been shown \cite{Adler:1969gk, Bell:1969ts} that the axial-vector current does not satisfy the usual divergence equation $\partial_{\mu}J^{\mu}_{5} = 0$ expected from naive use of equations of motion.\footnote{Whether the index `5' appears as a subscript or as a superscript is a matter of notational convenience: $J^{\mu}_{5}= \bar{\psi}\gamma^{\mu}\gamma_{5}\psi$, $J_{\mu}^{5}= \bar{\psi}\gamma_{\mu}\gamma_{5}\psi$.} Rather it satisfies the anomalous divergence equation given by Eq.~\eqref{anom.d}. The commutators\footnote{All the commutators appearing in this chapter are equal-time commutators. By $[J_{0}(x), J_{0}^5(y)]$ we mean $[J_{0}(\mathbf{x},t), J_{0}^5(\mathbf{y},t)]$, and so on. Likewise, $S_{00}(x,y)$ appearing in Eq.~\eqref{CS101} is to be understood as $S_{00}(\mathbf{x}, \mathbf{y}, t)$, and similarly for others.} involving the axial current which are relevant in the present context are \cite{Adler:1970qb, DeserOthers:1970}
\begin{gather}
\label{CS101}
S_{00}(x,y) \equiv \left[J_{0}(x), J_{0}^5(y)\right] = \frac{\ui}{4\pi^2}\varepsilon_{ijk}F_{jk}(y)\partial_{i}^x\delta^{3}(\mathbf{x}-\mathbf{y}),\displaybreak[0]%
\\
\label{CS102}
S_{i0}(x,y) \equiv \left[J_{i}(x), J_{0}^5(y)\right] = -\frac{\ui}{4\pi^2}\varepsilon_{ijk}F_{0j}(x)\partial_{k}^y\delta^{3}(\mathbf{x}-\mathbf{y}),\displaybreak[0]%
\\
\label{CS103}
S_{0i}(x,y) \equiv \left[J_{0}(x), J_{i}^5(y)\right] = \frac{\ui}{4\pi^2}\varepsilon_{ijk}F_{0j}(y)\partial_{k}^x\delta^{3}(\mathbf{x}-\mathbf{y}),\displaybreak[0]%
\\
\label{CA101}
L_{\sigma\mu}(x,y) \equiv \left[A_{\sigma}(x), J_{\mu}^5(y)\right] = 0,\displaybreak[0]%
\\
\label{CB101}
M_{0\mu}(x,y) \equiv \left[\partial_{0}A_{0}(x), J_{\mu}^5(y)\right] = 0,\displaybreak[0]%
\\
\label{CB102}
M_{i0}(x,y) \equiv \left[\partial_{0}A_{i}(x), J_{0}^5(y)\right] = \frac{\ui}{4\pi^2}\varepsilon_{ijk}F_{jk}\delta^{3}(\mathbf{x}-\mathbf{y}),\displaybreak[0]%
\\
\label{CB103}
M_{im}(x,y) \equiv \left[\partial_{0}A_{i}(x), J_{m}^5(y)\right] = \frac{\ui}{4\pi^2}\varepsilon_{imn}F_{0n}\delta^{3}(\mathbf{x}-\mathbf{y}).\end{gather}
All of the nonvanishing commutators given above are anomalous in the sense that if they are calculated by naive use of canonical commutation relations they vanish. These brackets are compatible with the axial anomaly \eqref{anom.d} as shown in \refscite{Adler:1970qb, DeserOthers:1970}. Some other commutators which will be useful later are
\begin{equation}\label{SCC}
\left[J_{\mu}(x), A_{\sigma}(y)\right]
= \left[J_{0}(x), \partial_{0}A_{\sigma}(y)\right] = 0.
\end{equation}


\subsection{Anomalous commutators in the noncommutative theory}

Now we are in a position to compute the anomalous commutators in the noncommutative theory. In the context of the ordinary theory it is well-known that the anomalous commutators are a different manifestation of the ABJ anomly. Since the standard ABJ anomaly is not modified in $\theta$-expanded theory, we argue that the set \eqref{CS101}--\eqref{CB103} of commutators remains valid in the $\theta$-expanded theory also. We further note that the equation of motion for the photon field in $\theta$-expanded theory will differ from \eqref{Seom2} by an $\uO(\theta)$ term. This will modify the canonical commutation relation $[A_{i}(x), \partial_{0}A_{j}(y)]$ given in Eq.~\eqref{Scr}, which will have an $\uO(\theta)$ extension. But we need not compute this $\uO(\theta)$ correction explicitly since later we shall use this particular commutation relation in such terms which will already involve $\theta$. The commutators $[A_{0}(x), \partial_{0}A_{\nu}(y)]$ and $[A_{\mu}(x), \partial_{0}A_{0}(y)]$ will not have any $\uO(\theta)$ extension.

Although our main interest is in the current--current commutators, we shall compute some other commutators as well which will later be useful when we discuss the consistency conditions. Now onwards we shall take $\theta$ to be of `magnetic' type so that $\theta^{0i} = 0$. Using the maps \eqref{J-hat2.m} and \eqref{J-hat5.m}, and Eq.~\eqref{SCC}, we find
\begin{equation}\label{CCA1011}\begin{split}
\xnc{S}_{00}(x,y) &\equiv \left[\xnc{J}_{0}(x), \xnc{J}_{0}^5(y)\right]\\
&= S_{00}(x,y) - \theta^{mn}\left[\partial_{n}^{y}\left(A_{m}(y)S_{00}(x,y)\right)\right.\\
&\qquad\qquad\qquad\qquad \left.{} + \partial_{n}^{x}\left(A_{m}(x)S_{00}(x,y) + J_{0}(x)L_{m0}(x,y)\right)\right] + \uO(\theta^2),
\end{split}\end{equation}
which may also be interpreted as a Seiberg--Witten-type map. Proceeding similarly, we obtain\footnote{To save space we omit arguments, writing $S_{\mu\sigma}$, $L_{\mu\sigma}$ and $M_{\mu\sigma}$ instead of $S_{\mu\sigma}(x,y)$, $L_{\mu\sigma}(x,y)$ and $M_{\mu\sigma}(x,y)$ respectively.}
\begin{gather}
\label{CCA102}\begin{split}
\xnc{S}_{i0}(x,y) &\equiv \left[\xnc{J}_{i}(x), \xnc{J}_{0}^5(y)\right]\\
&= S_{i0} - \theta^{mn}\left[\partial_{n}^{y}\left(A_{m}(y)S_{i0}\right) + \partial_{n}^{x}\left(A_{m}(x)S_{i0} + J_{i}(x)L_{m0}\right)\right]\\
&\quad {} - \theta^{im}\left[{F_{m}}^{\beta}(x)S_{\beta 0} + J_{0}(x)\left(\partial_{m}^{x}L_{00} - M_{m0}\right) - J_{n}(x)\left(\partial_{m}^{x}L_{n0} - \partial_{n}^{x}L_{m0}\right)\right]\\
&\quad {} + \uO(\theta^2),
\end{split}
\\
\label{CSA103}\begin{split}
\xnc{S}_{0i}(x,y) &\equiv \left[\xnc{J}_{0}(x), \xnc{J}_{i}^5(y)\right]\\
&= S_{0i} - \theta^{mn}\left[\partial_{n}^{y}\left(A_{m}(y)S_{0i}\right) + \partial_{n}^{x}\left(A_{m}(x)S_{0i} + J_{0}(x)L_{mi}\right)\right]\\
&\quad {} - \theta^{im}{F_{m}}^{\beta}(y)S_{0\beta} + \uO(\theta^2).
\end{split}
\end{gather}
The field--current algebra is likewise computed using Eqs.~\eqref{A.m}, \eqref{J-hat5.m} and \eqref{Scr}:
\begin{gather}
\label{CCA014}\begin{split}
\xnc{L}_{00}(x,y) &\equiv \left[\xnc{A}_{0}(x), \xnc{J}_{0}^5(y)\right]\\
&= L_{00} - \theta^{mn}\left[\partial_{n}^{y}\left(A_{m}(y)L_{00}\right) + \frac{1}{2}A_{m}(x)\left(2\partial_{n}^{x}L_{00} - M_{n0}\right)\right.\\
&\qquad\qquad\qquad\quad \left.{} + \frac{1}{2}L_{m0}\left(\partial_{n}A_{0}(x)+F_{n0}(x)\right)\right]+\uO(\theta^2),
\end{split}\displaybreak[0]%
\\
\label{CCA014bb}
\begin{split}
\xnc{L}_{0i}(x,y) &\equiv \left[\xnc{A}_{0}(x), \xnc{J}_{i}^5(y)\right]\\
&= L_{0i} - \theta^{im}{F_{m}}^{\beta}(y)L_{0\beta} - \theta^{mn}\left[\partial_{n}^{y}\left(A_{m}(y)L_{0i}\right) + \frac{1}{2}A_{m}(x)\left(2\partial_{n}^{x}L_{0i} - M_{ni}\right)\right.\\
&\qquad\qquad\qquad\qquad\qquad\qquad\qquad \left.{} + \frac{1}{2}L_{mi}\left(\partial_{n}A_{0}(x)+F_{n0}(x)\right)\right] + \uO(\theta^2),
\end{split}\displaybreak[0]%
\\
\label{CCA105}\begin{split}
\xnc{L}_{i0}(x,y) &\equiv \left[\xnc{A}_{i}(x), \xnc{J}_{0}^5(y)\right]\\
&= L_{i0} - \theta^{mn}\bigg[\partial_{n}^{y}\left(A_{m}(y)L_{i0}\right) + \frac{1}{2}A_{m}(x)\left(2\partial_{n}^{x}L_{i0} - \partial_{i}^{x}L_{n0}\right)\\
&\qquad\qquad\qquad {} + \frac{1}{2}L_{m0}\left(\partial_{n}A_{i}(x)+F_{ni}(x)\right)\bigg] + \uO(\theta^2),
\end{split}\displaybreak[0]%
\\
\label{CCA1052} 
\begin{split}
\xnc{L}_{im}(x,y) &\equiv \left[\xnc{A}_{i}(x), \xnc{J}_{m}^5(y)\right]\\
&= L_{im} + \ui\theta^{mi}J_{0}^{5}\delta^{3}(\mathbf{x}-\mathbf{y}) - \theta^{mj}{F_{j}}^{\beta}(y)L_{i\beta}\\
&\quad {} - \frac{1}{2}\theta^{jk}\left[2\partial_{k}^{y}\left(A_{j}(y)L_{im}\right) + L_{jm}\left(\partial_{k}A_{i}(x)+F_{ki}(x)\right)\right.\\
&\qquad\qquad\quad \left.{} + A_{j}(x)\left(2\partial_{k}^{x}L_{im}-\partial_{i}^{x}L_{km}\right)\right] + \uO(\theta^2),
\end{split}\displaybreak[0]%
\\
\label{CCA106} \raisetag{21pt}
\begin{split}
\xnc{M}_{00}(x,y) &\equiv \left[\partial_{0}\xnc{A}_{0}(x), \xnc{J}_{0}^5(y)\right]\\
&= M_{00} - \theta^{mn}\bigg\{\partial_{n}^{y}\left(A_{m}(y)M_{00}\right) + \frac{1}{2}A_{m}(x)\left(2\partial_{n}^{x}M_{00} - \left[\partial_{0}\partial_{0}A_{n}(x),J_{0}^{5}(y)\right]\right)\\
&\qquad\qquad\qquad\quad {} + \frac{1}{2}L_{m0}\partial_{0}\left(\partial_{n}A_{0}(x)+F_{n0}(x)\right) + \partial_{0}A_{m}(x)\partial_{n}^{x}L_{00}\\
&\qquad\qquad\qquad\quad {} - \partial_{m}A_{0}(x)M_{n0}\bigg\}  + \uO(\theta^2),
\end{split}\displaybreak[0]%
\\
\label{CCA107}\begin{split}
\xnc{M}_{i0}(x,y) &\equiv \left[\partial_{0}\xnc{A}_{i}(x), \xnc{J}_{0}^5(y)\right]\\
&= M_{i0} + \ui\theta^{in}\partial_{n}^y\left(J_{0}^5\delta^{3}(\mathbf{x}-\mathbf{y})\right)\\
&\quad {} - \theta^{mn}\bigg[\partial_{n}^{y}\left(A_{m}(y)M_{i0}\right) + \frac{1}{2}\partial_{0}A_{m}(x)\left(2\partial_{n}^{x}L_{i0}-\partial_{i}^{x}L_{n0}\right)\\
&\qquad\qquad\quad {}  + \frac{1}{2}L_{m0}\partial_{0}\left(\partial_{n}A_{i}(x)+F_{ni}(x)\right) + \frac{1}{2}\partial_{i}^{x}\left(A_{n}(x)M_{m0}\right)\\
&\qquad\qquad\quad {} + F_{ni}(x)M_{m0} + A_{m}(x)\partial_{n}^{x}M_{i0} \bigg] + \uO(\theta^2),
\end{split}\displaybreak[0]%
\\
\label{CCA108}\begin{split}
\xnc{M}_{ik}(x,y) &\equiv \left[\partial_{0}\xnc{A}_{i}(x), \xnc{J}_{k}^5(y)\right]\\
&= M_{ik} + \ui\theta^{in}\partial_{n}^y\left(J_{k}^5\delta^{3}(\mathbf{x}-\mathbf{y})\right) + \ui\theta^{ki}J_{n}^{5}(y)\partial_{n}^{y}\delta^{3}(\mathbf{x}-\mathbf{y})\\
&\quad {} - \theta^{km}\left(\ui J_{i}^{5}(y)\partial_{m}^{y}\delta^{3}(\mathbf{x}-\mathbf{y}) + {F_{m}}^{\beta}(y)M_{i\beta}\right)\\
&\quad {} - \theta^{mn}\bigg[\partial_{n}^{y}(A_{m}(y)M_{ik}) + \frac{1}{2}\partial_{0}A_{m}(x)\left(2\partial_{n}^{x}L_{ik}-\partial_{i}^{x}L_{nk}\right)\\
&\qquad\qquad\quad {}  + \frac{1}{2}L_{mk}\partial_{0}\left(\partial_{n}A_{i}(x)+F_{ni}(x)\right) + \frac{1}{2}\partial_{i}^{x}\left(A_{n}(x)M_{mk}\right)\\
&\qquad\qquad\quad {} + F_{ni}(x)M_{mk} + A_{m}(x)\partial_{n}^{x}M_{ik}\bigg] + \uO(\theta^2).
\end{split}
\end{gather}
Now we use the relations \eqref{CS101}--\eqref{CB103} to substitute for the commutators appearing on the right-hand sides in Eqs.~\eqref{CCA1011}--\eqref{CCA108}. In order to compute $[\partial_{0}\partial_{0}A_{n}(x), J_{0}^5(y)]$ appearing on the right-hand side of Eq.~\eqref{CCA106}, we make use of the equation of motion. The equation of motion \eqref{Seom2} of the usual theory in the Feynman gauge reads $\partial_{0}\partial_{0}A_{\mu}-{\nabla}^{2}A_{\mu}-J_{\mu} = 0$. Therefore the equation of motion of the noncommutative theory in terms of the usual variables,
\begin{equation}
\partial_{0}\partial_{0}A_{\mu}-{\nabla}^{2}A_{\mu}-J_{\mu}+\uO(\theta) = 0,
\end{equation}
implies
\begin{equation}\label{}
 \left[\partial_{0}\partial_{0}A_{n}(x), J_{0}^5(y)\right] = {\nabla}_{x}^{2}\left[A_{n}(x), J_{0}^5(y)\right] + \left[J_{n}(x), J_{0}^5(y)\right]+\uO(\theta),
\end{equation}
 which can be computed using Eqs.~\eqref{CS102} and \eqref{CA101}. Thus Eqs.~\eqref{CCA1011}--\eqref{CCA108} become
\begin{gather}
\label{CCB101}\begin{split}
\xnc{S}_{00}(x,y) &= \frac{\ui}{4\pi^2}\varepsilon_{ijk}F_{jk}(y)\partial_{i}^{x}\delta^{3}(\mathbf{x}-\mathbf{y})\\
&\quad {}-\frac{\ui}{4\pi^2}\theta^{mn}\varepsilon_{ijk}\left[\partial_{n}^{y}\left(A_{m}(y)F_{jk}(y)\partial_{i}^{x}\delta^{3}(\mathbf{x}-\mathbf{y})\right)\right.\\
&\qquad\qquad\qquad\qquad\left.{}+\partial_{n}^{x}\left(A_{m}(x)F_{jk}(y)\partial_{i}^{x}\delta^{3}(\mathbf{x}-\mathbf{y})\right)\right] + \uO(\theta^2),
\end{split}\displaybreak[0]%
\\
\label{CCB102} \raisetag{14pt}
\begin{split}
\xnc{S}_{i0}(x,y) &= -\frac{\ui}{4\pi^2}\varepsilon_{ijk}F_{0j}(x)\partial_{k}^{y}\delta^{3}(\mathbf{x}-\mathbf{y})\\
&\quad -\frac{\ui}{4\pi^2}\theta^{im}\left[\varepsilon_{njk}\left(F_{m0}(x)F_{jk}(y)\partial_{n}^{x}\delta^{3}(\mathbf{x}-\mathbf{y})+F_{mn}(x)F_{0j}(x)\partial_{k}^{y}\delta^{3}(\mathbf{x}-\mathbf{y})\right)\right.\\
&\qquad\qquad\qquad \left.{}-\varepsilon_{mjk}F_{jk}J_{0}\delta^{3}(\mathbf{x}-\mathbf{y})\right]\\
&\quad +\frac{\ui}{4\pi^2}\theta^{mn}\varepsilon_{ijk}\left[\partial_{n}^{y}\left(A_{m}(y)F_{0j}(x)\partial_{k}^{y}\delta^{3}(\mathbf{x}-\mathbf{y})\right)\right.\\
&\qquad\qquad\qquad\qquad \left.{}+\partial_{n}^{x}\left(A_{m}(x)F_{0j}(x)\partial_{k}^{y}\delta^{3}(\mathbf{x}-\mathbf{y})\right)\right] + \uO(\theta^2),
\end{split}\displaybreak[0]%
\\
\label{CCB103}\begin{split}
\xnc{S}_{0i}(x,y) &= \frac{\ui}{4\pi^2}\varepsilon_{ijk}F_{0j}(y)\partial_{k}^{x}\delta^{3}(\mathbf{x}-\mathbf{y})\\
&\quad - \frac{\ui}{4\pi^2}\theta^{im}\varepsilon_{njk}\left(F_{m0}(y)F_{jk}(y)\partial_{n}^{x}\delta^{3}(\mathbf{x}-\mathbf{y})-F_{mn}(y)F_{0j}(y)\partial_{k}^{x}\delta^{3}(\mathbf{x}-\mathbf{y})\right)\\
&\quad - \frac{\ui}{4\pi^2}\theta^{mn}\varepsilon_{ijk}\left[\partial_{n}^{y}\left(A_{m}(y)F_{0j}(y)\partial_{k}^{x}\delta^{3}(\mathbf{x}-\mathbf{y})\right)\right.\\
&\qquad\qquad\qquad\qquad \left.{} + \partial_{n}^{x}\left(A_{m}(x)F_{0j}(y)\partial_{k}^{x}\delta^{3}(\mathbf{x}-\mathbf{y})\right)\right] + \uO(\theta^2),
\end{split}\displaybreak[0]%
\\
\label{CCB104}
\xnc{L}_{00}(x,y)
= \frac{\ui}{8\pi^2}\theta^{mn}\varepsilon_{njk}A_{m}F_{jk}\delta^{3}(\mathbf{x}-\mathbf{y}) + \uO(\theta^2),\displaybreak[0]%
\\
\label{CCB105}
\xnc{L}_{0i}(x,y)
= \frac{\ui}{8\pi^2}\theta^{mn}\varepsilon_{nik}A_{m}F_{0k}\delta^{3}(\mathbf{x}-\mathbf{y}) + \uO(\theta^2),\displaybreak[0]%
\\
\label{CCB1051}
\xnc{L}_{i0}(x,y)
= \uO(\theta^2),\displaybreak[0]%
\\
\label{CCB1052}
\xnc{L}_{im}(x,y)
= \ui\theta^{mi}J_{0}^{5}\delta^{3}(\mathbf{x}-\mathbf{y}) + \uO(\theta^2),\displaybreak[0]%
\\
\label{CCB106}\begin{split}
\xnc{M}_{00}(x,y)
&= \frac{\ui}{4\pi^2}\theta^{mn}\varepsilon_{njk}\left(\partial_{m}A_{0}F_{jk}\delta^{3}(\mathbf{x}-\mathbf{y}) - \frac{1}{2}A_{m}(x)F_{0j}(x)\partial_{k}^{y}\delta^{3}(\mathbf{x}-\mathbf{y})\right)\\
&\quad {} + \uO(\theta^2),
\end{split}\displaybreak[0]%
\\
\label{CCB107}\begin{split}
\xnc{M}_{i0}(x,y)
&= \frac{\ui}{4\pi^2}\varepsilon_{ijk}F_{jk}\delta^{3}(\mathbf{x}-\mathbf{y}) + \ui\theta^{in}\partial_{n}^{y}\left(J_{0}^{5}\delta^{3}(\mathbf{x}-\mathbf{y})\right)\\
&\quad {} - \frac{\ui}{4\pi^2}\theta^{mn}\bigg[\varepsilon_{mjk}\left\{F_{ni}F_{jk}\delta^{3}(\mathbf{x}-\mathbf{y}) + \frac{1}{2}\partial_{i}^{x}\left(A_{n}F_{jk}\delta^{3}(\mathbf{x}-\mathbf{y})\right)\right\}\\
&\qquad\qquad\qquad {} + \varepsilon_{ijk}A_{m}\partial_{n}F_{jk}\delta^{3}(\mathbf{x}-\mathbf{y})\bigg] + \uO(\theta^2),
\end{split}\displaybreak[0]%
\\
\label{CCB108}
\begin{split}
\xnc{M}_{ik}(x,y)
&= \frac{\ui}{4\pi^2}\varepsilon_{ikj}F_{0j}\delta^{3}(\mathbf{x}-\mathbf{y}) + \ui\theta^{in}\partial_{n}^{y}\left(J_{k}^{5}\delta^{3}(\mathbf{x}-\mathbf{y})\right)+\ui\theta^{ki}J_{n}^{5}(y)\partial_{n}^{y}\delta^{3}(\mathbf{x}-\mathbf{y})\\
&\quad {} -\ui\theta^{km}\left\{J_{i}^{5}(y)\partial_{m}^{y}\delta^{3}(\mathbf{x}-\mathbf{y})+\frac{\ui}{4\pi^2}\varepsilon_{ijn}\left(F_{m0}F_{jn}+F_{mn}F_{0j}\right)\delta^{3}(\mathbf{x}-\mathbf{y})\right\}\\
&\quad {} - \frac{\ui}{4\pi^2}\theta^{mn}\bigg[\varepsilon_{mkj}\left\{F_{ni}F_{0j}\delta^{3}(\mathbf{x}-\mathbf{y}) + \frac{1}{2}\partial_{i}^{x}\left(A_{n}F_{0j}\delta^{3}(\mathbf{x}-\mathbf{y})\right)\right\}\\
&\qquad\qquad\qquad {} + \varepsilon_{ikj}A_{m}\partial_{n}F_{0j}\delta^{3}(\mathbf{x}-\mathbf{y})\bigg] + \uO(\theta^2).
\end{split}
\end{gather}
We have thus obtained various anomalous commutators up to the first order in a magnetic-type~$\theta$. These expressions are given in commutative variables. Using the inverse maps,
\begin{gather}
\label{inv1}
A_{\mu} = \xnc{A}_{\mu} + \frac{1}{2}\theta^{\alpha\beta}\xnc{A}_{\alpha}\left(\partial_{\beta}\xnc{A}_{\mu}+\xnc{F}_{\beta\mu}\right) + \uO(\theta^2),\displaybreak[0]\\
\label{inv2}
F_{\mu\nu} = \xnc{F}_{\mu\nu} + \theta^{\alpha\beta}\left(\xnc{A}_{\alpha}\partial_{\beta}\xnc{F}_{\mu\nu} + \xnc{F}_{\mu\alpha}\xnc{F}_{\beta\nu}\right) + \uO(\theta^2),\displaybreak[0]\\
\label{inv3}
J^{\mu} = \xnc{J}^{\mu} + \theta^{\alpha\beta}\left(\xnc{A}_{\alpha}\partial_{\beta}\xnc{J}^{\mu} - \frac{1}{2}\xnc{F}_{\alpha\beta}\xnc{J}^{\mu}\right)-\theta^{\mu\alpha}\xnc{F}_{\alpha\beta}\xnc{J}^{\beta} + \uO(\theta^2),
\end{gather}
with $\theta^{0i} = 0$, we can express them in terms of the noncommutative variables:
\begin{gather}
\label{CCC101} \raisetag{18pt}
\begin{split}
\xnc{S}_{00}(x,y)
&= \frac{\ui}{4\pi^2}\varepsilon_{ijk}\xnc{F}_{jk}(y)\partial_{i}^{x}\delta^{3}(\mathbf{x}-\mathbf{y})\\
&\quad {}+\frac{\ui}{4\pi^2}\theta^{mn}\varepsilon_{ijk}\left[\xnc{F}_{jm}(y)\xnc{F}_{nk}(y)\partial_{i}^{x}\delta^{3}(\mathbf{x}-\mathbf{y})-\xnc{F}_{jk}(y)\partial_{n}^{y}\left(\xnc{A}_{m}(y)\partial_{i}^{x}\delta^{3}(\mathbf{x}-\mathbf{y})\right)\right.\\
&\qquad\qquad\qquad\qquad \left.{}-\xnc{F}_{jk}(y)\partial_{n}^{x}\left(\xnc{A}_{m}(x)\partial_{i}^{x}\delta^{3}(\mathbf{x}-\mathbf{y})\right)\right] + \uO(\theta^2),
\end{split}\displaybreak[0]%
\\
\label{CCC102} \raisetag{18pt}
\begin{split}
\xnc{S}_{i0}(x,y)
&= -\frac{\ui}{4\pi^2}\varepsilon_{ijk}\xnc{F}_{0j}(x)\partial_{k}^{y}\delta^{3}(\mathbf{x}-\mathbf{y})\\
&\quad {} - \frac{\ui}{4\pi^2}\theta^{im}\left[\varepsilon_{njk}\left(\xnc{F}_{m0}(x)\xnc{F}_{jk}(y)\partial_{n}^{x}\delta^{3}(\mathbf{x}-\mathbf{y})+\xnc{F}_{mn}(x)\xnc{F}_{0j}(x)\partial_{k}^{y}\delta^{3}(\mathbf{x}-\mathbf{y})\right)\right.\\
&\qquad\qquad\qquad \left.{} - \varepsilon_{mjk}\xnc{J}_{0}\xnc{F}_{jk}\delta^{3}(\mathbf{x}-\mathbf{y})\right]\\
&\quad {} - \frac{\ui}{4\pi^2}\theta^{mn}\varepsilon_{ijk}\left[\xnc{F}_{0m}(x)\xnc{F}_{nj}(x)\partial_{k}^{y}\delta^{3}(\mathbf{x}-\mathbf{y})-\xnc{F}_{0j}(x)\partial_{n}^{y}\left(\xnc{A}_{m}(y)\partial_{k}^{y}\delta^{3}(\mathbf{x}-\mathbf{y})\right)\right.\\
&\qquad\qquad\qquad\qquad \left.{} - \xnc{F}_{0j}(x)\partial_{n}^{x}\left(\xnc{A}_{m}(x)\partial_{k}^{y}\delta^{3}(\mathbf{x}-\mathbf{y})\right)\right] + \uO(\theta^2),
\end{split}\displaybreak[0]%
\\
\label{CCC103} \raisetag{18pt}
\begin{split}
\xnc{S}_{0i}(x,y)
&= \frac{\ui}{4\pi^2}\varepsilon_{ijk}\xnc{F}_{0j}(y)\partial_{k}^{x}\delta^{3}(\mathbf{x}-\mathbf{y})\\
&\quad {} - \frac{\ui}{4\pi^2}\theta^{im}\varepsilon_{njk}\left(\xnc{F}_{m0}(y)\xnc{F}_{jk}(y)\partial_{n}^{x}\delta^{3}(\mathbf{x}-\mathbf{y}) - \xnc{F}_{mn}(y)\xnc{F}_{0j}(y)\partial_{k}^{x}\delta^{3}(\mathbf{x}-\mathbf{y})\right)\\
&\quad {} + \frac{\ui}{4\pi^2}\theta^{mn}\varepsilon_{ijk}\left[\xnc{F}_{0m}(y)\xnc{F}_{nj}(y)\partial_{k}^{x}\delta^{3}(\mathbf{x}-\mathbf{y}) - \xnc{F}_{0j}(y)\partial_{n}^{y}\left(\xnc{A}_{m}(y)\partial_{k}^{x}\delta^{3}(\mathbf{x}-\mathbf{y})\right)\right.\\
&\qquad\qquad\qquad\qquad \left.{} - \xnc{F}_{0j}(y)\partial_{n}^{x}\left(\xnc{A}_{m}(x)\partial_{k}^{x}\delta^{3}(\mathbf{x}-\mathbf{y})\right)\right] + \uO(\theta^2),
\end{split}\displaybreak[0]%
\\
\label{CCC104}
\xnc{L}_{00}(x,y)
= \frac{\ui}{8\pi^2}\theta^{mn}\varepsilon_{njk}\xnc{A}_{m}\xnc{F}_{jk}\delta^{3}(\mathbf{x}-\mathbf{y}) + \uO(\theta^2),\displaybreak[0]%
\\
\label{CCC105}
\xnc{L}_{0i}(x,y)
= \frac{\ui}{8\pi^2}\theta^{mn}\varepsilon_{nik}\xnc{A}_{m}\xnc{F}_{0k}\delta^{3}(\mathbf{x}-\mathbf{y}) + \uO(\theta^2),\displaybreak[0]\displaybreak[0]%
\\
\label{CCC1051}
\xnc{L}_{i0}(x,y)
= \uO(\theta^2),\displaybreak[0]%
\\
\label{CCC1052}
\xnc{L}_{im}(x,y)
= \ui\theta^{mi}\xnc{J}_{0}^{5}\delta^{3}(\mathbf{x}-\mathbf{y}) + \uO(\theta^2),\displaybreak[0]%
\\
\label{CCC106}\begin{split}
\xnc{M}_{00}(x,y)
&= \frac{\ui}{4\pi^2}\theta^{mn}\varepsilon_{njk}\left(\partial_{m}\xnc{A}_{0}\xnc{F}_{jk}\delta^{3}(\mathbf{x}-\mathbf{y}) - \frac{1}{2}\xnc{A}_{m}(x)\xnc{F}_{0j}(x)\partial_{k}^{y}\delta^{3}(\mathbf{x}-\mathbf{y})\right)\\
&\quad {} + \uO(\theta^2),
\end{split}\displaybreak[0]%
\\
\label{CCC107}\begin{split}
\xnc{M}_{i0}(x,y)
&= \frac{\ui}{4\pi^2}\varepsilon_{ijk}\xnc{F}_{jk}\delta^{3}(\mathbf{x}-\mathbf{y}) + \ui\theta^{in}\partial_{n}^{y}\left(\xnc{J}_{0}^{5}\delta^{3}(\mathbf{x}-\mathbf{y})\right)\\
&\quad {} - \frac{\ui}{4\pi^2}\theta^{mn}\bigg[\varepsilon_{mjk}\left\{\xnc{F}_{ni}\xnc{F}_{jk}\delta^{3}(\mathbf{x}-\mathbf{y}) + \frac{1}{2}\partial_{i}^{x}\left(\xnc{A}_{n}\xnc{F}_{jk}\delta^{3}(\mathbf{x}-\mathbf{y})\right)\right\}\\
&\qquad\qquad\qquad {} - \varepsilon_{ijk}\xnc{F}_{jm}\xnc{F}_{nk}\delta^{3}(\mathbf{x}-\mathbf{y})\bigg] + \uO(\theta^2),
\end{split}\displaybreak[0]%
\\
\label{CCC108} \raisetag{21pt}
\begin{split}
\xnc{M}_{ik}(x,y)
&= \frac{\ui}{4\pi^2}\varepsilon_{ikj}\xnc{F}_{0j}\delta^{3}(\mathbf{x}-\mathbf{y}) + \ui\theta^{in}\partial_{n}^{y}\left(\xnc{J}_{k}^{5}\delta^{3}(\mathbf{x}-\mathbf{y})\right) + \ui\theta^{ki}\xnc{J}_{n}^{5}(y)\partial_{n}^{y}\delta^{3}(\mathbf{x}-\mathbf{y})\\
&\quad {} - \ui\theta^{km}\left\{\xnc{J}_{i}^{5}(y)\partial_{m}^{y}\delta^{3}(\mathbf{x}-\mathbf{y})+\frac{\ui}{4\pi^2}\varepsilon_{ijn}\left(\xnc{F}_{m0}\xnc{F}_{jn}+\xnc{F}_{mn}\xnc{F}_{0j}\right)\delta^{3}(\mathbf{x}-\mathbf{y})\right\}\\
&\quad {} - \frac{\ui}{4\pi^2}\theta^{mn}\bigg[ \varepsilon_{mkj}\left\{\xnc{F}_{ni}\xnc{F}_{0j}\delta^{3}(\mathbf{x}-\mathbf{y}) + \frac{1}{2}\partial_{i}^{x}\left(\xnc{A}_{n}\xnc{F}_{0j}\delta^{3}(\mathbf{x}-\mathbf{y})\right)\right\}\\
&\qquad\qquad\qquad {} - \varepsilon_{ikj}\xnc{F}_{0m}\xnc{F}_{nj}\delta^{3}(\mathbf{x}-\mathbf{y})\bigg] + \uO(\theta^2).
\end{split}
\end{gather}
This completes our obtention of the anomalous commutators in both commutative as well as noncommutative variables.


\section{\label{sec:anomalous-consis}Consistency conditions and the anomalous commutators}

Just as the anomalous commutators in the usual theory are subjected to certain consistency conditions \cite{Adler:1970qb, DeserOthers:1970}, we now show that those in the noncommutative theory also obey certain consistency conditions, implying their compatibility with the noncommutative covariant anomaly~\eqref{anom-cov.d}.

To obtain the consistency criteria, we begin with
\begin{equation}\label{cons101}
\partial_{0}\xnc{S}_{00}(x,y) = \partial_{0}\left[\xnc{J}_{0}(x), \xnc{J}_{0}^5(y)\right]= \left[\partial_{0}\xnc{J}_{0}(x), \xnc{J}_{0}^5(y)\right]+\left[\xnc{J}_{0}(x), \partial_{0}\xnc{J}_{0}^5(y)\right].
\end{equation}
In view of Eq.~\eqref{DJ}, it follows from $\xnc{\uD}_{\mu}\star\xnc{J}^{\mu} = 0$, and  $\xnc{\uD}_{\mu}\star\xnc{J}^{\mu}_{5} = \xnc{\mathscr{A}}$ that (for $\theta^{0i}=0$)
\begin{gather}
\label{cons102}
\partial_{0}\xnc{J}_{0} = \partial_{m}\xnc{J}_{m}+\theta^{mn}\partial_{m}\xnc{J}^{\mu}\partial_{n}\xnc{A}_{\mu}+\uO(\theta^2),\\
\label{cons103}
\partial_{0}\xnc{J}_{0}^5 = \partial_{m}\xnc{J}_{m}^{5}+\theta^{mn}\partial_{m}\xnc{J}^{\mu}_{5}\partial_{n}\xnc{A}_{\mu}+\xnc{\mathscr{A}}+\uO(\theta^2).
\end{gather}
Using these to substitute for $\partial_{0}\xnc{J}_{0}$ and $\partial_{0}\xnc{J}_{0}^5$, Eq.~\eqref{cons101} yields a consistency relation among the anomalous commutators of the noncommutative theory:
\begin{equation}\label{cons104}\begin{split}
\partial_{0}\xnc{S}_{00}(x,y) &= \partial_{m}^{x}\xnc{S}_{m0}(x,y)+\partial_{m}^{y}\xnc{S}_{0m}(x,y)\\
&\quad +\theta^{mn}\left(\partial_{n}\xnc{A}^{\mu}(x)\partial_{m}^{x}\xnc{S}_{\mu0}(x,y)+\partial_{n}\xnc{A}^{\mu}(y)\partial_{m}^{y}\xnc{S}_{0\mu}(x,y)\right.\\
&\qquad\qquad \left.{} + \partial_{m}\xnc{J}^{\mu}(x)\partial_{n}^{x}\xnc{L}_{\mu 0}(x,y) + \partial_{m}\xnc{J}^{\mu}_{5}(y)\partial_{n}^{y}\left[\xnc{J}_{0}(x), \xnc{A}_{\mu}(y)\right]\right)\\
&\quad {} + \left[\xnc{J}_{0}(x), \xnc{\mathscr{A}}(y)\right]+\uO(\theta^2).
\end{split}\end{equation}
The essentially new ingredient is the last bracket involving the anomaly. Using the maps (with $\theta^{0i} = 0$) for $\xnc{J}_{0}$ and $\xnc{\mathscr{A}}$ given in Eqs.~\eqref{J-hat2.m} and \eqref{anom.m} respectively, we get
\begin{equation*}\begin{split}
\left[\xnc{J}_{0}(x), \xnc{\mathscr{A}}(y)\right] &= \left[J_{0}(x), \mathscr{A}(y)\right]\\
&\quad {} - \theta^{mn}\left(\partial_{n}^{y}\left[J_{0}(x), A_{m}(y)\mathscr{A}(y)\right] + \partial_{n}^{x}\left[A_{m}(x)J_{0}(x), \mathscr{A}(y)\right]\right)+\uO(\theta^2),
\end{split}
\end{equation*}
which, on substituting for the anomaly, $\mathscr{A} = ({1}/{16\pi^{2}})\varepsilon_{\mu\nu\lambda\rho}F^{\mu\nu}F^{\lambda\rho}$, 
and using the relations \eqref{Scr} and \eqref{SCC}, yields
\begin{equation}\label{cons105}\begin{split}
\left[\xnc{J}_{0}(x), \xnc{\mathscr{A}}(y)\right] = \frac{\ui}{4\pi^2}\theta^{mn}\varepsilon_{mjk}J_{0}(y)F_{jk}(y)\partial_{n}^{x}\delta^{3}(\mathbf{x}-\mathbf{y})+\uO(\theta^2).
\end{split}
\end{equation}
We observe that the $\theta\rightarrow 0$ limit of the condition \eqref{cons104} is
\begin{equation}\label{cons1041}
\partial_{0}S_{00}(x,y) = \partial_{m}^{x}S_{m0}(x,y) + \partial_{m}^{y}S_{0m}(x,y),
\end{equation}
which is easily verified using Eqs.~\eqref{CS101}--\eqref{CS103}. To show that Eq.~\eqref{cons104} indeed holds is also straightforward. Equation~\eqref{cons105} gives the last term on the right-hand side of Eq.~\eqref{cons104}. The commutator $[\xnc{J}_{0}(x), \xnc{A}_{\mu}(y)]$ occurs in an $\uO(\theta)$ term, and therefore it can be replaced by $[J_{0}(x), A_{\mu}(y)]$ which vanishes because of Eq.~\eqref{SCC}. The other terms in Eq.~\eqref{cons104} are also known in view of Eqs.~\eqref{CCB101}--\eqref{CCB108}. Substituting for all these commutators, we find that Eq.~\eqref{cons104} is satisfied. Alternatively, the verification of Eq.~\eqref{cons104} can be done in noncommutative variables by exploiting Eqs.~\eqref{CCC101}--\eqref{CCC108} and the one obtained by using the inverse maps \eqref{inv2} and \eqref{inv3} on the right-hand side of Eq.~\eqref{cons105} (this amounts to just replacing the usual variables by the noncommutative ones, since it is already an $\uO(\theta)$ term). This shows that our anomalous commutators are compatible with the noncommutative anomaly.

As another example of a consistency condition, we note that
\begin{equation*}
\partial_{0}\left[\xnc{A}_{\nu}(x), \xnc{J}_{0}^5(y)\right] = \left[\partial_{0}\xnc{A}_{\nu}(x), \xnc{J}_{0}^5(y)\right]+\left[\xnc{A}_{\nu}(x), \partial_{0}\xnc{J}_{0}^5(y)\right],
\end{equation*}
which, invoking the notations introduced earlier, can be rewritten compactly as
\begin{equation}\label{consc}
\partial_{0}\xnc{L}_{\nu 0}(x,y) = \xnc{M}_{\nu 0}(x,y) + \left[\xnc{A}_{\nu}(x), \partial_{0}\xnc{J}_{0}^5(y)\right].
\end{equation}
Using Eq.~\eqref{cons103} to substitute for $\partial_{0}\xnc{J}_{0}^5$ on the right-hand side gives a consistency condition
\begin{equation}\label{cons107}\begin{split}
\partial_{0}\xnc{L}_{\nu 0}(x,y) &= \xnc{M}_{\nu 0}(x,y) + \partial_{m}^{y}\xnc{L}_{\nu m}(x,y)\\
&\quad {}+\theta^{mn}\left(\partial_{n}\xnc{A}^{\mu}(y)\partial_{m}^{y}\xnc{L}_{\nu\mu}(x,y)
+\partial_{m}\xnc{J}^{\mu}_{5}(y)\partial_{n}^{y}\left[\xnc{A}_{\nu}(x), \xnc{A}_{\mu}(y)\right]\right)\\
&\quad {}+\left[\xnc{A}_{\nu}(x), \xnc{\mathscr{A}}(y)\right] + \uO(\theta^2).
\end{split}\end{equation}
Using the maps for $\xnc{A}_{0}$ and $\xnc{\mathscr{A}}$ given in Eqs.~\eqref{A.m} and \eqref{anom.m} we get
\begin{equation*}\begin{split}
\left[\xnc{A}_{0}(x), \xnc{\mathscr{A}}(y)\right] &= \left[A_{0}(x), \mathscr{A}(y)\right] - \theta^{mn}\bigg(\frac{1}{2}\left[A_{m}(x)(\partial_{n}A_{0}(x)+F_{n0}(x)), \mathscr{A}(y)\right]\\
&\qquad\qquad\qquad\qquad\qquad\quad {} + \partial_{n}^{y}\left[A_{0}(x), A_{m}(y)\mathscr{A}(y)\right]\bigg) + \uO(\theta^2).
\end{split}\end{equation*}
By substituting for the anomaly, $\mathscr{A}= ({1}/{16\pi^{2}})\varepsilon_{\mu\nu\lambda\rho}F^{\mu\nu}F^{\lambda\rho}$, and using Eq.~\eqref{Scr}, this is computed as
\begin{equation}\label{cons113}\begin{split}
\left[\xnc{A}_{0}(x), \xnc{\mathscr{A}}(y)\right] &= \frac{\ui}{4\pi^2}\theta^{mn}\varepsilon_{mjk}\bigg[\frac{1}{2}\left(\partial_{n}A_{0}+F_{n0}\right)F_{jk}\delta^{3}(\mathbf{x}-\mathbf{y})\\
&\qquad\qquad\qquad\quad {} - F_{0j}(y)A_{n}(x)\partial_{k}^{y}\delta^{3}(\mathbf{x}-\mathbf{y})\bigg] + \uO(\theta^2).
\end{split}\end{equation}
Similarly we get
\begin{equation}
\label{cons108}\begin{split}
\left[\xnc{A}_{i}(x), \xnc{\mathscr{A}}(y)\right] &= -\frac{\ui}{4\pi^2}\varepsilon_{ijk}F_{jk}\delta^{3}(\mathbf{x}-\mathbf{y})\\
&\quad {}+\frac{\ui}{4\pi^2}\theta^{mn}\bigg[\varepsilon_{mjk}\left\{F_{ni}F_{jk}\delta^{3}(\mathbf{x}-\mathbf{y}) + \frac{1}{2}\partial_{i}^{x}\left(A_{n}F_{jk}\delta^{3}(\mathbf{x}-\mathbf{y})\right)\right\}\\
&\qquad\qquad\qquad {} + \varepsilon_{ijk}\left(A_{m}\partial_{n}F_{jk}\delta^{3}(\mathbf{x}-\mathbf{y})\right)\bigg] + \uO(\theta^2).
\end{split}
\end{equation}
Also, in view of the map \eqref{A.m}, we observe that $[\xnc{A}_{\nu}(x), \xnc{A}_{\mu}(y)]$ will not have at least any $\theta$-independent part, which means that the term involving this commutator on the right-hand side of Eq.~\eqref{cons107} drops out. Using Eqs.~\eqref{A.m}, \eqref{CCB104}, \eqref{CCB105}, \eqref{CCB106} and \eqref{cons113}, the right-hand side of Eq.~\eqref{cons107} for $\nu = 0$ reduces to
\begin{equation}\label{cons114}
\frac{\ui}{8\pi^2}\theta^{mn}\varepsilon_{njk}\partial_{0}\left(A_{m}F_{jk}\right)\delta^{3}(\mathbf{x}-\mathbf{y}) + \uO(\theta^2),
\end{equation}
which is also what the left-hand side of Eq.~\eqref{cons107} for $\nu = 0$ reduces to upon substituting for the commutator from Eq.~\eqref{CCB104}. For $\mu = i$, the left-hand side of Eq.~\eqref{cons107}, up to $\uO(\theta)$, vanishes in view of the Eq.~\eqref{CCB1051}, and the right-hand side, using Eqs.~\eqref{A.m}, \eqref{CCB1051}, \eqref{CCB1052}, \eqref{CCB107} and \eqref{cons108}, also vanishes. This shows the compatibility of the noncommutative anomalous commutators with the noncommutative anomaly.


\paragraph{Ambiguities in anomalous commutators and the consistency conditions.}
As mentioned in \refscite{Adler:1970qb, DeserOthers:1970}, the commutators given in the set \eqref{CS101}--\eqref{CB103} for the ordinary theory have been deduced from the triangle graph alone, which is also responsible for the current-divergence anomaly. This does not rule out the possibility that higher orders of perturbation theory may modify the values of these commutators. However, the commutators $S_{00}(x,y)$ and $M_{i0}(x,y)$ can also be deduced from simpler, exact commutators and equations of motion, which suggests that their value is exact to all orders of perturbation theory. On the other hand, the values given in the set \eqref{CS101}--\eqref{CB103} for the commutators $S_{i0}(x,y)$, $S_{0i}(x,y)$ and $M_{ik}(x,y)$ cannot be deduced in a way similar to those of $S_{00}(x,y)$ and $M_{i0}(x,y)$, and the possibility of the presence of additional terms is not ruled out. It has been shown \cite{Adler:1970qb, DeserOthers:1970} that if values of these commutators are modified to
\begin{gather}
\label{CS102m}
S_{i0}(x,y)
= -\frac{\ui}{4\pi^2}\varepsilon_{ijk}F_{0j}(x)\partial_{k}^y\delta^{3}(\mathbf{x}-\mathbf{y})+\ui\partial_{k}^{y}\left(T^{ik}\delta^{3}(\mathbf{x}-\mathbf{y})\right),\displaybreak[0]%
\\
\label{CS103m}
S_{0i}(x,y)
= \frac{\ui}{4\pi^2}\varepsilon_{ijk}F_{0j}(y)\partial_{k}^x\delta^{3}(\mathbf{x}-\mathbf{y})-\ui\partial_{k}^{x}\left(T^{ki}\delta^{3}(\mathbf{x}-\mathbf{y})\right),\displaybreak[0]%
\\
\label{CB103m}
M_{im}(x,y)
= \frac{\ui}{4\pi^2}\varepsilon_{imn}F_{0n}\delta^{3}(\mathbf{x}-\mathbf{y})-\ui T^{im}\delta^{3}(\mathbf{x}-\mathbf{y}),
\end{gather}
with $T^{ik}(y)$ a pseudotensor operator, then the consistency conditions, Eq.~\eqref{cons1041} for example, are unchanged. The implications of these modifications will now be analysed in the present context.

The first point to note is that the various anomalous commutators might get altered due to the additional $T^{ij}$-dependent pieces. We explicitly compute these modifications. Equations~\eqref{CCA1011}--\eqref{CCA108} relate the anomalous commutators in the noncommutative theory with their commutative counterparts. It becomes clear from these equations that the modifications \eqref{CS102m}--\eqref{CB103m} will not alter  the values of the commutators $\xnc{S}_{00}(x,y)$, $\xnc{L}_{00}(x,y)$, $\xnc{L}_{i0}(x,y)$, $\xnc{L}_{im}(x,y)$ and $\xnc{M}_{i0}(x,y)$ as given in the set \eqref{CCB101}--\eqref{CCB108}. The values of the remaining commutators will be modified as
\begin{gather}
\label{amb101}\begin{split}
\xnc{S}_{i0}(x,y) &= [\text{right-hand side of Eq.~\eqref{CCB102}}] + \ui\partial_{k}^{y}\left(T^{ik}\delta^{3}(\mathbf{x}-\mathbf{y})\right)\\
&\quad {} + \ui\theta^{im}F_{mn}(x)\partial_{k}^{y}\left(T^{nk}\delta^{3}(\mathbf{x}-\mathbf{y})\right)\\
&\quad {} - \ui\theta^{mn}\left[\partial_{n}^{y}\left\{A_{m}(y)\partial_{k}^{y}\left(T^{ik}\delta^{3}(\mathbf{x}-\mathbf{y})\right)\right\}\right.\\
&\qquad\qquad\quad \left.{} + \partial_{n}^{x}\left\{A_{m}(x)\partial_{k}^{y}\left(T^{ik}\delta^{3}(\mathbf{x}-\mathbf{y})\right)\right\}\right],
\end{split}\displaybreak[0]%
\\
\label{amb102}\begin{split}
\xnc{S}_{0i}(x,y) &= [\text{right-hand side of Eq.~\eqref{CCB103}}] - \ui\partial_{k}^{x}\left(T^{ki}\delta^{3}(\mathbf{x}-\mathbf{y})\right)\\
&\quad {} - \ui\theta^{im}F_{mn}(y)\partial_{k}^{x}\left(T^{kn}\delta^{3}(\mathbf{x}-\mathbf{y})\right)\\
&\quad {}+\ui\theta^{mn}\left[\partial_{n}^{y}\left\{A_{m}(y)\partial_{k}^{x}\left(T^{ki}\delta^{3}(\mathbf{x}-\mathbf{y})\right)\right\}\right.\\
&\qquad\qquad\quad \left.{} + \partial_{n}^{x}\left\{A_{m}(x)\partial_{k}^{x}\left(T^{ki}\delta^{3}(\mathbf{x}-\mathbf{y})\right)\right\}\right],
\end{split}\displaybreak[0]%
\\
\label{amb103}
\xnc{L}_{0i}(x,y) = [\text{right-hand side of Eq.~\eqref{CCB105}}] -\frac{\ui}{2}\theta^{mn}A_{m}T^{ni}\delta^{3}(\mathbf{x}-\mathbf{y}),\displaybreak[0]\\
\label{amb104}
\xnc{M}_{00}(x,y) = [\text{right-hand side of Eq.~\eqref{CCB106}}] +\frac{\ui}{2}\theta^{mn}A_{m}(x)\partial_{i}^{y}\left(T^{ni}\delta^{3}(\mathbf{x}-\mathbf{y})\right),\displaybreak[0]%
\\
\label{amb105}
\xnc{M}_{ik}(x,y) = [\text{right-hand side of Eq.~\eqref{CCB108}}] + (\cdots),
\end{gather}
where $(\cdots)$ appearing on the right-hand side of Eq.~\eqref{amb105} represents the terms involving $T^{ik}(y)$ whose explicit structure is not needed for our purpose.

Next we show that the conditions \eqref{cons104} and \eqref{cons107} still hold. The left-hand side of the condition~\eqref{cons104} does not involve any of the modified commutators  given in the set \eqref{amb101}--\eqref{amb105}, its value therefore remains unaltered. The right-hand side does involve the modified commutators, but it is a matter of straightforward algebra to show that there is no change in its value. The consistency condition \eqref{cons107} for $\nu = i$ does not involve any of the modified commutators, and therefore it trivially remains valid. As far as the condition \eqref{cons107} with $\nu = 0$ is concerned, its left-hand side is $\partial_{0}\xnc{L}_{00}(x,y)$ whose value obviously remains unaffected. The right-hand side involves the modified commutators, but again after some algebra we find that its value remains unchanged.


\section{\label{sec:anomalous-conclu}Discussion}

One might be tempted to guess the structures of these anomalous commutators as those obtained by a naive covariant deformation of the ordinary results, just as the covariant divergence anomaly~\eqref{anom-cov.d} is obtained by a covariant deformation of the usual result~\eqref{anom.d}. But a simple inspection rules out this possibility. The point is that the covariant deformation of a gauge-invariant expression can only give a star-gauge-covariant expression. Since the currents $J^{\mu}_{5}$ and $\xnc{J}^{\mu}_{5}$ are, respectively, gauge invariant and star-gauge covariant, so are the divergences $\partial_{\mu}J^{\mu}_{5}$ and $\xnc{\uD}_{\mu}\star\xnc{J}^{\mu}_{5}$. One could therefore expect that the star-gauge-covariant anomaly is obtained by a covariant deformation of the usual gauge-invariant anomaly. Explicit calculations serve to verify this expectation~\cite{Ardalan:2000cy, Gracia-Bondia:2000pz, Armoni:2002fh}. On the other hand, although the commutator $[J_{0}(x), J_{0}^{5}(y)]$, for example, is gauge invariant, yet its noncommutative counterpart, $[\xnc{J}_{0}(x), \xnc{J}_{0}^{5}(y)]$, is not star-gauge covariant because it involves two distinct spacetime points, $x$ and $y$. Therefore it becomes clear that the non-covariant commutator, $[\xnc{J}_{0}(x), \xnc{J}_{0}^{5}(y)]$, cannot be obtained by just a standard covariant deformation of the usual gauge-invariant commutator. Equations \eqref{CCC101}--\eqref{CCC108} indeed show that there is a departure from the naive covariant deformation of the corresponding gauge-invariant expressions.

The implications of Seiberg--Witten maps were discussed in the previous chapter in the context of divergence anomalies. We found that these maps are also useful in obtaining commutator anomalies. Although we analysed the case of the star-gauge-covariant current, it should be possible to extend this analysis to the star-gauge-invariant current since corresponding Seiberg--Witten maps are known to exist~\cite{Banerjee:2001un, Banerjee:2003ce}.


%% file: chap_lorentz.tex

\chapter{\label{chap:lorentz}Noncommutative gauge theories and Lorentz symmetry}


The issue of Lorentz symmetry in a noncommutative field theory has been debated \cite{Carroll:2001ws, Bichl:2001yf, Iorio:2001qy, Carlson:2002wj, Kase:2002pi, Alvarez-Gaume:2003mb, Imai:2003ek, Morita:2003vt, Morita:2003vx, Ghosh:2003cu} seriously, but it still remains a challenge leading to fresh insights \cite{Chaichian:2004za, Chaichian:2004yh}. The problem stems from the fact that pointwise multiplication of operators is replaced by a star multiplication:
\begin{equation}\label{lo-1}
A(x)B(x) \to A(x) \star B(x),
\end{equation}
which was defined in Eq.~\eqref{star.dcurr}:\footnote{This is the so-called canonical definition. There are other realisations like the Lie-algebra valued structure or the $q$-deformed structure---see footnote 1 of Chapter~\ref{chap:intro}.}
\begin{equation}\label{star.dlor}
A(x)\star B(x)=\left.\exp\left(\frac{\mathrm{i}}{2}\theta^{\alpha\beta}\partial_{\alpha}\partial'_{\beta}\right)A(x)B(x')\right|_{x'=x},
\end{equation}
where $\theta^{\alpha\beta}$ is a constant antisymmetric object. Hence the ordinarily vanishing commutators among spacetime coordinates acquire a nontrivial form:
\begin{equation}\label{xxmoyal}
[x^\mu, x^\nu] \to [x^\mu, x^\nu]_{\star} \equiv x^\mu \star x^\nu - x^\nu \star x^\mu = \mathrm{i}\theta^{\mu\nu}.
\end{equation}
Since $\theta^{\mu\nu}$ is constant, theories defined on such a noncommutative spacetime are considered to violate Lorentz invariance.

Nevertheless, in spite of this vexing problem, the basic issues of noncommutative field theory, like unitarity \cite{Gomis:2000zz}, causality \cite{Seiberg:2000gc}, mixing of UV/IR divergences \cite{Minwalla:1999px}, anomalies \cite{Ardalan:2000cy, Banerjee:2001un, Banerjee:2003ce} are discussed in a formally Lorentz-invariant manner, using the representaion of Poincar\'e algebra. To achieve a reconciliation, therefore, it is essential to obtain a conceptually cleaner understanding of Lorentz symmetry and its interpretaion in the noncommutative context. Precisely such a study is provided in this chapter.

We adopt a Noether-like approach\footnote{A somewhat similar approach, but with a different viewpoint, was followed in Ref.~\cite{Iorio:2001qy}.} to analyse the various spacetime symmetries of noncommutative electrodynamics. Here we deal with the classical (non-quantised) electromagnetic field. Although the present study is confined to the $\mathrm{U}$(1) group, it can be extended to other (nonabelian) groups. Since $\theta^{\mu\nu}$ is a constant, it appears as a background field in noncommutative electrodynamics. The Noether analysis, which is usually done for dynamical variables, is reformulated to include background fields. Now there are two possibilities for a constant $\theta^{\mu\nu}$. It may either be the same constant in all frames or it may transform as a second-rank tensor, taking different constant values in different frames. It is found that although the criterion for preserving translational invariance is the same in both cases, the criterion for Lorentz invariance (invariance under rotations and boosts) is different. An explicit computation shows that the criterion for Lorentz symmetry is satisfied only when $\theta^{\mu\nu}$ transforms as a tensor. Translational invariance is always satisfied. We also show that the transformations are dynamically consistent since the Noether charges correctly generate the transformations of an arbitrary function of canonical variables. Also, these charges satisfy the appropriate Lie brackets among themselves.

As is well known, noncommutative electrodynamics can be studied in two formulations; either in terms of the original noncommutative variables or, alternatively, in terms of its commutative equivalents obtained by using the Seiberg--Witten maps \cite{Seiberg:1999vs}. Our analysis has been carried out in both formulations, up to first order in $\theta$. A complete equivalence among the results has also been established. This is rather nontrivial since there are examples where this equivalence does not hold. For example, the IR problem found in noncommutative field theory \cite{Matusis:2000jf, Chepelev:2000hm} is absent in the commutative-variable approach \cite{Bichl:2001nf}, revealing an inequivalence, at least on a perturbative level.

It is reassuring to note that an important feature \cite{Alvarez-Gaume:2003mb} of quantum field theory on 4-dimensional noncommutative spacetime, namely, the invariance for a constant nontransforming $\theta$ under the $\mathrm{SO}(1,1)\times\mathrm{SO}(2)$ subgroup of Lorentz group is reproduced by the criteria found here. This has been shown in both the commutative and noncommutative descriptions.

Although the noncommutativity of the spacetime coordinates violates relativistic invariance, it has been recently shown by using the (twisted) Hopf algebra that corresponding field theories possess deformed symmetries \cite{Wess:2003da, Chaichian:2004za, Chaichian:2004yh}. We shall discuss such deformed symmetries in Chapter \ref{chap:deform}.

In section \ref{sec:lorentz-review}, the occurrence of noncommutative algebra in various approaches and their possible connections is briefly reviewed. Section \ref{sec:lorentz-toy} deals with the implications of Lorentz symmetry in a toy model comprising a usual Maxwell field coupled to an external source, whereas 
section \ref{sec:lorentz-nced} provides
a detailed account of Lorentz symmetry in noncommutative electrodynamics, first in the commutative-variable approach and then in terms of noncommutative variables.


\section{\label{sec:lorentz-review}A brief review of noncommutative algebra}

We start by briefly reviewing Snyder's algebra \cite{Snyder:1946qz}. The special theory of relativity may be based on the invariance of the indefinite quadratic form
\begin{equation}\label{sr}
S^{2} = (x^{0})^{2}-(x^{1})^{2}-(x^{2})^{2}-(x^{3})^{2} = -x_{\mu}x^{\mu}
\end{equation}
for transformation from one inertial frame to another. We shall use $(-, +, +, +)$ signature for the flat Minkowski metric $\eta_{\mu\nu}$. It is usually assumed that the variables $x^{\mu}$ take on a continuum of values and that they may take on these values simultaneously. Snyder considered a different situation. He considered Hermitian operators, $\op{x}^{\mu}$, for the spacetime coordinates of a particular Lorentz frame. He further assumed that the spectra of spacetime coordinate operators $\op{x}^{\mu}$ are invariant under Lorentz transformations. The later assumption is evidently satisfied by the usual spacetime continuum, however it is not the only solution. Snyder showed that there exists a Lorentz-invariant spacetime in which there is a natural unit of length.

To find operators $\op{x}^{\mu}$ possessing Lorentz-invariant spectra, Snyder considered the homogeneous quadratic form
\begin{equation}\label{qf}
-(y)^{2} = (y_{0})^{2}-(y_{1})^{2}-(y_{2})^{2}-(y_{3})^{2}-(y_{4})^{2} = -y_{\mu}y^{\mu}-(y_{4})^{2},
\end{equation}
in which $y$'s are assumed to be real variables. Now $\op{x}^{\mu}$ are defined by means of the infinitesimal elements of the group under which the quadratic form \eqref{qf} is invariant. The $\op{x}^{\mu}$ are taken as
\begin{equation}\label{opx}
\op{x}^{\mu} = \mathrm{i}a\left(y_{4}\frac{\partial}{\partial y_{\mu}}-y^{\mu}\frac{\partial}{\partial y_{4}}\right),
\end{equation}
in which $a$ is the natural unit of length. These operators are assumed to be Hermitian and operate on the single-valued functions of $y_{\mu}, y_{4}$. The spectra of $\op{x}^{i}$, $i=1,2,3$, are discrete, but $\op{x}^{0}$ has a continuous spectrum extending from $-\infty$ to $+\infty$. Transformations which leave the quadratic form \eqref{qf} and $y_{4}$ invariant are covariant Lorentz transformations on the variables $y_{1}$, $y_{2}$, $y_{3}$ and $y_{0}$, and these transformations induce contravariant Lorentz transformations in $\op{x}^{\mu}$.

Now six additional operators are defined as
\begin{equation}\label{opM}
\op{M}^{\mu\nu} = -\mathrm{i}\left(y^{\mu}\frac{\partial}{\partial y_{\nu}}-y^{\nu}\frac{\partial}{\partial y_{\mu}}\right),
\end{equation}
which are the infinitesimal elements of the four-dimensional Lorentz group. The ten operators defined in Eqs.~\eqref{opx} and \eqref{opM} have the following commutation relations:
\begin{gather}
\label{xxlor}
\left[\op{x}^{\mu}, \op{x}^{\nu}\right] = \mathrm{i}a^{2}\op{M}^{\mu\nu},\\
\label{Mx}
\left[\op{M}^{\mu\nu}, \op{x}^{\lambda}\right] = \mathrm{i}\left(\op{x}^{\mu}\eta^{\nu\lambda}-\op{x}^{\nu}\eta^{\mu\lambda}\right),\\
\label{MM}
\left[\op{M}^{\mu\nu}, \op{M}^{\alpha\beta}\right] = \mathrm{i}\left(\op{M}^{\mu\beta}\eta^{\nu\alpha}-\op{M}^{\mu\alpha}\eta^{\nu\beta}+\op{M}^{\nu\alpha}\eta^{\mu\beta}-\op{M}^{\nu\beta}\eta^{\mu\alpha}\right).
\end{gather}
The Lorentz $\mathrm{SO}(3, 1)$ symmetry given in Eq.~\eqref{MM} is extended to $\mathrm{SO}(4, 1)$ symmetry specified by Eqs.~\eqref{xxlor}--\eqref{MM}.

Since the position operators $\op{x}^{i}$ have discrete spectra, we can understand it in terms of a nonzero minimal uncertainty in positions. It is possible to obtain the space part of Snyder algebra by considering the generalised Heisenberg algebra\footnote{The space part of Snyder algebra can also be obtained from another generalised Heisenberg algebra considered in \refcite{Kempf:1994su}.} (with $\hbar = 1$):
\begin{equation}\label{k-xp}
\left[\op{x}_i, \op{p}_j\right] = \mathrm{i}\delta_{ij}\sqrt{1+a^{2}p_{k}p_{k}},
\end{equation}
which implies nonzero minimal uncertainties in position coordinates, and preserves the rotational symmetry. Representing the generalised Heisenberg algebra on momentum wave functions $\psi(p) = \langle p|\psi\rangle$,
\begin{gather}
\label{k-ppsi}
\op{p}_i \psi(p) = p_i \psi(p),\\
\label{k-xpsi}
\op{x}_i \psi(p) = \mathrm{i}\sqrt{1+a^{2}p_k p_k} \partial_{p_i} \psi(p),
\end{gather}
we get the commutation relation among the position operators:
\begin{equation}\label{k-xx}
\left[\op{x}_i, \op{x}_j\right] = -a^{2}\left(p_{i}\partial_{p_{j}}-p_{j}\partial_{p_{i}}\right) \equiv \mathrm{i}a^{2}\op{M}_{ij},
\end{equation}
where we have defined
\begin{equation}\label{k-M.d}
\op{M}_{ij} = \mathrm{i}\left(p_{i}\partial_{p_{j}}-p_{j}\partial_{p_{i}}\right).
\end{equation}
Thus we have
\begin{gather}
\label{k-xM}
\left[\op{M}_{ij}, \op{x}_{k}\right] = \mathrm{i}\left(\op{x}_{i}\delta_{jk}-\op{x}_{j}\delta_{ik}\right),\\
\label{k-MM}
\left[\op{M}_{ij}, \op{M}_{kl}\right] = \mathrm{i}\left(\op{M}_{il}\delta_{jk}-\op{M}_{ik}\delta_{jl}+\op{M}_{jk}\delta_{il}-\op{M}_{jl}\delta_{ik}\right).
\end{gather}
The algebra \eqref{k-xx}, \eqref{k-xM} and \eqref{k-MM} exactly reproduces the space part of the Snyder algebra \eqref{xxlor}--\eqref{MM}.

Doplicher, Fredenhagen and Roberts \cite{Doplicher:1994zv, Doplicher:1994tu} proposed a new algebra (DFR algebra) of a noncommutative spacetime through considerations on the spacetime uncertainty relations derived from quantum mechanics and general relativity. This algebra defines a Lorentz-invariant noncommutative spacetime different from Snyder's quantised spacetime. Their algebra is given by
\begin{gather}
\label{xx-dfr}
\left[\op{x}^{\mu}, \op{x}^{\nu}\right] = \mathrm{i}\opgr{\theta}^{\mu\nu},\\
\label{thx}
\left[\opgr{\theta}^{\mu\nu}, \op{x}^{\lambda}\right] = 0,\\
\label{thth}
\left[\opgr{\theta}^{\mu\nu}, \opgr{\theta}^{\alpha\beta}\right] = 0.
\end{gather}

Recently, Carlson \emph{et~al.}~\cite{Carlson:2002wj} rederived this DFR algebra by `contraction' of Snyder's algebra. For that they considered
\begin{equation}\label{Mth}
\op{M}^{\mu\nu} = \frac{1}{b}{\opgr{\theta}^{\mu\nu}},
\end{equation}
and the limits $b \to 0$, $a \to 0$ with the ratio of $a^{2}$ and $b$ held fixed: $(a^{2}/b) \to 1$.
The result of this contraction is the algebra given by Eqs.~\eqref{xx-dfr}--\eqref{thth}. It also follows that
\begin{equation}\label{M-th}
\left[\op{M}^{\mu\nu}, \opgr{\theta}^{\alpha\beta}\right] = \mathrm{i}\left(\opgr{\theta}^{\mu\beta}\eta^{\nu\alpha}+\opgr{\theta}^{\nu\alpha}\eta^{\mu\beta}-\opgr{\theta}^{\mu\alpha}\eta^{\nu\beta}-\opgr{\theta}^{\nu\beta}\eta^{\mu\alpha}\right).
\end{equation}
Since $a \to 0$ is a part of the limit, the contracted algebra corresponds to a continuum limit of Snyder's quantised spacetime.\footnote{The validity of this contraction process is questionable. Let us recall the familiar contractions of the group $\mathrm{SO}(3)$ to the group $\mathrm{E}_2$, and of the Poincar\'e group to the Galilean group. In the limit of infinite radius, $\mathrm{SO}(3)$, which is the symmetry group of the surface of the sphere, contracts to $\mathrm{E}_2$, the symmetry group of a plane. Likewise, in the low-velocity limit, the Poincar\'e group contracts to the Galilean group. These contractions involve taking limit of one parameter only whereas the above mentioned contraction of Snyder algebra to DFR algebra is achieved by taking limits of two parameters, $a \to 0$ and $b \to 0$. Furthermore, in the standard group contraction we can identify a mapping among the generators of the two groups, but in the mapping \eqref{Mth}, $\opgr{\theta}^{\mu\nu}$ is not a generator associated with any symmetry group. In this context, therefore, we agree with Kase \emph{et~al.}~\cite{Kase:2002pi} that there is no connection between the two algebras.}

Here we shall consider noncommutative electrodynamics which is obtained by a standard deformation of the usual (commutative) Maxwell theory, replacing pointwise multiplication by a star multiplication defined by Eq.~\eqref{star.dlor}. We shall show in what precise sense Lorentz symmetry is interpreted to be valid, or otherwise. To facilitate our analysis we first develop the formulation in the context of a simple toy model.


\section{\label{sec:lorentz-toy}A toy model}

We know from Noether's theorem that the invariance of action under a symmetry group, and a spacetime transformation in particular, implies the existence of a current $J^{\mu}$ satisfying a continuity equation $\partial_{\mu}J^{\mu}=0$. We shall now investigate what happens when the action contains vector or tensor parameters which are not included in the configuration space, i.e., there are external vector or tensor parameters in the theory. Before we consider the noncommutative Maxwell theory, which contains a tensor parameter $\theta^{\alpha\beta}$, it will be advantageous to first start with a simpler case.

We consider ordinary Maxwell theory with the potential coupled to an external source:
\begin{equation}\label{901lor}
S \equiv \int\!\mathrm{d}^{4}x\,\mathscr{L} = -\int\!\mathrm{d}^{4}x\,\left(\frac{1}{4}F_{\mu\nu}F^{\mu\nu}+j^{\mu}A_{\mu}\right).
\end{equation}
Here $j_{\mu}$ is taken to be a constant vector, i.e., it is constant but transforms as a vector when we go from one coordinate frame to another.\footnote{Later we shall also consider the case where $j^{\mu}$ does not transform like a vector but is fixed for all frames. In that case, one expects that the Lorentz invariance of the action will not be preserved.} Here we would like to mention that for the realistic current sources, $j^{\mu}$ corresponds to a vector function which is localised in space. In this sense, therefore, $j^{\mu}$ should be treated as a hypothetical source as it has been taken to be constant throughout. We are just interested in studying the Lorentz-transformation property of this system.

Let us consider an infinitesimal transformation of the coordinate system:
\begin{equation}\label{902}
x^{\mu} \rightarrow x'^{\mu} = x^{\mu}+\delta x^{\mu},
\end{equation}
under which $A^{\mu}$ and $j^{\mu}$ transform as
\begin{gather}
\label{903-1}
A^{\mu}(x) \rightarrow A'^{\mu}(x') = A^{\mu}(x)+\delta A^{\mu}(x),\\
\label{903lor}
j^{\mu} \rightarrow j'^{\mu} = j^{\mu}+\delta j^{\mu}.
\end{gather}
The change in the action resulting from these transformations is
\begin{equation}\label{904lor}
\delta S = \int_{\Omega'}\!\mathrm{d}^{4}x'\,\mathscr{L}\left(A'_{\nu}(x'),\partial'_{\mu}A'_{\nu}(x');j'_{\nu}\right)-\int_{\Omega}\!\mathrm{d}^{4}x\,\mathscr{L}\left(A_{\nu}(x),\partial_{\mu}A_{\nu}(x);j_{\nu}\right),
\end{equation}
where $\Omega$ is an arbitrarily large closed region of spacetime and $\Omega'$ being the transform of $\Omega$ under the coordinate change \eqref{902}. The above change in action can be rewritten as
\begin{equation}\label{905lor}
\begin{split}
\delta S &= \int_{\Omega}\!\mathrm{d}^{4}x\,\left[\mathscr{L}\left(A'_{\nu}(x),\partial_{\mu}A'_{\nu}(x);j'_{\nu}\right)-\mathscr{L}\left(A_{\nu}(x),\partial_{\mu}A_{\nu}(x);j_{\nu}\right)\right]\\
& \quad{}+\int_{\Omega'-\Omega}\!\mathrm{d}^{4}x\,\mathscr{L}\left(A'_{\nu}(x),\partial_{\mu}A'_{\nu}(x);j'_{\nu}\right).
\end{split}
\end{equation}
The last term, an integral over the infinitesimal volume $\Omega'-\Omega$, can be written as an integral over the boundary $\partial\Omega$:
\begin{equation}\label{9051}
\begin{split}
\int_{\Omega'-\Omega}\!\mathrm{d}^{4}x\,\mathscr{L}\left(A'_{\nu},\partial_{\mu}A'_{\nu};j'_{\nu}\right) &= \int_{\partial\Omega}\!\mathrm{d}S_{\lambda}\,\delta x^{\lambda}\mathscr{L}(A_{\nu},\partial_{\mu}A_{\nu};j_{\nu})\\
&= \int_{\Omega}\!\mathrm{d}^{4}x\,\partial_{\lambda}\left[\delta x^{\lambda}\mathscr{L}(A_{\nu},\partial_{\mu}A_{\nu};j_{\nu})\right],
\end{split}
\end{equation}
where Gauss theorem has been used in the last step.
For any function $f(x)$, we can write
\begin{equation}\label{906lor}
\delta f = f'(x')-f(x) = \delta_{0}f + \delta x^{\mu}\partial_{\mu}f,
\end{equation}
where $\delta_{0}f = f'(x)-f(x)$ is the functional change. Since we have taken $j^{\mu}$ to be constant, $\delta_{0}j^{\mu}=\delta j^{\mu}$. Now we have
\begin{equation}\label{911lor}
\begin{split}
&\mathscr{L}\left(A'_{\nu}(x),\partial_{\mu}A'_{\nu}(x);j'_{\nu}\right)-\mathscr{L}\left(A_{\nu}(x),\partial_{\mu}A_{\nu}(x);j_{\nu}\right)\\
&\quad= \frac{\partial\mathscr{L}}{\partial A_{\nu}}\delta_{0}A_{\nu}+\frac{\partial\mathscr{L}}{\partial(\partial_{\mu}A_{\nu})}\delta_{0}\partial_{\mu}A_{\nu}+\frac{\partial\mathscr{L}}{\partial j_{\nu}}\delta j_{\nu}.
\end{split}
\end{equation}
Using the equation of motion
\begin{equation}\label{eom1}
\frac{\partial \mathscr{L}}{\partial A_{\nu}}-\partial_{\mu}\left(\frac{\partial \mathscr{L}}{\partial(\partial_{\mu}A_{\nu})}\right) = 0,
\end{equation}
and the relations \eqref{9051} and \eqref{911lor}, we can cast Eq.~\eqref{905lor} as
\begin{equation*}
\delta S = \int_{\Omega}\!\mathrm{d}^{4}x\,\left[\partial_{\mu}\left(\mathscr{L}\delta x^{\mu}+\frac{\partial\mathscr{L}}{\partial(\partial_{\mu}A_{\nu})}\delta_{0}A_{\nu}\right)+\frac{\partial\mathscr{L}}{\partial j_{\nu}}\delta j_{\nu}\right].
\end{equation*}
In view of Eq.~\eqref{906lor}, we can write\footnote{Now onwards we drop the explicit display of $\Omega$ as we take this to correspond to entire spacetime in a suitable limit.}
\begin{equation}\label{913lor}
\delta S = \int\!\mathrm{d}^{4}x\,\left[\partial_{\mu}\left(\frac{\partial\mathscr{L}}{\partial(\partial_{\mu}A_{\nu})}\delta A_{\nu}-T^{\mu\nu}\delta x_{\nu}\right)+\frac{\partial\mathscr{L}}{\partial j_{\nu}}\delta j_{\nu}\right],
\end{equation}
where $T^{\mu\nu}$ is the canonical energy--momentum tensor defined by
\begin{equation}\label{914lor}
T^{\mu\nu} = \frac{\partial\mathscr{L}}{\partial(\partial_{\mu}A_{\sigma})}\partial^{\nu}A_{\sigma}-\eta^{\mu\nu}\mathscr{L}.
\end{equation}

For spacetime translations, $\delta x^{\mu}=a^{\mu}$, a constant, while $\delta A_{\mu}=0$ and $\delta j_{\mu}=0$. So the invariance of the action under translations implies
\begin{equation*}
\int\!\mathrm{d}^{4}x\,(\partial_{\mu}T^{\mu\nu})a_{\nu} = 0.
\end{equation*}
Since it is true for arbitrary $a_{\nu}$, we must have
\begin{equation}\label{916lor}
\partial_{\mu}T^{\mu\nu} = 0.
\end{equation}
This is the criterion for translational invariance of the action.

In the case of infinitesimal Lorentz transformations (rotations and boosts), $\delta x_{\mu}=\omega_{\mu\nu}x^{\nu}$, $\delta A_{\mu}=\omega_{\mu\nu}A^{\nu}$ and $\delta j_{\mu}=\omega_{\mu\nu}j^{\nu}$, where $\omega_{\mu\nu}$ is constant and antisymmetric. So the invariance of the action implies
\begin{equation*}
\int\!\mathrm{d}^{4}x\,\left[\partial_{\mu}\bigg(\frac{\partial\mathscr{L}}{\partial(\partial_{\mu}A_{\lambda})}A^{\rho}-\frac{\partial\mathscr{L}}{\partial(\partial_{\mu}A_{\rho})}A^{\lambda}-T^{\mu\lambda}x^{\rho}+T^{\mu\rho}x^{\lambda}\bigg)+\frac{\partial \mathscr{L}}{\partial j_{\lambda}}j^{\rho}-\frac{\partial \mathscr{L}}{\partial j_{\rho}}j^{\lambda}\right]\omega_{\lambda\rho} = 0.
\end{equation*}
Since it is true for arbitrary $\omega_{\lambda\rho}$, we must have
\begin{equation}\label{918lor}
\partial_{\mu}M^{\mu\lambda\rho}+\frac{\partial \mathscr{L}}{\partial j_{\lambda}}j^{\rho}-\frac{\partial \mathscr{L}}{\partial j_{\rho}}j^{\lambda} = 0,
\end{equation}
where
\begin{equation}\label{9181}
M^{\mu\lambda\rho} = \frac{\partial\mathscr{L}}{\partial(\partial_{\mu}A_{\lambda})}A^{\rho}-\frac{\partial\mathscr{L}}{\partial(\partial_{\mu}A_{\rho})}A^{\lambda}-T^{\mu\lambda}x^{\rho}+T^{\mu\rho}x^{\lambda}.
\end{equation}
Therefore, the criterion for Lorentz invariance of the action is
\begin{equation}\label{918-22}
\partial_{\mu}M^{\mu\lambda\rho}-A^{\lambda}j^{\rho}+A^{\rho}j^{\lambda} = 0.
\end{equation}

Now we shall obtain the criteria for translational invariance and Lorentz invariance of the action when $j^{\mu}$ is not a genuine vector but has the same constant value in all frames. In that case we have $\delta j^{\mu} = 0$ not only under translations but also under Lorentz transformations. Therefore the last term inside the parentheses on the right-hand side of Eq.~\eqref{913lor} drops out and the criteria for the invariance of the action turn out to be
\begin{gather}
\label{crit1}
\partial_{\mu}T^{\mu\nu} = 0,\\
\label{crit2}
\partial_{\mu}M^{\mu\lambda\rho} = 0.
\end{gather}
Thus, the criterion for translational invariance is the same irrespective of whether $j^{\mu}$ is a genuine vector or not. However, this is not the case with the criterion for Lorentz invariance.

Now we shall explicitly evaluate $\partial_{\mu}T^{\mu\nu}$ and $\partial_{\mu}M^{\mu\lambda\rho}$ for our toy model \eqref{901lor}. This will obviously be independent of whether $j^{\mu}$ transforms like a vector or not. Using
\[
\partial^{\nu}\mathscr{L} = \frac{\partial \mathscr{L}}{\partial A_{\rho}}\partial^{\nu}A_{\rho}+\frac{\partial \mathscr{L}}{\partial(\partial_{\kappa} A_{\rho})}\partial^{\nu}\partial_{\kappa}A_{\rho},
\]
the equation of motion \eqref{eom1}, and the definition \eqref{914lor} of energy--momentum tensor, we find
\begin{equation}\label{find1}
\partial_{\mu}T^{\mu\nu} = 0.
\end{equation}
Also, using the equation of motion \eqref{eom1}, Eq.~\eqref{find1} and the defintion \eqref{9181} of $M^{\mu\lambda\rho}$, we find for our theory \eqref{901lor} that
\begin{equation}\label{dM}
\partial_{\mu}M^{\mu\lambda\rho} = A^{\lambda}j^{\rho}-A^{\rho}j^{\lambda}.
\end{equation}
As mentioned earlier, the results \eqref{find1} and \eqref{dM} do not depend whether $j^{\mu}$ transforms like a vector or not.

We have seen that the criterion for translational invariance is the same,  $\partial_{\mu}T^{\mu\nu} = 0$, in both the cases, independent of whether $j^{\mu}$ transforms like a vector or not. This is satisfied in view of Eq.~\eqref{find1}, thereby indicating that our toy model has translational invariance in both the cases. However, the criterion for Lorentz invariance is different in the two cases---see Eqs.~\eqref{918-22} and \eqref{crit2}---whereas what we have actually found is given by Eq.~\eqref{dM}. Since this agrees with the criterion \eqref{918-22}, our model has Lorentz invariance only when $j^{\mu}$ transforms like a vector, and not in the other case.

We shall now show that using the Noether charges
\begin{equation}\label{gen}
P^{\mu} = \int\!\mathrm{d}^{3}x\,T^{0\mu}, \qquad
J^{\mu\nu} = \int\!\mathrm{d}^{3}x\,M^{0\mu\nu},
\end{equation}
and the canonical equal-time Poisson brackets $\{A_{\mu}(t, \mathbf{x}), \pi^{\nu}(t, \mathbf{y})\} = \delta_{\mu}^{\nu}\delta^{3}(\mathbf{x}-\mathbf{y})$, we can generate the transformations of the dynamical variables $A_i$ and $\pi^{i}$:
\begin{equation}\label{AQL}
\left\{A_{i}, Q_{V}\right\} = \mathcal{L}_{V}A_{i}, \qquad
\left\{\pi^{i}, Q_{V}\right\} = \mathcal{L}_{V}\pi^{i},
\end{equation}
where $Q_{\partial_{\mu}}=P_{\mu}$, $Q_{x_{[\mu}\partial_{\nu]}}=J_{\mu\nu}$ and  $\mathcal{L}_{V}A_{i}$ stands for the Lie derivative\footnote{If $W{}^{\alpha\ldots\beta}_{\mu\ldots\nu}(x) \to W'{}^{\alpha\ldots\beta}_{\mu\ldots\nu}(x')$ for an arbitrary tensor field under the infinitesimal transformation $x^{\mu} \to x'^{\mu} = x^{\mu}-bV^{\mu}$, then the Lie derivative of $W(x)$ with respect to the vector field  $V(x) = V^{\mu}(x)\partial_{\mu}$ is defined as
\[
\left(\mathcal{L}_{V}W\right)^{\alpha\ldots\beta}_{\mu\ldots\nu}(x) = \lim_{b \to 0}\frac{1}{b}\left(W'{}^{\alpha\ldots\beta}_{\mu\ldots\nu}(x)-W{}^{\alpha\ldots\beta}_{\mu\ldots\nu}(x)\right).
\]
} of the field $A_i$ with respect to the vector field $V$ associated with the charge $Q_{V}$.

The canonical momenta of the theory are
\begin{gather}
\label{pi0}
\pi^{0} = \frac{\partial\mathscr{L}}{\partial(\partial_{0}A_{0})} = 0,\\
\label{pii}
\pi^{i} = \frac{\partial\mathscr{L}}{\partial(\partial_{0}A_{i})} = F^{i0}.
\end{gather}
It follows from the definitions \eqref{914lor} and \eqref{9181} that
\begin{gather}
\label{T00}
T^{00} = \pi^{i}\partial_{i}A^{0}-\frac{1}{2}\pi^{i}\pi_{i}-\frac{1}{4}F_{ij}F^{ij}-j^{\mu}A_{\mu},\\
\label{T0i}
T^{0i} = \pi^{j}\partial^{i}A_{j},\\
\label{M00i}
M^{00i} = -T^{00}x^{i}-\pi^{i}A^{0}+x^{0}\pi^{j}\partial^{i}A_{j},\\
\label{M0ij}
M^{0ij} = \pi^{i}A^{j}-x^{j}\pi^{k}\partial^{i}A_{k}-\pi^{j}A^{i}+x^{i}\pi^{k}\partial^{j}A_{k},
\end{gather}
where we have used Eq.~\eqref{pii} to eliminate velocities in favour of momenta. Now we compute the Poisson brackets of the field $A_{i}$ with the charges:
\begin{gather}
\label{AP1}
\left\{A_{i}, P_{j}\right\} = \partial_{j}A_{i},\\
\label{AP2}
\left\{A_{i}, P_{0}\right\} = \partial_{i}A_{0}+\pi_{i} = \partial_{0}A_{i},\\
\label{AJ1}
\left\{A_{i}, J_{kl}\right\} = \eta_{ik}A_{l}-x_{l}\partial_{k}A_{i}-\eta_{il}A_{k}+x_{k}\partial_{l}A_{i},\\
\label{AJ2}
\left\{A_{i}, J_{0l}\right\} = -x_{l}\left(\partial_{i}A_{0}+\pi_{i}\right)-\eta_{il}A_{0}+x_{0}\partial_{l}A_{i} = -x_{l}\partial_{0}A_{i}-\eta_{il}A_{0}+x_{0}\partial_{l}A_{i},
\end{gather}
where the definition \eqref{pii} of momenta has been used in the second steps of Eqs.~\eqref{AP2} and \eqref{AJ2}. Since
\begin{gather}
\label{ld1}
\mathcal{L}_{\partial_{\mu}}A_{i} = \partial_{\mu}A_{i},\\
\label{ld2}
\mathcal{L}_{x_{[\mu}\partial_{\nu]}}A_{i} = \eta_{i\mu}A_{\nu}-x_{\nu}\partial_{\mu}A_{i}-\eta_{i\nu}A_{\mu}+x_{\mu}\partial_{\nu}A_{i},
\end{gather}
it follows that
\begin{equation}
\label{APL}
\left\{A_{i}, P_{\mu}\right\} = \mathcal{L}_{\partial_{\mu}}A_{i}, \qquad
\left\{A_{i}, J_{\mu\nu}\right\} = \mathcal{L}_{x_{[\mu}\partial_{\nu]}}A_{i}.
\end{equation}
The brackets of the momenta $\pi_{i}$ with the charges are
\begin{gather}
\label{piP1}
\left\{\pi_{i}, P_{j}\right\} = \partial_{j}\pi_{i},\\
\label{piP2}
\left\{\pi_{i}, P_{0}\right\} = \partial_{k}{F^{k}}_{i}-j_{i} = \partial_{0}\pi_{i},\\
\label{piJ1}
\left\{\pi_{i}, J_{kl}\right\} = \eta_{ik}\pi_{l}-x_{l}\partial_{k}\pi_{i}-\eta_{il}\pi_{k}+x_{k}\partial_{l}\pi_{i},\\
\label{piJ2}
\left\{\pi_{i}, J_{0l}\right\} = -x_{l}\left(\partial_{k}{F^{k}}_{i}-j_{i}\right)+x_{0}\partial_{l}\pi_{i}-F_{li} = -x_{l}\partial_{0}\pi_{i}+x_{0}\partial_{l}\pi_{i}-F_{li},
\end{gather}
where, in the second steps of Eqs.~\eqref{piP2} and \eqref{piJ2}, we have used $\partial_{0}\pi^{i}=\partial_{k}F^{ki}-j^{i}$ which is a consequence of the equation of motion \eqref{eom1}:
\begin{gather}
\label{motion}
\partial_{\mu}F^{\mu\nu}-j^{\nu}=0\\
\nonumber
\Rightarrow \quad \partial_{0}F^{0i}+\partial_{k}F^{ki}-j^{i} = -\partial_{0}\pi^{i}+\partial_{k}F^{ki}-j^{i}=0.
\end{gather}
Since\footnote{It is perhaps worthwhile to mention that while computing the Lie derivative of $\pi^{i}$, one should keep in mind that $\pi^{i}$ are not the components of a 4-vector. Rather, $\pi^{i}$ are the components of a tensor, $\pi^{i}=F^{i0}$.}
\begin{gather}
\label{pild1}
\mathcal{L}_{\partial_{\mu}}\pi_{i} = \partial_{\mu}\pi_{i},\\
\label{pild2}
\mathcal{L}_{x_{[k}\partial_{l]}}\pi_{i} = \eta_{ik}\pi_{l}-x_{l}\partial_{k}\pi_{i}-\eta_{il}\pi_{k}+x_{k}\partial_{l}\pi_{i},\\
\label{pild3}
\mathcal{L}_{(x_{0}\partial_{l}-x_{l}\partial_{0})}\pi_{i} = -x_{l}\partial_{0}\pi_{i}+x_{0}\partial_{l}\pi_{i}-F_{li},
\end{gather}
it follows that
\begin{equation}
\label{piPL}
\left\{\pi_{i}, P_{\mu}\right\} = \mathcal{L}_{\partial_{\mu}}\pi_{i}, \qquad
\left\{\pi_{i}, J_{\mu\nu}\right\} = \mathcal{L}_{x_{[\mu}\partial_{\nu]}}\pi_{i}.
\end{equation}
Hence we have shown that Eq.~\eqref{AQL} is indeed satisfied.

We also find that
\begin{gather}
\label{PP}
\left\{P_{i}, P_{j}\right\} = 0,\\
\label{PJ}
\left\{P_{i}, J_{kl}\right\} = \eta_{ik}P_{l}-\eta_{il}P_{k},\\
\label{JJ}
\left\{J_{ij}, J_{kl}\right\} = \eta_{jk}J_{il}+\eta_{il}J_{jk}-\eta_{ik}J_{jl}-\eta_{jl}J_{ik}.
\end{gather}
Now it follows that restricting to kinematical generators ($P_{i}$ and $J_{ij}$) only, we have
\begin{equation}\label{QQ}
\left\{Q_{U}, Q_{V}\right\} = Q_{[U,V]}.
\end{equation}
Thus we see that, although $\partial_{0}J^{\mu\nu}\neq 0$ (in view of Eq.~\eqref{dM} and the definition of $J^{\mu\nu}$ in \eqref{gen}), we still have Eqs.~\eqref{AQL} and \eqref{QQ}. This is necessary for establishing the dynamical consistency of the transformations.

It should be stressed that the Hamiltonian approach violates manifest Lorentz invariance. The fact that it gets restored is thus quite nontrivial. A possible way to see the manifest violation is through Eq.~\eqref{pi0}. Within the Hamiltonian formulation, however, this equation really is a primary constraint and the equality is only `weakly' valid \cite{Dirac:1964}. Time-conserving the primary constraint leads to a secondary (Gauss) constraint. This is basically the zero-component of the equation of motion \eqref{motion}, expressed in phase-space variables:
\begin{equation}\label{gauss1}
\partial_{i}\pi^{i}-j^{0} \approx 0.
\end{equation}
There are no further constraints. These constraints do not affect the realisation of the three-dimensional Euclidean symmetry \eqref{PP}--\eqref{JJ}.


\section{\label{sec:lorentz-nced}Noncommutative electrodynamics}


\subsection{Commutative-variable approach}

We now generalise the case of vector source considered in the previous section to antisymmetric tensor `source' $\theta^{\mu\nu}$. We take the noncommutative Maxwell theory:
\begin{equation}\label{l-105}
\xnc{S}=-\frac{1}{4}\int\!\mathrm{d}^{4}x\,\left(\xnc{F}_{\mu\nu}\star\xnc{F}^{\mu\nu}\right).
\end{equation}
On applying the Seiberg--Witten maps,
\begin{gather}
\label{101-lo}
\xnc{A}_{\mu} = A_{\mu}-\frac{1}{2}\theta^{\alpha\beta}A_{\alpha}\left(\partial_{\beta}A_{\mu}+F_{\beta\mu}\right)+ \uO(\theta^{2}),\\
\label{107-lo}
\xnc{F}_{\mu\nu} = F_{\mu\nu}-\theta^{\alpha\beta}\left(A_{\alpha}\partial_{\beta}F_{\mu\nu}+F_{\mu\alpha}F_{\beta\nu}\right)+ \uO(\theta^{2}),
\end{gather}
we get the effective theory in terms of usual (commutative) variables:
\begin{equation}\label{921lor}
S = -\int\!\mathrm{d}^{4}x\,\left[\frac{1}{4}F_{\mu\nu}F^{\mu\nu}+\theta^{\alpha\beta}\left(\frac{1}{2}F_{\mu\alpha}F_{\nu\beta}+\frac{1}{8}F_{\beta\alpha}F_{\mu\nu}\right)F^{\mu\nu}\right]+ \uO(\theta^2),
\end{equation}
where a boundary term has been dropped in order to express it solely in terms of the field strength. Although we have kept only linear terms in $\theta$, our conclusions are expected to hold for the full theory. The Euler--Lagrange equation of motion for this theory (in view of the fact that $\mathscr{L}$ does not have explicit dependence on $A_{\mu}$) is 
\begin{equation}\label{eomm}
\partial_{\rho}\left(\frac{\partial\mathscr{L}}{\partial(\partial_{\sigma}A_{\rho})}\right) = 0.
\end{equation}

Popular noncommutative spacetime is characterised by a constant and fixed (same value in all frames) noncommutativity parameter but here first we take $\theta^{\alpha\beta}$ to be a constant tensor parameter, i.e., it is constant but transforms as a tensor under Poincar\'e transfomations. Proceeding as in the previous section, we find that for spacetime translations, invariance of the action implies, as before,
\begin{equation}\label{916-2}
\partial_{\mu}T^{\mu\nu} = 0,
\end{equation}
with $T^{\mu\nu}$ defined as in \eqref{914lor}, i.e.,
\begin{equation}\label{914'}
T^{\mu\nu} = \frac{\partial\mathscr{L}}{\partial(\partial_{\mu}A_{\sigma})}\partial^{\nu}A_{\sigma}-\eta^{\mu\nu}\mathscr{L}.
\end{equation}

In case of infinitesimal Lorentz transformations, $\delta x_{\mu}=\omega_{\mu\nu}x^{\nu}$, $\delta A_{\mu}=\omega_{\mu\nu}A^{\nu}$ and $\delta \theta_{\mu\nu}=\omega_{\mu\alpha}{\theta^{\alpha}}_{\nu}-\omega_{\nu\alpha}{\theta^{\alpha}}_{\mu}$. With $M^{\mu\lambda\rho}$ defined as in \eqref{9181},
\begin{equation}\label{9181'}
M^{\mu\lambda\rho} = \frac{\partial\mathscr{L}}{\partial(\partial_{\mu}A_{\lambda})}A^{\rho}-\frac{\partial\mathscr{L}}{\partial(\partial_{\mu}A_{\rho})}A^{\lambda}-T^{\mu\lambda}x^{\rho}+T^{\mu\rho}x^{\lambda},
\end{equation}
the analogue of Eq.~\eqref{918lor} turns out to be
\begin{equation}\label{918-2}
\partial_{\mu}M^{\mu\lambda\rho}+2\frac{\partial \mathscr{L}}{\partial \theta_{\alpha\rho}}{\theta^{\lambda}}_{\alpha}-2\frac{\partial \mathscr{L}}{\partial \theta_{\alpha\lambda}}{\theta^{\rho}}_{\alpha} = 0,
\end{equation}
which, upon substituting
\begin{equation*}
\frac{\partial\mathscr{L}}{\partial \theta_{\alpha\rho}} = -\frac{1}{2}\left(F^{\mu\alpha}F^{\nu\rho}+\frac{1}{4}F^{\rho\alpha}F^{\mu\nu}\right)F_{\mu\nu},
\end{equation*}
gives us the criterion for Lorentz invariance of the action as
\begin{equation}\label{407n}
\partial_{\mu}M^{\mu\lambda\rho} - {\theta^{\lambda}}_{\alpha}F_{\mu\nu}\left(F^{\mu\alpha}F^{\nu\rho}+\frac{1}{4}F^{\mu\nu}F^{\rho\alpha}\right)+{\theta^{\rho}}_{\alpha}F_{\mu\nu}\left(F^{\mu\alpha}F^{\nu\lambda}+\frac{1}{4}F^{\mu\nu}F^{\lambda\alpha}\right) = 0.
\end{equation}

In the case when $\theta^{\mu\nu}$ does not transform like a tensor but is fixed in all frames, we have $\delta \theta_{\mu\nu} = 0$ under translations and Lorentz transformations. In that case, the criteria for the invariance of the action turn out to be
\begin{gather}
\label{crit1n}
\partial_{\mu}T^{\mu\nu} = 0,\\
\label{crit2n}
\partial_{\mu}M^{\mu\lambda\rho} = 0,
\end{gather}
which are the exact analogues of the criteria \eqref{crit1} and \eqref{crit2}.

Now we shall explicitly evaluate $\partial_{\mu}T^{\mu\nu}$ and $\partial_{\mu}M^{\mu\lambda\rho}$ for our model \eqref{921lor}. We have
\begin{equation}\label{1191}
\frac{\partial \mathscr{L}}{\partial(\partial_{\sigma}A_{\rho})} = F^{\rho\sigma}+\theta^{\alpha\sigma}F^{\mu\rho}F_{\mu\alpha}-\theta^{\alpha\rho}F^{\mu\sigma}F_{\mu\alpha}-\frac{1}{4}\theta^{\rho\sigma}F^{\mu\nu}F_{\mu\nu}+\theta^{\alpha\beta}\left({F^{\rho}}_{\alpha}{F^{\sigma}}_{\beta}+\frac{1}{2}F_{\beta\alpha}F^{\rho\sigma}\right).
\end{equation}
Taking the derivative of Eq.~\eqref{914'} and using the equation of motion \eqref{eomm}, yields
\begin{equation}\label{find1n}
\partial_{\mu}T^{\mu\nu} = 0.
\end{equation}
Similarly, taking the derivative of Eq.~\eqref{9181'}, using Eqs.~\eqref{eomm} and \eqref{find1n}, and finally substituting \eqref{1191}, we find
\begin{equation}\label{find2n}
\partial_{\mu}M^{\mu\lambda\rho} = {\theta^{\lambda}}_{\alpha}F_{\mu\nu}\left(F^{\mu\alpha}F^{\nu\rho}+\frac{1}{4}F^{\mu\nu}F^{\rho\alpha}\right)-{\theta^{\rho}}_{\alpha}F_{\mu\nu}\left(F^{\mu\alpha}F^{\nu\lambda}+\frac{1}{4}F^{\mu\nu}F^{\lambda\alpha}\right).
\end{equation}
The results \eqref{find1n} and \eqref{find2n} do not depend on whether $\theta^{\mu\nu}$ transforms like a tensor or not.

We have seen that the criterion for translational invariance is the same,  $\partial_{\mu}T^{\mu\nu} = 0$, in both the cases when $\theta^{\mu\nu}$ transforms like a tensor and when it does not. This is satisfied in view of Eq.~\eqref{find1n}. However, the criterion for Lorentz invariance is different in the two cases---see Eqs.~\eqref{407n} and \eqref{crit2n}---and what we have actually found is given by Eq.~\eqref{find2n}. Therefore, as expected, our theory has Lorentz invariance only when $\theta^{\mu\nu}$ transforms like a tensor, and not in the other case. The Seiberg--Witten maps \eqref{101-lo} and \eqref{107-lo} have an explicit Lorentz-invariant form provided that $\theta$ transforms like a Lorentz tensor, in accordance with the result found here.

As in the toy model, we now show that the Poisson bracket of the dynamical fields $A_i$ and $\pi^i$ with the charge is equal to the Lie derivative of the field with respect to the vector field associated with the charge. As usual, the Hamiltonian formulation \cite{Banerjee:2002qh} is commenced by computing the canonical momenta of the theory:
\begin{gather}
\label{pi0n}
\pi^{0} = 0,\\
\label{piin}
\begin{split}
\pi^{i} &= F^{i0}-\theta^{mn}\left(F{^{i}}_{n}F{^{0}}_{m}+\frac{1}{2}F_{nm}F^{0i}\right)-\theta^{in}F_{kn}F^{0k}-\theta^{0n}\left(F^{0i}F_{0n}+F^{mi}F_{mn}\right)\\&\quad{}+\theta^{0i}\left(\frac{1}{4}F^{mn}F_{mn}-\frac{1}{2}F^{0m}F_{0m}\right).
\end{split}
\end{gather}
As before, Eq.~\eqref{pi0n} is interpreted as a primary constraint. Since the definition \eqref{piin} of momenta $\pi^{i}$ contains terms quadratic in `velocities', it is highly nontrivial to invert this relation to express velocities in terms of phase-space variables. Therefore, we now implement the condition\footnote{The simplifications achieved by this condition are well known in the Hamiltonian formulation of noncommutative gauge theories. It eliminates the higher-order time-derivatives so that the standard Hamiltonian prescription can be adopted.} $\theta^{0i}=0$,  which enables us to write down the velocities in terms of phase-space variables:
\begin{equation}\label{piin2}
F^{i0}=\pi^{i}-\theta^{mn}\left({F^{i}}_{n}\pi_{m}+\frac{1}{2}F_{nm}\pi^{i}\right)-\theta^{in}F_{kn}\pi_{k}.
\end{equation}

It follows from the definitions of $T^{\mu\nu}$ \eqref{914'} and $M^{\mu\lambda\rho}$ \eqref{9181'} that
\begin{gather}
\label{T00n}
\begin{split}
T^{00} &= \pi^{i}\partial_{i}A^{0}-\frac{1}{2}\pi^{i}\pi_{i}-\frac{1}{4}F_{ij}F^{ij}\\
& \quad{}-\theta^{ij}\left(\frac{1}{2}F_{ki}F_{mj}F^{km}+\frac{1}{8}F_{ji}F_{km}F^{km}-\frac{1}{4}F_{ji}\pi_{k}\pi^{k}-F_{kj}\pi_{i}\pi^{k}\right),
\end{split}\\
\label{T0in}
T^{0i} = \pi^{j}\partial^{i}A_{j},\\
\label{M00in}
M^{00i} = -T^{00}x^{i}-\pi^{i}A^{0}+x^{0}\pi^{j}\partial^{i}A_{j},\\
\label{M0ijn}
M^{0ij} = \pi^{i}A^{j}-x^{j}\pi^{k}\partial^{i}A_{k}-\pi^{j}A^{i}+x^{i}\pi^{k}\partial^{j}A_{k},
\end{gather}
where we have used Eq.~\eqref{piin2} to eliminate velocities in favour of momenta. Time-conserving the primary constraint with the Hamiltonian $\int \mathrm{d}^{3}x\,{T^{0}}_{0}$ yields the Gauss constraint
\begin{equation}\label{gauss2}
\partial_{i}\pi^{i} \approx 0.
\end{equation}
There are no further constraints.

Now we find
\begin{gather}
\label{AP1n}
\left\{A_{i}, P_{j}\right\} = \partial_{j}A_{i},\\
\label{AP2n}
\left\{A_{i}, P_{0}\right\} = \partial_{i}A_{0}+\pi_{i}-{\theta_{i}}^{n}F_{mn}\pi^{m}-\theta^{mn}\left(F_{in}\pi_{m}+\frac{1}{2}F_{nm}\pi_{i}\right),
\\
\label{AJ1n}
\left\{A_{i}, J_{kl}\right\} = \eta_{ik}A_{l}-\partial_{k}A_{i}x_{l}-\eta_{il}A_{k}+\partial_{l}A_{i}x_{k},\\
\label{AJ2n}
\begin{split}
\left\{A_{i}, J_{0k}\right\} &= -x_{k}\left[\partial_{i}A_{0}+\pi_{i}-{\theta_{i}}^{n}F_{mn}\pi^{m}-\theta^{mn}\left(F_{in}\pi_{m}+\frac{1}{2}F_{nm}\pi_{i}\right)\right]
+\partial_{k}A_{i}x_{0}\\&\quad{}-\eta_{ik}A_{0}.
\end{split}
\end{gather}
As in the toy model, here also we obtain
\begin{equation}\label{AQLn}
\left\{A_{i}, Q_{V}\right\} = \mathcal{L}_{V}A_{i}, \qquad
\left\{\pi^{i}, Q_{V}\right\} = \mathcal{L}_{V}\pi^{i}.
\end{equation}
We find that algebra \eqref{PP}--\eqref{JJ} is satisfied here also, which in turn implies that the condition \eqref{QQ} holds, i.e., restricting to $P_{i}$ and $J_{ij}$, we have
\begin{equation}\label{QQn}
\left\{Q_{U}, Q_{V}\right\} = Q_{[U,V]}.
\end{equation}

Finally, we would like to mention that there are certain choices of constant nontransforming $\theta$ for which the Lorentz invariance can be partially restored. Let us get back to Eq.~\eqref{xxmoyal} which characterises the noncommutativity. Under Lorentz transformation, $\delta x^{\mu} = {\omega^{\mu}}_{\lambda}x^{\lambda}$, this equation imposes the following restriction on nontransforming $\theta$: 
\begin{equation}\label{lo-111}
\Omega^{\mu\nu} \equiv {\omega^{\mu}}_{\lambda}\theta^{\lambda\nu}-{\omega^{\nu}}_{\lambda}\theta^{\lambda\mu} = 0.
\end{equation}
There is no nontrivial solution of this set of equations. However, some subsets of this set of equations are soluble. It can be easily seen that the equation
\begin{equation*}
\Omega^{01} \equiv {\omega^{0}}_{2}\theta^{21}+{\omega^{0}}_{3}\theta^{31}-{\omega^{1}}_{2}\theta^{20}-{\omega^{1}}_{3}\theta^{30} = 0
\end{equation*}
is satisfied for $\theta^{02}=\theta^{03}=\theta^{12}=\theta^{13}=0$. This choice of $\theta$ also solves $\Omega^{23}=0$. Thus, invariance under a rotation in $23$-plane and under a boost in $1$-direction can be restored (for nontransforming $\theta$) by choosing
\begin{equation}\label{lo-113}
\left\{\theta^{\mu\nu}\right\} = \begin{pmatrix}0&\theta_{e}&0&0\\-\theta_{e}&0&0&0\\0&0&0&\theta_{m}\\0&0&-\theta_{m}&0
\end{pmatrix}.
\end{equation}
Likewise it can be seen that the invariance under a rotation in $13$-plane and under a boost in $2$-direction is restored for $\theta^{01}=\theta^{03}=\theta^{12}=\theta^{23}=0$, whereas for $\theta^{01}=\theta^{02}=\theta^{13}=\theta^{23}=0$, the invariance under a rotation in $12$-plane and under a boost in $3$-direction is restored. The spacetime symmetry group for these choices of $\theta$ is $[\mathrm{SO(1,1)}\times\mathrm{SO(2)}]\rtimes \mathrm{T}_{4}$, where $\rtimes$ represents semi-direct product.

We now show that these results also follow from our analysis. We have shown that the criterion for Lorentz invariance when $\theta$ does not transform is $\partial_{\mu}M^{\mu\lambda\rho}=0$, Eq.~\eqref{crit2n}. For the choice \eqref{lo-113} of $\theta$, Eq.~\eqref{find2n} indeed gives $\partial_{\mu}M^{\mu 23}=0$ and $\partial_{\mu}M^{\mu 01}=0$. Similarly, our analysis gives consistent results for the other choices of $\theta$. It is worthwhile to mention that the choice \eqref{lo-113} has recently been studied \cite{Alvarez-Gaume:2003mb, Morita:2003vx} and CPT theorem in noncommutative field theories has been proved~\cite{Alvarez-Gaume:2003mb}.

Noncommutative gauge theories in two dimensions are always Lorentz invariant, since, in two dimensions, the noncommutativity parameter becomes proportional to the antisymmetric tensor $\varepsilon^{\mu\nu}$, which has the same value in all frames. Our analysis is also consistent with this fact; in two dimensions, Eq.~\eqref{find2n} gives $\partial_{\mu}M^{\mu 01}=0$.


\subsection{Noncommutative-variable approach}

Here we shall reconsider the analysis just presented, but in noncommutative variables. However, as earlier, we again restrict ourselves to the first order in $\theta$. In this approximation, the original theory~\eqref{l-105} reads
\begin{equation}\label{nl-105}
\xnc{S} = -\frac{1}{2}\int\!\mathrm{d}^4x\,\left[\partial_{\mu}\xnc{A}_{\nu}\left(\partial^{\mu}\xnc{A}^{\nu}-\partial^{\nu}\xnc{A}^{\mu}\right)+2\theta^{\alpha\beta}\partial_{\alpha}\xnc{A}^{\mu}\partial_{\beta}\xnc{A}^{\nu}\partial_{\mu}\xnc{A}_{\nu}\right].
\end{equation}
The change of $\xnc{A}_{\mu}$ under Poincar\'e transformation is dictated by the noncommutativity parameter $\theta^{\mu\nu}$ through the Seiberg--Witten map \eqref{101-lo}; $\xnc{A}_{\mu}$ will transform differently depending on whether $\theta^{\mu\nu}$ transforms like a tensor or not. For spacetime translations, however, it does not matter; $\delta A_{\mu}=0$ and $\delta \theta^{\mu\nu}=0$ imply $\delta \xnc{A}_{\mu}=0$. Under Lorentz transformation, $\delta A_{\mu}=\omega_{\mu\nu}A^{\nu}$, $\delta F_{\beta\mu}=\omega_{\beta\lambda}{F^{\lambda}}_{\mu}-\omega_{\mu\lambda}{F^{\lambda}}_{\beta}$, and $\delta \theta_{\mu\nu}=\omega_{\mu\alpha}{\theta^{\alpha}}_{\nu}-\omega_{\nu\alpha}{\theta^{\alpha}}_{\mu}$ if $\theta_{\mu\nu}$ transforms as a tensor, otherwise $\delta \theta_{\mu\nu}=0$ if it does not transform. Therefore, for transforming $\theta$,
 map \eqref{101-lo} gives
\begin{equation}\label{delA1-lo}
\delta \xnc{A}_{\mu}=\omega_{\mu\lambda}\xnc{A}^{\lambda},
\end{equation}
which is the expected noncommutative deformation of the standard transformation for a covariant vector. For nontransforming $\theta$,
\begin{equation}\label{delA2-lo}
\delta \xnc{A}_{\mu} = \omega_{\mu\lambda}\xnc{A}^{\lambda}-\frac{1}{2}\theta^{\alpha\beta}\omega_{\beta\lambda}\left[\xnc{A}^{\lambda}\partial_{\mu}\xnc{A}_{\alpha}-\xnc{A}_{\alpha}\partial_{\mu}\xnc{A}^{\lambda}-2\left(\xnc{A}^{\lambda}\partial_{\alpha}\xnc{A}_{\mu}-\xnc{A}_{\alpha}\partial^{\lambda}\xnc{A}_{\mu}\right)\right].
\end{equation}

Proceeding as in the case of toy model, we find that the change in action under spacetime transformations is given by
\begin{equation}\label{913-lo}
\delta \xnc{S} = \int\!\mathrm{d}^{4}x\,\left[\partial_{\mu}\left(\frac{\partial\xnc{\mathscr{L}}}{\partial(\partial_{\mu}\xnc{A}_{\nu})}\delta \xnc{A}_{\nu}-\xnc{T}^{\mu\nu}\delta x_{\nu}\right)+\frac{\partial\xnc{\mathscr{L}}}{\partial \theta^{\mu\nu}}\delta \theta^{\mu\nu}\right],
\end{equation}
where the canonical energy--momentum tensor is defined as
\begin{equation}\label{914-lo}
\xnc{T}^{\mu\nu} = \frac{\partial\xnc{\mathscr{L}}}{\partial(\partial_{\mu}\xnc{A}_{\sigma})}\partial^{\nu}\xnc{A}_{\sigma}-\eta^{\mu\nu}\xnc{\mathscr{L}}.
\end{equation}

Therefore, the criterion for translational invariance of the action, irrespective of whether $\theta$ is a tensor or not, is
\begin{equation}\label{916-lo}
\partial_{\mu}\xnc{T}^{\mu\nu} = 0,
\end{equation}
since $\delta A_{\nu} = \delta \theta^{\mu\nu} = 0$. It follows from the definition \eqref{914-lo} that the criterion \eqref{916-lo} is indeed satisfied once we use the equation of motion (Lagrangian density does not have explicit dependence on $\xnc{A}_{\mu}$)
\begin{equation}\label{eom1-lo}
\partial_{\mu}\left(\frac{\partial \xnc{\mathscr{L}}}{\partial(\partial_{\mu}\xnc{A}_{\nu})}\right) = 0.
\end{equation}
This implies that the action~\eqref{nl-105} is invariant under translations.

In the case of transforming $\theta$, the criterion of Lorentz invariance, using the transformation \eqref{delA1-lo}, turns out to be
\begin{equation}\label{crit2nc}
\partial_{\mu}\xnc{M}^{\mu\lambda\rho}-\left(\partial_{\mu}\xnc{A}_{\nu}-\partial_{\nu}\xnc{A}_{\mu}\right)\left({\theta^{\lambda}}_{\alpha}\partial^{\alpha}\xnc{A}^{\mu}\partial^{\rho}\xnc{A}^{\nu}-{\theta^{\rho}}_{\alpha}\partial^{\alpha}\xnc{A}^{\mu}\partial^{\lambda}\xnc{A}^{\nu}\right) = 0,
\end{equation}
where
\begin{equation}\label{9181-lo}
\xnc{M}^{\mu\lambda\rho} = \frac{\partial\xnc{\mathscr{L}}}{\partial(\partial_{\mu}\xnc{A}_{\lambda})}\xnc{A}^{\rho}-\frac{\partial\xnc{\mathscr{L}}}{\partial(\partial_{\mu}\xnc{A}_{\rho})}\xnc{A}^{\lambda}-\xnc{T}^{\mu\lambda}x^{\rho}+\xnc{T}^{\mu\rho}x^{\lambda}.
\end{equation}
On the other hand, using the transformation \eqref{delA2-lo} for nontransforming $\theta$, the invariance of the action under Lorentz transformations demands
\begin{equation}\label{crit22nc}
\begin{split}
\partial_{\mu}\xnc{M}^{\mu\lambda\rho}-\frac{1}{2}{\theta^{\lambda}}_{\alpha}\left(\partial_{\mu}\xnc{A}_{\nu}-\partial_{\nu}\xnc{A}_{\mu}\right)\partial^{\mu}\left[\xnc{A}^{\alpha}(2\partial^{\rho}\xnc{A}^{\nu}-\partial^{\nu}\xnc{A}^{\rho})-\xnc{A}^{\rho}\left(2\partial^{\alpha}\xnc{A}^{\nu}-\partial^{\nu}\xnc{A}^{\alpha}\right)\right]
\\+\frac{1}{2}{\theta^{\rho}}_{\alpha}\left(\partial_{\mu}\xnc{A}_{\nu}-\partial_{\nu}\xnc{A}_{\mu}\right)\partial^{\mu}\left[\xnc{A}^{\alpha}(2\partial^{\lambda}\xnc{A}^{\nu}-\partial^{\nu}\xnc{A}^{\lambda})-\xnc{A}^{\lambda}\left(2\partial^{\alpha}\xnc{A}^{\nu}-\partial^{\nu}\xnc{A}^{\alpha}\right)\right] = 0.
\end{split}
\end{equation}

Next we compute $\partial_{\mu}\xnc{M}^{\mu\lambda\rho}$ from the definition \eqref{9181-lo}. Using the equation of motion \eqref{eom1-lo} and
\begin{equation}\label{lo-101}
\frac{\partial \xnc{\mathscr{L}}}{\partial(\partial_{\mu}\xnc{A}_{\lambda})} = \partial^{\lambda}\xnc{A}^{\mu}-\partial^{\mu}\xnc{A}^{\lambda}-\theta^{\alpha\beta}\partial_{\alpha}\xnc{A}^{\mu}\partial_{\beta}\xnc{A}^{\lambda}-\theta^{\mu\beta}\partial_{\beta}\xnc{A}^{\nu}\left(\partial^{\lambda}\xnc{A}_{\nu}-\partial_{\nu}\xnc{A}^{\lambda}\right),
\end{equation}
it follows from \eqref{9181-lo} that
\begin{equation}\label{find2nc}
\partial_{\mu}\xnc{M}^{\mu\lambda\rho} = \left(\partial_{\mu}\xnc{A}_{\nu}-\partial_{\nu}\xnc{A}_{\mu}\right)\left({\theta^{\lambda}}_{\alpha}\partial^{\alpha}\xnc{A}^{\mu}\partial^{\rho}\xnc{A}^{\nu}-{\theta^{\rho}}_{\alpha}\partial^{\alpha}\xnc{A}^{\mu}\partial^{\lambda}\xnc{A}^{\nu}\right),
\end{equation}
which shows that the criterion \eqref{crit2nc} is satisfied and not \eqref{crit22nc}. Thus, the action \eqref{nl-105} is invariant under Lorentz transformations only when $\theta$ transforms as a tensor, which is like the case of noncommutative electrodynamics in usual variables, considered in the previous section.

We shall now establish a connection between the two descriptions of noncommutative electrodynamics considered here and in the previous section. The Lagrangian densities in the two formulations are related by the map
\begin{equation}\label{LLhat}
\xnc{\mathscr{L}} = \mathscr{L}+\frac{1}{4}\theta^{\alpha\beta}\partial_{\beta}\left(A_{\alpha}F_{\mu\nu}F^{\mu\nu}\right).
\end{equation}
Since $\xnc{\mathscr{L}}$ and $\mathscr{L}$ differ by a total-derivative term, we have $\xnc{S} = S$.

Now we shall find the maps between $T^{\mu\nu}$ and $\xnc{T}^{\mu\nu}$ as well as between $M^{\mu\lambda\rho}$ and $\xnc{M}^{\mu\lambda\rho}$. First we apply the Seiberg--Witten map \eqref{101-lo} on the right-hand side of Eq.~\eqref{lo-101} and take into account Eq.~\eqref{1191} to get
\begin{equation}\label{lo-101a}
\frac{\partial \xnc{\mathscr{L}}}{\partial(\partial_{\mu}\xnc{A}_{\lambda})} = \frac{\partial\mathscr{L}}{\partial(\partial_{\mu}A_{\lambda})}+\theta^{\alpha\mu}F^{\lambda\sigma}\partial_{\sigma}A_{\alpha}+\theta^{\alpha\lambda}F^{\sigma\mu}F_{\sigma\alpha}-\theta^{\alpha\beta}\partial_{\beta}(A_{\alpha}F^{\lambda\mu})+\frac{1}{4}\theta^{\lambda\mu}F_{\kappa\sigma}F^{\kappa\sigma}.
\end{equation}
Using the maps \eqref{101-lo}, \eqref{LLhat} and \eqref{lo-101a}, we get a map\footnote{A similar map among the symmetric energy--momentum tensors is defined in \refcite{Banerjee:2003vc}, the energy--momentum tensors considered here follow from Noether's prescription.} between $\xnc{T}^{\mu\nu}$ \eqref{914-lo} and  $T^{\mu\nu}$ \eqref{914'}:
\begin{equation}\label{TThat}
\begin{split}
\xnc{T}^{\mu\nu}&= T^{\mu\nu}+\theta^{\alpha\mu}\left(F^{\lambda\sigma}\partial_{\sigma}A_{\alpha}\partial^{\nu}A_{\lambda}+\frac{1}{4}F_{\lambda\rho}F^{\lambda\rho}\partial^{\nu}A_{\alpha}\right)\\
&\quad{}+\theta^{\alpha\beta}\left[\frac{1}{2}F^{\lambda\mu}\partial_{\lambda}A_{\alpha}\partial^{\nu}A_{\beta}-\partial_{\beta}\left(A_{\alpha}F^{\lambda\mu}\right)\partial^{\nu}A_{\lambda}-\frac{1}{4}\eta^{\mu\nu}\partial_{\beta}\left(A_{\alpha}F_{\lambda\rho}F^{\lambda\rho}\right)\right.\\
&\qquad\qquad\left.{}-\frac{1}{2}A_{\alpha}F^{\lambda\mu}\partial^{\nu}\left(\partial_{\beta}A_{\lambda}+F_{\beta\lambda}\right)\right].
\end{split}
\end{equation}
Similarly, using the maps \eqref{101-lo}, \eqref{lo-101a} and \eqref{TThat}, we get a map between $\xnc{M}^{\mu\lambda\rho}$ \eqref{9181-lo} and  $M^{\mu\lambda\rho}$ \eqref{9181'}:
\begin{equation}\label{MMhat}
\xnc{M}^{\mu\lambda\rho} = M^{\mu\lambda\rho}+M_{(\theta)}^{\mu\lambda\rho}-M_{(\theta)}^{\mu\rho\lambda},
\end{equation}
where
\begin{equation}\label{Mth-mlr}\raisetag{21pt}
\begin{split}
M_{(\theta)}^{\mu\lambda\rho}&=\theta^{\lambda\alpha}F^{\mu\sigma}F_{\sigma\alpha}A^{\rho}+\frac{1}{4}\theta^{\lambda\mu}A^{\rho}F_{\kappa\sigma}F^{\kappa\sigma}\\
&\quad{}+\theta^{\alpha\mu}\left[F^{\lambda\sigma}A^{\rho}\partial_{\sigma}A_{\alpha}-x^{\rho}\left(F^{\kappa\sigma}\partial_{\sigma}A_{\alpha}\partial^{\lambda}A_{\kappa}+\frac{1}{4}F_{\kappa\sigma}F^{\kappa\sigma}\partial^{\lambda}A_{\alpha}\right)\right]\\&\quad{}-\theta^{\alpha\beta}\left[A^{\rho}\partial_{\beta}\left(A_{\alpha}F^{\lambda\mu}\right)+\frac{1}{2}F^{\lambda\mu}A_{\alpha}\left(\partial_{\beta}A^{\rho}+{F_{\beta}}^{\rho}\right)\right.\\&\qquad\qquad\left.{}+x^{\rho}\left(\frac{1}{2}F^{\sigma\mu}\partial_{\sigma}A_{\alpha}\partial^{\lambda}A_{\beta}-\partial^{\lambda}A_{\sigma}\partial_{\beta}\left(A_{\alpha}F^{\sigma\mu}\right)-\frac{1}{4}\eta^{\mu\lambda}\partial_{\beta}\left(A_{\alpha}F_{\kappa\sigma}F^{\kappa\sigma}\right)\right.\right.\\&\qquad\qquad\qquad\quad\left.\left.{}-\frac{1}{2}A_{\alpha}F^{\sigma\mu}\partial^{\lambda}\left(\partial_{\beta}A_{\sigma}+F_{\beta\sigma}\right)\right)\right].
\end{split}
\end{equation}

It follows from Eq.~\eqref{TThat} that
\begin{equation}\label{lo-102}
\partial_{\mu}\xnc{T}^{\mu\nu} = \partial_{\mu}T^{\mu\nu},
\end{equation}
where we have used the equation of motion, $\partial_{\mu}F^{\mu\nu}+ \uO(\theta) = 0$. This shows the compatibility of the criteria for translational invariance in the two descriptions, Eqs.~\eqref{916-2}, \eqref{crit1n} and \eqref{916-lo}.

Next we show the compatibility of the criteria for Lorentz invariance. It follows from Eq.~\eqref{MMhat} that
\begin{equation}\label{lo-103}
\begin{split}
\partial_{\mu}\xnc{M}^{\mu\lambda\rho} &= \partial_{\mu}M^{\mu\lambda\rho}+\theta^{\alpha\lambda}\left(F^{\sigma\mu}F_{\sigma\alpha}\partial_{\mu}A^{\rho}-F^{\sigma\mu}\partial_{\sigma}A_{\alpha}\partial^{\rho}A_{\mu}+\frac{1}{4}F_{\kappa\sigma}F^{\kappa\sigma}{F^{\rho}}_{\alpha}\right)\\&\quad{}-\theta^{\alpha\rho}\left(F^{\sigma\mu}F_{\sigma\alpha}\partial_{\mu}A^{\lambda}-F^{\sigma\mu}\partial_{\sigma}A_{\alpha}\partial^{\lambda}A_{\mu}+\frac{1}{4}F_{\kappa\sigma}F^{\kappa\sigma}{F^{\lambda}}_{\alpha}\right),
\end{split}
\end{equation}
where again the equation of motion, $\partial_{\mu}F^{\mu\nu}+ \uO(\theta) = 0$, has been used. Now we use the maps \eqref{101-lo} and \eqref{lo-103} on the left-hand side of Eq.~\eqref{crit2nc} to obtain
\begin{equation}\label{lo-104}
\begin{split}
&\partial_{\mu}\xnc{M}^{\mu\lambda\rho}-\left(\partial_{\mu}\xnc{A}_{\nu}-\partial_{\nu}\xnc{A}_{\mu}\right)\left({\theta^{\lambda}}_{\alpha}\partial^{\alpha}\xnc{A}^{\mu}\partial^{\rho}\xnc{A}^{\nu}-{\theta^{\rho}}_{\alpha}\partial^{\alpha}\xnc{A}^{\mu}\partial^{\lambda}\xnc{A}^{\nu}\right)\\ &= \partial_{\mu}M^{\mu\lambda\rho} - {\theta^{\lambda}}_{\alpha}F_{\mu\nu}\left(F^{\mu\alpha}F^{\nu\rho}+\frac{1}{4}F^{\mu\nu}F^{\rho\alpha}\right)+{\theta^{\rho}}_{\alpha}F_{\mu\nu}\left(F^{\mu\alpha}F^{\nu\lambda}+\frac{1}{4}F^{\mu\nu}F^{\lambda\alpha}\right).
\end{split}
\end{equation}
Thus, the left-hand side of criterion \eqref{crit2nc} goes over to the left-hand side of criterion \eqref{407n} under the Seiberg--Witten maps, which shows the compatibility of the two criteria for Lorentz invariance when $\theta$ transforms as a tensor. Turning to the case when $\theta$ does not transform, we now apply the maps \eqref{101-lo} and \eqref{lo-103} on the left-hand side of the criterion \eqref{crit22nc}:
\begin{equation}\label{lo-105}
\begin{split}
&\partial_{\mu}\xnc{M}^{\mu\lambda\rho}-\frac{1}{2}{\theta^{\lambda}}_{\alpha}\left(\partial_{\mu}\xnc{A}_{\nu}-\partial_{\nu}\xnc{A}_{\mu}\right)\partial^{\mu}\left[\xnc{A}^{\alpha}\left(2\partial^{\rho}\xnc{A}^{\nu}-\partial^{\nu}\xnc{A}^{\rho}\right)-\xnc{A}^{\rho}\left(2\partial^{\alpha}\xnc{A}^{\nu}-\partial^{\nu}\xnc{A}^{\alpha}\right)\right]\\
&\quad{}+\frac{1}{2}{\theta^{\rho}}_{\alpha}\left(\partial_{\mu}\xnc{A}_{\nu}-\partial_{\nu}\xnc{A}_{\mu}\right)\partial^{\mu}\left[\xnc{A}^{\alpha}\left(2\partial^{\lambda}\xnc{A}^{\nu}-\partial^{\nu}\xnc{A}^{\lambda}\right)-\xnc{A}^{\lambda}\left(2\partial^{\alpha}\xnc{A}^{\nu}-\partial^{\nu}\xnc{A}^{\alpha}\right)\right]\\
&= \partial_{\mu}M^{\mu\lambda\rho}+\frac{1}{4}{\theta^{\lambda}}_{\alpha}\left[\partial^{\alpha}\left(A^{\rho}F_{\kappa\sigma}F^{\kappa\sigma}\right)-\partial^{\rho}\left(A^{\alpha}F_{\kappa\sigma}F^{\kappa\sigma}\right)\right]\\
&\quad{}-\frac{1}{4}{\theta^{\rho}}_{\alpha}\left[\partial^{\alpha}\left(A^{\lambda}F_{\kappa\sigma}F^{\kappa\sigma}\right)-\partial^{\lambda}\left(A^{\alpha}F_{\kappa\sigma}F^{\kappa\sigma}\right)\right].
\end{split}
\end{equation}
Thus, the left-hand side of criterion \eqref{crit22nc} goes over to the left-hand side of criterion \eqref{crit2n} up to total-derivative terms. The origin of these total-derivative terms is presumably due to the fact that $\xnc{\mathscr{L}}$ and $\mathscr{L}$ are not exactly equal but differ by a total-derivative term, Eq.~\eqref{LLhat}.

We shall now show that using the Noether charges
\begin{equation}\label{gen-lo}
\xnc{P}^{\mu} = \int\!\mathrm{d}^{3}x\,\xnc{T}^{0\mu}, \qquad
\xnc{J}^{\mu\nu} = \int\!\mathrm{d}^{3}x\,\xnc{M}^{0\mu\nu},
\end{equation}
and the canonical equal-time Poisson brackets $\{\xnc{A}_{\mu}(t, \mathbf{x}), \xnc{\pi}^{\nu}(t, \mathbf{y})\} = \delta_{\mu}^{\nu}\delta^{3}(\mathbf{x}-\mathbf{y})$, we can generate the transformations of the dynamical variables $\xnc{A}_i$ and $\xnc{\pi}_i$:
\begin{equation}\label{AQL-lo}
\left\{\xnc{A}_{i}, \xnc{Q}_{V}\right\} = \mathcal{L}_{V}\xnc{A}_{i}, \qquad
\left\{\xnc{\pi}^{i}, \xnc{Q}_{V}\right\}= \mathcal{L}_{V}\xnc{\pi}^{i}.
\end{equation}

The canonical momenta of the theory are
\begin{gather}
\label{pi02-lo}
\xnc{\pi}^{0} = -\theta^{0i}\partial_{i}\xnc{A}_{j}\left(\partial^{0}\xnc{A}^{j}-\partial^{j}\xnc{A}^{0}\right),\\
\label{pii2-lo}
\begin{split}
\xnc{\pi}^{i} &=\partial^{i}\xnc{A}^{0}-\partial^{0}\xnc{A}^{i}-\theta^{kl}\partial_{k}\xnc{A}^{0}\partial_{l}\xnc{A}^{i}\\
&\quad{}-\theta^{0l}\left(\partial_{0}\xnc{A}^{0}\partial_{l}\xnc{A}^{i}-2\partial_{l}\xnc{A}^{0}\partial_{0}\xnc{A}^{i}+\partial_{l}\xnc{A}_{0}\partial^{i}\xnc{A}^{0}+\partial^{i}\xnc{A}^{k}\partial_{l}\xnc{A}_{k}-\partial^{k}\xnc{A}^{i}\partial_{l}\xnc{A}_{k}\right).
\end{split}
\end{gather}
As in the previous section, here also we set $\theta^{0i}=0$, so that the above definitions simplify to
\begin{gather}
\label{pi0-lo}
\xnc{\pi}^{0} = 0,\\
\label{pii-lo}
\xnc{\pi}^{i} =\partial^{i}\xnc{A}^{0}-\partial^{0}\xnc{A}^{i}-\theta^{kl}\partial_{k}\xnc{A}^{0}\partial_{l}\xnc{A}^{i}.
\end{gather}
It follows from the definitions \eqref{914-lo}, \eqref{9181-lo} and \eqref{pii-lo} that
\begin{gather}
\label{T00-lo}
\begin{split}
\xnc{T}^{00} &= \xnc{\pi}^{i}\partial_{i}\xnc{A}^{0}-\frac{1}{2}\xnc{\pi}^{i}\xnc{\pi}_{i}-\frac{1}{2}\partial_{i}\xnc{A}_{j}\left(\partial^{i}\xnc{A}^{j}-\partial^{j}\xnc{A}^{i}\right)\\
&\quad{}-\theta^{kl}\left[\xnc{\pi}_{i}\partial_{k}\xnc{A}^{0}\partial_{l}\xnc{A}^{i}+\partial_{k}\xnc{A}^{i}\partial_{l}\xnc{A}^{j}\partial_{i}\xnc{A}_{j}\right],
\end{split}\\
\label{T0i-lo}
\xnc{T}^{0i} = \xnc{\pi}^{j}\partial^{i}\xnc{A}_{j},\\
\label{M00i-lo}
\xnc{M}^{00i} = -\xnc{T}^{00}x^{i}-\xnc{\pi}^{i}\xnc{A}^{0}+x^{0}\xnc{\pi}^{j}\partial^{i}\xnc{A}_{j},\\
\label{M0ij-lo}
\xnc{M}^{0ij} = \xnc{\pi}^{i}\xnc{A}^{j}-x^{j}\xnc{\pi}^{k}\partial^{i}\xnc{A}_{k}-\xnc{\pi}^{j}\xnc{A}^{i}+x^{i}\xnc{\pi}^{k}\partial^{j}\xnc{A}_{k}.
\end{gather}
After some algebra, we find that
\begin{equation}
\label{APL-lo}
\left\{\xnc{A}_{i}, \xnc{P}_{\mu}\right\} = \mathcal{L}_{\partial_{\mu}}\xnc{A}_{i},\qquad
\left\{\xnc{A}_{i}, \xnc{J}_{\mu\nu}\right\} = \mathcal{L}_{x_{[\mu}\partial_{\nu]}}\xnc{A}_{i},
\end{equation}
and likewise for $\xnc{\pi}_{i}$, which proves Eq.~\eqref{AQL-lo}. We also find that
\begin{gather}
\label{PP-lo}
\left\{\xnc{P}_{i}, \xnc{P}_{j}\right\} = 0,\\
\label{PJ-lo}
\left\{\xnc{P}_{i}, \xnc{J}_{kl}\right\} = \eta_{ik}\xnc{P}_{l}-\eta_{il}\xnc{P}_{k},\\
\label{JJ-lo}
\left\{\xnc{J}_{ij}, \xnc{J}_{kl}\right\} = \eta_{jk}\xnc{J}_{il}+\eta_{il}\xnc{J}_{jk}-\eta_{ik}\xnc{J}_{jl}-\eta_{jl}\xnc{J}_{ik},
\end{gather}
from where it follows that
\begin{equation}\label{QQ-lo}
\left\{\xnc{Q}_{U}, \xnc{Q}_{V}\right\} = \xnc{Q}_{[U,V]},
\end{equation}
where we have restricted to kinematical generators ($\xnc{P}_{i}$ and $\xnc{J}_{ij}$) only. Thus we see that, although $\partial_{0}\xnc{J}^{\mu\nu} \neq 0$, we still have Eqs.~\eqref{AQL-lo} and \eqref{QQ-lo}. This is necessary for establishing the dynamical consistency of the transformations.

Finally, we would like to mention that for the choice \eqref{lo-113} of $\theta$, Eq.~\eqref{find2nc} gives $\partial_{\mu}\xnc{M}^{\mu 23}=0$ and $\partial_{\mu}\xnc{M}^{\mu 01}=0$. The criterion \eqref{crit22nc} for Lorentz invariance when $\theta$ does not transform is not compatible with Eq.~\eqref{find2nc} in general. However, for this particular choice of $\theta$ the criterion \eqref{crit22nc} also gives $\partial_{\mu}\xnc{M}^{\mu 23}=0$ and $\partial_{\mu}\xnc{M}^{\mu 01}=0$. Thus, Lorentz invariance is partially restored.


\section{\label{sec:lorentz-conclu}Discussion}

The present analysis fits in with the general notions of observer versus particle Lorentz transformations. As is known, usually (without a background) these two approaches to Lorentz symmetry agree. In the presence of a background, this equivalence fails since the background (here $\theta_{\mu\nu}$) transforms as a tensor under observer Lorentz transformations but as a set of scalars under particle Lorentz transformations. The effect of observer and particle Lorentz transformations was captured here by the distinct set of criteria---Eqs.~\eqref{407n} and \eqref{crit2n} in the commutative description and Eqs.~\eqref{crit2nc} and \eqref{crit22nc} in the noncommutative description---obtained for a transforming or a nontransforming $\theta$. Lorentz symmetry was preserved only for a transforming $\theta$ which conforms to observer Lorentz transformations.

The analysis of Lorentz symmetry in the presence of the background field $\theta$ seems to parallel the discussion of gauge symmetry\footnote{For a detailed study of the connection between Lorentz and gauge symmetries in the Maxwell theory, see \refcite{Weinberg:1995}.} in the presence of a background magnetic field $B$.\footnote{Indeed $\theta$ can be regarded as the inverse of $B$.} In the present treatment, Lorentz symmetry of the action is preserved although there may not be a conserved generator.\footnote{The generators, however, are dynamically consistent as shown, for instance, in Eqs.~\eqref{AQLn}, \eqref{QQn}, \eqref{AQL-lo} and \eqref{QQ-lo}.} Likewise, gauge symmetry of the action, say for a particle moving in the presence of background magnetic field, is preserved although a generator, like the Gauss operator, does not exist, since there is no dynamical piece for the gauge field.

Finally, we mention that the present analysis refers to the standard realisation of Poincar\'e symmetry over trivial co-commutative Hopf algebra of fields. Recently it has been shown \cite{Chaichian:2004za, Chaichian:2004yh} that for constant $\theta$, an explicit twisted Poincar\'e symmetry is realised within the twisted Hopf algebra of fields. This is discussed in the next chapter.


%% file: chap_deform.tex

\chapter{\label{chap:deform}Deformed symmetries on noncommutative spaces}


The introduction of noncommuting relativistic coordinate spacetime,
\begin{equation*}\label{101}
\left[\ncx^\mu, \ncx^\nu\right] =  \ui\theta^{\mu\nu}, \quad \mu, \nu = 0, i,
\end{equation*}
for constant $\theta^{\mu\nu}$ implies, among other things, a breakdown of Lorentz invariance. However, it has been shown by using the (twisted) Hopf algebra \cite{Chaichian:2004za} that corresponding field theories possess \emph{deformed} Lorentz invariance. This suggests above all to use the representation theory of the deformed Poincar\'e algebra as a basis for systematic field theoretic discussions of these theories. In the related developments, Wess~\cite{Wess:2003da} and collaborators \cite{Dimitrijevic:2004rf, Aschieri:2005yw, Koch:2004ud} have discussed the deformation of various symmetries on noncommutative spaces. A deformation of the algebra of diffeomorphisms is constructed for noncommutative spaces with a constant $\theta$ parameter. The deformation of the Poincar\'e algebra naturally follows as a subgroup of the deformed diffeomorphism algebra. It has been shown that the algebraic relations remain unaffected but the coproduct rule changes. The modified coproduct rule obtained for the Poincar\'e generators is found to agree with an alternative (quantum-group-theoretic) derivation \cite{Chaichian:2004za, Chaichian:2004yh, Matlock:2005zn} based on the application of twist functions \cite{Oeckl:2000eg}. The extension of these ideas to field theory and possible implications for Noether symmetry are discussed in \refscite{Banerjee:2004ev, Gonera:2005hg, Calmet:2004ii}. An attempt to extend such notions to supersymmetry has been done in Refs.~\cite{Kobayashi:2004ep, Zupnik:2005ut, Ihl:2005zd, Banerjee:2005ig}. Very recently, the deformed Poincar\'e generators for  Lie-algebraic $\theta$ (rather than a constant $\theta$) \cite{Lukierski:2005fc} and Snyder \cite{Snyder:1946qz} noncommutativity \cite{Banerjee:2006wf} have also been analysed.

There are principally two approaches for discussing the deformed symmetries and these give equivalent results. In the first  method \cite{Wess:2003da, Dimitrijevic:2004rf, Aschieri:2005yw, Koch:2004ud} higher-order differential operators are constructed which are compatible with the star-product for a constant (canonical) noncommutative parameter. The deformations brought about by the presence of these operators are such that the comultiplication rules are modified but the algebra remains undeformed. In the second method \cite{Chaichian:2004za, Chaichian:2004yh} the modified comultiplication rules are obtained by an application of an abelian twist function on the primitive coproducts.

In this chapter we develop an algebraic method for analysing the deformed relativistic and nonrelativistic symmetries in noncommutative spaces with a constant noncommutativity parameter. By requiring the twin conditions of consistency with the noncommutative space and  closure of the Lie algebra, we obtain deformed generators with arbitrary free parameters. For relativistic conformal-Poincar\'e symmetries a specific choice of these parameters yields the undeformed algebra, although the generators are still deformed. For the nonrelativistic (Schr\"odinger \cite{Niederer:1972, Hagen:1972pd, Burdet:1972xd}) case two possibilities are discussed for introducing the free parameters. In one of these there is no choice of the parameters that yields the undeformed algebra while in the other way, this possibility exists. 

A differential-operator realisation of the deformed generators is given in the coordinate and momentum representations. The various expressions naturally contain the free parameters. For the particular choice of these parameters that yields the undeformed algebra, the deformations in the generators drop out completely in the momentum representation.

The modified comultiplication rules (in the coordinate representation) and the associated Hopf algebra are calculated. For the choice of parameters that leads to the undeformed algebra we show that these rules agree with those obtained by an application of the abelian twist function on the primitive comultiplication rule.\footnote{For the conformal-Poincar\'e case this computation of modified coproduct rules using the twist function already exists in the literature \cite{Chaichian:2004za, Chaichian:2004yh, Matlock:2005zn, Oeckl:2000eg}, but a similar analysis for the nonrelativistic symmetries is new and presented here.} For other choices of the free parameters the deformations cannot be represented by twist functions. The possibility that there can be such deformations  also arises in the context of $\kappa$-deformed symmetries \cite{Lukierski:2006fv}.
  
Coordinate transformations mapping the undeformed generators with the deformed ones have been given, once again for the particular choice of parameters when the algebra remains undeformed. Consequently such transformations are meaningful only when the deformations are expressed through twist functions. Also, these transformations are valid both for the relativistic and nonrelativistic treatments.

In section \ref{sec:deform-conf-rel} we discuss the deformed conformal-Poincar\'e symmetries. The special conformal generator contains an arbitrary free parameter. New algebraic structures are obtained. Section~\ref{sec:deform-conf-nonrel} has a detailed analysis of the  Schr\"odinger symmetry \cite{Niederer:1972, Hagen:1972pd, Burdet:1972xd} (Schr\"odinger group contains, in addition to the centrally extended Galilean group, two conformal generators, namely dilatations and special conformal transformations or expansions). Two generalisations are possible, both of which contain free parameters. We show that if only $\uO(\theta)$ deformations are considered, then the closure of the algebra is such that no choice of the free parameters yields the undeformed algebra. This is feasible only if $\uO(\theta^2)$ deformations are included. In either case the algebra closes nontrivially leading to new structures. Also, a  deformed conformal-Galilean algebra is obtained in this section by a contraction of the deformed conformal-Poincar\'e algebra.


\section{\label{sec:deform-conf-rel}Deformed conformal-Poincar\'e algebra}

In this section we analyse the deformations in the full conformal-Poincar\'e generators compatible with a canonical (constant) noncommutative spacetime. First, confining to the Poincar\'e sector only, we find that it is possible to obtain a generalisation (by including, apart from the translations and rotations, a symmetric second-rank tensor operator) of the Poincar\'e algebra containing two arbitrary parameters. Fixing these parameters yields the usual undeformed algebra. This result is in comformity with that obtained in \refcite{Koch:2004ud}. Including the conformal sector yields further novel algebraic structures. We find that there exists a one-parameter class of deformed special conformal generators that yields a closed algebra whose structure is completely new. A particular value of the parameter leads to the undeformed algebra.

We begin by presenting an algebraic approach whereby compatibility is achieved with noncommutative spacetime by the various Poincar\'e generators. This spacetime is characterised by the algebra
\begin{equation} \label{brac:nc101}
\left[ \ncx^{\mu}, \ncx^{\nu} \right] = \ui \theta^{\mu \nu} \, , \qquad
\left[ \ncp_{\mu}, \ncp_{\nu} \right] = 0 \, , \qquad
\left[ \ncx^{\mu}, \ncp_{\nu} \right] = \ui {\delta^{\mu}}_{\nu} \, .
\end{equation}
For constant $\theta$, it follows that, for any spacetime transformation,
\begin{equation} \label{nc102}
\left[ \delta \ncx^{\mu}, \ncx^{\nu} \right] + \left[ \ncx^{\mu}, \delta \ncx^{\nu} \right] = 0 \, .
\end{equation}
It is obvious that  translations, $\delta \ncx^{\mu} = a^{\mu}$, with constant $a^{\mu}$, are compatible with the condition \eqref{nc102}. The generator of the transformation, consistent with $\delta \ncx^{\mu} = \ui a^{\sigma} [ \ncopP_{\sigma}, \ncx^{\mu} ]$, is
\begin{equation} \label{nc103}
\ncopP_{\mu} = \ncp_{\mu} \, .
\end{equation}

For an undeformed Lorentz transformation, $\delta \ncx^{\mu} = \omega^{\mu \nu} \ncx_{\nu}$, $\omega^{\mu \nu} = - \omega^{\nu \mu}$, the requirement \eqref{nc102} implies ${\omega^\mu}_\lambda \theta^{\lambda \nu} - {\omega^\nu}_\lambda \theta^{\lambda \mu} = 0$, which is not satisfied except for two dimensions, when $\omega_{\mu\nu}$ and $\theta_{\mu\nu}$ become proportional to the antisymmetric tensor $\varepsilon_{\mu\nu}$. Therefore, in general, the usual Lorentz transformation is not consistent with the condition \eqref{nc102}. A deformation of the Lorentz transformation is therefore mandatory. We consider the minimal deformation so that the transformation law is modified by terms proportional to $\theta$:
\begin{equation*} \label{nc104}
\delta \ncx^{\mu} = \omega^{\mu \nu} \ncx_{\nu} + n_1 {\omega^{\mu}}_{\nu} \theta^{\nu \sigma} \ncp_{\sigma} + n_2 {\omega_{\nu}}^{\sigma} \theta^{\mu \nu} \ncp_{\sigma} + n_3 \omega_{\nu \sigma} \theta^{\nu \sigma} \ncp^{\mu} \, ,
\end{equation*}
where $n_1$, $n_2$ and $n_3$ are coefficients to be determined by consistency arguments. The  generator,\footnote{Whenever convenient, we shall use the symbol $\langle \mu \nu \rangle$ to denote the preceding terms with $\mu$ and $\nu$ interchanged. For example, $Z^{\cdots \mu \cdots \nu \cdots} - \langle \mu \nu \rangle = Z^{\cdots \mu \cdots \nu \cdots} - Z^{\cdots \nu \cdots \mu \cdots} = Z^{\cdots \mu \cdots \nu \cdots} - \langle \nu \mu \rangle$, and $-Z^{\cdots \mu \cdots \nu \cdots} - \langle \mu \nu \rangle = -Z^{\cdots \mu \cdots \nu \cdots} + Z^{\cdots \nu \cdots \mu \cdots}$.}
\begin{equation} \label{nc105}
\begin{split}
\ncopJ^{\mu \nu}
&= \ncx^{\mu} \ncp^{\nu} - \ncx^{\nu} \ncp^{\mu} + \lambda_{1} \left( \theta^{\mu \sigma} \ncp_{\sigma} \ncp^{\nu} - \theta^{\nu \sigma} \ncp_{\sigma} \ncp^{\mu} \right) + \lambda_{2} \theta^{\mu \nu} \ncp^{2} \\
&\equiv \ncx^{\mu} \ncp^{\nu} + \lambda_{1} \theta^{\mu \sigma} \ncp_{\sigma} \ncp^{\nu} + \tfrac{1}{2} \lambda_{2} \theta^{\mu \nu} \ncp^{2} - \langle \mu \nu \rangle \, ,
\end{split}
\end{equation}
reproduces the above transformation as
\begin{equation*}
\delta \ncx^{\mu} = -\frac{\ui}{2} \omega_{\rho \sigma} \left[ \ncopJ^{\rho \sigma}, \ncx^{\mu} \right]
\end{equation*}
for $n_1 = n_2 + 1 = \lambda_1$, $n_3 = -\lambda_2$, a result which follows on using the basic noncommutative algebra \eqref{brac:nc101}. It is therefore clear that $n_1 = n_2 = 0$ is not possible, which necessitates the modification of the transformation as well as the generator. It turns out that\footnote{The symbol $\langle \mu \nu \rho \sigma \rangle$ means the following: $Z^{\cdots} - \langle \mu \nu \rho \sigma \rangle = \left( Z^{\cdots} - \langle \mu \nu \rangle \right) - \langle  \rho \sigma \rangle$.}
\begin{equation} \label{nc106}
\begin{split}
\left[ \ncopJ^{\mu \nu}, \ncopJ^{\rho \sigma} \right]
&= \ui \Big[ \eta^{\mu \rho} \ncopJ^{\nu \sigma} - \eta^{\nu \rho} \ncopJ^{\mu \sigma} - \eta^{\mu \sigma} \ncopJ^{\nu \rho} + \eta^{\nu \sigma} \ncopJ^{\mu \rho} \\
&\qquad {} - \theta^{\mu \rho}   \left\{ (2\lambda_1 - 1) \ncp^{\nu} \ncp^{\sigma} + \lambda_2 \ncp^2 \eta^{\nu \sigma} \right\} + \theta^{\nu \rho}   \left\{ (2\lambda_1 - 1) \ncp^{\mu} \ncp^{\sigma} + \lambda_2 \ncp^2 \eta^{\mu \sigma} \right\} \\
&\qquad {} + \theta^{\mu \sigma} \left\{ (2\lambda_1 - 1) \ncp^{\nu} \ncp^{\rho} + \lambda_2 \ncp^2 \eta^{\nu \rho} \right\} - \theta^{\nu \sigma} \left\{ (2\lambda_1 - 1) \ncp^{\mu} \ncp^{\rho} + \lambda_2 \ncp^2 \eta^{\mu \rho} \right\} \Big] \\
&\equiv \ui \Big[ \eta^{\mu \rho} \ncopJ^{\nu \sigma} - \theta^{\mu \rho} \left\{ (2\lambda_1 - 1) \ncp^{\nu} \ncp^{\sigma} + \lambda_2 \ncp^2 \eta^{\nu \sigma} \right\} \Big] - \langle \mu \nu \rho \sigma \rangle \, .
\end{split}
\end{equation}
The closure of the normal Lorentz algebra is obtained only for $\lambda_1 = 1/2$ and $\lambda_2 = 0$ \cite{Koch:2004ud}.

As a curiosity we remark that it is possible to have a generalised type of Poincar\'e algebra with generators $\ncopP^\mu$, $\ncopJ^{\mu \nu}$, $\ncopS^{\mu \nu} \equiv \ncp^\mu \ncp^\nu$. Since the $\ncopP$--$\ncopP$ and $\ncopJ$--$\ncopP$ algebras retain their undeformed structures, it is clear that the closure of this algebra with an extended generator holds. It is worthwhile to mention here that a symmetric second-rank tensor as a generator occurs in the example of the 3-dimensional isotropic harmonic oscillator: $\opH = \mathbf{p}^2 / 2m + m \omega^2 \mathbf{x}^2 / 2$. The dynamical symmetry generators, $\opJ_i = \varepsilon_{ijk} x_j p_k$, $\opQ_{ij} = x_i x_j - \delta_{ij} \mathbf{x}^2 / 3$, satisfy an $\mathrm{SU}(3)$ algebra. The quadrupole operator $\opQ_{ij}$ is obviously symmetric and traceless.

Similarly, the usual scale transformation, $\delta \ncx^{\mu} =  \alpha \ncx^{\mu}$, is not consistent with the condition \eqref{nc102}. A minimal deformed form of the transformation may be written as
\begin{equation*} \label{nc107}
\delta \ncx^{\mu} = \alpha \ncx^{\mu} + \alpha n \theta^{\mu \nu} \ncp_{\nu} \, .
\end{equation*}
The consistency, $\delta \ncx^{\mu} = \ui \alpha [ \ncopD, \ncx^{\mu} ]$, is achieved only for $n = 1$ by
\begin{equation} \label{nc108}
\ncopD = \ncx^\mu \ncp_\mu \, .
\end{equation}

Likewise, to achieve consistency with the condition \eqref{nc102}, we start with the minimally deformed form of the special conformal transformation:
\begin{equation*} \label{nc109}
\begin{split}
\delta \ncx^{\mu}
&= 2 \omega_{\rho} \ncx^{\rho} \ncx^{\mu} - \omega^{\mu} x^2 \\
&\quad {} + \omega_{\rho} \left( m_1 \theta^{\sigma \mu} \ncx^{\rho} \ncp_{\sigma} + m_2 \theta^{\sigma \mu} \ncx_{\sigma} \ncp^{\rho} + m_3 \theta^{\mu \rho} \ncx^{\sigma} \ncp_{\sigma} + m_4 \theta^{\mu \rho} + m_5 \theta^{\rho \sigma} \ncx^{\mu} \ncp_{\sigma} \right) \\
&\quad {} + m_6 \omega^{\mu} \theta^{\rho \sigma} \ncx_{\rho} \ncp_{\sigma} \, .
\end{split}
\end{equation*}
The  generator,
\begin{equation} \label{nc110}
\ncopK^{\rho} = 2 \ncx^{\rho} \ncx^{\sigma} \ncp_{\sigma} - \ncx^2 \ncp^\rho + \eta_1 \theta^{\rho \sigma} \ncp_{\sigma} + \eta_2 \theta^{\rho \sigma} \ncx^{\beta} \ncp_{\beta} \ncp_{\sigma} + \eta_3 \theta^{\sigma \beta} \ncx^{\sigma} \ncp_{\beta} \ncp^{\rho}
\end{equation}
is consistent with $\delta \ncx^{\mu} = \ui \omega_{\rho} [ \ncopK^{\rho}, \ncx^{\mu} ]$ for $\eta_2 = \eta_3 = 0 = m_6 = m _5 = m_2 -2 = m_3 - 2 = m_1 + 2$ and $m_4 = - \eta_1$.

This completes our demonstration of the compatibility of the various transformation laws with the basic noncommutative algebra. However, achieving consistency with the transformation and closure of the algebra are two different things. It can and does turn out that the minimal $\uO(\theta)$ deformation, while preserving consistency, does not yield a closed algebra. Indeed we find that the conformal algebra
\begin{equation*} \label{nc111}
\left[ \ncopK^\rho, \ncopD \right]
= \ui \left[ \ncopK^\rho + 2 \theta^{\rho \mu} \left( \ncp_\mu \ncopD - \eta_1 \ncp_\mu \right) - \theta^{\sigma \mu} \left( \ncx_\sigma \ncp_\mu \ncp^\rho + \ncp_\mu \ncx_\sigma \ncp^\rho \right) \right]
\end{equation*}
does not close, necessitating the inclusion of $\uO(\theta^2)$ terms in the deformed transformation and the deformed generator. Therefore, instead of the form appearing above Eq.~\eqref{nc110} we now start with
\begin{equation*} \label{nc112}
\begin{split}
\delta \ncx^{\mu}
&= 2 \omega_{\rho} \ncx^{\rho} \ncx^{\mu} - \omega^{\mu} x^2 \\
&\quad {} + \omega_{\rho} \left( m_1 \theta^{\sigma \mu} \ncx^{\rho} \ncp_{\sigma} + m_2 \theta^{\sigma \mu} \ncx_{\sigma} \ncp^{\rho} + m_3 \theta^{\mu \rho} \ncx^{\sigma} \ncp_{\sigma} + m_4 \theta^{\mu \rho} + m_5 \theta^{\rho \sigma} \ncx^{\mu} \ncp_{\sigma} \right) \\
&\quad {} + m_6 \omega^{\mu} \theta^{\rho \sigma} \ncx_{\rho} \ncp_{\sigma} + \omega^{\mu} \left( m_7 \theta^{\alpha \beta} \theta_{\alpha \beta} \ncp^2 + m_8 \theta^{\alpha \beta} {\theta_{\alpha}}^{\sigma} \ncp_{\sigma} \ncp_{\beta} \right) \\
&\quad {} + \omega_{\rho} \Big( m_9 \theta^{\alpha \beta} \theta_{\alpha \beta} \ncp^{\rho} \ncp^{\mu} + m_{10} \theta^{\rho \alpha} {\theta_{\alpha}}^{\sigma} \ncp_{\sigma} \ncp^{\mu} \\
&\qquad\qquad {} + m_{11} \theta^{\mu \alpha} {\theta_{\alpha}}^{\sigma} \ncp_{\sigma} \ncp^{\rho} + m_{12} \theta^{\rho \alpha} {\theta_{\alpha}}^{\mu} \ncp^2 + m_{13} \theta^{\rho \alpha} \theta^{\sigma \mu} \ncp_{\alpha} \ncp_{\sigma} \Big) \, .
\end{split}
\end{equation*}
An appropriately deformed form of the generator containing  $\uO(\theta^2)$ terms is given by
\begin{equation} \label{nc113}
\begin{split}
\ncopK^{\rho}
&= 2 \ncx^{\rho} \ncx^{\sigma} \ncp_{\sigma} - \ncx^2 \ncp^\rho + \eta_1 \theta^{\rho \sigma} \ncp_{\sigma} + \eta_2 \theta^{\rho \sigma} \ncx^{\beta} \ncp_{\beta} \ncp_{\sigma} + \eta_3 \theta^{\sigma \beta} \ncx^{\sigma} \ncp_{\beta} \ncp^{\rho} \\
&\quad {} + \eta_4 \theta^{\alpha \beta} {\theta_{\alpha}}^{\sigma} \ncp_{\sigma} \ncp_{\beta} \ncp^{\rho} + \eta_5 \theta^{\alpha \beta} \theta_{\alpha \beta} \ncp^2 \ncp^{\rho} + \eta_6 \theta^{\rho \alpha} {\theta_{\alpha}}^{\sigma} \ncp_{\sigma} \ncp^2 \, .
\end{split}
\end{equation}
Consistency with the transformation law now requires $m_1 = -2$, $m_2 = m_6 + 2 = \eta_3 + 2$, $m_3 = 2 - m_5 = 2 + m_{13} = 2 - \eta_2$, $m_4 = -\eta_1$, $m_7 = m_9/2 = \eta_5$, $m_8 = (m_2 - 2 -m_{11})/2 = \eta_4$ and $m_{10} = 2 m_{12} = 2 \eta_6$, implying 6 free parameters in the generator and in the transformation. However, the closure of the algebra
\begin{equation} \label{nc114}
\begin{split}
\left[ \ncopK^\rho, \ncopD \right]
&= \ui \left[ \ncopK^\rho + 2 (1 - \eta_2) \theta^{\rho \mu} \ncp_\mu \ncopD - 2 (\ui \eta_2 + \eta_1) \theta^{\rho \mu} \ncp_\mu - 2 (1 + \eta_3) \theta^{\alpha \beta} \ncx_\alpha \ncp_\beta \ncp^\rho \right. \\
&\qquad \, {} + \left. (\eta_3 - 4 \eta_4) \theta^{\alpha \beta} {\theta_\alpha}^\sigma \ncp_\sigma \ncp_\beta \ncp^\rho - 4 \eta_5 \theta^{\alpha \beta} \theta_{\alpha \beta} \ncp^2 \ncp^\rho - 4 \eta_6 \theta^{\rho \alpha} {\theta_\alpha}^\sigma \ncp_\sigma \ncp^2 \right]
\end{split}
\end{equation}
fixes 5 parameters, $\eta_2 = -\eta_3 = -4 \eta_4 = 1$, $\eta_5 = \eta_6 = 0$, leaving only one, $\eta_1$, as free.

The final form of the deformed generators, therefore, is given by
\begin{equation} \label{nc115}
\begin{aligned}
&\ncopP_{\mu} = \ncp_{\mu} \, , \\
&\ncopJ^{\mu \nu} = \ncx^{\mu} \ncp^{\nu} + \tfrac{1}{2} \theta^{\mu \sigma} \ncp_{\sigma} \ncp^{\nu} - \langle \mu \nu \rangle \, , \\
&\ncopD = \ncx^\mu \ncp_\mu \, , \\
&\ncopK^{\rho} = 2 \ncx^{\rho} \ncx^{\sigma} \ncp_{\sigma} - \ncx^2 \ncp^\rho + \eta_1 \theta^{\rho \sigma} \ncp_{\sigma} + \theta^{\rho \sigma} \ncx^{\beta} \ncp_{\beta} \ncp_{\sigma} - \theta^{\sigma \beta} \ncx_{\sigma} \ncp_{\beta} \ncp^{\rho} - \tfrac{1}{4} \theta^{\alpha \beta} {\theta_{\alpha}}^{\sigma} \ncp_{\sigma} \ncp_{\beta} \ncp^{\rho} \, ,
\end{aligned}
\end{equation}
which involves one free parameter. Observe that the free parameters in the Lorentz generator are ruled out as a consequence of the closure of the $\ncopJ$--$\ncopK$ algebra. The various  generators satisfy the deformed algebra:
\begin{equation} \label{nc116}
\begin{aligned}
&\left[ \ncopP^{\mu}, \ncopP^{\nu} \right] = 0 \, , \rule{4.66cm}{0cm}
\left[ \ncopP^{\mu}, \ncopJ^{\rho \sigma} \right] = - \ui \eta^{\mu \rho} \ncopP^{\sigma} - \langle \rho\sigma \rangle \, , \\
&\left[ \ncopJ^{\mu \nu}, \ncopJ^{\rho \sigma} \right] = \ui \eta^{\mu \rho} \ncopJ^{\nu \sigma} - \langle \mu \nu \rho \sigma \rangle \, , \qquad \quad
\left[ \ncopD, \ncopP^{\mu} \right] = \ui \ncopP^{\mu} \, , \\
&\left[ \ncopD, \ncopJ^{\mu \nu} \right] = 0 \, , \rule{4.65cm}{0cm}
\left[ \ncopK^{\rho}, \ncopP^{\mu} \right] = 2 \ui \left( \eta^{\rho \mu} \ncopD + \ncopJ^{\rho \mu} \right), \\
&\left[ \ncopK^{\rho}, \ncopJ^{\mu \nu} \right] = - \ui \Big[ \eta^{\rho \mu} \ncopK^{\nu} + (\ui + \eta_1) \left( \theta^{\rho \mu} \ncopP^{\nu} - \eta^{\rho \mu} \theta^{\nu \sigma} \ncopP_{\sigma} \right) \Big] - \langle \mu\nu \rangle \, , \\
&\left[ \ncopK^{\rho}, \ncopD \right] = \ui \left[ \ncopK^{\rho} - 2 (\ui + \eta_1) \theta^{\rho \mu} \ncopP_{\mu} \right], \\
&\left[ \ncopK^{\rho}, \ncopK^{\mu} \right] = -2 \ui (\ui + \eta_1) \left( \theta^{\rho \mu} \ncopD - \theta^{\mu \sigma} {\ncopJ^{\rho}{}}_{\sigma} \right) - \langle \rho\mu \rangle \, .
\end{aligned}
\end{equation}
We observe that the Poincar\'e sector remains unaffected, but the conformal sector changes. A one-parameter class of closed algebras is found. We therefore obtain new algebraic structures in the conformal sector. Also, unlike the Poincar\'e sector discussed earlier, it is not necessary to extend the set of generators to obtain these new structures. Fixing $\eta_1 = - \ui$ yields the usual (undeformed) Lie algebra. In that case the deformed special conformal generator also agrees with the result given in \refcite{Banerjee:2005ig}.


\subsection{Coordinate transformations and generators}

The form of the generators in Eq.~\eqref{nc115} with $\eta_1 = -\ui$ obeys the usual conformal-Poincar\'e algebra. It is possible to obtain this form of the generators starting from the generators in the commutative space and then using the appropriate transformation from the commutative $(x, p)$ to the noncommutative $(\ncx, \ncp)$ description. To this end, we note that the transformation $\ncp_{\mu} = p_{\mu}$, $\ncx^{\mu} = x^{\mu} - \tfrac{1}{2} \theta^{\mu \nu} p_{\nu}$
preserves the basic commutation relations:
\begin{align*}
\left[ \ncp_{\mu}, \ncp_{\nu} \right] &= \left[ p_{\mu}, p_{\nu} \right] = 0 \, , \\
\left[ \ncx^{\mu}, \ncp_{\nu} \right] &= \left[ x^{\mu} - \tfrac{1}{2} \theta^{\mu \sigma} p_{\sigma}, p_{\nu} \right] = \ui {\delta^{\mu}}_{\nu} \, , \\
\left[ \ncx^{\mu}, \ncx^{\nu} \right] &= \left[ x^{\mu} - \tfrac{1}{2} \theta^{\mu \sigma} p_{\sigma}, x^{\nu} - \tfrac{1}{2} \theta^{\nu \lambda} p_{\lambda} \right] = \ui \theta^{\mu \nu} \, .
\end{align*}
Now taking the generators in the commutative space and applying the inverse transformation, $p_{\mu} = \ncp_{\mu}$, $x^{\mu} = \ncx^{\mu} + \tfrac{1}{2} \theta^{\mu \nu} \ncp_{\nu}$, yields the generators in the noncommutative space:
\begin{equation} \label{def16031}
\begin{aligned}
&\ncopP^\mu = \opP^\mu \left(x(\ncx,\ncp), p(\ncx,\ncp)\right) = p^\mu (\ncx,\ncp) = \ncp^\mu \, , \\
&\ncopJ^{\mu \nu}
= \opJ^{\mu \nu}\left(x(\ncx,\ncp), p(\ncx,\ncp)\right)
 = x^\mu p^\nu - x^\nu p^\mu
 = \left( \ncx^{\mu} + \tfrac{1}{2} \theta^{\mu \sigma} \ncp_{\sigma} \right) \ncp^{\nu} - \langle \mu \nu \rangle \\
&\qquad\! = \ncx^{\mu} \ncp^{\nu} + \tfrac{1}{2} \theta^{\mu \sigma} \ncp^{\nu} \ncp_{\sigma} - \langle \mu \nu \rangle \, ,\\
&\ncopD = \opD \left(x(\ncx,\ncp), p(\ncx,\ncp)\right) = x^\mu p_\mu =\left( \ncx^{\mu} + \tfrac{1}{2} \theta^{\mu \sigma} \ncp_{\sigma} \right) \ncp_{\mu} = \ncx^{\mu} \ncp_{\mu} \, , \\
&\ncopK^{\rho}
= \opK^\rho \left(x(\ncx,\ncp), p(\ncx,\ncp)\right)= 2x^\rho x^\sigma p_\sigma - x^2 p^\rho \\
&\quad\, = 2 \left( \ncx^{\rho} + \tfrac{1}{2} \theta^{\rho \alpha} \ncp_{\alpha} \right) \left( \ncx^{\sigma} + \tfrac{1}{2} \theta^{\sigma \beta} \ncp_{\beta} \right) \ncp_{\sigma} - \left( \ncx^{\mu} + \tfrac{1}{2} \theta^{\mu \alpha} \ncp_{\alpha} \right) \left( \ncx_{\mu} + \tfrac{1}{2} \theta_{\mu \beta} \ncp^{\beta} \right) \ncp^\rho \\
&\quad\, = 2 \ncx^{\rho} \ncx^{\sigma} \ncp_{\sigma} - \ncx^2 \ncp^\rho - \ui \theta^{\rho \sigma} \ncp_{\sigma} + \theta^{\rho \sigma} \ncx^{\beta} \ncp_{\beta} \ncp_{\sigma} - \theta^{\sigma \beta} \ncx_{\sigma} \ncp_{\beta} \ncp^{\rho} - \tfrac{1}{4} \theta^{\alpha \beta} {\theta_{\alpha}}^{\sigma} \ncp_{\sigma} \ncp_{\beta} \ncp^{\rho} \, .
\end{aligned}
\end{equation}
This also explains the fact that these deformed generators satisfy the usual undeformed algebra. Nontrivial distinctions arise when $\eta_1 \neq \ui$ in which case new structures are obtained. These cannot be reproduced by simple coordinate transformations.


\subsection{Representations}

In the usual commutative space a symmetry exists between the coordinates $x$ and momenta $p$. Each is an observable with eigenvalues extending from $-\infty$ to $+\infty$ and the usual commutation relations involving $x$ and $p$ remain invariant if $x$ and $p$ are interchanged and `$\ui$' is replaced by `$-\ui$'. One may then set up the coordinate representation in which $x$ is diagonal and $p =-\ui \frac{\partial}{\partial x}$ with $\hbar=1$. Alternatively it is also feasible to write the momentum representation where $p$ is diagonal and $x = \ui \frac{\partial}{\partial p}$.

In the noncommutative space, on the other hand, the symmetry between $x$ and $p$ is lost. As will soon be shown, this leads to nontrivial distinctions between the coordinate and momentum representations. The relations in Eq.~\eqref{brac:nc101} are easily reproduced by representing
\begin{equation} \label{repr-x}
\ncx^\mu = \ncx^\mu, \qquad \ncp_\mu = - \ui \ncpartial_\mu \equiv -\ui \frac{\partial}{\partial \ncx^\mu} \, ,
\end{equation}
in view of the relations $[ \ncx^\mu, \ncx^\nu ] = \ui \theta^{\mu \nu}$, $[ \ncpartial_\mu, \ncpartial_\nu ] = 0$, $[ \ncpartial_\mu, \ncx^\nu ] = {\delta_\mu}^\nu$. This is the coordinate representation. One may also choose the momentum representation:
\begin{equation} \label{ncrepr-p}
\ncp_{\mu} = \ncp_{\mu} \, , \qquad
\ncx^{\mu} = \ui \nceth^{\mu} - \tfrac{1}{2} \theta^{\mu \nu} \ncp_{\nu}
\equiv \ui \frac{\partial}{\partial \ncp_{\mu}} - \tfrac{1}{2} \theta^{\mu \nu} \ncp_{\nu} \, .
\end{equation}
The relations in Eq.~\eqref{brac:nc101} are now reproduced in view of $[ \ncp_\mu, \ncp_\nu ] = 0$, $[ \nceth^\mu, \nceth^\nu ] = 0$, $[ \nceth^\mu, \ncp_\nu ] = {\delta^\mu}_\nu$. The deformed generators in coordinate representation read
\begin{equation} \label{nc-cor115}
\begin{aligned}
&\ncopP_{\mu} = -\ui \ncpartial_{\mu} \, , \\
&\ncopJ^{\mu \nu} = -\ui \ncx^{\mu} \ncpartial^{\nu} - \tfrac{1}{2} \theta^{\mu \sigma} \ncpartial_{\sigma} \ncpartial^{\nu} - \langle \mu \nu \rangle \, , \\
&\ncopD = -\ui \ncx^\mu \ncpartial_\mu \, , \\
&\ncopK^{\rho} = -2 \ui \ncx^{\rho} \ncx^{\sigma} \ncpartial_{\sigma} + \ui \ncx^2 \ncpartial^\rho - \ui  \eta_1 \theta^{\rho \sigma} \ncpartial_{\sigma} - \theta^{\rho \sigma} \ncx^{\beta} \ncpartial_{\beta} \ncpartial_{\sigma} \\
&\qquad\,\, {} + \theta^{\sigma \beta} \ncx_{\sigma} \ncpartial_{\beta} \ncpartial^{\rho} - \frac{\ui}{4} \theta^{\alpha \beta} {\theta_{\alpha}}^{\sigma} \ncpartial_{\sigma} \ncpartial_{\beta} \ncpartial^{\rho} \, .
\end{aligned}
\end{equation}
It is a matter of straightforward calculation to show that the algebra \eqref{nc116} is indeed satisfied.

\paragraph{Momentum representation.}
From a purely algebraic point of view one may use either coordinate or momentum representation. However it appears that, for noncommutative space, momentum representation is more favoured since the momenta still continue to commute. This is even true from an algebraic point of view, as we now demonstrate by writing the generators in the momentum representation.

Translations are trivially represented by $\ncopP^{\mu} = \ncp^{\mu}$. Let us write down the generator of Lorentz transformations in momentum representation:
\begin{equation*}
\begin{split}
\ncopJ^{\mu \nu}
&= \ncp^\nu \ncx^\mu + \tfrac{1}{2} \theta^{\mu \sigma} \ncp^\nu \ncp_\sigma - \langle \mu \nu \rangle
 = \ncp^\nu \left( \ui \nceth^\mu - \tfrac{1}{2} \theta^{\mu \sigma} \ncp_{\sigma} \right) + \tfrac{1}{2} \theta^{\mu \sigma} \ncp^\nu \ncp_\sigma - \langle \mu \nu \rangle \\
&= \ui \left( \ncp^\nu \nceth^\mu - \ncp^\mu \nceth^\nu \right).
\end{split}
\end{equation*}
We note that the extra (deformed) pieces exactly cancel out. The definition of the Lorentz generator, as compared to the commutative space description, is thus form-invariant. This is a generic feature, it is also true for dilatations:
\[
\ncopD = \ncx^{\mu} \ncp_{\mu}
= \ncp_{\mu} \ncx^{\mu} + \ui N
= \ncp_{\mu} \left( \ui \nceth^\mu - \tfrac{1}{2} \theta^{\mu \sigma} \ncp_{\sigma} \right) + \ui N
= \ui \ncp_{\mu} \nceth^\mu + \ui N \, ,
\]
where $N = {\delta^\mu}_\mu$ is the number of spacetime dimensions. For special conformal transformations we have
\[
\ncopK^\rho = \ncp^\rho \nceth^2 -  2 \ncp_\sigma \nceth^\rho \nceth^\sigma - 2 N \nceth^\rho + (\eta_1 + \ui) \theta^{\rho \sigma} \ncp_\sigma \, .
\]
Although there is deformation in the generator for the general case, for $\eta_1 = -\ui$, when the generators satisfy the usual (undeformed) algebra, the deformation in $\ncopK^\rho$ drops out in the momentum representation. 

Thus all the generators have exactly the same structure as in the commutative description. It shows the naturalness of the momentum representation. This is also intuitively understandable since noncommutative-space momenta still commute among themselves, as they do in the commutative space.


\subsection{Coproducts and Hopf algebra}

The deformed generators lead to new comultiplication rules. To obtain these rules we apply the operator to a product of two functions. Using the coordinate representation it follows that
\begin{equation*}
\ncopP_{\mu} \left(\xnc{f} {\,} \xnc{g} \right) = - \ui \ncpartial_{\mu} \left(\xnc{f} {\,} \xnc{g} \right) = \left(- \ui \ncpartial_{\mu} \xnc{f} \right) \xnc{g} + \xnc{f} \left( - \ui \ncpartial_{\mu} \xnc{g} \right) = \left( \ncopP_{\mu} \xnc{f} \right) \xnc{g} + \xnc{f} \left( \ncopP_{\mu} \xnc{g} \right),
\end{equation*}
which yields
\begin{equation} \label{nc-coP}
\Delta ( \ncopP_{\mu} ) =  \ncopP_{\mu} \otimes \opid + \opid \otimes \ncopP_{\mu} \, .
\end{equation}
Similarly we find
\begin{gather}
\label{nc-coJ}
\Delta ( \ncopJ^{\mu \nu} ) = \tfrac{1}{2} \left[ \ncopJ^{\mu \nu} \otimes \opid + \opid \otimes \ncopJ^{\mu \nu} + \theta^{\mu \sigma} \left( \ncopP^{\nu} \otimes \ncopP^{\sigma} - \ncopP^{\sigma} \otimes \ncopP^{\nu} \right) \right] - \langle \mu \nu \rangle \, , \\
\label{nc-coD}
\Delta ( \ncopD ) = \ncopD \otimes \opid + \opid \otimes \ncopD + \theta^{\mu \nu} \ncopP_{\mu} \otimes \ncopP_{\nu} \, , \\
\label{nc-coK} \begin{split}
\Delta ( \ncopK^{\rho} ) &=  \ncopK^{\rho} \otimes \opid + \opid \otimes \ncopK^{\rho} + \theta^{\rho \sigma} \left( \ncopD \otimes \ncopP_{\sigma} - \ncopP_{\sigma} \otimes \ncopD \right) \\
&\quad {} + \theta^{\alpha \sigma} \left( {\ncopJ^{\rho}{}}_{\alpha} \otimes \ncopP_{\sigma} - \ncopP_{\sigma} \otimes {\ncopJ^{\rho}{}}_{\alpha} \right)\\
&\quad {} + \tfrac{1}{2} \theta^{\alpha \sigma} \theta^{\rho \beta} \left( \ncopP_{\alpha} \otimes \ncopP_{\beta} \ncopP_{\sigma} + \ncopP_{\beta} \ncopP_{\sigma} \otimes \ncopP_{\alpha} \right) \\
&\quad {} - \tfrac{1}{4} \theta^{\alpha \sigma} {\theta_{\alpha}}^{\beta} \left( \ncopP^{\rho} \otimes \ncopP_{\beta} \ncopP_{\sigma} + \ncopP_{\beta} \ncopP_{\sigma} \otimes \ncopP^{\rho} \right).
\end{split}
\end{gather}
The free parameter appearing in $\ncopK^{\rho}$ does not appear explicitly in $\Delta (\ncopK^{\rho})$. The coproduct rules for the Poincar\'e sector were earlier derived in Refs.~\cite{Wess:2003da, Koch:2004ud, Chaichian:2004za, Chaichian:2004yh} and for the conformal sector in Refs.~\cite{Banerjee:2005ig, Matlock:2005zn}. Now we compute the basic Hopf algebra. It turns out that the Hopf algebra can be read off from Eq.~\eqref{nc116} by just replacing the generators by the coproducts. For example,
\begin{equation*}
\left[ \Delta ( \ncopK^{\rho} ), \Delta ( \ncopD ) \right] = \ui \left[ \Delta ( \ncopK^{\rho} ) - 2 (\ui + \eta_1) \theta^{\rho \mu} \Delta ( \ncopP_{\mu} ) \right].
\end{equation*}


\section{\label{sec:deform-conf-nonrel}Deformed Schr\"odinger and conformal-Galilean algebras}

The analysis of the previous section is now done for the nonrelativistic symmetries. We consider separately the Schr\"odinger symmetry and the conformal-Galilean symmetry, both of which are extensions of the Galilean symmetry.


\subsection{Deformed Galilean symmetry}

The undeformed $n$-dimensional Galilean algebra, which involves Hamiltonian ($\opH$), translations ($\opP^{i}$), rotations ($\opJ^{ij}$) and boosts ($\opG^{i}$), is given by
\begin{equation} \label{alg:gal}
\begin{aligned}
&\left[ \opP^{i}, \opP^{j} \right] = 0 \, , &
&\left[ \opJ^{ij}, \opJ^{k\ell} \right] = \ui \delta^{ik} \opJ^{j\ell} - \langle i j k \ell \rangle \, , \\
&\left[ \opG^{i}, \opG^{j} \right] = 0 \, , &
&\left[ \opH, \opP^{i} \right] = 0 \, , \\
&\left[ \opH, \opJ^{ij} \right] = 0 \, , &
&\left[ \opH, \opG^{i} \right] = - \ui \opP^{i} \, , \\
&\left[ \opP^{i}, \opJ^{jk} \right] = \ui \delta^{ik} \opP^{j} - \langle j k \rangle \, , & \qquad \quad
&\left[ \opP^{i}, \opG^{j} \right] = - \ui m \delta^{ij} \, , \\
&\left[ \opG^{i}, \opJ^{jk} \right] = \ui \delta^{ik} \opG^{j} - \langle j k \rangle \, .
\end{aligned}
\end{equation}
The standard free-particle representation of this algebra is given by
\begin{equation} \label{gen:gal}
\begin{aligned}
&\opH = \frac{1}{2m} \vec{p}^{2} \, , &
&\opP^{i} = p^{i} \, , \\
&\opJ^{ij} = x^{i} p^{j} - x^{j} p^{i} \, , & \qquad
&\opG^{i} = m x^{i} - t p^{i} \, .
\end{aligned}
\end{equation}
Using the usual commutation relations $[ x^{i}, x^{j} ] = [ p^{i}, p^{j} ] = 0$, $[ x^{i}, p^{j} ] = \ui \delta^{ij}$, the algebra \eqref{alg:gal} is easily reproduced from generators \eqref{gen:gal}.

Now we introduce noncommutativity in space:
\begin{equation} \label{brac:nc}
\left[ \ncx^{i}, \ncx^{j} \right] = \ui \theta^{ij} \, , \qquad
\left[ \ncp^{i}, \ncp^{j} \right] = 0 \, , \qquad
\left[ \ncx^{i}, \ncp^{j} \right] = \ui \delta^{ij} \, .
\end{equation}
Exactly as done for the deformed Poincar\'e generators, we follow a two-step algebraic process. First, by requiring the compatibility of transformations with Eq.~\eqref{brac:nc}, a general deformation of the generators is obtained. A definite structure emerges after demanding the closure of the algebra. Let us first consider the minimal deformation in the generators. The linear momentum $\ncp^{i}$ and the Hamiltonian, $\opH=\vec{\ncp}^{2}/2m$, retain their original forms, basically because the algebra of $\ncp^{i}$ is identical to $p^{i}$. For rotations and boosts a deformation is necessary. Considering the minimal (i.e.\ least order in $\theta$) deformation, we obtain the following structure:
\begin{equation} \label{gen-nc:gal}
\begin{aligned}
&\ncopH = \frac{1}{2m} \vec{\ncp}^{2} \, , \\
&\ncopP^{i} = \ncp^{i} \, , \\
&\ncopJ^{ij} = \ncx^{i} \ncp^{j} - \ncx^{j} \ncp^{i} + \lambda_{1} \left( \theta^{ik} \ncp^{k} \ncp^{j} - \theta^{jk} \ncp^{k} \ncp^{i} \right) + \lambda_{2} \theta^{ij} \vec{\ncp}^{2} \, , \\
&\ncopG^{i} = m \ncx^{i} - t \ncp^{i} + \lambda_{3} m \theta^{ij} \ncp^{j} + \lambda_{4} m^{3} \theta^{ij} \ncx^{j} \, .
\end{aligned}
\end{equation}
The transformations derived from these generators are consistent with the noncommuting algebra \eqref{brac:nc}. Till now the $\lambda$ parameters are arbitrary. These will be determined by requiring the closure of the algebra. Using the brackets \eqref{brac:nc} we find
\begin{gather*}
\left[ \ncopJ^{ij}, \ncopJ^{k\ell} \right]
= \ui \Big[ \delta^{ik} \ncopJ^{j\ell} - \theta^{ik} \left\{ (2\lambda_1 - 1) \ncp^{j} \ncp^{\ell} + 2 \lambda_2 m \ncopH \delta^{j\ell} \right\} \Big] - \langle i j k \ell \rangle, \\
\left[ \ncopG^i, \ncopG^j \right]
= \ui \left[ m^2 (1- 2\lambda_3) \theta^{ij} - 2 t m^3 \lambda_4 \theta^{ij} - m^6 \lambda_4^2 \theta^{ik} \theta^{k\ell} \theta^{\ell j} \right], \\
\begin{split}
\left[ \ncopG^i, \ncopJ^{jk} \right]
&= \ui \Big[ \delta^{ik} \ncopG^j + m (1 - \lambda_1 - \lambda_3) \theta^{ij} \ncopP^k + m (\lambda_1 - \lambda_3) \delta^{ik} \theta^{jm} \ncopP^m + m \lambda_2 \theta^{jk} \ncopP^i \\
&\qquad {} + \lambda_4 m^3 \Big\{ \theta^{ik} \ncx^j - \delta^{ik} \theta^{j\ell} \ncx^\ell + (1 - \lambda_1) \theta^{i\ell} \theta^{\ell j} \ncopP^k \\
&\qquad \qquad \qquad {} + \left( \lambda_1 \theta^{ik} \theta^{j\ell} + \lambda_2 \theta^{i\ell} \theta^{jk} \right) \ncopP^\ell \Big\} \Big] - \langle j k \rangle.
\end{split}
\end{gather*}
If we conform to the usual type of algebra, in the sense that any bracket between the generators should not involve product of generators, then the first equation requires $\lambda_1$ to be set to $1/2$, in order to get rid of the term involving $\ncp^{k} \ncp^{j}$. Also, as is clear from the last equation, the closure of the algebra requires $\lambda_4$ to vanish. For this reason, we set $\lambda_1 = 1/2$ and $\lambda_4 = 0$, so that the above equations simplify to
\begin{equation} \label{alg-nc:gal-11}
\begin{aligned}
&\left[ \ncopJ^{ij}, \ncopJ^{k\ell} \right] = \ui \left( \delta^{ik} \ncopJ^{j\ell} - 2 \theta^{ik} \lambda_2 m \ncopH \delta^{j\ell} \right) - \langle i j k \ell \rangle, \\
&\left[ \ncopG^i, \ncopG^j \right] = \ui m^2 (1- 2\lambda_3) \theta^{ij}, \\
&\left[ \ncopG^i, \ncopJ^{jk} \right] = \ui \left[ \delta^{ik} \ncopG^j + m \left( \tfrac{1}{2} - \lambda_3 \right) \left( \theta^{ij} \ncopP^k + \delta^{ik} \theta^{jm} \ncopP^m \right) + m \lambda_2 \theta^{jk} \ncopP^i \right] - \langle j k \rangle.
\end{aligned}
\end{equation}
The structure of the other brackets remains unaltered:
\begin{equation} \label{alg-nc2:gal}
\begin{aligned}
&\left[ \ncopP^{i}, \ncopP^{j} \right] = 0, &
&\left[ \ncopH, \ncopP^{i} \right] = 0, \\
&\left[ \ncopH, \ncopJ^{ij} \right] = 0, &
&\left[ \ncopH, \ncopG^{i} \right] = -\ui \ncopP^{i}, \\
&\left[ \ncopP^{i}, \ncopJ^{jk} \right] = \ui \delta^{ik} \ncopP^{j} - \langle j k \rangle, & \qquad
&\left[ \ncopP^{i}, \ncopG^{j} \right] = -\ui m \delta^{ij}.
\end{aligned}
\end{equation}
The deformed generators \eqref{gen-nc:gal} now read
\begin{equation} \label{gen-nc2:gal}
\begin{aligned}
&\ncopH = \frac{1}{2m} \vec{\ncp}^{2}, \\
&\ncopP^{i} = \ncp^{i}, \\
&\ncopJ^{ij} = \ncx^{i} \ncp^{j} - \ncx^{j} \ncp^{i} + \tfrac{1}{2} \left( \theta^{ik} \ncp^{k} \ncp^{j} - \theta^{jk} \ncp^{k} \ncp^{i} \right) + \lambda_{2} \theta^{ij} \vec{\ncp}^{2}, \\
&\ncopG^{i} = m \ncx^{i} - t \ncp^{i} + \lambda_{3} m \theta^{ij} \ncp^{j}.
\end{aligned}
\end{equation}
We thus have the deformed Galilean algebra \eqref{alg-nc:gal-11} satisfied by the generators \eqref{gen-nc2:gal}. As happens for the relativistic case, here also we find new algebraic structures. There are two arbitrary parameters $\lambda_2$ and $\lambda_3$. Fixing $\lambda_2 = 0$ and $\lambda_3 = 1/2$ yields the standard (undeformed) algebra; the generators are still deformed, however.

Now we can give the operators some differential representation. The deformed generators in coordinate representation (Eq.~\eqref{repr-x} with $\mu = i$) read
\begin{equation} \label{gen-ncx:gal}
\begin{aligned}
&\ncopH = - \frac{1}{2m} \vec{\ncnabla}^2, \\
&\ncopP^i = - \ui \ncpartial^i, \\
&\ncopJ^{ij} = - \ui \left( \ncx^i \ncpartial^j - \ncx^j \ncpartial^i \right) - \tfrac{1}{2} \left( \theta^{ik} \ncpartial^k \ncpartial^j - \theta^{jk} \ncpartial^k \ncpartial^i \right) - \lambda_2 \theta^{ij} \vec{\ncnabla}^2, \\
&\ncopG^i = m \ncx^i + \ui t \ncpartial^i - \ui \lambda_3 m \theta^{ij} \ncpartial^j.
\end{aligned}
\end{equation}
In momentum representation (Eq.~\eqref{ncrepr-p} with $\mu = i$), on the other hand, they read
\begin{equation} \label{gen-ncp:gal}
\begin{aligned}
&\ncopH = \frac{1}{2m} \vec{\ncp}^2, \\
&\ncopP^i = \ncp^i, \\
&\ncopJ^{ij} = - \ncp^{i} \ncx^{j} + \ncp^{j} \ncx^{i} + \tfrac{1}{2} \left( \theta^{ik} \ncp^{k} \ncp^{j} - \theta^{jk} \ncp^{k} \ncp^{i} \right) + \lambda_{2} \theta^{ij} \vec{\ncp}^{2} \\
&\qquad\!\! = {} - \ui \left( \ncp^i \nceth^j - \ncp^j \nceth^i \right) + \lambda_2 \theta^{ij} \vec{\ncp}^2, \\
&\ncopG^i = \ui m \nceth^i - t \ncp^i + \left( \lambda_3 - \tfrac{1}{2} \right) m \theta^{ij} \ncp^j.
\end{aligned}
\end{equation}
Expectedly, for $\lambda_2 = 0$, $\lambda_3 = 1/2$, which corresponds to the standard (undeformed) Galilean algebra, there is no deformation in the generators in the momentum representation. The same thing also happened for the relativistic treatment.


\subsection{Deformed Schr\"odinger algebra}

The standard Schr\"odinger algebra is given by extending the Galilean algebra with the algebra of dilatation ($\opD$) and expansion or special conformal transformation ($\opK$). The relations \eqref{alg:gal} are augmented by
\begin{equation} \label{alg:DK}
\begin{aligned}
&\left[ \opH, \opD \right] = - 2 \ui \opH \, , & \qquad
&\left[ \opH, \opK \right] = - \ui \opD \, , & \qquad
&\left[ \opD, \opP^i \right] = \ui \opP^i \, , \\
&\left[ \opK, \opP^i \right] = \ui \opG^i \, , &
&\left[ \opJ^{ij}, \opD \right] = 0 \, , &
&\left[ \opJ^{ij}, \opK \right] = 0 \, , \\
&\left[ \opD, \opG^i \right] = - \ui \opG^i \, , &
&\left[ \opK, \opG^i \right] = 0 \, , &
&\left[ \opD, \opK \right] = - 2 \ui \opK \, .
\end{aligned}
\end{equation}
The free-particle representation of this algebra is given by the relations \eqref{gen:gal} along with
\begin{equation} \label{gen:DK}
\opD = p^i x^i - \frac{t}{m} \vec{p}^2 \, , \qquad
\opK = \frac{m}{2} \left( \vec{x} - \frac{t}{m} \vec{p} \right)^2.
\end{equation}
Introducing noncommutativity and starting with the minimal deformation, we write down for dilatation and expansion:
\begin{align}
\label{gen-nc:D}
\ncopD &= \ncp^i \ncx^i - \frac{t}{m} \vec{\ncp}^2 + \lambda_5 \theta^{ij} m^2 \ncx^i \ncp^j, \\
\label{gen-nc:K}
\ncopK &= \frac{m}{2} \left( \vec{\ncx} - \frac{t}{m} \vec{\ncp} \right)^2 + \lambda_6 m \theta^{ij} \ncx^i \ncp^j.
\end{align}
These modifications are compatible with the noncommutative algebra \eqref{brac:nc}. Next, the Lie algebra is considered. Using Eqs.~\eqref{gen-nc2:gal} and \eqref{gen-nc:D}, we obtain
\begin{equation} \label{alg-nc:DJ}
\left[ \ncopJ^{ij}, \ncopD \right] = \ui \left[ -2 m \lambda_2 \theta^{ij} \ncopH + \lambda_5 m^2 \theta^{ik} \left( \ncx^j \ncp^k - \ncx^k \ncp^j + \tfrac{1}{2} \theta^{k\ell} \ncp^{\ell} \ncp^j \right) \right]- \langle i j \rangle.
\end{equation}
The closure of the algebra requires $\lambda_5 = 0$. Then the brackets of $\ncopD$ with other generators are found to be
\begin{equation} \label{alg-nc:D}
\begin{aligned}
&\left[ \ncopH, \ncopD \right] = - 2 \ui \ncopH, &
&\left[ \ncopD, \ncopP^i \right] = \ui \ncopP^i, \\
&\left[ \ncopJ^{ij}, \ncopD \right] = - 4 \ui m \lambda_2 \theta^{ij} \ncopH, & \qquad
&\left[ \ncopD, \ncopG^i \right] = -\ui \left[ \ncopG^i + m (1 - 2 \lambda_3) \theta^{ij} \ncopP^j \right],
\end{aligned}
\end{equation}
leading to a non-standard closure of the algebra.

Turning to expansion now, we find
\begin{equation*}
\left[ \ncopD, \ncopK \right] = \ui \left[ - 2 \ncopK + \left( \tfrac{1}{2} - \lambda_6 \right) m \theta^{ij} \left( 2 \lambda_2 m \theta^{ij} \ncopH - \ncopJ^{ij} \right) + 2 \left( \tfrac{1}{4} - \lambda_6 \right) m \theta^{ij} \theta^{im} \ncp^m \ncp^j \right],
\end{equation*}
which fixes $\lambda_6 = 1/4$. Then the brackets involving $\ncopK$ are seen to be
\begin{equation} \label{alg-nc:K}
\begin{aligned}
&\left[ \ncopH, \ncopK \right] = -\ui \ncopD, \\
&\left[ \ncopK, \ncopP^i \right] = \ui \left[ \ncopG^i + \left( \tfrac{1}{4} - \lambda_3 \right) m \theta^{ij} \ncopP^j \right], \\
&\left[ \ncopJ^{ij}, \ncopK \right] = \ui \left[ \frac{m}{4} \left( \theta^{ik} \ncopJ^{kj} - \theta^{jk} \ncopJ^{ki} \right) - 2 \lambda_2 m \theta^{ij} \ncopD \right], \\
&\left[ \ncopK, \ncopG^i \right] = -\ui m \theta^{ij} \left[ \left( \tfrac{3}{4} - \lambda_3 \right) \ncopG^j + \left( \lambda_3^2 - \lambda_3 + \tfrac{1}{4} \right) m \theta^{jk} \ncopP^k \right], \\
&\left[ \ncopD, \ncopK \right] = -\ui \left( 2 \ncopK + \frac{m}{4} \theta^{ij} \ncopJ^{ij} - \tfrac{1}{2} \lambda_2 m^2 \theta^{ij} \theta^{ij} \ncopH \right).
\end{aligned}
\end{equation}
Thus the dilatation and expansion have the final form
\begin{gather}
\label{gen-nc2:D}
\ncopD = \ncp^i \ncx^i - \frac{t}{m} \vec{\ncp}^2, \\
\label{gen-nc2:K}
\ncopK = \frac{m}{2} \left( \vec{\ncx} - \frac{t}{m} \vec{\ncp} \right)^2 + \frac{m}{4} \theta^{ij} \ncx^i \ncp^j.
\end{gather}

Some comments are in order. We have obtained the deformed Schr\"odinger algebra involving two parameters, $\lambda_2$ and $\lambda_3$. The closure of the algebra is highly nontrivial and yields new structures. For $\theta\rightarrow 0$, the deformed algebra reduces to the undeformed one. A distinctive feature is that there is no choice of the free parameters for which the standard (undeformed) algebra can be reproduced. This is an obvious (and important) difference from the Poincar\'e treatment.

It is however possible to obtain an alternative deformation which, for a particular choice of parameters, yields the undeformed algebra. First, notice that as far as the Galilean part is concerned, fixing $\lambda_2 = 0$ and $\lambda_3 = 1/2$ gives the standard algebra, although the generators are deformed. With this choice, the brackets involving $\ncopD$ and Galilean generators, given in Eq.~\eqref{alg-nc:D}, also reduce to the standard ones. The same is, however, not true for brackets involving $\ncopK$, given in Eq.~\eqref{alg-nc:K}. So let us remodify the form \eqref{gen-nc:K}, allowing the possiblity of  $\theta^2$ terms:
\begin{equation} \label{gen-nc:K-2}
\ncopK = \frac{m}{2} \left( \vec{\ncx} - \frac{t}{m} \vec{\ncp} \right)^2 + \lambda_6 m \theta^{ij} \ncx^i \ncp^j + \theta^{ij} \theta^{jk} \left( \lambda_7 m \ncp^i \ncp^k + \lambda_8 m^5 \ncx^i \ncx^k + \lambda_9 m^3 \ncx^i \ncp^k \right).
\end{equation}
Now we get
\begin{equation} \label{alg-nc:HK-2}
\left[ \ncopH, \ncopK \right] = - \ui \left[ \ncopD + \theta^{ik} \theta^{kj} \left( 2 \lambda_8 m^4 \ncp^i \ncx^j + \lambda_9 m^2 \ncp^i \ncp^j \right) \right], \\
\end{equation}
which fixes $\lambda_8 = \lambda_9 = 0$. Further we also have
\begin{equation} \label{alg-nc:DK-2}
\begin{split}
\left[ \ncopD, \ncopK \right] &= \ui \Big[ - 2 \ncopK + \left( \tfrac{1}{2} - \lambda_6 \right) m \theta^{ij} \left( 2 \lambda_2 m \theta^{ij} \ncopH - \ncopJ^{ij} \right) \\
&\qquad \, {}+ 2 \left( \tfrac{1}{4} - \lambda_6 - 2 \lambda_7 \right) m \theta^{ij} \theta^{im} \ncp^m \ncp^j \Big],
\end{split}
\end{equation}
which necessitates a relation between $\lambda_6$ and $\lambda_7$ so as to make the last term on the right-hand side  vanish. We therefore set $\lambda_7 = (1/8) - \lambda_6/2$. Then the brackets involving $\ncopK$ turn out to be
\begin{equation} \label{alg-nc:K-2}
\begin{aligned}
&\left[ \ncopH, \ncopK \right] = -\ui \ncopD \, , \\
&\left[ \ncopK, \ncopP^i \right] = \ui \left[ \opG^i + \left( \lambda_6 - \lambda_3 \right) m \theta^{ij} \ncopP^j \right], \\
&\left[ \ncopJ^{ij}, \ncopK \right] = \ui \left[ \left( \tfrac{1}{2} - \lambda_6 \right) m \left( \theta^{ik} \ncopJ^{kj} - \theta^{jk} \ncopJ^{ki} \right) - 2 \lambda_2 m \theta^{ij} \ncopD \right], \\
&\left[ \ncopK, \ncopG^i \right] = -\ui m \theta^{ij} \left[ \left( 1 - \lambda_3 - \lambda_6 \right) \ncopG^j + \left( \lambda_3^2 - \lambda_3 + \tfrac{1}{4} \right) m \theta^{jk} \ncopP^k \right], \\
&\left[ \ncopD, \ncopK \right] = \ui \left[ - 2 \ncopK + \left( \tfrac{1}{2} - \lambda_6 \right) m \theta^{ij} \left( 2 \lambda_2 m \theta^{ij} \ncopH - \ncopJ^{ij} \right) \right],
\end{aligned}
\end{equation}
which are the analogue of the set \eqref{alg-nc:K}, for $\ncopK$ involving $\theta^2$ deformation:
\begin{equation} \label{gen-nc2:K-2}
\ncopK = \frac{m}{2} \left( \vec{\ncx} - \frac{t}{m} \vec{\ncp} \right)^2 + \lambda_6 m \theta^{ij} \ncx^i \ncp^j + m \left( \frac{1}{8} - \frac{\lambda_6}{2} \right) \theta^{ij} \theta^{jk} \ncp^i \ncp^k \, .
\end{equation}
We have thus obtained another deformed Schr\"odinger algebra, involving three parameters, $\lambda_2$, $\lambda_3$ and $\lambda_6$. It is easily seen from \eqref{alg-nc:K-2} that the particular choice of parameters, $\lambda_2 = 0$ and $\lambda_3 = \lambda_6 = 1/2$, reproduces the standard algebra. This agrees with \refcite{Banerjee:2005zt}.

Rewriting Eq.~\eqref{gen-nc2:D} as $\ncopD = \ncx^i \ncp^i - ({t}/{m}) \vec{\ncp}^2 - \ui N$, now $N = {\delta^i}_i$ being the number of space dimensions, the coordinate representation of $\ncopD$ becomes obvious:
\begin{equation} \label{gen-nc2x:D}
\ncopD = - \ui \ncx^i \ncpartial^i + \frac{t}{m} \vec{\ncnabla}^2 - \ui N \, .
\end{equation}
For momentum representation Eq.~\eqref{gen-nc2:D} yields
\begin{equation} \label{gen-nc2p:D}
\ncopD = \ui \ncp^i \nceth^i - \frac{t}{m} \vec{\ncp}^2 \, .
\end{equation}
For $\ncopK$, first we write Eqs.~\eqref{gen-nc2:K} and \eqref{gen-nc2:K-2} in the expanded form:
\begin{align}
\label{gen-nc2q:K-1}
\ncopK_{(1)} &= \frac{m}{2} \vec{\ncx}^2 + \frac{t^2}{2m} \vec{\ncp}^2 - \frac{t}{2} \left( \ncx^i \ncp^i + \ncp^i \ncx^i \right) + \frac{m}{4} \theta^{ij} \ncx^i \ncp^j \, , \\
\label{gen-nc2q:K-2}
\ncopK_{(2)} &= \frac{m}{2} \vec{\ncx}^2 + \frac{t^2}{2m} \vec{\ncp}^2 - \frac{t}{2} \left( \ncx^i \ncp^i + \ncp^i \ncx^i \right) + \lambda_6 m \theta^{ij} \ncx^i \ncp^j + m \left( \frac{1}{8} - \frac{\lambda_6}{2} \right) \theta^{ij} \theta^{jk} \ncp^i \ncp^k \, .
\end{align}
Using $\ncx^i \ncp^i + \ncp^i \ncx^i = 2 \ncx^i \ncp^i - \ui N$ in the above equations, yields the coordinate representation for $\ncopK$ as
\begin{align}
\label{gen-nc2x:K-1}
\ncopK_{(1)} &= \frac{m}{2} \vec{\ncx}^2 - \frac{t^2}{2m} \vec{\ncnabla}^2 + \ui t \ncx^i \ncpartial^i + \ui \frac{tN}{2} - \ui \frac{m}{4} \theta^{ij} \ncx^i \ncpartial^j \, , \\
\label{gen-nc2x:K-2}
\ncopK_{(2)} &= \frac{m}{2} \vec{\ncx}^2 - \frac{t^2}{2m} \vec{\ncnabla}^2 + \ui t \ncx^i \ncpartial^i + \ui \frac{tN}{2} - \ui \lambda_6 m \theta^{ij} \ncx^i \ncpartial^j - m \left( \frac{1}{8} - \frac{\lambda_6}{2} \right) \theta^{ij} \theta^{jk} \ncpartial^i \ncpartial^k \, .
\end{align}
For momentum representation we use $\ncx^i \ncp^i + \ncp^i \ncx^i = 2 \ncp^i \ncx^i + \ui N$:
\begin{align}
\label{gen-nc2p:K-1}
\ncopK_{(1)} &= - \frac{m}{2} \nceth^i \nceth^i + \frac{t^2}{2m} \vec{\ncp}^2 - \ui t \ncp^i \nceth^i - \ui \frac{tN}{2} - \ui \frac{m}{4} \theta^{ij} \ncp^j \nceth^i \, , \\
\label{gen-nc2p:K-2}
\ncopK_{(2)} &= - \frac{m}{2} \nceth^i \nceth^i + \frac{t^2}{2m} \vec{\ncp}^2 - \ui t \ncp^i \nceth^i - \ui \frac{tN}{2} + \ui m \left( \lambda_6 - \tfrac{1}{2} \right) \theta^{ij} \ncp^j \nceth^i \, .
\end{align}
We notice that in the momentum representation, there is no deformation in $\ncopK_{(2)}$ for the special case of $\lambda_6 = 1/2$, which corresponds to the standard algebra.

Now onwards we shall restrict to $\ncopK_{(2)}$ whenever expansions are considered.


\subsection{Coproducts and Hopf algebra}

The comultiplication rules, using the coordinate representation, for generators given in Eqs.~\eqref{gen-ncx:gal}, \eqref{gen-nc2x:D} and \eqref{gen-nc2x:K-2}, turn out to be
\begin{gather}
\label{co-ncH}
\Delta ( \ncopH ) = \ncopH \otimes \opid + \opid \otimes \ncopH + \frac{1}{m} \ncopP^i \otimes \ncopP^i, \\
\label{co-ncP}
\Delta ( \ncopP^i ) = \ncopP^i \otimes \opid + \opid \otimes \ncopP^i, \\
\label{co-ncJ}
\begin{split}
\Delta ( \ncopJ^{ij} ) &= \tfrac{1}{2} \left[ \ncopJ^{ij} \otimes \opid + \opid \otimes \ncopJ^{ij} + \theta^{im} \left( \ncopP^j \otimes \ncopP^m - \ncopP^m \otimes \ncopP^j \right) \right] \\
&\qquad\quad {} + \lambda_2 \theta^{ij} \ncopP^m \otimes \ncopP^m - \langle i j \rangle,
\end{split} \\
\label{co-ncG}
\begin{split}
\Delta ( \ncopG^i ) &= \tfrac{1}{2} \Big[ \ncopG^i \otimes \opid + \opid \otimes \ncopG^i - t \left( \ncopP^i \otimes \opid + \opid \otimes \ncopP^i \right) \\
&\qquad\quad {} + m \theta^{ij} \left\{ (\lambda_3 - 1) \ncopP^j \otimes \opid + \lambda_3 \opid \otimes \ncopP^j \right\} \Big],
\end{split} \\
\label{co-ncD}
\Delta ( \ncopD ) =  \ncopD \otimes \opid + \opid \otimes \ncopD - \frac{2t}{m} \ncopP^i \otimes \ncopP^i + \ui \frac{N}{2} \opid \otimes \opid + \theta^{ij} \ncopP^i \otimes \ncopP^j, \\
\label{co-ncK}
\begin{split}
\Delta ( \ncopK ) &=  \ncopK \otimes \opid \\
&\quad{} + \opid \otimes \ncopK + \frac{1}{2m} \left[ t^2 \ncopP^i \otimes \ncopP^i - \ncopG^i \otimes \ncopG^i - t \left( \ncopP^i \otimes \ncopG^i + \ncopG^i \otimes \ncopP^i \right) \right] \\
&\quad{} - \ui \frac{tN}{2} \opid \otimes \opid - \tfrac{1}{2} \theta^{ij} \left[ t \ncopP^i \otimes \ncopP^j + (\lambda_3 - 1) \ncopP^i \otimes \ncopG^j + \lambda_3 \ncopG^j \otimes \ncopP^i \right] \\
&\quad{} - \frac{m}{2} \theta^{ij} \theta^{ik} \left( \lambda_3^2 - \lambda_3 + \tfrac{1}{2} \right) \ncopP^j \otimes \ncopP^k.
\end{split}
\end{gather}
Note that among the free parameters $\lambda_2$, $\lambda_3$ and $\lambda_6$ appearing in the definition of the deformed generators, only the first two occur in the expressions for the deformed coproducts. The parameter $\lambda_6$, which is present in $\ncopK$, however, does not occur in $\Delta(\ncopK)$. Now we compute the basic Hopf algebra. Expectedly, it turns out that the Hopf algebra can be read off from Eqs.~\eqref{alg-nc2:gal}, \eqref{alg-nc:gal-11}, \eqref{alg-nc:D} and \eqref{alg-nc:K-2} by just replacing the generators by the coproducts.

As is known there is an alternative method, based on quantum-group-theoretic arguments, of computing the coproducts \cite{Chaichian:2004za, Chaichian:2004yh, Matlock:2005zn}. This is obtained for the particular case when the deformed generators satisfy the undeformed algebra. In our analysis it corresponds to the choice $\lambda_2 = 0$, $\lambda_3 = \lambda_6 = 1/2$. The essential ingredient is the application of the abelian twist function, $\opF = \exp ( \frac{\ui}{2} \theta^{ij} \opP^i \otimes \opP^j )$, as a similarity transformation on the primitive coproduct rule to abstract the deformed rule.

Consider first the Baker--Campbell--Hausdorff relation,
\begin{equation*}
\ue^\opA {\,} \opB {\,} \ue^{-\opA} = \opB + [ \opA, \opB ] + \frac{1}{2!} \big[ \opA, [\opA, \opB] \big] + \cdots,
\end{equation*}
which implies
\begin{equation} \label{bch}
\begin{split}
\opF {\,} \opB {\,} \opF^{-1} &= \opB + \frac{\ui}{2} \theta^{ij} \left[ \opP^i \otimes \opP^j, \opB \right] \\
&\quad {} + \frac{1}{2!} \left(\frac{-1}{4}\right) \theta^{ij} \theta^{k\ell} \left[ \opP^k \otimes \opP^\ell, \left[ \opP^i \otimes \opP^j, \opB \right] \right] + \cdots.
\end{split}
\end{equation}
Let us now take the specific example of Galilean boosts. Therefore, taking the primitive coproduct (Eq.~\eqref{co-ncG} with $\theta =0$, the commutative-space analogue),
\begin{equation*}
\Delta ( \opG^i ) =  \tfrac{1}{2} \left[ \opG^i \otimes \opid + \opid \otimes \opG^i - t \left( \opP^i \otimes \opid + \opid \otimes \opP^i \right) \right],
\end{equation*}
we find, after an application of the twist function,
\begin{equation} \label{co-twG}
\opF {\,} \Delta ( \opG^i ) {\,} \opF^{-1} =  \Delta ( \opG^i ) - \frac{m}{4} \theta^{ij} \left( \opP^j \otimes \opid - \opid \otimes \opP^j \right),
\end{equation}
where use has been made of Eq.~\eqref{bch}. This is the deformed coproduct rule \eqref{co-ncG} (for the specific values of the free parameters already stated) obtained by identifying
\begin{equation} \label{co-nctG}
\Delta ( \ncopG^i ) = \left[ \opF {\,} \Delta ( \opG^i ) {\,} \opF^{-1} \right]_{\opG^i \rightarrow \ncopG^i, \opP^j \rightarrow \ncopP^j} \, .
\end{equation}
Similarly the coproducts for other generators can also be obtained from the same twist element.


\subsection{Deformed conformal algebra through contraction}
 
Strictly speaking, the algebra obtained by enlarging the Galilean algebra by including dilatations and expansions, as discussed in the previous subsections, is not a conformal algebra since it does not inherit some basic characteristics like vanishing of the mass, equality of the number of translations and the special conformal transformations, etc. However since it is a symmetry of the Schr\"odinger equation, this enlargement of the Galilean algebra is appropriately referred to as the Schr\"odinger algebra. It is possible to discuss the conformal extension of the Galilean algebra by means of a nonrelativistic contraction of the relativistic conformal-Poincar\'e algebra. Recently this was discussed for the particular case of three dimensions \cite{Lukierski:2005xy}. This algebra is different from the Schr\"odinger algebra discussed earlier. We scale the generators and the noncommutativity parameter as
\begin{equation}
\begin{aligned}
&\ncopD = \scopD, \\
&\ncopK^{\rho} = \left( \ncopK^0, \ncopK^i \right) = \left( c \scopK, c^2 \scopK^i \right), \\
&\ncopP^{\mu} = \left( \ncopP^0, \ncopP^i \right) = \left( \scopH /c, \scopP^i \right), \\
&\ncopJ^{\mu \nu} = \left( \ncopJ^{0i}, \ncopJ^{ij} \right) = \left( c \scopG^i, \scopJ^{ij} \right), \\
&\theta^{\mu \nu} = \left( \theta^{0i}, \theta^{ij} \right) = \left( 0, c^2 \overline{\theta}^{ij} \right),
\end{aligned}
\end{equation}
where $c$ is the velocity of light. We use this scaling in Eq.~\eqref{nc116} and take the limit $c \rightarrow \infty$. Finally we redefine to choose the same symbols for the nonrelativistic case; i.e. we do the replacements $\scopD \rightarrow \ncopD$, etc. Then we get the deformed algebra
\begin{equation} \label{nc-cont:DK}
\begin{aligned}
&\begin{aligned}
&\left[ \ncopD, \ncopH \right] = \ui \ncopH, &
&\left[ \ncopD, \ncopP^i \right] = \ui \ncopP^i, \\
&\left[ \ncopD, \ncopJ^{ij} \right] = 0, & 
&\left[ \ncopD, \ncopG^i \right] = 0 \, , \\
&\left[ \ncopK, \ncopH \right] = 2 \ui \eta^{00} \ncopD \, , &
&\left[ \ncopK, \ncopP^i \right] = 2 \ui \ncopG^i \, , \\
&\left[ \ncopK, \ncopD \right] = \ui \ncopK \, , &
&\left[ \ncopK, \ncopK \right] = 0 \, , \\
&\left[ \ncopK, \ncopG_i \right] = - \ui \eta^{00} \left[ \ncopK^i - (\ui + \eta_1) \theta^{ij} \ncopP_j \right], & \qquad
&\left[ \ncopK, \ncopK^i \right] = 2 \ui (\ui + \eta_1) \theta^{ij} \ncopG_j \, , \\
&\left[ \ncopK, \ncopJ^{ij} \right] = 0 \, , &
&\left[ \ncopK^i, \ncopH \right] = - 2 \ui \ncopG^i \, , \\
&\left[ \ncopK^i, \ncopP^j \right] = 0 \, , &
&\left[ \ncopK^i, \ncopG^j \right] = 0 \, ,
\end{aligned} \\
&\begin{aligned}
&\left[ \ncopK^i, \ncopJ^{jk} \right] = - \ui \Big[ \eta^{ij} \ncopK^k + (\ui + \eta_1) \left( \theta^{ij} \ncopP^k - \eta^{ij} \theta^{k\ell} \ncopP_\ell \right) \Big] - \langle j k \rangle, \\
&\left[ \ncopK^i, \ncopD \right] = \ui \left[ \ncopK^i - 2 (\ui + \eta_1) \theta^{ij} \ncopP_j \right].
\end{aligned}
\end{aligned}
\end{equation}
This algebra also contains a free parameter. Restricting to three dimensions and the specific choice $\eta_1 = -\ui$ reproduces the results obtained recently in \refcite{Lukierski:2005xy}.


\section{\label{sec:deform-conlu}Discussion}

We have considered in full generality the most simple solutions to Eq.~\eqref{nc102} subject to the condition of a noncommutative spacetime. These solutions are first-order in the noncommutativity parameter $\theta$. For the Poincar\'e symmetry, our results agree with an alternative approach provided in \refcite{Koch:2004ud}. Inclusion of the conformal sector leads to the first nontrivial effect. We find that there is no first-order solution that yields a closed algebra. It becomes mandatory to include second-order terms in the conformal generator to get this closure. For the Schr\"odinger symmetry there is a first-order solution that satisfies the closure property. However, as already stated, the intriguing point here is that there is no solution for the free parameters that reproduces the standard (primitive) closure. It becomes essential to include second-order terms to have this property.

The present analysis can be extended to other (non-constant) types of noncommutativity. Some results in this direction have already been provided for the Snyder space \cite{Banerjee:2006wf}.


%% file: chap_conclu.tex

\chapter{\label{chap:discussions}Concluding remarks}


Although it has a longer history, the idea that configuration-space coordinates may not commute has arisen recently from string theory. Noncommuting spatial coordinates and fields can be realised in actual physical situations \cite{Jackiw:2001dj}. Therefore, there is enough motivation to investigate what follows just from the idea that coordinates are operators that do not commute. Noncommutative field theores, which are the field theories in which the coordinates do not commute, have many novel features. Today we have enough literature on the subject. The aim of this thesis was to further these investigations. We studied some aspects of noncommutativity in field theory, strings and membranes.

We started, in Chapter \ref{chap:intro}, with a brief introduction to noncommutative spaces. Then we discussed briefly the Landau problem, an important physically realised example of noncommuting coordinates.

In Chapter \ref{chap:membrane}, we first presented a review of noncommutativity in an open string moving in a background Neveu--Schwarz field in a gauge-independent Hamiltonian approach. The noncommutativity was seen to be a direct consequence of the nontrivial boundary conditions, which, contrary to several approaches, were not treated as constraints. The origin of any modification in the usual Poisson algebra was the presence of boundary conditions. In a gauge-independent formulation of a free Polyakov string, the boundary conditions naturally led to a noncommutative structure among the coordinates. This noncommutativity  vanished in the conformal gauge, as expected. For the interacting string, a more involved boundary condition led to a more general type of noncommutativity. Contrary to the standard conformal-gauge expressions, this noncommutative algebra survived at all points of the string and not just at the boundaries. In contrast to the free theory, this noncommutativity could not be removed in any gauge. In the conformal gauge, noncommutativity survived only at the string endpoints.

We then analysed an open membrane, with square and cylindrical topology, ending on $p$-branes. Both the free case as well as the theory where the membrane is coupled to a background three-form potential were considered. 

For the free theory, the world-volume action was taken to be either the Nambu--Goto type or the Polyakov type. For the Nambu--Goto action, a gauge-independent formulation, similar to that adopted in \refcite{HansonReggeTeitelboim:1976} for the string theory, was presented. The reparametrisation invariances were manifested by the freedom in the choice of the multipliers enforcing the constraints of the theory. The implications of the boundary conditions in preserving the stability of the free membrane were discussed, highlighting the parallel with the string treatment. A set of quasi-orthonormal gauge-fixing conditions was systematically obtained, which simplified the structure of the Hamiltonian.

A constrained analysis of the Polyakov action, contrary to the Nambu--Goto action,  led to the presence of second-class constraints. However, by an iterative prescription of computing Dirac brackets, the first-class sector was identified. The Dirac brackets of this sector were identical to the Poisson brackets and exactly matched with the involutive algebra found in the Nambu--Goto case. The analogue of the quasi-orthonormal gauge was also discussed in the Polyakov formulation. It naturally led to the choice of the metric which is used to perform calculations in the light-front variables \cite{Taylor:2001vb}.
Moreover, in this gauge, the energy--momentum tensor was expressed as a combination of the constraints. On the constraint shell this tensor was seen to have a vanishing trace.

A fundamental difference of the quasi-orthonormal gauge fixing in the two cases was pointed out. In the Polyakov case, gauge fixing entailed certain restrictions on the metric. Since the metric is regarded as an independent field, the gauge fixing does not affect the constraints of the theory which generate the reparametrisation invariances. The discussion was thus confined to the Poisson algebra only. A similar gauge fixing in the Nambu--Goto case obviously restricts the target-space coordinates. The first-class constraints get converted into second-class ones, thereby necessitating the use of Dirac brackets. Their evaluation is quite complicated due to nonlinear terms. 

Since Dirac brackets were avoided in the Polyakov formulation, we proceeded to discuss noncommutativity only in this formulation. Also, cylindrical topology of the membrane was considered. Contrary to standard approaches \cite{Kawamoto:2000zt, Das:2001mg, Tezuka:2002wn, Chu:1998qz, Romero:2002vg},  boundary conditions were not treated as primary constraints of the theory. Our approach was in line with the treatment for string theory discussed in \refcite{Banerjee:2002ky}. Thus, noncommutative algebra, if any, would be a manifestation of the Poisson brackets and not Dirac brackets. The noncommutative algebra was required to establish algebraic consistency of the boundary conditions with the basic Poisson brackets. For the free theory it was found that there was no clash between the boundary conditions and the Poisson brackets, hence there was no noncommutativity.

For the membrane interacting with a three-form potential a nontrivial algebraic relation was found that revealed the occurrence of noncommutativity, independent of any gauge choice or any approximations. Since this equation could not be solved, we passed on to its low-energy limit. Now this limit, which takes a membrane to a string, has been known for quite some time \cite{Duff:1987bx} and has been studied or exploited in several circumstances \cite{Lindstrom:1989xa, Bergshoeff:1987qx, Smolin:1997ai}. The cylindrical membrane is assumed to wrap around a circle, whose radius is taken to be vanishingly small. This enforces a double dimensional reduction with the eleven-dimensional compactified target space passing over to the ten-dimensional space while the membrane effectively reduces to an open string. We studied this limit and showed  how the membrane boundary conditions, action and the world-volume metric were transformed into the corresponding expressions for the string. The equation governing noncommutativity in the membrane was likewise shown to reduce to the string example. Since every point in D-brane can correspond to the endpoints of the cylindrical membrane, we get noncommutativity in D-brane coordinates also---albeit in this low-energy limit. Of course, this feature of noncommutativity will persist even if this limit is not considered, otherwise the basic equation \eqref{514} becomes inconsistent.


In Chapter \ref{chap:current} we already took a noncommutative spacetime and discussed its implications. The Seiberg--Witten map, which provides an alternative method of studying noncommutative gauge theories by recasting these in terms of their commutative equivalents (by replacing the noncommuting vector potential by a function of a commuting potential), was discussed.

Then we provided a Seiberg--Witten-type map relating the sources in the noncommutative and commutative descriptions. For investigating quantum aspects of the mapping, we applied it to the divergence anomalies for the abelian theory in the two descriptions. For the slowly-varying-field approximation, the anomalies indeed got identified. Thus the classical map correctly accounted for the quantum effects inherent in the calculations of the anomalies. The results were checked up to $\uO(\theta^{2})$. We also provided an indirect method of extending the calculations and found an agreement up to $\uO(\theta^{3})$. The analysis strongly suggests that the classical mapping would hold for all orders in $\theta$, albeit in the slowly-varying-field approximation. In the nonabelian theory, the classical maps for the currents and their covariant divergences were given up to $\uO(\theta)$. Our findings may also be compared with \refscite{Kaminsky:2003qq, Kaminsky:2003mn} where the classical equivalence of the Chern--Simons theories in different descriptions was found to persist even in the quantum case.

For arbitrary field configurations, derivative corrections to the classical source map were explicitly computed up to $\uO(\theta^{2})$. Indeed, it is known that if one has to go beyond the slowly-varying-field approximation, derivative corrections are essential. For instance, Dirac--Born--Infeld actions with derivative corrections have been discussed \cite{Wyllard:2000qe, Wyllard:2001ye, Das:2001xy}.

In Chapter \ref{chap:anomalous} we obtained the $\uO(\theta)$ structure of all the anomalous commutators involving the covariant axial-vector current in noncommutative electrodynamics for a magnetic-type $\theta$. The basic step in our approach was to exploit the Seiberg--Witten maps for currents and fields that relate the noncommutative and usual (commutative) descriptions. The commutators in the noncommutative theory were thereby expressed in terms of their commutative counterparts which are known. Substituting for these known commutators we obtained the commutators in the noncommutative theory. The results were displayed both in terms of the commutative (usual) and noncommutative variables.

We showed that the commutators we obtained were compatible with the noncommutative covariant anomaly. For this we derived certain consistency conditions involving this anomaly and then showed that the commutators indeed satisfied these conditions. It may be remarked that such consistency conditions were used in usual electrodynamics to reveal the compatibility of the various anomalous commutators with the Adler--Bell--Jackiw anomaly. In the usual quantum electrodynamics without axial-vector currents, anomalies in potential--current commutators (`seagulls') and in current--current commutators (`Schwinger terms') are related and cancel exactly when the divergence of covariant matrix element is taken, reproducing the familiar current conservation. The distinguishing feature of the commutator anomalies associated with the triangle diagram is that when the axial-vector divergence is taken, the seagulls and Schwinger terms do not cancel~\cite{Jackiw:1968xt}. Rather, they combine to give the divergence anomaly (Adler--Bell--Jackiw anomaly), giving an alternative interpretation of the divergence anomaly as the result of non-cancellation of seagulls and Schwinger terms. Our analysis thus suggests that the star-gauge-covariant anomaly can also be regarded as consequence of a similar effect in noncommutative electrodynamics. Finally, we analysed the implications of certain ambiguities present in the ordinary commutators on our scheme, and showed that the commutators satisfy the consistency conditions irrespective of these ambiguities.

Most popular noncommutative field theories are characterised by a constant noncommutativity parameter $\theta$ that violates Lorentz invariance. Violations of Lorentz symmetry are intrinsic to noncommutative theories by virtue of nonzero $\theta_{\mu\nu}$. The aim of Chapter \ref{chap:lorentz} was to provide a conceptually cleaner understanding of Lorentz symmetry and its interpretaion in the noncommutative context. Here we derived, starting from a first-principle Noether-like approach, criteria for preserving Poincar\'e invariance in a noncommutative gauge theory with a constant noncommutativity parameter $\theta$. The criterion for translational invariance was the same irrespective of whether $\theta$ transformed as a second-rank tensor or was the same constant in all frames. This criterion was then shown to hold by performing an explicit check. Thus, as expected, translational invariance was valid.

The issue of Lorentz invariance (invariance under rotations and boosts) was quite subtle. We found distinct criteria depending on the nature of transformation of $\theta$. An explicit check using the equations of motion confirmed the particular criterion for Lorentz invariance when $\theta_{\mu\nu}$ transformed as covariant second-rank tensor. Thus Lorentz invariance was preserved only for a transforming $\theta$.

We showed that all the transformations are dynamically consistent. The Noether charges generated the appropriate transformations on the phase-space variables. These charges also satisfied the desired Lie brackets among themselves.

The complete analysis was done in both the commutative and noncommutative descriptions. By the use of suitable Seiberg--Witten-type maps, compatibility among the results found in the two descriptions was established.

The criteria for Lorentz invriance found here were also consistent with the fact that, for a constant nontransforming $\theta$ having special values, the symmetry group breaks down to $\mathrm{SO}(1,1)\times\mathrm{SO}(2)$, a subgroup of the Lorentz group.

Although the noncommutativity of the spacetime coordinates violates relativistic invariance, it has been shown by using the (twisted) Hopf algebra that corresponding field theories possess deformed symmetries. Chapter \ref{chap:deform} is devoted to the study of deformed relativistic and nonrelativistic symmetries on canonical noncommutative spaces. Here we analysed the deformed conformal-Poincar\'e, Schr\"odinger and conformal-Galilean symmetries compatible with the canonical (constant) noncommutative spacetime and found new algebraic structures. We followed a two-step algebraic process. First, by requiring the compatibility of transformations with noncommutativity, a general deformation of the generators was obtained. Then a definite structure emerged after demanding the closure of the algebra satisfied by the deformed generators.

For the Poincar\'e sector, we obtained a generalisation (by including, apart from the translations and rotations, a symmetric second-rank tensor operator) of the Poincar\'e algebra containing two arbitrary parameters. Fixing these parameters reproduced the usual undeformed algebra.

For the full conformal-Poincar\'e case we obtained new algebraic structures. We found a one-parameter class of deformed special conformal generators that yielded a closed algebra whose structure was completely new. Unlike the Poincar\'e sector, it was not necessary to extend the set of generators to obtain these new structures. Fixing the arbitrary parameter reproduced the usual (undeformed) Lie algebra. In this case the deformed special conformal generator also agreed with the result given in \refcite{Banerjee:2005ig}.

We derived the structures of the generators in the coordinate and momentum representations and demonstrated that momentum representation is more favoured for the noncommutative space \cite{Banerjee:2005zt}. Although there was deformation in the generator for the general case, for a particular value of the parameter for which the generators satisfied the usual (undeformed) algebra, the deformation in generators dropped out in the momentum representation. 

Next we considered the Schr\"odinger symmetry and obtained the deformed Schr\"odinger algebra involving two parameters. The generators involved $\uO(\theta)$ deformations. For $\theta\rightarrow 0$, the deformed algebra reduced to the undeformed one. However a distinctive feature was that there was no choice of the free parameters for which the standard (undeformed) algebra could be reproduced.

Exploring other possibilities, then we obtained an alternative deformation which, for a particular choice of parameters, indeed reproduced the undeformed algebra. In this case the modified special conformal generator involved $\uO(\theta^2)$ terms while the other genrators involved at most $\uO(\theta)$ terms only. The deformed Schr\"odinger algebra now involved three parameters, a particular choice of which reproduced the standard algebra.

In all these examples we computed the modified comultiplication rules associated with the deformed generators. These rules also contained the free parameters entering in the definition of the generators. As a consistency, we showed that the comultiplication rules, for the particular values of the free parameters yielding the undeformed algebra, agreed with those obtained by an application of the abelian twist function on the primitive coproduct.

We also discussed the conformal extension of the Galilean algebra by means of a nonrelativistic contraction of the relativistic conformal-Poincar\'e algebra. Recently this was discussed for the particular case of three dimensions \cite{Lukierski:2005xy}. This algebra is different from the Schr\"odinger algebra, both in the commutative and noncommutative descriptions.

\paragraph{Future directions.}
In this thesis we studied certain aspects of noncommutativity in field theory, strings and membranes. Noncommutative field theories have many novel properties which are not exhibited by conventional quantum field theories and we shall continue to further these studies. The fact that quantum field theory on a noncommutative space arises naturally in string theory and Matrix theory strongly suggests that spacetime noncommutativity is a general feature of a unified theory of quantum gravity. Noncommutative field theories should be properly understood as lying somewhere between ordinary field theory and string theory. From these models we may learn something about string theory and the classification of its backgrounds, using the somewhat simpler techniques of quantum field theory. Extension of our results of noncommutative electrodynamics to higher orders and further studies of deformed symmetries, including supersymmetric extension, are among the possible near-future directions.